\pdfoutput=1 

\documentclass[12pt,twoside,singlespace]{mitthesis}

\usepackage{lgrind}
\usepackage{cmap}
\usepackage[T1]{fontenc}
\usepackage[utf8]{inputenc}
\usepackage[english]{babel}
\usepackage{lmodern}
\usepackage[babel=true]{microtype}
\usepackage{graphicx}
\usepackage{bm} 
\usepackage{array} 
\usepackage{xspace}

\newcolumntype{L}[1]{>{\raggedright\let\newline\\\arraybackslash\hspace{0pt}}m{#1}}
\newcolumntype{C}[1]{>{\centering\let\newline\\\arraybackslash\hspace{0pt}}m{#1}}
\newcolumntype{R}[1]{>{\raggedleft\let\newline\\\arraybackslash\hspace{0pt}}m{#1}}

\usepackage{url} 
\usepackage{hhline}

\usepackage[normalem]{ulem}
\usepackage{booktabs}
\usepackage{makecell} 
\usepackage{titlesec} 
\usepackage{indentfirst} 
\usepackage{float} 
\usepackage{dblfloatfix} 
\usepackage{changepage} 
\usepackage{algorithm}
\usepackage[noend]{algpseudocode}

\newcommand{\delete}[1]{{\color{red}\sout{}}}

\usepackage{multirow}
\usepackage{comment}

\usepackage[nolessnomore, italic]{mathastext}
\usepackage{amsmath,amssymb,amsfonts}

\usepackage{textcomp}

\usepackage{fancyhdr}
\def\BibTeX{{\rm B\kern-.05em{\sc i\kern-.025em b}\kern-.08em
    T\kern-.1667em\lower.7ex\hbox{E}\kern-.125emX}}
\usepackage{upgreek} 
\usepackage{hhline}
\usepackage{enumitem}
\newcommand{\js}[1]{{\color{black}{#1}}}
\newcommand{\nh}[1]{{\color{black}{#1}}}
\newcommand{\juan}[1]{{\color{black}{#1}}}

\newcommand{\fig}[1]{{Figure~#1}\xspace} 
\newcommand{\tab}[1]{{Table~#1}\xspace} 
\newcommand{\head}[1]{{\noindent\textbf{#1.}\xspace}} 

\usepackage{listings}
\usepackage[font=small,labelfont=bf,textfont={bf}]{caption}
\usepackage[labelfont=bf,textfont=normalfont,singlelinecheck=off,justification=centering]{subcaption}

 \usepackage{amsmath,amssymb,amsfonts}
 \usepackage{cleveref}

\usepackage{booktabs} 
\usepackage{textcomp}

\usepackage{balance}
\usepackage[rightcaption]{sidecap} 
\usepackage{fancyhdr}
\usepackage[us,12hr]{datetime}

\newcommand{\ignore}[1]{}

\usepackage{tikz}
\newcommand*\circled[1]{\tikz[baseline=(char.base)]{
            \node[shape=circle,draw,inner sep=0pt,fill=black, text=white] (char) {#1};}}
\newcommand*\circledWhite[1]{\tikz[baseline=(char.base)]{
            \node[shape=circle,draw,inner sep=0pt,fill=white, text=black] (char) {#1};}}

\usepackage{grffile}
\usepackage{float}
\usepackage{svg}
\usepackage{pifont}
\usepackage{threeparttable}
\usepackage{graphics}
\usepackage{multirow}
\usepackage{tabularx}
\usepackage{verbatim}
\usepackage{booktabs}

\usepackage{dblfloatfix}

\usepackage{adjustbox}
\usepackage{diagbox}
\definecolor{LightCyan}{rgb}{0.88,1,1}
\definecolor{Gray}{gray}{0.9}

\definecolor{dred}{rgb}{255.00, 0.00, 0.00}
\newcommand*{\Scale}[2][4]{\scalebox{#1}{$#2$}}%
\definecolor{dred}{rgb}{0.75, 0.00, 0.00}

\definecolor{dgreen}{rgb}{0.00, 0.75, 0.00}
\definecolor{dpink}{rgb}{0.75, 0.0, 0.75}
\definecolor{dblack}{rgb}{0.00, 0.00, 0.00}
\definecolor{dblue}{rgb}{0.00, 0.00, 0.75}
\definecolor{gyell}{rgb}{0.5, 0.5, 0.0}
\definecolor{dbleudefrance}{rgb}{0.19, 0.55, 0.91}
\definecolor{navyblue}{rgb}{0.0, 0.0, 0.5}
\definecolor{forestgreen}{rgb}{0.13, 0.55, 0.13}

\newcommand{\gthesis}[1]{{\color{dblack}#1}}
\newcommand{\ggthesis}[1]{{\color{dblack}#1}}
\newcommand{\fc}[1]{{\color{dblack}#1}}
\newcommand{\ph}[1]{{\color{dblack}#1}}
\newcommand{\dd}[1]{{\color{dblack}#1}}
\newcommand{\ou}[1]{{\color{dblack}#1}}

\newcommand{\gcheck}[1]{{\color{dblack}#1}}
\newcommand{\gagan}[1]{{\color{dblack}#1}}
\newcommand{\gagann}[1]{{\color{black}#1}}
\newcommand{\gagannn}[1]{{\color{black}#1}}

\newcommand{\rakesh}[1]{{\color{black}#1}}
\newcommand{\juang}[1]{{\color{black}#1}}
\newcommand{\juangg}[1]{{\color{black}#1}}
\newcommand{\juanggg}[1]{{\color{black}#1}}
\newcommand{\jgl}[1]{}
\usepackage{notoccite}
\newcommand{\hps}{\textsf{HPS}\xspace}
\newcommand{\cde}{\textsf{CDE}\xspace}
\newcommand{\slow}{\textsf{Slow-Only}\xspace}
\newcommand{\fast}{\textsf{Fast-Only}\xspace}
\newcommand{\oracle}{\textsf{Oracle}\xspace}

\newcommand{\etal}{\textit{et al.}\xspace}

\include{pythonlisting}
\newcommand{\namePaper}{NERO\xspace} 
\newcommand{\nameNERO}{NERO\xspace} 

\newcommand{\myGlobalTransformation}[2]
{
    \pgftransformcm{1}{0}{0.4}{0.5}{\pgfpoint{#1cm}{#2cm}}
}

\newcommand{\gridThreeD}[3]
{
    \begin{scope}
        \myGlobalTransformation{#1}{#2};
        \draw [#3] grid (5,5);
    \end{scope}
}
\newcommand\crule[3][black]{\textcolor{#1}{\rule{#2}{#3}}}

\newcommand{\gridThreeDSecond}[3]
{
    \begin{scope}
        \myGlobalTransformation{#1}{#2};
        \draw [#3] grid (3,3);
    \end{scope}
}

\newcommand{\gridThreeDThird}[3]
{
    \begin{scope}
        \myGlobalTransformation{#1}{#2};
        \draw [#3] grid (1,1);
    \end{scope}
}

\tikzstyle myBG=[line width=3pt,opacity=1.0]

\newcommand{\dotesUpper}[2]
{
    \begin{scope}
        \myGlobalTransformation{#1}{#2};
   
     \node at (2.5,0.5) [circle,fill=black] {};
     \node at (1.5,1.5) [circle,fill=black] {};
    \node at (3.5,1.5) [circle,fill=black] {};

     \node at (0.5,2.5) [circle,fill=black] {};
      \node at (2.5,2.5) [circle,fill=black] {};
      \node at (4.5,2.5) [circle,fill=black] {};
      \node at (1.5,3.5) [circle,fill=black] {};
         \node at (3.5,3.5) [circle,fill=black] {};
           \node at (2.5,4.5) [circle,fill=black] {};

    \end{scope}
}

\newcommand{\lowerDots}[2]
{
    \begin{scope}
        \myGlobalTransformation{#1}{#2};
 
     \node at (0.5,1.5) [circle,fill=black] {};
       \node at (-1.5,1.5) [circle,fill=black] {};
          \node at (-0.5,0.5) [circle,fill=black] {};
           \node at (-0.5,2.5) [circle,fill=black] {};

    \end{scope}
}

\newcommand{\lowestDot}[2]
{
    \begin{scope}
        \myGlobalTransformation{#1}{#2};
 
    \node at (0.5,1.5) [circle,fill=black] {};

    \end{scope}
}
\usepackage[backend=biber,maxbibnames=99,style=ieee,citestyle=numeric,autocite=plain, giveninits,maxcitenames=99,sortfirstinits= true,sorting=nyt,dashed=false,sortcites=true]{biblatex} 
\DeclareNameAlias{sortname}{family-given}
\DeclareNameAlias{author}{last-first} 
\bibliography{ref}

\newcommand{\tikzcircle}[2][red,fill=red]{\tikz[baseline=-0.5ex]\draw[#1,radius=#2] (0,0) circle ;}%

\newcommand{\hdiff}{\texttt{hdiff}\xspace}
\newcommand{\sevpoint}{\texttt{7-point}\xspace}
\newcommand{\tfivepoint}{\texttt{25-point}\xspace}
\newcommand{\bitwidth}{{bitwidth}\xspace }
\newcommand{\integer}{\texttt{integer}\xspace } 
\newcommand{\fractional}{\texttt{fractional}\xspace } 
\newcommand{\cmark}{\ding{51}\xspace}%
\newcommand{\xmark}{\ding{55}\xspace}%

\usepackage[nottoc,notlot,notlof]{tocbibind}
\usepackage{array}
\newcolumntype{H}{>{\setbox0=\hbox\bgroup}c<{\egroup}@{}}

\newlength\mylen

\newcommand{\chapternote}[1]{{%
  \let\thempfn\relax
  \footnotetext[0]{{#1}}
}}

\usepackage[]{hyperref}
\hypersetup{
    pdftitle={Designing, Modeling, and Optimizing Data-Intensive Computing Systems},
    pdfauthor={Gagandeep Singh},
    bookmarksnumbered=true,     
    pdfstartview=Fit,           
    pdfpagemode=UseOutlines    
}

\usepackage{nicematrix}

\usepackage{bigstrut}
\makeatletter
\newlength\mylena
\newlength\mylenb
\newcommand\mystrut[1][2]{%
    \setlength\mylena{#1\ht\@arstrutbox}%
    \setlength\mylenb{#1\dp\@arstrutbox}%
    \rule[\mylenb]{0pt}{\mylena}}
\makeatother

\DeclareSortingTemplate{ydndt}{
  \sort{
    \field{presort}
  }
  \sort[final]{
    \field{sortkey}
  }
  \sort[direction=descending]{
    \field{sortyear}
    \field{year}
    \literal{9999}
  }
  \sort[direction=descending]{
    \field{sortname}
    \field{author}
    \field{editor}
    \field{translator}
    \field{sorttitle}
    \field{title}
  }
  \sort{
    \field{sorttitle}
    \field{title}
  }
}

\usepackage{color, colortbl}
\definecolor{Gray}{gray}{0.9}
\usepackage[first=0,last=9]{lcg}

\newcommand{\gh}[1]{{\color{black}#1}}
\newcommand{\go}[1]{{\color{black}#1}}

\newcommand{\gup}[1]{{\color{black}#1}}
\newcommand{\gca}[1]{{\color{black}#1}}
\newcommand{\rakeshisca}[1]{{\color{black}#1}}
\crefformat{section}{\S#2#1#3} 

\newcommand{\gsa}[1]{{\color{black}#1}}
\newcommand{\gor}[1]{{\color{black}#1}}
\newcommand{\gf}[1]{{\color{black}#1}}

\newcommand{\gon}[1]{{\color{black}#1}}
\newcommand{\gonn}[1]{{\color{black}#1}}
\newcommand{\gont}[1]{{\color{black}#1}}
\newcommand{\gonff}[1]{{\color{black}#1}}
\newcommand{\gonz}[1]{{\color{black}#1}}
\newcommand{\gonx}[1]{{\color{black}#1}}
\newcommand{\gaganF}[1]{[{\color{magenta}GAGAN:#1}]}

\newcommand{\ghpca}[1]{{\color{black}#1}}

\newcommand{\grakesh}[1]{{\color{black}#1}}

\newcommand{\gonf}[1]{{\color{black}#1}}
\newcommand{\rcam}[1]{{\color{black}#1}}
\newcommand{\rcamfix}[1]{{\color{black}#1}}
\newcommand{\rncam}[1]{{\color{black}#1}}
\newcommand{\rnlast}[1]{{\color{black}#1}}
\newcommand{\rbc}[1]{{\color{black}#1}}

\newcommand{\gonzz}[1]{{\color{black}#1}}
\newcommand{\jgll}[1]{{\color{ddgreen}\textbf{\textit{JGL: #1}}}}

\newcommand{\juancr}[1]{{\color{black}{#1}}}

\newcommand{\hlssd}{\textsf{$H\&L$}\xspace}
\newcommand{\hmssd}{\textsf{$H\&M$}\xspace}
\definecolor{NavyBlue}{rgb}{0.19, 0.55, 0.91}
\newcommand{\comm}[1]{{\color{NavyBlue}#1}}
\renewcommand{\algorithmiccomment}[1]{\bgroup\hfill\fontsize{5.8}{6}\selectfont$\triangleright$ {\textcolor{blue}{#1}}\egroup}
\newcommand{\arcc}{\textsf{Archivist}\xspace}
\newcommand{\kleio}{\textsf{RNN-HSS}\xspace}
\newcommand{\sibyldef}{\textsf{Sibyl$_{Def}$}\xspace}
\newcommand{\sibylopt}{\textsf{Sibyl$_{Opt}$}\xspace}

\newcommand{\thold}{\textsf{$H\&M\&L$}}
\newcommand{\thnew}{\textsf{$H\&M\&L_{SSD}$}}

\newcommand\scalemath[2]{\scalebox{#1}{\mbox{\ensuremath{\displaystyle #2}}}}
\newcommand*\circleds[1]{\tikz[baseline=(char.base)]{
            \node[shape=circle,draw,inner sep=-0.5pt,fill=black, text=white] (char) {#1};}}

\newcommand\mix{\textsf{$mix\_1$}\xspace}
\newcommand\hmone{\textsf{$hm\_1$}\xspace}

\newcommand\proxy{\textsf{$prxy\_0$}\xspace}

\usepackage[toc,nopostdot,style=alttree,nogroupskip,acronym,nonumberlist]{glossaries}

\makeglossaries

\newglossaryentry{latex}
{
        name=latex,
        description={Is a mark up language specially suited for 
scientific documents}
}

\newglossaryentry{maths}
{
        name=mathematics,
        description={Mathematics is what mathematicians do}
}

\newglossaryentry{formula}
{
        name=formula,
        description={A mathematical expression}
}

\newacronym{nmc}{NMC}{Near-Memory Computing}
\newacronym{scm}{SCM}{Storage Class Memory}
\newacronym{soc}{SoC}{System-on-a-Chip}
\newacronym{ssd}{SSD}{Solid-State Drive}
\newacronym{llvm}{LLVM}{Low Level Virtual Machine}
\newacronym{tco}{TCO}{Total Cost of Ownership}
\newacronym{asic}{ASIC}{Application-Specific Integrated Circuits}
\newacronym{cgra}{CGRA}{Coarse-Grained Reconfigurable Architecture}
\newacronym{cim}{CIM}{Computation-In Memory}
\newacronym{hpc}{HPC}{High-Performance Computing}
\newacronym{cosmo}{COSMO}{Consortium for Small-Scale Modeling}
\newacronym{fpga}{FPGA}{Field-Programmable Gate Array}
\newacronym{ml}{ML}{Machine Learning}
\newacronym{rl}{RL}{Reinforcement Learning}
\newacronym{pe}{PE}{Processing Element}
\newacronym{hbm}{HBM}{High-Bandwidth Memory}
\newacronym{hmc}{HMC}{Hybrid Memory Cube}
\newacronym{doe}{DoE}{Design of Experiments}
\newacronym{lhs}{LHS}{Latin Hypercube Sampling}
\newacronym{uram}{URAM}{UltraRAM}
\newacronym{bram}{BRAM}{Block RAM}
\newacronym{dse}{DSE}{Design Space Exploration}
\newacronym{lut}{LUT}{Lookup Table}
\newacronym{ff}{FF}{Flip Flop}
\newacronym{gpu}{GPU}{Graphics Processing Unit}
\newacronym{dsp}{DSP}{Digital Signal Processor}
\newacronym{clb}{CLB}{Configurable Logic Block}
\newacronym{api}{API}{Application Programming Interface }
\newacronym{hls}{HLS}{High-Level Synthesis}
\newacronym{capi}{CAPI}{Coherent Accelerator Processor Interface}
\newacronym{psl}{PSL}{Power-Service Layer}
\newacronym{ilp}{ILP}{Instruction Level Parallelism}
\newacronym{afu}{AFU}{Accelerator Functional Unit}
\newacronym{ccd}{CCD}{Central Composite Design}
\newacronym{nn}{NN}{Neural Network}
\newacronym{rf}{RF}{Random Forest}
\newacronym{ipc}{IPC}{Instructions Per Cycle}
\newacronym{edp}{EDP}{Energy-Delay Product}
\newacronym{nvm}{NVM}{Non-volatile Memory}
\newacronym{slc}{SLC}{Single-level Cell}
\newacronym{mlc}{MLC}{Multi-level Cell}
\newacronym{tlc}{TLC}{Triple-level Cell}
\newacronym{tlb}{TLB}{Translation Lookaside Buffer}
\newacronym{hss}{HSS}{Hybrid Storage System}

\begin{document}
\title{Near-Data Computing and \\ Data-Driven Architectures} 
\title{Performance modeling and enhancement for data-centric systems} 
\title{Designing, modeling, and optimizing \\data-centric architectures} 
\covertitle{Designing, modeling, and optimizing \\data-centric architectures} 
\title{Designing, Modeling, and Optimizing \\Data-Intensive Computing Systems}
\covertitle{Designing, Modeling, and Optimizing Data-Intensive Computing Systems}
\author{Gagandeep Singh} 
\birthplace{Chandigarh, India}
\defencedate{maandag 29  maart 2021 om 16:00 uur}
\backoffront{\noindent Dit proefschrift is goedgekeurd door de promotoren en de samenstelling van de \linebreak promotiecommissie is als volgt:\\[2em]

\noindent\begin{tabular}{ll}
voorzitter:     & prof.dr.ir.\ A. B.\ Smolders \\
1$^e$ promotor: & prof.dr.\ H.\ Corporaal\\
2$^e$ promotor:  & prof.dr.\ O.\ Mutlu    (ETH Z\"urich)\\
copromotor:     & dr.ir.\ S.\ Stuijk \\
 leden: & prof.dr.\ H.P.\ Hofstee    (IBM Research Austin/TU Delft)\\
                  & prof.dr.\ F.\ Catthoor (IMEC/KU Leuven)\\
                  & prof.dr.ir.\ C.H.\ van Berkel\\
 adviseurs:   & dr.\ O.\ Unsal (BSC-CNS) \\
 
 & dr.\ D.\ Diamantopoulos (IBM Research Europe) \\
 				
\end{tabular}

\vfill
\noindent Het onderzoek dat in dit proefschrift wordt beschreven is uitgevoerd in overeenstemming met de TU/e Gedragscode Wetenschapsbeoefening.
}

\backofcover{\noindent Doctorate committee:\\

\noindent\begin{tabular}{ll}
prof.dr.\ H.\ Corporaal         & Eindhoven University of Technology, 1$^{st}$  promotor\\
prof.dr.\ O.\ Mutlu    & ETH Z\"urich, 2$^{nd}$ promotor \\
dr.ir.\ S.\ Stuijk        & Eindhoven University of Technology, copromotor\\
prof.dr.ir.\ A.B.\ Smolders  & Eindhoven University of Technology, chairman \\
prof.dr.\ H.P.\ Hofstee  & IBM Research Austin/TU Delft\\
 prof.dr.\ F.\ Catthoor & IMEC/KU Leuven\\
 prof.dr.ir.\ C.H.\ van Berkel & Eindhoven University of Technology\\
dr.\ O.\ Unsal & BSC-CNS\\
dr.\ D.\ Diamantopoulos &IBM Research Europe \\

 \end{tabular}
\vfill \small

\noindent This work is supported in part by the European Commission under Marie Sklodowska-Curie Innovative Training Networks European Industrial Doctorate (Project ID: 676240).
\vspace{1em}

\noindent \textcopyright{} Copyright 2021, Gagandeep Singh\\All rights reserved.  Reproduction in whole or in part is prohibited without the written consent of the copyright owner.
\vspace{1em}

\noindent Cover design by Harpreet Kaur
\vspace{1em}

\noindent Printed by ProefschriftMaken || www.proefschriftmaken.nl 
\vspace{1em}

\noindent A catalogue record is available from the Eindhoven University of Technology Library.\\
ISBN: 978-90-386-5227-6
 }
\maketitle

\section*{Abstract}
The cost of moving data between the memory units and the compute units is a major contributor to the execution time and energy consumption of modern workloads in computing systems. At the same time, we are witnessing an enormous amount of data being generated across multiple application domains.  
Moreover, the end of Dennard scaling, the slowing of Moore's law, and the emergence of dark silicon limit the attainable performance on current computing systems. 
These trends suggest a need for a paradigm shift towards a \emph{data-centric} approach where computation is performed close to where the data resides. This approach allows us to overcome our current systems' performance and energy limitations by minimizing the \emph{data movement overhead} \ggthesis{by ensuring that data does not overwhelm system components.} Further, a {data-centric} approach can enable a \emph{data-driven} view where we take advantage of vast amounts of available data to improve architectural decisions. Our current systems are designed to follow rigid and simple policies that lack adaptability. Therefore, current system policies fail to 
provide robust improvement across varying workloads and system conditions.

As a step towards modern computing architectures, this dissertation contributes to various aspects of the \emph{data-centric} approach and further proposes several \emph{data-driven} mechanisms.

First, we design NERO, a data-centric accelerator for a real-world weather prediction application. NERO overcomes the memory bottleneck of weather prediction stencil kernels by exploiting near-memory computation capability on specialized field-programmable gate array (FPGA) accelerators with high-bandwidth memory (HBM) that are attached  through a cache-coherent interconnect to a host CPU system. Second, we explore the applicability of different number formats, including fixed-point, floating-point, and posit, for memory-bound stencil kernels. 
We search for the appropriate bit-width that reduces the memory footprint and improves the performance and energy efficiency with minimal loss in the accuracy. 

Third, we propose NAPEL, an ML-based application performance and energy prediction framework for data-centric architectures. NAPEL uses ensemble learning to build a model that, once trained for a fraction of programs, 
 can predict the performance and energy consumption of {different} applications. Fourth, we present LEAPER, the first use of {few-shot learning} to transfer FPGA-based computing models across different hardware platforms and applications. LEAPER provides the ability to reuse a prediction model  built on an {{inexpensive}} low-end local system to a new, unknown, high-end FPGA-based system. 

Fifth, we propose Sibyl, the first reinforcement learning-based mechanism for data placement in hybrid storage systems.  Sibyl is a data-driven mechanism. It observes different features of the running workload as well as the storage devices to make system-aware data placement decisions. For every decision it makes, Sibyl receives a reward from the system that it uses to evaluate the long-term performance impact of its decision and continuously optimizes its data placement policy online. Our extensive real-system evaluation demonstrates that Sibyl provides adaptivity and extensibility by continuously learning from and autonomously adapting to the workload characteristics, storage configuration and device characteristics, and system-level feedback to maximize the overall long-term performance of a hybrid storage system. We interpret Sibyl's policy through our explainability analysis and conclude that Sibyl provides an effective and robust approach to data placement in current and future hybrid storage systems.


Overall, this thesis provides two key conclusions: (1) hardware acceleration on an FPGA+HBM fabric is a promising solution to overcome the data movement bottleneck of our current computing systems; (2) data should drive system and design decisions by leveraging inherent data characteristics to make our computing systems more efficient. 
Thus, we conclude that the mechanisms proposed by this dissertation provide promising solutions to handle data well by following a \emph{data-centric} approach and further demonstrates the importance of leveraging data to devise \emph{data-driven} policies.

We hope that the proposed architectural techniques and detailed experimental results presented in this dissertation will enable the development of energy-efficient data-intensive computing systems and drive the exploration of new mechanisms to improve the performance and energy efficiency of future computing systems. 

\section*{Acknowledgements}
At the end of this journey, I believe Ph.D. is more than just a thesis. It is who you have become. I would like to reflect on the many people who have supported and helped me to become who I am today. Certainly, I would not have gotten this far without the support of many excellent people around me.

First and foremost, I express my gratitude to my promotor, Dr. Henk Corporaal, for giving me the opportunity and for his unwavering support and guidance throughout my Ph.D. Henk has been more than a supervisor and always guided me on any matter I needed his help. Despite his busy schedule, he always made time to discuss and brainstorm about any idea and always provided rigorous feedback. I would like to thank my co-promotor, Dr. Sander Stuijk, for his supervision and guidance. Sander was instrumental in my Ph.D. as he always believed in my skills and helped me to set ambitious but realistic goals.  Moreover, a special mention for helping with incessant administrative issues.  

My time at ETH with the SAFARI research group took my research career to the next level. I am deeply indebted to Dr. Onur Mutlu for his invaluable scientific discussions, feedback, and support. His passion for Computer Architecture intrigued me to pursue  research in this domain. I am grateful to have experienced the rigor, thought, and care Onur puts into every scientific task. \gthesis{Further, he always makes sure that extraneous issues do not constraint us in any way, so we can focus on high-impact research and realize our true potential.} I cannot begin without expressing my gratitude to Dr. Juan G{\'o}mez-Luna, who provided me with the encouragement and patience throughout my Ph.D. He always believed in my abilities and motivated me to keep going. 
I also thank Jisung Park, Nastaran Hajinazar, and Rakesh Nadig for their discussions, feedback, collaboration, and support. I would like to recognize the help I received from Kaan Kara,  Geraldo Oliviera, Jeremie Kim, Damla Senol Cali, Mohammed Alser, Rahul Bera, Konstantinos Kanellopoulas, Abdullah Giray Ya\u{g}l{\i}k\c{c}{\i}, Can Firtina, Taha Shahroodi, Tracy Ewen, and Christian Rossi.

During my secondment at IBM Research, I had invaluable experiences that have shaped me both personally and professionally.  I am incredibly grateful to Dionysios Diamantopoulos. I spent more time in his office than in mine discussing and brainstorming while using his coffee machine.  I would like to extend my sincere thanks to Christoph Hagleitner for the guidance and for providing access to the IBM systems.  I always enjoyed our interactions, both technical and lunch-time. Special thanks to Ronald Luijten for our numerous discussions on weather forecasting and personal guidance. One of the greatest joys of working at IBM is the presence of many talented individuals who I called upon numerous times for their expertise during my research, and thus I would like to express my appreciation to Robert Haas,  Giovanni Mariani, Andreea Anghel, Abu Sebastian, Florian Auernhammer, Gero Dittmann,  Raphael Polig, Jan van Lunteren,  Teodoro Laino, Miguel Prada, Martino Dazzi, Irem Boybat, Iason Giannopoulos, Mitra Purandare, Francois Abel, Beat Weiss, Pier Andrea Francese, Celestine D\"unner, Burkhard Ringlein, Judith Blanc, Lilli-Marie Pavka, and numerous other IBMers and interns for many interesting discussions and assistance with my work. I also had the great pleasure of working with Lorenzo Chelini, Ahsan Jawed Awan (Ericsson Research), and Stefano Corda.

I would like to thank the members of the doctoral committee for agreeing to participate in my defense during such difficult times, as well as for taking the time to read my drafts and provide constructive feedback, which helped improve the quality of this thesis.

I am deeply indebted to my family, who has done everything in their power to support me in both my academic and personal life. My father, Harinder Singh's steadfastness and humbleness, my mother Ravinder Kaur's kindness and encouragement, and my sister Jasmeet Kaur's unique perspective and guidance have helped me focus on what is important. Their support underlies everything that I do, and I am grateful to have such a wonderful family.

  I am grateful to Jan and Evelyn for making my time in the Netherlands memorable. Last but not least, I cannot begin to express my gratitude to Lenka, Ilde,  Mladen,  Savvas, Victor, Luc, Mark, Barry, Amr, Marja,  Kanishkan, Sajid, Alessandro, Sayandip, Paul, Hamideh, Mojtaba, Kamlesh, Umar, Roel, and other friends and colleagues from Eindhoven who always invited me to every event and never made me feel that I am away from them. I appreciate their continued  motivation and support throughout my stay away from home in the Netherlands and Switzerland.\\

 \begin{flushright}
\noindent \textbf{Gagandeep Singh}\\
 \textit{Z\"urich, Switzerland}\\
\textit{March 12, 2021}
\end{flushright}

\begingroup
\setstretch{1.05}
\setlength\bibitemsep{0.1pt}
\pdfbookmark{\contentsname}{toc}
\tableofcontents
\endgroup

\newpage
\pagenumbering{roman}


\begingroup
\setstretch{1}
\setlength\bibitemsep{0.1pt}
\listoffigures
\addcontentsline{toc}{chapter}{\listfigurename}
\endgroup


\begingroup
\setstretch{1}
\setlength\bibitemsep{0.1pt}
\listoftables
\addcontentsline{toc}{chapter}{\listtablename}
\endgroup

\begingroup
\setstretch{0.95}
\setlength\bibitemsep{0.1pt}
\glsfindwidesttoplevelname
\printglossary[title=List of Acronyms, toctitle=List of Acronyms, type=\acronymtype]
\endgroup

\pagestyle{fancy}
\fancyhf{}
\fancyhead[LE,RO]{\small \thepage}
\fancyhead[RE]{\small \selectfont  \textit{ \uppercase{\leftmark}} }
\fancyhead[LO]{\small \selectfont  \textit{ \uppercase{\rightmark}} }
\renewcommand{\headrulewidth}{0pt}
\pagenumbering{arabic}
\setcounter{page}{1}


 \chapter{Introduction}
\label{chapter:intro}

A wide range of application domains have emerged with the ubiquity of computing platforms in every aspect of our daily lives. These modern workloads (e.g., machine
learning, graph processing, and bioinformatics) demand high compute capabilities within strict power constraints~\cite{ghose2019processing}. However, today's computing systems are getting constrained by current technological capabilities, making them incapable of delivering the required performance. We highlight three major trends that call for a paradigm shift in the computing system landscape.

\noindent\textbf{(1) The Memory Wall:} Over the years, the improvements in memory access latency have not been able to keep up with the improvements in processor latency, which is referred to as the \emph{memory wall}~\cite{Wulf1995}.
Earlier, system architects tried to bridge this gap by introducing multiple levels of the memory hierarchy that can cache data to avoid unnecessary off-chip data movement. Yet, the system remained \emph{compute-centric}, where the data has to move a long distance from the memory subsystem over the power-hungry off-chip data bus through multiple levels of memory hierarchy to the compute units for processing. This compute-centric nature has lead to a fundamental \emph{data movement bottleneck}, which incurs a significant amount of energy and latency overhead in our current computing systems~\cite{mutlu2020modern}. \fc{In addition, for a given application,  there can be a strong mismatch between the nature of the data access patterns and the layout of the data in memory.  
Such a mismatch leads to limited spatial locality, which causes frequent data movement between the memory subsystem and the processing units.} 



\noindent\textbf{(2) The Slowdown of Moore's Law:} \gthesis{The continuation of Moore's law allowed more transistors per chip for each transistor node technology generation. This transistor scaling enabled us to have CPUs composed of a multi-core architecture with multiple levels of caches. In 1974, Dennard~\cite{1050511} postulated that the total chip power for a given chip area stayed constant from one process generation to another.} This trend allowed CPU vendors to have higher clock frequencies without drastically impacting the overall system power consumption. However, with the demise of Dennard scaling~\cite{1050511}, we witness  that with every process generation, total chip power  does not remain constant due to the leakage effects. \ph{The leakage effects limit the shrinking of gate oxide in a transistor that forces higher voltage than required for Dennard scaling, which in turn increases the power density.} \fc{As a result,  the system-level performance gains are not enough to motivate a further increase in the single-core clock frequency compared to the significant increase in energy consumption.} This increase in power consumption has led to the onset of dark silicon~\cite{6175879} \fc{in servers}, whereby a large portion of transistors on a chip has to be switched off or run at a lower frequency to avoid the effects of high leakage~\cite{martin2014postdennard}. \fc{While in system-on-a-chip (\acrshort{soc})-based designs, specialized heterogeneous cores have emerged to overcome the disadvantages of dark silicon and slowdown of Moore's law~\cite{hardavellas2012rise}, where each core is optimized for a specific task.}   Therefore, in the Post-Dennard scaling era, the single-core CPU performance has stagnated because of its inability to operate within the given power budget. 

Over the years, advancements in manufacturing technology have been slowing due to device physics limitations in transistor node scaling. \fc{On the other hand, the memory scaling is becoming even more challenging~\cite{mutlu2017rowhammer,ghose2018your,frigo2020trrespass}}. However, the demand for performance with high energy-efficiency has continued to grow. In the high-performance computing (\acrshort{hpc}) domain, current scaling issues \fc{and high communication overheads} limit HPC systems to realize exascale computers needed for modern and future data-intensive workloads~\cite{nair2015active}.

Therefore, future architectural innovations are expected to come from optimizations across the entire hardware/software computing stack~\cite{mutlu2020intelligent,mutlu2021intelligent_DATE}.

\noindent\textbf{(3) The Data Avalanche:} At the same time, we are witnessing an enormous amount of data being generated across multiple application domains~\cite{nair2015active} like weather prediction modeling, radio astronomy, bioinformatics, material science, chemistry, health sciences, etc.  In the domain of climate and weather modeling, for example, there is a data avalanche due to large atmospheric simulations~\cite{schar2020kilometer}. Major efforts are currently underway towards refining the resolution grid of climate models that would generate \emph{zettabytes} of data~\cite{schar2020kilometer}. These high-resolution simulations are useful to predict and address events like severe storms. However, the sheer amount of generated data is one of the biggest challenges to overcome. 

The \acrlong{cosmo} (\acrshort{cosmo})~\cite{doms1999nonhydrostatic} 
\juang{built} one such weather model 
to meet the high-resolution forecasting requirements of weather services. \gthesis{We use COSMO as a major case-study in this thesis.}   The main computational pipeline of COSMO consists of compound stencil kernels that operate on a three-dimensional grid~\cite{gysi2015modesto}. 
The performance of these compound stencil kernels is dominated by memory-bound operations with irregular memory access patterns and low arithmetic intensity that often results in $<$10\% sustained floating-point performance on current CPU-based systems~\cite{chris} that standard CPU-based optimization techniques cannot overcome (see Chapter~\ref{chapter:nero}). 

We find another relevant example in radio astronomy. The first phase of the Square Kilometre Array (SKA) aims to process over 100 terabytes of raw data samples per second, yielding of the order of 300 petabytes of SKA data produced annually~\cite{6898703}. \fc{ Recent biological disciplines such as genomics have also emerged as one of the most data-intensive workloads across all different sciences wherein just a single human genome sequence produces hundreds of gigabytes of raw data. With the rapid advancement in sequencing technology, the data volume in genomics is projected to surpass the data volume in all other application domains~\cite{navarro2019genomics}.}  \\ 

\emph{{\gthesis{The above trends suggest that} computer architects and system designers need to\linebreak develop novel architectural solutions to overcome the aforementioned major technological challenges and effectively handle the overwhelming amount of data. }}\\

This chapter serves as an introduction to this dissertation. The chapter is structured as follows. Section~\ref{sec:intro/problem} highlights the two guiding principles that we follow in this dissertation to  overcome our current system challenges. Based on these guiding principles, we provide the thesis statement. Section~\ref{sec:intro/overview} describes our approach to solve the thesis statement, while Section~\ref{sec:intro/contributions} lists the contributions of this dissertation. Finally, Section~\ref{sec:intro/outline} provides the dissertation outline.

\section{Problem and Thesis Statement}
\label{sec:intro/problem}

\gthesis{Future architectural innovations are expected to come from optimizations across the entire hardware/software computing stack~\cite{mutlu2020intelligent}. We highlight two guiding principles that can be applied in \dd{complementary} aspects of computer architecture to overcome the current system challenges and improve the overall performance: (1) \emph{data-centric \dd{computing}},  where we bring processing closer to the memory and use different techniques to reduce the data movement bottleneck (such as hardware specialization, data quantization, and  domain specific-memory hierarchies); (2) \emph{data-driven \dd{system optimization}}, where we exploit the available data to perform architectural decisions or predictions. }
\subsection{Data-Centric \dd{Computing}}
\label{subsection:intro/data_centric}
\begin{figure*}[h]
\centering
\includegraphics[width=1\textwidth]{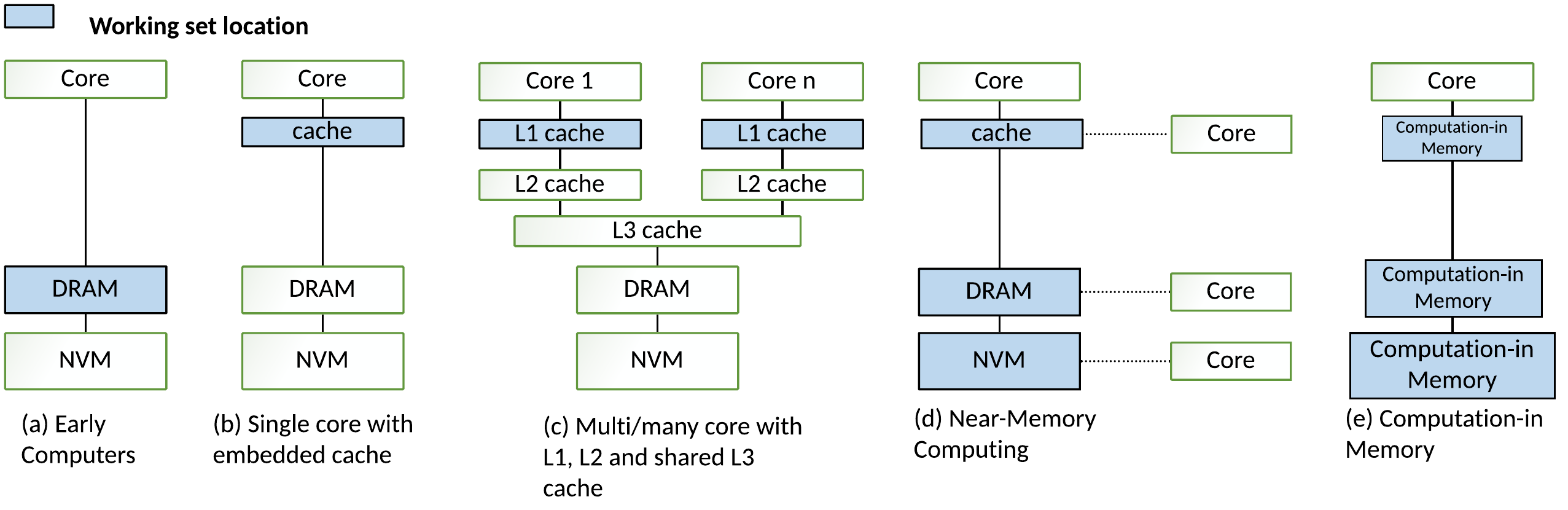}
\caption[Classification of computing systems based on working set location~\cite{7092668}.]{Classification of computing systems based on working set location~\cite{7092668}. Prior systems were based on a compute-centric approach where data is moved to the core for processing (Figure 1-1 (a)-(c)), whereas now, with near-memory computing (Figure 1-1 (d)), the processing cores are brought closer to the place where data resides. Therefore, in a data-centric approach, the data resides much closer to the processing units than in a compute-centric approach. Computation-in-memory (Figure 1-1 (e)) further reduces data movement by using memories with built-in compute capability (e.g., phase-change memory (PCM)~\cite{li2016pinatubo})
\label{fig:nmc_review_allsystems}}
\end{figure*}

Today's memory hierarchy usually consists of multiple levels of cache, the main memory, and the storage. The traditional approach is to move data up to caches from the storage and then process it. Figure~\ref{fig:nmc_review_allsystems} depicts the system evolution based on the information referenced by a program during execution, which is referred to as a working set~\cite{7092668}. Prior systems were based on a \emph{compute-centric} approach where data is moved to the core for processing (Figure 1 (a)-(c)). In contrast, the \emph{data-centric} approach aims at processing close to where the data resides. This approach couples compute units close to the data and seek to minimize the expensive data movement. The system ensures that data does not overwhelm its components. 

As shown in Figure 1 (d-e), near-memory computing (\acrshort{nmc})~\cite{nair2015active,teserract,7446059,ahn2015pim,hsieh2016accelerating,tom,ke2020recnmp,gomezluna2021upmem,cali2020genasm,fernandez2020natsa,singh2020nero} and computation-in memory (\acrshort{cim})~\cite{seshadri2013rowclone,chang2016low,simon2020blade,rezaei2020nom,gao2019computedram,seshadri2019dram,seshadri2017ambit,li2016pinatubo,computecache,li2017drisa} are two data-centric paradigms.  In near-memory computing (Figure 1 (d)), we place the processing cores closer to the memory units. On the other hand, the computation-in-memory (Figure 1 (e)) paradigm is more disruptive as it aims at reducing data movement completely by using memories with \ggthesis{built-in} compute capability (e.g., resistive random-access memory (ReRAM) and  phase-change memory (PCM)). Processing right at the ``home'' of data can significantly diminish the data movement problem of data-intensive applications. \gthesis{Thus, data-centric architectures have the potential to overcome our current data movement bottleneck.} 

\fc{
\textbf{Heterogeneous computing} has emerged as an answer to continue to improve performance beyond the limits imposed by the slow down in Moore’s Law and the end of Dennard scaling~\cite{sundararajan2010hpcwithfpga}. Heterogeneous computing entails complementing processing elements with different compute capabilities, each to perform the tasks to which it is best suited. In the HPC domain, coupling specialized compute units with general-purpose cores while following a data-centric approach can meet the high-performance computing demands and provides the ability to realize exascale systems  needed to process data-intensive workloads~\cite{nair2015active}. }


The graphics processing unit (\acrshort{gpu}) is one of the most popular acceleration platforms. GPUs have been used to accelerate workloads like computer graphics and linear algebra~\cite{volkov2008benchmarking} because of their many-core architecture. However, GPUs are power-hungry due to high transistor density and, depending on the power constraints, may not always be the ideal platform for implementation. Recently, the use of field-programmable gate array (\acrshort{fpga}) in accelerating machine learning workloads with high energy-efficiency has inspired researchers to explore the use of FPGAs instead of GPUs for various high-performance computing applications~\cite{caulfield2016cloud,duarte2018fast}.  

FPGAs provide a unique combination of flexibility and performance without the cost, complexity, and risk of developing custom application-specific integrated circuits (\acrshort{asic}s). 
The researchers at CERN, for example, are using FPGAs to accelerate physics workload in CERN's \dd{exploration} for dark matter~\cite{duarte2018fast}. Microsoft's Project Catapult~\cite{caulfield2016cloud} is another example of how FPGAs can be used in the data center infrastructure. \dd{Driven by Catapult's promising research results, Microsoft further deployed the architecture on the Azure cloud marketplace~\cite{FPGA_in_azure}. } Such integration \fc{for certain workloads} can even offer more energy efficiency than CPU or GPU-based systems. However, taking full advantage of 
FPGAs for accelerating a workload is not a trivial task. Compared to CPUs or GPUs, {an} FPGA must exploit an order of magnitude more parallelism in a target workload to compensate for the lower clock frequency. 

The above trend raises a question: \textbf{how can we accelerate other scientific applications in an energy-efficient way on specialized hardware?}\\

\noindent Modern FPGAs show four key trends. 
\begin{itemize} [leftmargin=*,topsep=0pt,itemsep=1ex,partopsep=1ex,parsep=1ex] 
    \item The integration of high-bandwidth memory (\acrshort{hbm})~\cite{hbm} on the same package as an FPGA allows us to implement our accelerator logic much closer to the memory with an order of magnitude more bandwidth than the traditional DDR4-based FPGA boards. Thus, these modern FPGAs \gthesis{adopt} a \emph{data-centric} approach. 
    \item The introduction of UltraRAM (URAM)~\cite{uram} along with the BlockRAM (BRAM) that offers massive \emph{scratchpad}-based on-chip memory next to the logic. 
    \item New cache-coherent interconnects, such as IBM Coherent Accelerator Processor Interface (\acrshort{capi})~\cite{openCAPI}, Cache Coherent Interconnect for Accelerators (CCIX)~\cite{benton2017ccix}, and Compute Express Link (CXL)~\cite{cxlwhitepaper},  allow tight integration of FPGAs with CPUs at high bidirectional bandwidth {({on} the order of tens of GB/s)}. This integration \ggthesis{reduces programming effort and} enables us to coherently access the host system's memory through a pointer rather than having multiple copies of the data.   
    \item FPGAs are being manufactured with an advanced technology node of 7-14nm FinFET technology~\cite{gaide2019xilinx} that offers higher performance. \\
\end{itemize}


\noindent These four trends suggest that \textbf{modern FPGA architectures deliver \linebreak unprecedented levels of integration and compute capability due to new technological advances, which \fc{further} provides an opportunity to overcome {the} \emph{memory bottleneck} of real-world data-intensive applications.}


\subsection{Data-Driven System Optimization}
\label{subsection:intro/data_driven}
On a computing system, we generally run a diverse set of workloads that generate a large amount of data.  However, our present systems do not take advantage of vast amounts of data available to them~\cite{mutlu2020modern, mutlu2020intelligent,mutlu2021intelligent_DATE}.  
Current systems are built around rigid design rules that follow fixed heuristic-driven policies or perform an exhaustive exploration with a change in system scenario rather than leveraging previous knowledge.  Moreover, these heuristic-driven approaches cannot fully capture complex relations present in various aspects of the computer architecture, such as data placement, memory management, and task scheduling. These approaches favor certain workloads and/or system configurations over others without considering an application's inherent behavior or the underlying device characteristics. For example, the storage subsystem keeps executing the same data placement policy during the entire lifetime of a system regardless of the impact of the resulting decisions on the system. A storage subsystem sees a vast amount of data, yet it cannot learn from that data and adapt its policy because the policy is rigid and hard-coded by a human. 

To overcome the inefficiency of our current computing system, we need to devise \emph{data-driven} mechanisms that take advantage of the vast amount of available data by exploiting inherent data characteristics~\cite{ipek2008self,jimenez2001dynamic,jimenez2003fast,teran2016perceptron,jimenez2005piecewise,jimenez2011optimized,bhatia2019perceptron,peled2019neural,peled2015semantic,rl_GC_TECS,rl_NOC_AIDArc_2018}. To this end, a data-driven approach can enable machine learning (\acrshort{ml}) techniques in different aspects of computer architecture design and use. ML models are trained to make predictions or decisions without explicit programming by discovering inherent patterns or relationships in the data. 
Traditionally, computer architects have \ph{focused more} on accelerating ML algorithms. 
In contrast, \fc{only in the past few years have we seen a growth in works using ML for architecture design and use~\cite{MLsurvey2019}. }


Therefore, there is \textbf{an opportunity to develop novel mechanisms that allow computers to learn from experiences and reuse those experiences to make future decisions. In turn, closing a loop where computer architects enable ML and ML improves computer architectures. Moreover, we need to assess if these machine learning-driven approaches can outperform our current human-driven approaches.}

\gthesis{Data-driven techniques using ML can assist us in many aspects of computer architecture, including  architectural evaluation and design space exploration (\acrshort{dse}).} In the early design stage, architects often use various evaluation techniques to navigate the design space of new architectures, avoiding the cost of chip fabrication.  Usually, we employ analytic models or simulation techniques to provide performance and energy consumption estimates while using different workloads or architectural configurations. Analytic models are typically based on simple mathematical equations that provide fast estimates but at the cost of accuracy. Therefore, we often resort to simulation-based techniques that can model architectural interactions more accurately. However, simulation techniques \gthesis{can be} extremely slow because a single simulation for a real-world application with a representative dataset can take hours or even days (see Chapter~\ref{chapter:napel}). This slow speed cannot meet the modern design productivity demands, despite the efforts of accelerating simulations with both hardware~\cite{pellauer2011hasim} and software-based~\cite{sanchez2013zsim} techniques. 

In the later phase of the design cycle, architects often use FPGAs to prototype their design to get the accurate performance and power estimates. An FPGA is highly configurable and allows us to reconfigure its circuitry to implement any algorithm. However, an FPGA's large configuration space and the complex interactions among configuration options lead many developers to explore individual optimization options in an ad hoc manner. Moreover, FPGAs have infamously low productivity due to the time-consuming FPGA \fc{mapping} process~\cite{o2018predictive}. 

Therefore, a common challenge that past works {have also faced} is \textbf{how to evaluate the performance of an application or a new architecture in a reasonable amount~of~time~\cite{o2018hlspredict} \gthesis{such that a large design space can be covered}? 
}

\subsection*{Thesis Statement}
\gthesis{Based on the above-discussed aspects,} the thesis problem statement is as follows: \emph{Design system architectures to effectively handle data by: (1) overcoming the data movement bottleneck in data-intensive applications through a data-centric approach of bringing processing close to the memory and ensuring that data does not overwhelm system components; thus, enabling high performance in an energy-efficient way; and (2) further leverage the enormous amount of data to model and optimize computing systems by exploiting the inherent data characteristics to perform data-driven decisions. }


\section{Overview of Our Approach}
\label{sec:intro/overview}
In line with the thesis statement, this dissertation provides five contributions based on nine methods (concepts) to handle and leverage data effectively.  In Table~\ref{tab:intro/methods}, we discuss nine different methods that we use in this dissertation.  Methods are architectural concepts that we use in our contribution to reach a specific goal. \ggthesis{ We use six data-centric (DC) methods that put data and its processing at the center of the design. These methods minimize data movement and maximize efficiency while processing, accessing, and storing data. We also adopt three data-driven (DD) methods that take advantage of the vast amount of data that flow through the system.  We briefly discuss these nine methods.
\begin{table}[h]
\caption[List of contributions and methods used to achieve the thesis statement]{We highlight across five contributions nine different methods (concepts) used in this dissertation to achieve the thesis statement of handling data well. We make use of two guiding principles: (1) \textit{data-centric} (DC) is bringing processing closer to where data resides and ensuring that data does not overwhelm system components; (2) \textit{data-driven} (DD) is leveraging data to perform architectural decisions or predictions
\label{tab:intro/methods}}
\centering
\HUGE
\renewcommand{\arraystretch}{0.8}
\setlength{\arrayrulewidth}{2pt}
\setlength{\tabcolsep}{10pt}
\resizebox{1\linewidth}{!}{
\begin{tabular}{l|l|l|Hl|l|l}
\toprule 
\rule{0pt}{32pt}& \multicolumn{5}{c}{\textbf{CONTRIBUTIONS}} \\ \cmidrule{2-7} 
\textbf{METHODS} & \textbf{NERO~\cite{singh2020nero}}                                                                                                   & \textbf{Low Precision~\cite{singh2019low}}                                                                                        & \textbf{PreciseFPGA~\cite{not_published_arpan2021preciseNN}}                                                                                      & \textbf{NAPEL~\cite{NAPEL}}                                                                                       & \textbf{LEAPER~\cite{singh2021leaper_fpga}}                                                                                                                 & \textbf{Sibyl~\cite{not_published_singh2021qrator}}                                                                                   \\ \hline 
\rowcolor[HTML]{DCDCDC} 
\mystrut  
\textbf{Specialization} (DC)                                                                          &  \begin{tabular}[c]{@{}l@{}}FPGA-based \\ accelerator\end{tabular}    &                                             \begin{tabular}[c]{@{}l@{}}Implementation on\\an FPGA for fixed-\\point and floating-\\point representation\end{tabular}                                                                                                                 & \begin{tabular}[c]{@{}l@{}}FPGA-based \\ accelerator\end{tabular}                                         &                                                                                                      & \begin{tabular}[c]{@{}l@{}}FPGA-based \\ accelerator\end{tabular}                                                               &                                                                                                   \\ \hline 
\rowcolor[HTML]{DCDCDC} 
\mystrut  
\begin{tabular}[c]{@{}l@{}}\textbf{Revisit memory}\\\textbf{hierarchy} (DC)       \end{tabular}                                                                & \begin{tabular}[c]{@{}l@{}}Scratchpad-based\\ hybrid memory\end{tabular}                                 &                                                                                                               &                                                                                                           &                                                                                                      &                                                                                                                                 & \begin{tabular}[c]{@{}l@{}}Tiered hybrid\\storage system\end{tabular}                             \\ \hline 
\rowcolor[HTML]{DCDCDC}
\mystrut  
\begin{tabular}[c]{@{}l@{}}\textbf{Reducing copies}\\\textbf{between host and}\\\textbf{accelerator} (DC)\end{tabular} & \begin{tabular}[c]{@{}l@{}}Shared memory \\space    \end{tabular}                                                                                             & Quantize data                                                                                                 & Quantize data                                                                                             &                                                                                                      &                                                                                                                                 &                                                                                                   \\ \hline 
\rowcolor[HTML]{DCDCDC} 
\mystrut   
\begin{tabular}[c]{@{}l@{}}\textbf{Dataflow} \\ \textbf{architecture} (DC)     \end{tabular}                                                                                   & Task pipelining                                                                                                 &                                                                                                               &                                                                                                           &                                                                                                      &                                                                                                                                 &                                                                                                   \\ \hline 
\rowcolor[HTML]{DCDCDC} 
\mystrut  
\begin{tabular}[c]{@{}l@{}}\textbf{Near-memory}\\\textbf{computing} (DC)    \end{tabular}                                                               & \begin{tabular}[c]{@{}l@{}}{Processing near-}\\{high-bandwidth} \\{memory}   \end{tabular}                                                                                            &                                                                                                               &                                                                                                           & \begin{tabular}[c]{@{}l@{}}Processing near-\\3D stacked memory \end{tabular}                           &                                                                                                                                 &                                                                                                   \\ \hline 
\rowcolor[HTML]{DCDCDC} 
\mystrut   
\begin{tabular}[c]{@{}l@{}}\textbf{Reducing memory}\\ \textbf{footprint} (DC)     \end{tabular}                                                          &                                                                                                          \begin{tabular}[c]{@{}l@{}}Single and half\\floating-point\\precision \end{tabular}        & \begin{tabular}[c]{@{}l@{}}Different number \\ representations--\\ posit, fixed-point,\\and floating-point\end{tabular}                           & \begin{tabular}[c]{@{}l@{}}Automatic\\ quantization of \\ fixed-point\\ representation\end{tabular} &                                                                                                      &                                                                                                                                 &                                                                                                   \\ \hline 
\rowcolor{gray!45} 
\mystrut  
\textbf{Static ML} (DD)                                                                 &                                                                                                                 &                                                                                                               &                                             \begin{tabular}[c]{@{}l@{}}Learns resource\\ and power\\utilization\end{tabular}                                                                &\begin{tabular}[c]{@{}l@{}}Learns application\\ performance and\\ energy consumption\end{tabular}                                                                                   & \begin{tabular}[c]{@{}l@{}}Learns resource\\ utilization and\\ performance\\ models\end{tabular}                                    &         
\\\hline 
\rowcolor{gray!45} 
\mystrut  
\begin{tabular}[c]{@{}l@{}}\textbf{Speed up design} \\ \textbf{space exploration}\\  (DD)\end{tabular}         &    \begin{tabular}[c]{@{}l@{}}Auto-tuning\\for data transfer\\window size\end{tabular}                                                                                                                      &                                                                                                               & Supervised ML                                                                                                  & \begin{tabular}[c]{@{}l@{}}Supervised ML,\\Design of \\ experiment\end{tabular}                                      & \begin{tabular}[c]{@{}l@{}}Few-shot\\learning\end{tabular}                                                                                                           &                           \begin{tabular}[c]{@{}l@{}}Design of\\experiment\end{tabular}                                                                            \\\hline 
\rowcolor{gray!45}
\mystrut  
\textbf{Dynamic ML} (DD)                                                          &                                                                                                                 &                                                                                                               &                                                                                                           &                                                                                                      &                                                                                                                                 & \begin{tabular}[c]{@{}l@{}}Reinforcement\\ Learning\end{tabular}                                                                             \\ \hline 
\mystrut  
\textbf{Goal}                                                                                & {\begin{tabular}[c]{@{}l@{}}Overcome memory\\bottleneck of\\weather prediction\\ application\end{tabular}} & {\begin{tabular}[c]{@{}l@{}}Investigate\\ computationally\\ cheaper number\\ representations\end{tabular}} &   {\begin{tabular}[c]{@{}l@{}}Automate DSE\\ for fixed-point\\ representation\end{tabular}}                                                                                  & {\begin{tabular}[c]{@{}l@{}}Quick performance\\ and energy\\ estimates of new \\ applications\end{tabular}} & {\begin{tabular}[c]{@{}l@{}}Quick area and\\ performance\\ estimates on\\ new high-end \\FPGA-based \\platforms\end{tabular}} & {\begin{tabular}[c]{@{}l@{}}Efficient and \\ high performance\\data placement \\mechanism\end{tabular}} \\ \bottomrule
\end{tabular}
}
\end{table}

  \begin{enumerate} [leftmargin=*,topsep=0pt,itemsep=0.3ex] 
     \item \textbf{Specialization  (DC):} Using specialized hardware platforms such as an FPGA to accelerate an application in an energy-efficient way.
     \item \textbf{Revisit memory hierarchy  (DC):} Design memory hierarchies to provide low latency access to data.
     \item \textbf{Reducing copies between host and accelerator (DC):} Minimize the redundant amount of data copies between the host and the specialized hardware.
     \item\textbf{Near-memory computing (DC):} Process data closer to the main memory to reduce the data movement overhead. 
     \item\textbf{Dataflow architecture (DC):} Enable task-level parallelism \fc{while exploiting data-level parallelism} to maximize the utilization of compute resources.
     \item \textbf{Reducing the memory footprint (DC):} Reduce an application's required memory space to efficiently access and process data.
     \item \textbf{Static machine learning  (DD):} Leveraging ML during the design phase of architecture.
     \item \textbf{Speedup design space exploration  (DD):} Quick and accurate architectural evaluation and exploration.
     \item \textbf{Dynamic machine learning  (DD):} Leveraging reinforcement learning (\acrshort{rl}) during the run-time of a system to continuously adapt system policies.\\
 \end{enumerate}

 }

\noindent We demonstrate these above nine methods across the following five mechanisms.\\ \linebreak
First,  we design NERO, a data-centric accelerator for a real-world weather prediction application. As mentioned above, the sheer amount of atmospheric simulation data generated is one of the biggest challenges in the domain of weather prediction.  We use a heterogeneous system comprising of IBM\textsuperscript{\textregistered} POWER9 CPU with field-programmable gate array (FPGA) as our target platform. An FPGA can provide both flexibility and energy efficiency, and moreover, is a cost-effective \gthesis{alternative} to an application-specific integrated circuit (ASIC). We create a heterogeneous domain-specific memory hierarchy using on-chip \acrshort{uram}s and  \acrshort{bram}s on an FPGA. Unlike traditionally fixed CPU memory hierarchies, which perform poorly with irregular access patterns and suffer from cache pollution effects, application-specific memory hierarchies are shown to improve energy and latency by tailoring the cache levels and cache sizes to an application's memory access patterns~\cite{jenga}.  
NERO overcomes the memory bottleneck {of weather {prediction} stencil kernels} by {exploiting near-memory {computation} capabilities on} specialized FPGA accelerators with high-bandwidth memory (HBM), {which are attached} {to the host CPU}. 

Second, we explore the applicability of different number formats and exhaustively search for the appropriate bit-width for memory-bound stencil kernels to improve performance and energy-efficiency with minimal loss in the accuracy.  Stencils are one of the most widely used kernels in real-world applications. Based on an exhaustive exploration of a broad range of number systems -- fixed-point, floating-point, and posit~\cite{gustafson2017beating} -- we provide the precision and the corresponding accuracy deviation. \gthesis{Each number representation offers a different dynamic range, the usability of which depends upon the target workload.} 

Third, we propose NAPEL, a machine learning-based application performance and energy prediction framework for data-centric architectures. NAPEL uses ensemble learning to build a model that, once trained for a fraction of programs on a number of architecture configurations, can predict the performance and energy consumption of {different} applications. 

Fourth, we present LEAPER, the first use of \emph{few-shot learning} to transfer FPGA-based computing models across different hardware platforms and applications. \gthesis{Machine learning (ML)-based modeling has emerged as an alternative to traditional, slow simulation (or evaluation) techniques~\cite{o2018hlspredict}. ML modeling provides the capability to both quickly evaluate various architectural design choices and perform suitability analysis for many workloads. Thus, quick exploration and large prediction time savings compared to simulation are possible. To this end, we develop NAPEL and LEAPER, which are ML-based model solutions. 

However, ML needs a large amount of data to train models, which requires running time-consuming simulations. To alleviate this problem, we use a technique called the \emph{design of experiments} (\acrshort{doe})~\cite{montgomery2017design} to extract representative data with a small number of experimental runs.  DoE is a set of statistical techniques meant to locate a small set of points in a parameter space with the goal of representing the whole parameter space. The traditional brute-force approach to collecting training data is time-consuming: the sheer number of experiments renders detailed simulations intractable.  } 

Fifth, we propose Sibyl, the first technique that uses  reinforcement learning (\acrshort{rl}) \gca{for data placement in hybrid storage systems}. Sibyl observes different features \gon{of} the \gon{running workload}  \gon{as well as the}  storage devices to make system-aware data placement decisions. For every decision \gon{it makes}, Sibyl receives a reward from the system that it uses to evaluate the long-term \gon{performance} impact of its decision and continuously optimizes its data placement policy online.  \gthesis{Compared to supervised learning, reinforcement learning provides the following three benefits. \ggthesis{First, supervised learning requires large amounts of labeled data. In some scenarios, collecting labeled data can be difficult or even infeasible.} Second, unlike supervised learning, which is purely driven by prediction accuracy, an RL-agent is objective-driven, making RL a great fit for objective-driven policies. Third, RL does not require separate training and testing phases. Instead, RL continuously learns and adapts based on the changes in the environment.} We implement Sibyl on  \emph{real} \gonn{systems with various} \go{HSS configurations, including dual- and tri-hybrid storage systems}. Our in-depth evaluation of Sibyl shows that it outperforms \gonn{four} state-of-the-art 
    techniques over a wide variety of applications with a low implementation overhead.



\section{Contributions}
\label{sec:intro/contributions}
\noindent This dissertation makes the following \textbf{key contributions}:

\begin{enumerate}[leftmargin=*,topsep=0pt,itemsep=1.3ex]
\renewcommand{\namePaper}{NERO\xspace} 
\item In \textbf{Chapter~\ref{chapter:nero}}, we propose \namePaper, {the first} near-HBM FPGA-based accelerator for representative kernels from a real-world weather prediction application.  Weather prediction is one such high-performance computing application that generates a large amount of data. It consists of   compound stencil kernels that operate on a three-dimensional grid. Such compound kernels are dominated by memory-bound operations with complex memory access patterns and low arithmetic intensity. This poses a fundamental challenge to acceleration.
 \begin{enumerate}[leftmargin=*, noitemsep, topsep=-1pt]
    \item We perform a detailed roofline analysis to show that representative weather prediction kernels are constrained by memory bandwidth on state-of-the-art CPU systems.
    \item We optimize \namePaper~with a data-centric caching scheme with precision-optimized tiling for a heterogeneous memory hierarchy (consisting of URAM, BRAM, and HBM).
    \item We evaluate the performance and energy consumption of our accelerator and perform a scalability analysis. We show that an FPGA+HBM-based design 
    outperforms a complete 16-core POWER9 system~(running 64~threads)~by~$4.2\times$ for the vertical advection (\texttt{vadvc}) and $8.3\times$ for 
    the horizontal diffusion (\texttt{hdiff}) kernels with energy reductions of $22\times$ and $29\times$, respectively. 
    \end{enumerate}
\item In \textbf{Chapter~\ref{chapter:low_precision_stencil}}, we perform a precision exploration of the three-dimensional stencil kernels for future mixed-precision systems using a wide range of number systems, including fixed-point, floating-point, and posit.
       \begin{enumerate}[leftmargin=*, noitemsep, topsep=-1pt]
    \item  We provide the precision and the corresponding accuracy deviation for a broad range of number systems -- fixed-point, floating-point, and posit.
    \item We tune stencil-based kernels on a state-of-the-art IBM POWER9 CPU and further evaluate them on an FPGA, which is coherently attached to the host memory. Thus, this chapter fills the gap between the current hardware capabilities and future hardware design.
    \item As an extension of this chapter, in Appendix~\ref{chapter:precise_fpga}, we demonstrate our approach to automate the exploration for fixed-point configurations. We show our results for tuning the precision of weights for a neural network.
    \end{enumerate}
\item In \textbf{Chapter~\ref{chapter:napel}}, we propose NAPEL, a new, {fast} high-level performance and energy estimation framework for NMC architectures. NAPEL {is the first such model to} leverage ensemble learning techniques, specifically random forest, to {quickly estimate} the performance and energy consumption of previously-unseen applications in the early stages of design space exploration for NMC architectures.

     \begin{enumerate}[leftmargin=*, noitemsep, topsep=-1pt]
    \item We {reduce} the simulation time needed to gather training data for NAPEL by employing a DoE technique~\cite{mariani2017predicting}, which selects a {small} number of application input configurations that well represent the entire space of input configurations.
    \item {We} show that NAPEL can provide performance and energy estimates {$220\times$ faster than a state-of-the-art microarchitecture simulator} with an {average error rate of 8.5\% (performance) and 11.6\% (energy)} compared to the simulator.
    \item We show that we can use NAPEL to {accurately} determine if, {and by how much,} executing a certain workload on a specific NMC architecture can improve performance and reduce energy consumption {versus execution on a CPU}.
    \end{enumerate}

\item In \textbf{Chapter~\ref{chapter:leaper}}, we present LEAPER, the first use of \emph{few-shot learning} to transfer FPGA-based computing models across different hardware platforms and applications. This approach dramatically reduces (up to $10\times$) the training overhead by adapting a \emph{base model} trained on a low-end {edge} {FPGA} platform to a {new}, unknown high-end environment ({a cloud environment in our case}) rather than building a new model from scratch).
  \begin{enumerate}[leftmargin=*, noitemsep, topsep=-1pt]
    \item We create an {ensemble of transfer learning models} to accurately transfer learning from multiple base learners to avoid a negative transfer, i.e.,  severe degradation of the predictive power of the transferred model. 
    \item We demonstrate our approach across \textit{five} {state-of-the-art,} high-end, {on-premise cloud} FPGA-based platforms with \textit{three} different {interconnect} technologies, between host CPU and FPGA, on \textit{six} real-world applications. 
    For \textit{5-shot} learning, we achieve an average performance and area prediction accuracy of 80--90\%. 
    \end{enumerate}

 \renewcommand{\namePaper}{Sibyl\xspace} 
 \item In \textbf{Chapter~\ref{chapter:sibyl}}, we propose \namePaper
  , a new self-\gon{optimizing}  mechanism 
  \gon{that uses} reinforcement learning \gon{to make data placement decisions in hybrid storage systems}. \namePaper{} 
  \go{dynamically}
  \emph{learns}\gont{,} \gonn{ using both multiple workload features and system-level feedback information\gont{, how} to continuously adapt its policy to improve \rnlast{its} long-term performance \rnlast{for} a workload.}
    \begin{enumerate}[leftmargin=*, noitemsep, topsep=-1pt]
    \item  We show on real \gonn{hybrid storage systems (HSSs)}  that prior state-of-the-art \gont{HSS data placement mechanisms} fall 
  short of the oracle placement due to: lack of  \gon{(1) adapt\gonn{i}vity to workload changes and \gonn{storage device characteristics}, and (2) extensibility. } 

    \item We conduct an in-depth evaluation of \namePaper on \gon{real \gonn{systems with various} HSS configurations\gont{,}} showing that it outperforms \gonn{four} state-of-the-art 
    techniques over a wide variety of applications with a low implementation overhead.
    \item We provide an \rbc{in-depth} explanation of \namePaper's actions that show that \namePaper performs dynamic data placement decisions by learning \gonn{ changes in the level of asymmetry in the read/write latenc\gonzz{ies} and the number and types of storage devices.} 
    \item \gca{ We \gon{freely} open-source \namePaper to aid 
 {future research \gon{in} data placement for storage systems~\cite{sibylLink}}.} 
    \end{enumerate}
\item In addition to the above contributions, in \textbf{Chapter~\ref{chapter:background}}, we analyze and organize the extensive body of literature on near-memory computing  architectures across various dimensions: starting from the memory level where this paradigm is applied to the granularity of an application that could be executed on these architectures. 
    \begin{enumerate}[leftmargin=*, noitemsep, topsep=-1pt]
    \item A survey of existing near-memory computing architectures. We review more than 30 architectures in detail and identify the strengths and weaknesses of the existing architectures in Appendix~\ref{chapter:appendixA}.
    \item We highlight the opportunities and the challenges in the domain of near-memory computing. 
    \end{enumerate}
    
\end{enumerate}
\subsection*{Thesis Conclusion}
\noindent Overall, we make the following two conclusions for this thesis. 
 \begin{enumerate} 
\item \emph{Hardware acceleration on an FPGA+HBM fabric is a promising solution to reduce the data movement bottleneck of our current computing systems in an energy-efficient way. }

\item \emph{Data should drive system and design decisions by exploiting the inherent \linebreak characteristics of data to perform efficient architectural decisions or predictions in various design aspects of the computer architecture.} 
  \end{enumerate}
Therefore, we conclude that the mechanisms proposed by this dissertation provide \linebreak promising solutions to handle data well by following a \emph{data-centric} approach and further demonstrate the importance of leveraging data to devise \emph{data-driven} policies. 

\section{Dissertation Structure}
\label{sec:intro/outline}
This thesis is organized into \gcheck{eight} chapters. Chapter~\ref{chapter:background} provides background into near-memory computing, where we classify and evaluate various state-of-the-art \emph{data-centric} architectures. In Appendix~\ref{chapter:appendixA}, we describe in detail all the evaluated architectures. Additionally,  Chapter~\ref{chapter:background} also highlights various challenges that need to be addressed. Chapter~\ref{chapter:nero} presents NERO, a \emph{data-centric} architecture of weather prediction application. Chapter~\ref{chapter:low_precision_stencil} explores the applicability of different number systems \ou{for stencil kernels}. As an extension to Chapter~\ref{chapter:low_precision_stencil}, Appendix~\ref{chapter:precise_fpga} presents PreciseFPGA. It provides an automated exploration framework for fixed-point representation. Chapter~\ref{chapter:napel} introduces NAPEL, a fast high-level performance, and energy estimation framework. Chapter~\ref{chapter:leaper} presents LEAPER, our approach to quickly model different FPGA-based hardware platforms and applications. Chapter~\ref{chapter:sibyl} introduces Sibyl, the first RL-based \emph{data-driven} mechanism for data placement in a hybrid storage system. Chapter~\ref{chapter:conclusion} concludes this dissertation and provides future directions that are enabled by \ggthesis{its results} -- both in the domain of \emph{data-centric computing}  and \emph{data-driven optimization}  that can help us overcome present computing system challenges. In addition to the works presented in this thesis, Appendix~\ref{chapter:appendixB} highlights several other contributions of the author.
 







\chapter{Near-Memory Computing}
\label{chapter:background}
\chapternote{The content of this chapter was published as \emph{``Near-Memory Computing: Past, Present, and Future''} in MICPRO 2019.}




In the literature, data-centric computing has manifested with names such as \textit{processing-in memory} (PIM),  \textit{near-data processing} (NDP), \textit{near-memory processing} (NMP), or in the case of non-volatile memories as \textit{in-storage processing} (ISP). However, all these terms fall under the same umbrella of \textit{near-memory computing} (NMC), with the core principle of performing processing closer to the memory in contrast to the traditional \textit{compute-centric} approach. In this dissertation, we focus on NMC rather than \textit{in-situ} data-centric computing called \textit{computation-in-memory} that performs logical operations using memory itself  by exploiting physical properties of memory devices, such as phase-change memory. 

This chapter deals with analyzing and organizing the extensive body of literature on NMC architectures across various dimensions: starting from the memory level where this paradigm is applied to the granularity of an application that could be executed on these architectures. \fc{We provide representative architectures in each category of our NMC taxonomy.} 
The remainder of this chapter is structured as follows. Section~\ref{sec:nmc_review/background} provides background on near-memory computing. Section~\ref{sec:nmc_review/classificationmodel} outlines the evaluation and classification scheme that we use.  Section~\ref{sec:nmc_review/challenges}  highlights the present challenges with NMC-based systems, which include 
lack of evaluation tools, virtual memory, memory coherence, task scheduling, and data mapping.  Finally, Section~\ref{sec:nmc_review/concludes} concludes the chapter.

\section{Background and Related Work}
\label{sec:nmc_review/background}
The idea of processing close to the memory dates back to the 1960s~\cite{ALogicinMemoryComputer}. However, the first appearance of data-centric systems can be traced back to the early 1990s~\cite{5727436,Kogge1994,fbram1994,375174,592312,808425}. As an example, \textit{Vector IRAM} (VIRAM)~\cite{612252}, where the researchers develop a vector processor with an on-chip embedded DRAM. They use VIRAM to exploit data parallelism in multimedia applications. Although such works obtained promising results, these earlier systems did not penetrate the market, and their adoption remained limited. One of the main reasons was attributed to the technological limitations because the amount of memory we could integrate with the processor was limited due to the difference in logic and memory technology processes.

\begin{figure}[h]
\centering
\includegraphics[bb=87 146 666 399,width=0.7\linewidth,trim={3cm 5cm 10.2cm 4.8cm},clip]{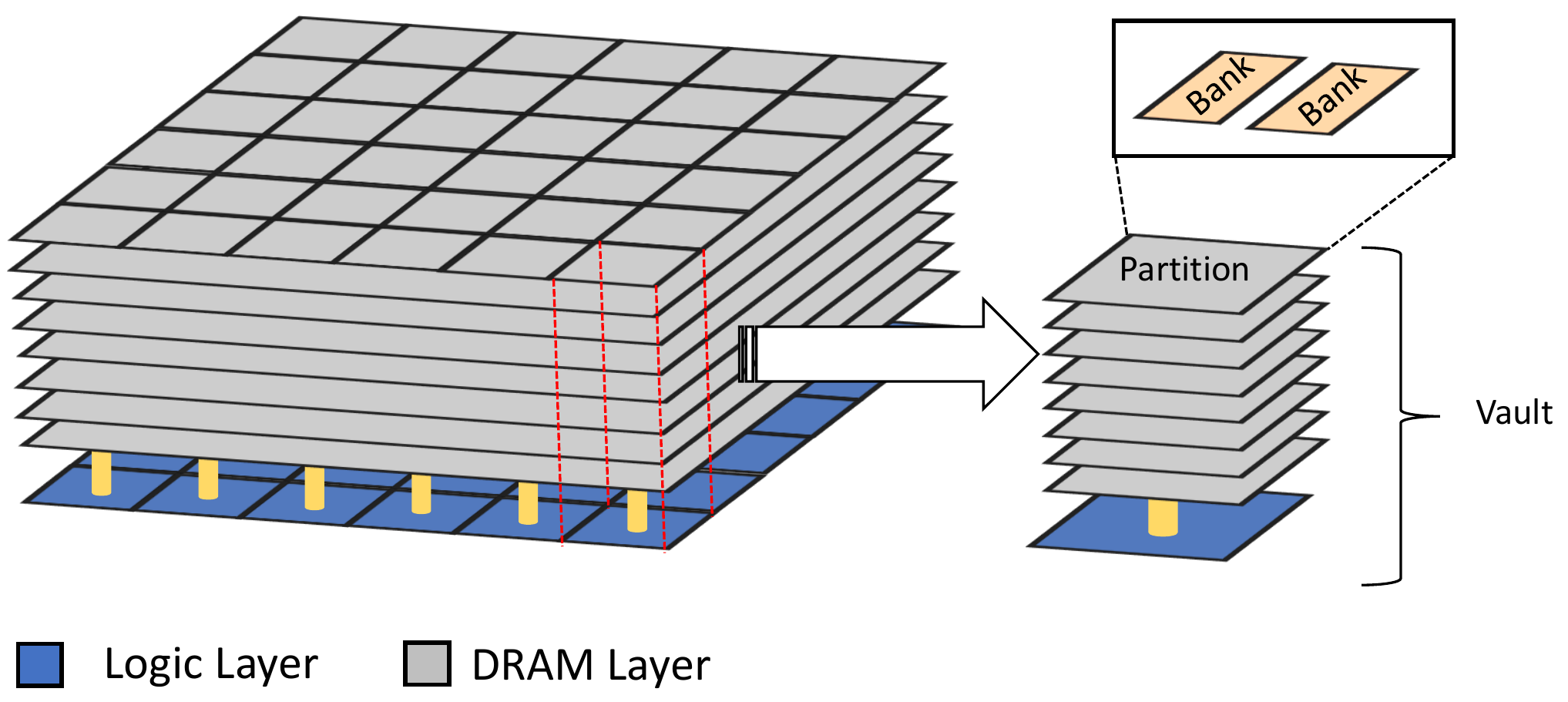}
\caption{Micron's Hybrid Memory Cube (\acrshort{hmc})~\cite{7477494} comprising of several DRAM layers stacked on top of a logic layer connected by through-silicon vias (TSVs). The memory organization is divided into vaults, with each vault consisting of multiple DRAM banks
\label{fig:nmc_review/HMCLayout}}
\end{figure}

Today, after almost two decades of dormancy, research in NMC architectures is regaining attention. We can largely attribute this resurgence to the following three reasons. First, technological advancements in the stacking technology -- 3D ( e.g., hybrid memory cube (HMC)~\cite{7477494} see Figure~\ref{fig:nmc_review/HMCLayout}) and 2.5D (e.g., high-bandwidth memory (HBM)~\cite{hbm}) stacking that blends logic and memory in the same package. Second, moving the computation closer to where the data reside allows for sidestepping the performance and energy bottlenecks due to data movement by circumventing memory-package pin-count limitations. Third, \ph{the increase in data volumes produced in various} application domains, such as weather prediction modeling, radio astronomy, and bioinformatics, calls for newer architectures designed to handle the overwhelming amount of data. As a result, in recent years, researchers have proposed various NMC designs and proved their potential in enhancing performance in many application domains~\cite{teserract, ahn2015pim,nair2015active,7446059,hsieh2016accelerating,tom,ke2020recnmp,gomezluna2021upmem,cali2020genasm,fernandez2020natsa,singh2020nero}. \fc{For CIM-based architectures, prior works demonstrate that CIM can be achieved using various memory technologies such as static random-access memory (SRAM)~\cite{simon2020blade,computecache,kang2014energy,neuralcache2018}, dynamic random-access memory (DRAM)~\cite{seshadri2017ambit,chang2016low,seshadri2013rowclone}, PCM~\cite{li2016pinatubo}, and ReRAM~\cite{levy2014logic,kvatinsky2014magic,shafiee2016isaac}.}

Loh~\etal~\cite{loh2013processing}, in their position paper, present an initial taxonomy for NMC. This taxonomy is based on the computing interface with software. Siegl~\etal~\cite{siegl2016data}, in an overview paper, gave a historical evolution of NMC. Similar \dd{to the approach of this thesis}, Mutlu~\etal\cite{mutlu2020modern},  provide a thorough overview of the mechanisms and challenges in the field of near-memory computing. Unlike \dd{the survey of this thesis}, the paper~\cite{mutlu2020modern} does not focus on providing systematization to the literature. Our review characterizes near-memory computing literature in various dimensions starting from the memory level, where we apply the paradigm of near-memory computing to the type of near-memory processing unit, memory integration, and type of workloads/applications.

\section{Classification and Evaluation}
\label{sec:nmc_review/classificationmodel}

\begin{figure}[h]
\centering
  \hbox{\hspace{7em} \includegraphics[width=0.7\linewidth]{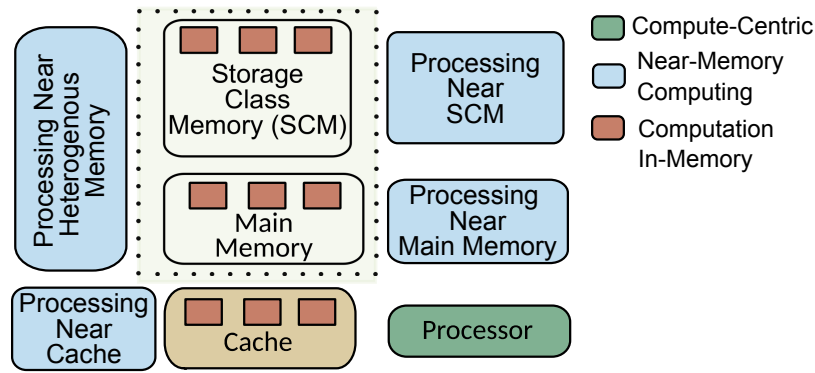}}
\caption{Processing options in the memory hierarchy highlighting three computation paradigms: (1) Compute-centric approach where data is moved through various levels of memory to the processor for computing; (2) Near-memory computing approach where the processing elements are placed closer to memory; and (3) Computation-in memory approach that uses inherent properties of memory to perform computation
\label{fig:nmc_review/taxonomy}}
\end{figure}

Figure~\ref{fig:nmc_review/taxonomy} shows a high-level view of our classification based on the level in the memory hierarchy. We further split our classification into the type of processing unit (programmable, fixed-function, or reconfigurable). Conceptually the approach  of near-memory computing can be applied to any level or type of memory to improve the overall system performance. 
\dd{Our taxonomy does not include magnetic disk-based systems because nowadays, it is only used as a long-term \emph{cold data} storage medium, i.e., for long-term and rarely accessed data~\cite{fuji_tape}. }
Nevertheless, there have been various research efforts towards providing processing capabilities in the disk. However, the industry did not adopt it widely due to the marginal performance improvement that could not justify the associated cost~\cite{Keeton:1998:CID:290593.290602,Riedel:1998:ASL:645924.671345}. Instead, we include emerging non-volatile memories termed  storage class memory (\acrshort{scm})~\cite{7151782}, which are trying to fill the latency gap between DRAM and disk. 

This section introduces the classification and evaluation metrics (see Table~\ref{tab:nmc_review/arcLegend}) used to analyze various architectures in Appendix~\ref{sec:nmc_review/procmainmem} and Appendix~\ref{sec:nmc_review/procstorage}. We summarize different architectures in Table~\ref{tab:nmc_review/arcLegend}. 

For each architecture, we evaluate and classify across five main categories:
\begin{itemize}
 \setlength{\topsep}{-3pt}
\setlength{\itemsep}{2pt}
\setlength{\parskip}{0pt}
\setlength{\parsep}{0pt}
\item{\textbf{Memory}} - The type of memory technology is one of the most fundamental questions on which the near-memory architecture depends.
\item{\textbf{Processing}} - The type of processing unit and the granularity of processing it performs plays a critical role in our analysis.
\item{\textbf{Tool}} - Any system's success depends heavily on the available tool support. The effectiveness of the tool infrastructure indicates the maturity of the architecture. 
\item{\textbf{Interoperability}} - \ggthesis{It is the integration of NMC processing units into the overall computer system architecture. }Interoperability deals with aspects such as virtual memory support, memory coherence,  efficient task scheduling, and data mapping. Interoperability is one of the key enablers for the adoption of any new system.
\item{\textbf{Application}} - NMC follows a data-centric principle and is usually specialized for a particular workload or a set of workloads. Therefore, in our evaluation, we include the domain of the application.  
\end{itemize}

\begin{table}[h]
\renewcommand{\arraystretch}{0.9}
\setlength{\tabcolsep}{10pt}
\caption{Classification metrics that we use to analyze some notable NMC architectures (see Table~\ref{tab:nmc_review/arcClassification}).}
\label{tab:nmc_review/arcLegend}
\centering
\footnotesize
\resizebox{0.85\columnwidth}{!}{
\begin{tabular}{l|l|c|l}
\toprule
\textbf{} & \textbf{Property} & \textbf{Abbreviation} & \textbf{Description} \\ \hline
\multirow{11}{*}{\textbf{Memory}} & \multirow{3}{*}{Hierarchy} & MM & Main memory \\
 &  & SCM & Storage class memory \\
 &  & HM & Heterogenous memory \\ \cline{2-4} 
 & \multirow{6}{*}{Type} & C3D & Commercial 3D memory \\
 &  & PCM & Phase change memory \\
 &  & DRAM & Dynamic random-access memory \\
 &  & \acrshort{ssd} & Solid-state drive \\
 \cline{2-4} 
 & \multirow{2}{*}{Integration} & US & Conventional unstacked \\
 &  & S & Stacked using 2.5D or 3D \\ \hline
\multirow{12}{*}{\textbf{Processing}} & \multirow{5}{*}{\begin{tabular}[c]{@{}l@{}}NMC/Host\\ Unit\end{tabular}} & CPU & Central processing unit \\
 &  & GPU & Graphics processing unit \\
 &  & FPGA & Field programmable gate array \\
 &  & \acrshort{cgra} & \begin{tabular}[c]{@{}l@{}}Coarse-grained reconfigurable \\ architecture\end{tabular} \\
 &  & ACC & Application-specific accelerator \\ \cline{2-4} 
 & \multirow{3}{*}{\begin{tabular}[c]{@{}l@{}}Implemen-\\ tation\end{tabular}} & P & Programmable unit \\
 &  & F & Fixed-function unit \\
 &  & R & Reconfigurable unit \\ \cline{2-4} 
 & \multirow{3}{*}{Granularity} & I & Instruction \\
 &  & K & Kernel \\
 &  & A & Application \\ \cline{2-4} 
 & Host Unit &  & Type of the host unit \\ \hline
\multirow{3}{*}{\textbf{Tool}} & \multirow{3}{*}{\begin{tabular}[c]{@{}l@{}}Evaluation \\ Technique\end{tabular}} & A & Analytic \\
 &  & S & Simulation \\
 &  & P & Prototype/Hardware \\ \hline
\multirow{3}{*}{\textbf{\begin{tabular}[c]{@{}l@{}}Interop-\\ erability\end{tabular}}} & \begin{tabular}[c]{@{}l@{}}Programming \\ Model\end{tabular} & - & \begin{tabular}[c]{@{}l@{}}Programming model for \\ NMC unit \end{tabular} \\
 & \begin{tabular}[c]{@{}l@{}}Memory Coherence\end{tabular} & Y/N & Mechanism for memory coherence \\
 & \begin{tabular}[c]{@{}l@{}}Virtual Memory\end{tabular} & Y/N & Virtual memory support \\ \hline
\textbf{\begin{tabular}[c]{@{}l@{}}Application \\ Domain\end{tabular}} & Workload & - & \begin{tabular}[c]{@{}l@{}}Target application domain \\ for the architecture\end{tabular} \\ \bottomrule
\end{tabular}
}

\end{table}

\subsection*{Discussion}

Based on the above classification, we highlight some of the notable architectures in the domain of near-memory computing. All solutions discussed in this section are summarized in Table~\ref{tab:nmc_review/arcClassification} and described in Appendix~\ref{chapter:appendixA}.  From Table~\ref{tab:nmc_review/arcClassification}, we make the following five observations. 
\begin{table*}[h]
\renewcommand{\arraystretch}{1.2}

\caption{Classification and evaluation of \fc{representative architectures in category of our NMC taxonomy}, refer to Table~\ref{tab:nmc_review/arcLegend} for the legend}
\label{tab:nmc_review/arcClassification}
\centering
\tabcolsep=0.22cm
\resizebox{\textwidth}{!}{
\begin{tabular}{@{}cccclccllclllcl@{}|}
\hline
\multicolumn{2}{|c}{\textbf{NMC Architecture}}& \multicolumn{3}{|c|}{\textbf{Memory}}& \multicolumn{4}{c|}{\textbf{Processing}} &\textbf{Tool}&\multicolumn{3}{|c|}{\textbf{Interoperability}}& \multicolumn{1}{|c|}{\textbf{App. Domain}} \\ \hline
\multicolumn{1}{|l}{\rotatebox[origin=c]{90}{\textbf{Architecture}}} & \multicolumn{1}{c|}{\rotatebox[origin=c]{90}{\textbf{Year}}}&
\multicolumn{1}{c|}{\rotatebox[origin=c]{90}{\textbf{Hierarchy}}}   & \multicolumn{1}{c|}{\rotatebox[origin=c]{90}{\textbf{Type}}} & \multicolumn{1}{l|}{\rotatebox[origin=c]{90}{\textbf{Integration}}} & \multicolumn{1}{c|}{\rotatebox[origin=c]{90}{\textbf{NMC Unit}}} & \multicolumn{1}{l|}{\rotatebox[origin=c]{90}{\textbf{Implementation}}} &
\multicolumn{1}{c|}{\rotatebox[origin=c]{90}{\textbf{Granularity}}} & \multicolumn{1}{l|}{\rotatebox[origin=c]{90}{\textbf{Host Unit}}} & \multicolumn{1}{c|}{\rotatebox[origin=c]{90}{\textbf{Evaluation}}} & 
\multicolumn{1}{c|}{\rotatebox[origin=c]{90}{\textbf{Programming Model} }} &
\multicolumn{1}{l|}{\rotatebox[origin=c]{90}{\textbf{Cache Coherence}}} & \multicolumn{1}{l|}{\rotatebox[origin=c]{90}{\textbf{Virtual Memory}}} &
\multicolumn{1}{c|}{\rotatebox[origin=c]{90}{\textbf{Workload}}}\\ 

\multicolumn{1}{|c}{}& 
\multicolumn{1}{c|}{}& 
\multicolumn{1}{c|}{}& 
\multicolumn{1}{c|}{}&  
\multicolumn{1}{l|}{}& 
\multicolumn{1}{c|}{}& 
\multicolumn{1}{l|}{}& 
\multicolumn{1}{l|}{}& 
\multicolumn{1}{l|}{}& 
\multicolumn{1}{l|}{}& 
\multicolumn{1}{l|}{}&
\multicolumn{1}{l|}{}&  
\multicolumn{1}{l|}{}&  
\multicolumn{1}{l|}{}\\ \hline 

\multicolumn{1}{|l}{XSD~\cite{cho2013xsd}}&
\multicolumn{1}{c|}{2013} &
\multicolumn{1}{c|}{SCM} &
\multicolumn{1}{c|}{SSD} & 
\multicolumn{1}{l|}{US}  & 
\multicolumn{1}{c|}{GPU}&
\multicolumn{1}{l|}{P} &
\multicolumn{1}{l|}{A} & 
\multicolumn{1}{l|}{CPU} &
\multicolumn{1}{c|}{S}&
\multicolumn{1}{l|}{MapReduce}&
\multicolumn{1}{l|}{-}& 
\multicolumn{1}{l|}{-} & 
\multicolumn{1}{l|}{MapReduce}\\ \hline

\multicolumn{1}{|l}{SmartSSD~\cite{kang2013enabling}}& 
\multicolumn{1}{c|}{2013} & 
\multicolumn{1}{c|}{SCM} & 
\multicolumn{1}{c|}{SSD} &  
\multicolumn{1}{l|}{US}  & 
\multicolumn{1}{c|}{CPU}& 
\multicolumn{1}{l|}{P} & 
\multicolumn{1}{l|}{A} & 
\multicolumn{1}{l|}{CPU} & 
\multicolumn{1}{c|}{P}& 
\multicolumn{1}{l|}{MapReduce}&
\multicolumn{1}{l|}{Y}&  
\multicolumn{1}{l|}{N} &  
\multicolumn{1}{l|}{Database}\\ \hline 

\multicolumn{1}{|l}{WILLOW~\cite{seshadri2014willow}}& 
\multicolumn{1}{c|}{2014} & 
\multicolumn{1}{c|}{SCM} & 
\multicolumn{1}{c|}{SSD} &  
\multicolumn{1}{l|}{US}  & 
\multicolumn{1}{c|}{CPU}& 
\multicolumn{1}{l|}{P} & 
\multicolumn{1}{l|}{K} & 
\multicolumn{1}{l|}{CPU} & 
\multicolumn{1}{c|}{P}& 
\multicolumn{1}{l|}{API}&
\multicolumn{1}{l|}{Y}&  
\multicolumn{1}{l|}{-}&  
\multicolumn{1}{l|}{Generic}\\ \hline 

\multicolumn{1}{|l}{NDC~\cite{6844483}}&
\multicolumn{1}{c|}{2014} &
\multicolumn{1}{c|}{MM} &
\multicolumn{1}{c|}{C3D} & 
\multicolumn{1}{l|}{S}  & 
\multicolumn{1}{c|}{CPU}&
\multicolumn{1}{l|}{P} &
\multicolumn{1}{l|}{K} & 
\multicolumn{1}{l|}{CPU} &
\multicolumn{1}{c|}{S}&
\multicolumn{1}{l|}{MapReduce}&
\multicolumn{1}{l|}{R} & 
\multicolumn{1}{l|}{N}& 
\multicolumn{1}{l|}{MapReduce}\\ \hline

\multicolumn{1}{|l}{TOP-PIM~\cite{zhang2014top}}& 
\multicolumn{1}{c|}{2014} & 
\multicolumn{1}{c|}{MM} & 
\multicolumn{1}{c|}{C3D} &  
\multicolumn{1}{l|}{S}  & 
\multicolumn{1}{c|}{GPU}& 
\multicolumn{1}{l|}{P} & 
\multicolumn{1}{l|}{K} & 
\multicolumn{1}{l|}{CPU} & 
\multicolumn{1}{c|}{S}& 
\multicolumn{1}{l|}{OpenCL}&
\multicolumn{1}{l|}{Y}&  
\multicolumn{1}{l|}{-} &  
\multicolumn{1}{l|}{Graph and HPC}\\ \hline 

\multicolumn{1}{|l}{AMC~\cite{nair2015active}}&
\multicolumn{1}{c|}{2015} &
\multicolumn{1}{c|}{MM} &
\multicolumn{1}{c|}{C3D} & 
\multicolumn{1}{l|}{S}  & 
\multicolumn{1}{c|}{CPU}&
\multicolumn{1}{l|}{P} &
\multicolumn{1}{l|}{K} & 
\multicolumn{1}{l|}{CPU} &
\multicolumn{1}{c|}{S}&
\multicolumn{1}{l|}{OpenMP}&
\multicolumn{1}{l|}{Y}& 
\multicolumn{1}{l|}{Y} & 
\multicolumn{1}{l|}{HPC}\\ \hline

\multicolumn{1}{|l}{JAFAR~\cite{xi2015beyond}}&
\multicolumn{1}{c|}{2015} &
\multicolumn{1}{c|}{MM} &
\multicolumn{1}{c|}{DRAM} & 
\multicolumn{1}{l|}{US}  & 
\multicolumn{1}{c|}{ACC}&
\multicolumn{1}{l|}{F} &
\multicolumn{1}{l|}{K} & 
\multicolumn{1}{l|}{CPU} &
\multicolumn{1}{c|}{S}&
\multicolumn{1}{l|}{API}&
\multicolumn{1}{l|}{-}& 
\multicolumn{1}{l|}{Y}& 
\multicolumn{1}{l|}{Database}\\ \hline

\multicolumn{1}{|l}{TESSERACT~\cite{teserract}}& 
\multicolumn{1}{c|}{2015} & 
\multicolumn{1}{c|}{MM} & 
\multicolumn{1}{c|}{C3D} &  
\multicolumn{1}{l|}{S}  & 
\multicolumn{1}{c|}{CPU}& 
\multicolumn{1}{l|}{F} & 
\multicolumn{1}{l|}{A} & 
\multicolumn{1}{l|}{CPU} & 
\multicolumn{1}{c|}{S}& 
\multicolumn{1}{l|}{API}&
\multicolumn{1}{l|}{Y}&  
\multicolumn{1}{l|}{N} &  
\multicolumn{1}{l|}{Graph processing}\\ \hline 

\multicolumn{1}{|l}{Gokhale~\cite{gokhale2015near}}&
\multicolumn{1}{c|}{2015} &
\multicolumn{1}{c|}{MM} &
\multicolumn{1}{c|}{C3D} & 
\multicolumn{1}{l|}{S}  & 
\multicolumn{1}{c|}{ACC}&
\multicolumn{1}{l|}{F} &
\multicolumn{1}{l|}{K} & 
\multicolumn{1}{l|}{CPU} &
\multicolumn{1}{c|}{S}&
\multicolumn{1}{l|}{API}&
\multicolumn{1}{l|}{Y}& 
\multicolumn{1}{l|}{Y} & 
\multicolumn{1}{l|}{Generic}\\ \hline

\multicolumn{1}{|l}{HRL~\cite{7446059}}&
\multicolumn{1}{c|}{2015} &
\multicolumn{1}{c|}{MM} &
\multicolumn{1}{c|}{C3D} & 
\multicolumn{1}{l|}{S}  & 
\multicolumn{1}{c|}{CGRA+FPGA}& 
\multicolumn{1}{l|}{R} & 
\multicolumn{1}{l|}{A} & 
\multicolumn{1}{l|}{CPU} &
\multicolumn{1}{c|}{S}&
\multicolumn{1}{l|}{MapReduce}& 
\multicolumn{1}{l|}{Y}& 
\multicolumn{1}{l|}{N} & 
\multicolumn{1}{l|}{Data analytics}\\ \hline

\multicolumn{1}{|l}{ProPRAM~\cite{wang2015propram}}& 
\multicolumn{1}{c|}{2015} & 
\multicolumn{1}{c|}{SCM} & 
\multicolumn{1}{c|}{PCM} &  
\multicolumn{1}{l|}{US}  & 
\multicolumn{1}{c|}{CPU}& 
\multicolumn{1}{l|}{P} & 
\multicolumn{1}{l|}{I} & 
\multicolumn{1}{l|}{-} & 
\multicolumn{1}{c|}{S}& 
\multicolumn{1}{l|}{ISA Extension}&
\multicolumn{1}{l|}{-}&  
\multicolumn{1}{l|}{-} &  
\multicolumn{1}{l|}{Data analytics}\\ \hline 

\multicolumn{1}{|l}{BlueDBM~\cite{jun2015bluedbm}}& 
\multicolumn{1}{c|}{2015} & 
\multicolumn{1}{c|}{SCM} & 
\multicolumn{1}{c|}{SSD} &  
\multicolumn{1}{l|}{US}  & 
\multicolumn{1}{c|}{FPGA}& 
\multicolumn{1}{l|}{R} & 
\multicolumn{1}{l|}{K} & 
\multicolumn{1}{l|}{-} & 
\multicolumn{1}{c|}{P}& 
\multicolumn{1}{l|}{API}&
\multicolumn{1}{l|}{-}&  
\multicolumn{1}{l|}{-} &  
\multicolumn{1}{l|}{Data analytics} \\ \hline 


\multicolumn{1}{|l}{NDA~\cite{7056040}}& 
\multicolumn{1}{c|}{2015} & 
\multicolumn{1}{c|}{MM} & 
\multicolumn{1}{c|}{DRAM} &  
\multicolumn{1}{l|}{S}  & 
\multicolumn{1}{c|}{CGRA}& 
\multicolumn{1}{l|}{R} & 
\multicolumn{1}{l|}{K} & 
\multicolumn{1}{l|}{CPU} & 
\multicolumn{1}{c|}{S}& 
\multicolumn{1}{l|}{OpenCL}&
\multicolumn{1}{l|}{Y}&  
\multicolumn{1}{l|}{Y} &  
\multicolumn{1}{l|}{MapReduce}\\ \hline 

\multicolumn{1}{|l}{PIM-enabled~\cite{ahn2015pim}}& 
\multicolumn{1}{c|}{2015} & 
\multicolumn{1}{c|}{MM} & 
\multicolumn{1}{c|}{C3D} &  
\multicolumn{1}{l|}{S}  & 
\multicolumn{1}{c|}{ACC}& 
\multicolumn{1}{l|}{F} & 
\multicolumn{1}{l|}{I} & 
\multicolumn{1}{l|}{CPU} & 
\multicolumn{1}{c|}{S}& 
\multicolumn{1}{l|}{ISA extension}&
\multicolumn{1}{l|}{Y}&  
\multicolumn{1}{l|}{Y} &  
\multicolumn{1}{l|}{Generic}\\ \hline 


\multicolumn{1}{|l}{IMPICA~\cite{hsieh2016accelerating}}& 
\multicolumn{1}{c|}{2016} & 
\multicolumn{1}{c|}{MM} & 
\multicolumn{1}{c|}{C3D} &  
\multicolumn{1}{l|}{S}  & 
\multicolumn{1}{c|}{ACC}& 
\multicolumn{1}{l|}{F} & 
\multicolumn{1}{l|}{K} & 
\multicolumn{1}{l|}{CPU} & 
\multicolumn{1}{c|}{S}& 
\multicolumn{1}{l|}{API}&
\multicolumn{1}{l|}{Y}&  
\multicolumn{1}{l|}{Y} &  
\multicolumn{1}{l|}{Pointer chasing}\\ \hline 

\multicolumn{1}{|l}{TOM~\cite{tom}}&
\multicolumn{1}{c|}{2016} &
\multicolumn{1}{c|}{MM} &
\multicolumn{1}{c|}{C3D} & 
\multicolumn{1}{l|}{S}  & 
\multicolumn{1}{c|}{GPU}&
\multicolumn{1}{l|}{P} &
\multicolumn{1}{l|}{K} & 
\multicolumn{1}{l|}{GPU} &
\multicolumn{1}{c|}{S}&
\multicolumn{1}{l|}{CUDA}&
\multicolumn{1}{l|}{Y}& 
\multicolumn{1}{l|}{Y} & 
\multicolumn{1}{l|}{Generic}\\ \hline


\multicolumn{1}{|l}{BISCUIT~\cite{gu2016biscuit}}& 
\multicolumn{1}{c|}{2016} & 
\multicolumn{1}{c|}{SCM} & 
\multicolumn{1}{c|}{SSD} &  
\multicolumn{1}{l|}{US}  & 
\multicolumn{1}{c|}{ACC}& 
\multicolumn{1}{l|}{F} & 
\multicolumn{1}{l|}{K} & 
\multicolumn{1}{l|}{CPU} & 
\multicolumn{1}{c|}{P}& 
\multicolumn{1}{l|}{API}&
\multicolumn{1}{l|}{-}&  
\multicolumn{1}{l|}{-} &  
\multicolumn{1}{l|}{Database}\\ \hline 

\multicolumn{1}{|l}{Pattnaik~\cite{7756764}}& 
\multicolumn{1}{c|}{2016} & 
\multicolumn{1}{c|}{MM} & 
\multicolumn{1}{c|}{C3D} &  
\multicolumn{1}{l|}{S}  & 
\multicolumn{1}{c|}{GPU}& 
\multicolumn{1}{l|}{P} & 
\multicolumn{1}{l|}{K} & 
\multicolumn{1}{l|}{GPU} & 
\multicolumn{1}{c|}{S}& 
\multicolumn{1}{l|}{CUDA}&
\multicolumn{1}{l|}{Y}&  
\multicolumn{1}{l|}{-} &  
\multicolumn{1}{l|}{Generic}\\ \hline 

\multicolumn{1}{|l}{CARIBOU~\cite{istvan2017caribou}}& 
\multicolumn{1}{c|}{2017} & 
\multicolumn{1}{c|}{SCM} & 
\multicolumn{1}{c|}{DRAM} &  
\multicolumn{1}{l|}{US}  & 
\multicolumn{1}{c|}{FPGA}& 
\multicolumn{1}{l|}{R} & 
\multicolumn{1}{l|}{K} & 
\multicolumn{1}{l|}{CPU} & 
\multicolumn{1}{c|}{P}& 
\multicolumn{1}{l|}{API}&
\multicolumn{1}{l|}{-}&  
\multicolumn{1}{l|}{-} &  
\multicolumn{1}{l|}{Database}\\ \hline 

\multicolumn{1}{|l}{Vermij~\cite{vermij2017sorting}}&
\multicolumn{1}{c|}{2017} &
\multicolumn{1}{c|}{MM} &
\multicolumn{1}{c|}{C3D} & 
\multicolumn{1}{l|}{S}  & 
\multicolumn{1}{c|}{ACC}&
\multicolumn{1}{l|}{F} &
\multicolumn{1}{l|}{A} & 
\multicolumn{1}{l|}{CPU} &
\multicolumn{1}{c|}{S}&
\multicolumn{1}{l|}{API}&
\multicolumn{1}{l|}{Y}& 
\multicolumn{1}{l|}{Y} & 
\multicolumn{1}{l|}{Sorting}\\ \hline

\multicolumn{1}{|l}{SUMMARIZER~\cite{koo2017summarizer}}& 
\multicolumn{1}{c|}{2017} & 
\multicolumn{1}{c|}{SCM} & 
\multicolumn{1}{c|}{SSD} &  
\multicolumn{1}{l|}{US}  & 
\multicolumn{1}{c|}{CPU}& 
\multicolumn{1}{l|}{P} & 
\multicolumn{1}{l|}{K} & 
\multicolumn{1}{l|}{CPU} & 
\multicolumn{1}{c|}{P}& 
\multicolumn{1}{l|}{API}&
\multicolumn{1}{l|}{-}&  
\multicolumn{1}{l|}{-} &  
\multicolumn{1}{l|}{Database}\\ \hline 

\multicolumn{1}{|l}{MONDRIAN~\cite{de2017mondrian}}& 
\multicolumn{1}{c|}{2017} & 
\multicolumn{1}{c|}{MM} & 
\multicolumn{1}{c|}{C3D} &  
\multicolumn{1}{l|}{S}  & 
\multicolumn{1}{c|}{CPU}& 
\multicolumn{1}{l|}{P} & 
\multicolumn{1}{l|}{K} & 
\multicolumn{1}{l|}{CPU} & 
\multicolumn{1}{c|}{A+S}& 
\multicolumn{1}{l|}{API}&
\multicolumn{1}{l|}{-}&  
\multicolumn{1}{l|}{Y} &  
\multicolumn{1}{l|}{Data analytics}\\ \hline 

\multicolumn{1}{|l}{GraphPIM~\cite{nai2017graphpim}}&
\multicolumn{1}{c|}{2017} &
\multicolumn{1}{c|}{MM} &
\multicolumn{1}{c|}{DRAM} & 
\multicolumn{1}{l|}{US}  & 
\multicolumn{1}{c|}{ACC}&
\multicolumn{1}{l|}{F} &
\multicolumn{1}{l|}{I} & 
\multicolumn{1}{l|}{CPU} &
\multicolumn{1}{c|}{S}&
\multicolumn{1}{l|}{API}&
\multicolumn{1}{l|}{Y}& 
\multicolumn{1}{l|}{N} & 
\multicolumn{1}{l|}{Graph}\\ \hline

\multicolumn{1}{|l}{MCN~\cite{alian2018application}}&
\multicolumn{1}{c|}{2018} &
\multicolumn{1}{c|}{MM} &
\multicolumn{1}{c|}{DRAM} & 
\multicolumn{1}{l|}{US}  & 
\multicolumn{1}{c|}{CPU}&
\multicolumn{1}{l|}{P} &
\multicolumn{1}{l|}{K} & 
\multicolumn{1}{l|}{CPU} &
\multicolumn{1}{c|}{P}&
\multicolumn{1}{l|}{TCP/IP}&
\multicolumn{1}{l|}{Y}& 
\multicolumn{1}{l|}{Y} & 
\multicolumn{1}{l|}{Generic}\\ \hline


\multicolumn{1}{|l}{DNN-PIM~\cite{liu2018processing}}&
\multicolumn{1}{c|}{2018} &
\multicolumn{1}{c|}{MM} & 
\multicolumn{1}{c|}{C3D} & 
\multicolumn{1}{l|}{S}  & 
\multicolumn{1}{c|}{CPU + ACC}& 
\multicolumn{1}{l|}{P+F} & 
\multicolumn{1}{l|}{K} & 
\multicolumn{1}{l|}{CPU} & 
\multicolumn{1}{c|}{P+S}& 
\multicolumn{1}{l|}{OpenCL}& 
\multicolumn{1}{l|}{Y}&  
\multicolumn{1}{l|}{N} & 
\multicolumn{1}{l|}{DNN training}\\ \hline 

\multicolumn{1}{|l}{Boroumand~\cite{googleWorkloads}}& 
\multicolumn{1}{c|}{2018} & 
\multicolumn{1}{c|}{MM} & 
\multicolumn{1}{c|}{C3D} &  
\multicolumn{1}{l|}{S}  & 
\multicolumn{1}{c|}{CPU+ACC}& 
\multicolumn{1}{l|}{P+F} & 
\multicolumn{1}{l|}{K} & 
\multicolumn{1}{l|}{CPU} & 
\multicolumn{1}{c|}{S}& 
\multicolumn{1}{l|}{-}& 
\multicolumn{1}{l|}{Y}&  
\multicolumn{1}{l|}{-}&  
\multicolumn{1}{l|}{Google workloads}\\ \hline

\multicolumn{1}{|l}{GRIM-Filter~\cite{kim2018grim}}& 
\multicolumn{1}{c|}{2018} & 
\multicolumn{1}{c|}{MM} & 
\multicolumn{1}{c|}{C3D} &  
\multicolumn{1}{l|}{S}  & 
\multicolumn{1}{c|}{ACC}& 
\multicolumn{1}{l|}{F} & 
\multicolumn{1}{l|}{K} & 
\multicolumn{1}{l|}{CPU} & 
\multicolumn{1}{c|}{S}& 
\multicolumn{1}{l|}{-}& 
\multicolumn{1}{l|}{Y}&  
\multicolumn{1}{l|}{-}&  
\multicolumn{1}{l|}{Read mapping}\\ \hline

\multicolumn{1}{|l}{CompStor~\cite{torabzadehkashi2018compstor}}& 
\multicolumn{1}{c|}{2018} & 
\multicolumn{1}{c|}{SCM} & 
\multicolumn{1}{c|}{SSD} &  
\multicolumn{1}{l|}{US}  & 
\multicolumn{1}{c|}{CPU}& 
\multicolumn{1}{l|}{P} & 
\multicolumn{1}{l|}{A} & 
\multicolumn{1}{l|}{CPU} & 
\multicolumn{1}{c|}{P}& 
\multicolumn{1}{l|}{API}&
\multicolumn{1}{l|}{-}&  
\multicolumn{1}{l|}{Y} &  
\multicolumn{1}{l|}{Text search}\\ \hline

\multicolumn{1}{|l}{RecNMP~\cite{ke2020recnmp}}& 
\multicolumn{1}{c|}{2020} & 
\multicolumn{1}{c|}{MM} & 
\multicolumn{1}{c|}{DRAM} &  
\multicolumn{1}{l|}{US}  & 
\multicolumn{1}{c|}{ACC}& 
\multicolumn{1}{l|}{F} & 
\multicolumn{1}{l|}{K} & 
\multicolumn{1}{l|}{CPU} & 
\multicolumn{1}{c|}{S}& 
\multicolumn{1}{l|}{API}&
\multicolumn{1}{l|}{Y}&  
\multicolumn{1}{l|}{Y} &  
\multicolumn{1}{l|}{Recommendation system}\\ \hline

\multicolumn{1}{|l}{GenASM~\cite{cali2020genasm}}& 
\multicolumn{1}{c|}{2020} & 
\multicolumn{1}{c|}{MM} & 
\multicolumn{1}{c|}{C3D} &  
\multicolumn{1}{l|}{S}  & 
\multicolumn{1}{c|}{ACC}& 
\multicolumn{1}{l|}{F} & 
\multicolumn{1}{l|}{K} & 
\multicolumn{1}{l|}{CPU} & 
\multicolumn{1}{c|}{S}& 
\multicolumn{1}{l|}{API}&
\multicolumn{1}{l|}{Y}&  
\multicolumn{1}{l|}{-} &  
\multicolumn{1}{l|}{String matching}\\ \hline

\multicolumn{1}{|l}{NATSA~\cite{fernandez2020natsa}}& 
\multicolumn{1}{c|}{2020} & 
\multicolumn{1}{c|}{MM} & 
\multicolumn{1}{c|}{C3D} &  
\multicolumn{1}{l|}{S}  & 
\multicolumn{1}{c|}{ACC}& 
\multicolumn{1}{l|}{F} & 
\multicolumn{1}{l|}{K} & 
\multicolumn{1}{l|}{-} & 
\multicolumn{1}{c|}{S}& 
\multicolumn{1}{l|}{API}&
\multicolumn{1}{l|}{-}&  
\multicolumn{1}{l|}{-} &  
\multicolumn{1}{l|}{Time series analysis}\\ \hline

\multicolumn{1}{|l}{{NERO}~\cite{singh2020nero}}& 
\multicolumn{1}{c|}{{2020}} & 
\multicolumn{1}{c|}{{MM}} & 
\multicolumn{1}{c|}{{C3D}} &  
\multicolumn{1}{l|}{{S}}  & 
\multicolumn{1}{c|}{{ACC}}& 
\multicolumn{1}{l|}{{R}} & 
\multicolumn{1}{l|}{{K}} & 
\multicolumn{1}{l|}{{CPU}} & 
\multicolumn{1}{c|}{{P}}& 
\multicolumn{1}{l|}{{API}}&
\multicolumn{1}{l|}{{Y}}&  
\multicolumn{1}{l|}{{Y}} &  
\multicolumn{1}{l|}{{Weather prediction modeling}}\\ \hline

\multicolumn{1}{|l}{{FIMDRAM}~\cite{fimdram2021ISSCC}}& 
\multicolumn{1}{c|}{{2021}} & 
\multicolumn{1}{c|}{{MM}} & 
\multicolumn{1}{c|}{{C3D}} &  
\multicolumn{1}{l|}{{S}}  & 
\multicolumn{1}{c|}{{ACC}}& 
\multicolumn{1}{l|}{{P}} & 
\multicolumn{1}{l|}{{I}} & 
\multicolumn{1}{l|}{{-}} & 
\multicolumn{1}{c|}{{P}}& 
\multicolumn{1}{l|}{{API}}&
\multicolumn{1}{l|}{{Y}}&  
\multicolumn{1}{l|}{{Y}} &  
\multicolumn{1}{l|}{{Machine learning}}\\ \bottomrule
\end{tabular}
}
\end{table*}


First, the efficacy of architectural proposals has mostly been tested using simulators. A few works emulate~\cite{gu2016biscuit, awan2017performance, koo2017summarizer, istvan2017caribou,kang2013enabling,seshadri2014willow} their proposal on an FPGA. Recently, HBM~\cite{hbm} has been adopted by GPU and FPGA vendors. In the future, we expect more evaluation studies on these  HBM-equipped platforms and upcoming platforms with NMC capabilities, such as  FIMDRAM~\cite{fimdram2021ISCA} and UPMEM~\cite{gomezluna2021upmem}. 

Second, the majority of research papers published over the years have proposed homogeneous processing units near the main memory. However, the logic considered varies in its compute capabilities, e.g., simple in-order cores~\cite{torabzadehkashi2018compstor,lee2018application,alian2018application,de2017mondrian,koo2017summarizer,7927081,teserract,nair2015active,6844483}, graphics processing units~\cite{zhang2014top,7756764,tom,cho2013xsd}, field-programmable gate arrays~\cite{7446059, jun2015bluedbm,istvan2017caribou}, and application-specific accelerators~\cite{xi2015beyond,gokhale2015near,ahn2015pim,hsieh2016accelerating,gu2016biscuit,vermij2017sorting,nai2017graphpim,liu2018processing,fernandez2020natsa,ke2020recnmp}. 

Third, the majority of the NMC proposals are targeted towards data-intensive applications e.g., graph processing~\cite{zhang2014top,teserract,nai2017graphpim}, MapReduce~\cite{cho2013xsd,6844483,7056040}, machine learning~\cite{liu2018processing,lee2018application}, and database~\cite{kang2013enabling,xi2015beyond,gu2016biscuit,istvan2017caribou,7927081,koo2017summarizer}.  

Fourth, most of the architectures propose adding compute capabilities in the logic layer of HMC-based memory~\cite{ke2020recnmp,cali2020genasm,fernandez2020natsa,googleWorkloads,liu2018processing,lee2018application,de2017mondrian,vermij2017sorting,7756764,tom,hsieh2016accelerating,ahn2015pim,7446059,gokhale2015near,teserract,nair2015active,zhang2014top,6844483,fimdram2021ISCA}. However, memory vendors such as Micron have announced to pursue HBM instead of focusing on HMC~\cite{noHMC}. Fifth, despite the promises made by existing proposals on NMC, the support for virtual memory, memory coherence, and compiler support is fairly limited. In most of the works~\cite{torabzadehkashi2018compstor, lee2018application, nai2017graphpim, de2017mondrian, koo2017summarizer,vermij2017sorting,istvan2017caribou,gu2016biscuit}, the programmers are expected to re-write their code using specialized \acrshort{api}s in order to reap the benefits of NMC. 

Based on the above observations, we make the following three conclusions. 

\begin{enumerate}[topsep=0pt,itemsep=0ex,partopsep=1ex,parsep=1ex]
    \item HBM-equipped specialized hardware has the potential to reduce the memory bandwidth bottleneck, but a study of their advantages for a real-world data-intensive application is still missing.  

\item The idea of populating homogeneous processing units near-memory to accelerate a specific class of workloads is limited in the sense that NMC enabled servers that would be deployed in the data centers are expected to host a wide variety of workloads. Hence, these systems would need heterogeneous processing units near the memory~\cite{googleWorkloads,liu2018processing,7446059} to support the complex mix of data center workloads. 

\item For broader adoption of the NMC by the application programmers, we would require methods that enable transparent offloading to the NMC units. Transparent offloading requires the compiler or the run-time system to identify NMC-suitable code regions based on some application characteristics, such as the number of last-level cache misses~\cite{Hadidi:2017:CCT:3154814.3155287, awan2017identifying,awan2017performance, gokhale2015near, ahn2015pim} and bandwidth utilization~\cite{zhang2014top, liu2018processing, hsieh2016accelerating}. Unfortunately, integrating a profiler (such as Linux perf~\cite{perf} or Intel Pin~\cite{reddi2004pin}) in a compiler or run-time system is still a challenging task~\cite{Hadidi:2017:CCT:3154814.3155287} due to its dynamic nature. Therefore, current solutions rely on commercial profiling tools, such as Intel VTune~\cite{Vtune}, to detect the offloading kernels~\cite{awan2017performance, awan2017identifying, hsieh2016accelerating}.
\end{enumerate}

\section{Challenges of Near-Memory Computing}
\label{sec:nmc_review/challenges}
In this section, we highlight five critical challenges in the domain of NMC. The challenges include evaluation tools,  virtual memory support, memory coherency, task scheduling, and data mapping. We need to address these challenges before NMC can be established as a de facto solution for modern data-intensive workloads.  
\subsection{Performance Evaluation Tools and Benchmarks}
\label{subsec:nmc_review/performanceevaluation}
As mentioned in Section~\ref{sec:intro/problem}, architects often use various evaluation techniques to navigate the design space of a new architecture. Based on the level of detail required, architects make use of analytic models or more detailed simulation-based techniques. 
 
\noindent\textbf{(1) Analytic modeling} abstracts low-level system details and provide quick performance estimates at the cost of accuracy. In the early design stage, system architects are faced with large design choices that range from semiconductor physics and circuit level to micro-architectural properties and cooling concerns~\cite{jongerius2017analytic}. Thus, during the first stage of design-space exploration, analytic models can provide quick estimates.


\noindent \textbf{(2) Simulation-based modeling} allows us to achieve more accurate performance numbers by precisely modeling various micro-architectural mechanisms. This approach, however, can be quite slow compared to analytic techniques. There have been various academic efforts~\cite{7336224,7529923,7544479,ramulator-pim-repo} to build open-source NMC simulators. However, there is a large room for improvement for developing a cycle-accurate simulator that can allow us to explore a wide range of near-memory compute configurations. In Table~\ref{tab:nmc_review/simulator}, we mention some of the academic efforts to create NMC simulation infrastructure.


\begin{table}[ht]
\renewcommand{\arraystretch}{1.2}
\caption{Academic NMC simulators}
\label{tab:nmc_review/simulator}
\centering
{\footnotesize
\begin{tabular}{lcHc}
\toprule
\textbf{Simulator} & \textbf{Year} & \textbf{Category} & \textbf{NMC capabilities} \\ \midrule
Sinuca~\cite{7336224} & 2015 & Cycle-Accurate & Yes \\
HMC-SIM~\cite{7529923} & 2016 & Cycle-Accurate & Limited \\
CasHMC~\cite{7544479} & 2016 & Cycle-Accurate & No \\
SMC~\cite{Azarkhish:2016:DEP:2963802.2963805} & 2016 & Cycle-Accurate & Yes \\
CLAPPS~\cite{junior2017generic} & 2017 & Cycle-Accurate & Yes \\ 
Ramulator-PIM~\cite{ramulator-pim-repo}      & 2019    & Cycle-Accurate & Yes  \\
\bottomrule
\end{tabular}
}
\end{table}

As the field of NMC does not have very sophisticated tools and techniques, researchers often \dd{spend} a significant amount of time building the appropriate evaluation environment~\cite{junior2017generic,NAPEL}. Additionally, there is a critical need for near-memory specific benchmarks of workloads that could benefit from NMC~\cite{oliveira2021pimbench}. Such a benchmark suite can allow researchers to evaluate different architectural proposals and faithfully reproduce results.

\subsection{Virtual Memory Support}
\label{subsec:nmc_review/virtualmemndp}
To access data inside the main memory, the CPU performs address translation from a data's virtual address to the actual physical address in the main memory. The address translation can be achieved by using the following two mechanisms: \textit{segmentation} or \textit{paging}. Segmentation~\cite{7459501,7459537} consists of a simplified approach where part of the linear virtual address space is mapped to physical memory using a direct segment. However, segmentation requires frequent swapping of segments between the main memory and the storage, leading to fragmentation. Therefore, \dd{the use of a paging mechanism is gaining wider adoption.}

Paging is a memory management mechanism that entails dividing virtual address space into blocks of addresses referred to as pages. A page table stores mapping between virtual to physical address and cache recently used mapping into a translation lookaside buffer (\acrshort{tlb}). A miss in the TLB would lead to a long-latency table walk, which can degrade the application performance. \fc{Several studies have been proposed to improve the efficiency of address translation, such as by speeding up address translation~\cite{barr2011spectlb,papadopoulou2015prediction}, increasing the TLB reach~\cite{ausavarungnirun2017mosaic,cox2017efficient}, and introducing caches to store page table address~\cite{barr2010translation,bhattacharjee2013large,bhargava2008accelerating}. }

In an NMC-based system, if an NMC accelerator requires many page table walks for the host CPU, it would substantially reduce the overall performance. Therefore, we need an effective address translation mechanism for NMC architectures. As an example, Hsieh~\etal~\cite{hsieh2016accelerating} design an NMC-side page table for their NMC accelerator, which avoids the use of CPU-side address translation. Past works adopt either a software-based~\cite{gao2015practical,Sura:2015:DAO:2742854.2742863,tom,Wei05anear-memory} or a hardware-based~\cite{Azarkhish:2016:DEP:2963802.2963805} approach to map between virtual and physical addresses.

\subsection{Memory Coherency}
\label{subsec:nmc_review/cachecoherencendp}
Coherency is one of the most critical challenges in the adoption of NMC. An NMC processing unit could modify the data, which the host CPU might require. Therefore, we need to maintain a coherence protocol between the shared memory. A fine-grain coherence mechanism might lead to a large number of coherence messages between the NMC cores and the host cores.  Therefore, the employed coherence mechanism can drastically affect the performance and the programming model. In NMC, researchers try to overcome this issue by following the \dd{two approaches listed below}.

\noindent\textbf{(1) Restricted memory region-based} techniques such as the one used by Farmahini \etal~\cite{7056040} divide the memory into two parts: one for the host processor and another for the accelerator, which is uncacheable. Ahn~\etal~\cite{teserract} use a similar approach for graph processing algorithms. 
Another strategy proposed by Ahn~\etal\cite{ahn2015pim} provides a simple hardware-based solution in which the NMC operations are restricted to only one last-level cache block, due to which they can monitor the cache block and request for invalidation or write-back if required.

\noindent\textbf{(2) Non-restricted memory region-based} techniques allow NMC units to access the entire memory space.  Pattnaik~\etal\cite{7756764} propose maintaining coherence between the host GPU and near-memory compute units by flushing the L2 cache in the host GPU after kernel execution. However, this approach could evict potentially useful data from the cache. Another way is to implement a look-up table-based mechanism, as Hsieh~\etal~\cite{tom}. The NMC units record the cache line address that the offloaded block has updated, and once the offloaded block is processed, the NMC units send this address back to the host system. Subsequently, the host system gets the latest data from memory by invalidating the reported cache lines. 
More recently, Boroumand~\etal\cite{boroumand2019conda} overcome the coherence issue with a specialized coherence protocol that batches and compresses multiple coherence requests from NMC units. As a result of this, the authors can achieve a near-ideal coherence mechanism.

\subsection{Task Scheduling}
\label{subsec:nmc_review/programmingmodel}

A critical challenge in adopting NMC is to support a heterogeneous processing environment comprising a host system and NMC processing units. It is not trivial to determine which part of an application should run on the NMC processing units. Works such as~\cite{Hadidi:2017:CCT:3154814.3155287, tom} leave this effort on the compiler, while others~\cite{ghose2018enabling, xi2015beyond, ahn2015pim} assume the programmer would manage the scheduling of tasks. Another approach~\cite{ahn2015pim, lee2018application, nai2017graphpim} uses some special set of NMC instructions, which invokes NMC processing units. This approach, however, calls for a sophisticated mechanism as it affects most of the software stack from the application down to the instruction set architecture. 

To this end, a run-time system capable of dynamically profiling applications to identify the potential offloads candidates for NMC processing units~\cite{liu2018processing, tom} can significantly help with task scheduling. Therefore, there is still a lot of research required in coming up with an efficient approach to ease the programming burden.

\subsection{Data Mapping}
\label{subsec:nmc_review/datamappNDPaPIM}
\fc{The problem of data mapping and data layout have been analyzed in various contexts to improve the spatial locality of a workload~\cite{wolf1991data,mckinley1996improving,chandra1994scheduling,kelm2010waypoint}. }The absence of an adequate data mapping mechanism can severely hamper the benefits of processing close to memory. A data mapping scheme should map data in such a way that the data required by the NMC processing units is readily available in the vicinity (data and code co-location). Hence, it is crucial to look into effective data mapping schemes. Hsieh~\etal~\cite{tom} propose a hardware/software co-design method to predict which pages of the memory would be used by the offloaded code segment, following which they place those pages in the memory stack closest to the offloaded code segment. 

Yitbarek~\etal~\cite{7459537} propose a data mapping scheme to place contiguous addresses in the same vault of an HMC-based memory allowing accelerators to access data directly from their local vault. Xiao~\etal~\cite{xiao2018prometheus} propose to model an application as a two-layer graph through the \acrshort{llvm}-intermediate representation (LLVM-IR), to distinguish between memory and computation operations. On building one such graph, their framework detects groups of vertices, called \textit{community}, that have a higher probability of connection with each other. Each community is mapped to a different vault of an HMC-based memory. Thus, \dd{this technique allows multiple} NMC units to perform computation in parallel on multiple data elements.

\section{Conclusion}
\label{sec:nmc_review/concludes}
Data-centric computing aims to reduce the data movement overhead by implementing processing capabilities close to where the data resides. With ``close'' being a relative term, there is a wide range of possibilities to bring computation closer to the data, resulting in various architectures being investigated today. Data-centric computing is attributed as one of the few real solutions to address the current scaling issues in HPC systems to realize exascale computers needed for modern and future data-intensive workloads.  There are two different approaches to enable data-centric computing. First, near-memory computing (NMC), which adds processing capabilities closer to the existing memory architectures. Second, computation-in memory (CIM), which  exploits the memory architecture and intrinsic properties of emerging technologies to perform operations using memory itself. \dd{This thesis focuses on} NMC-based architectures and techniques to overcome the data movement bottleneck.

This chapter analyzes NMC-based architectures across various dimensions and highlights that NMC is still in its infancy. We need to address multiple architectural challenges before NMC can be established as an essential component of HPC systems to accelerate data-intensive workloads. Besides designing and evaluating NMC processing capabilities for data-intensive workloads,  we stress the demand for sophisticated tools and techniques to enable the design space exploration for these novel architectures. Further, we need to solve various challenges related to the overall system integration. 
\ou{To overcome these challenges, we should consider data as a paramount resource and
provide various mechanisms to handle and leverage the vast amount of data. In this dissertation, we tackle the above challenges in three different ways.
First, we demonstrate NMC processing capabilities for a real-world data-intensive weather prediction application. Second, we provide data-driven machine learning-based solutions that allow quick and accurate performance estimation during the design-time. Third, we leverage the vast amount of available data to drive run-time system-level decisions.  }

\chapter[\texorpdfstring{NERO: A Near-High Bandwidth Memory Stencil Accelerator for Weather Prediction Modeling}{NERO: A Near-High Bandwidth Memory Stencil Accelerator for Weather Prediction Modeling}
]{\chaptermark{header} NERO: A Near-High Bandwidth Memory Stencil Accelerator for Weather Prediction Modeling}
\chaptermark{NERO}
\label{chapter:nero}
\chapternote{The content of this chapter was published as \emph{``NERO: A Near High-Bandwidth Memory Stencil Accelerator for Weather Prediction Modeling''} in FPL 2020. (\emph{Nominated for the Stamatis Vassiliadis Memorial Best Paper Award})\\
Our earlier work was published as \emph{``{NARMADA}: {N}ear-Memory Horizontal Diffusion Accelerator for Scalable Stencil Computations''} in FPL 2019 where we accelerate only horizontal diffussion kernel from the COSMO (Consortium for Small-Scale Modeling) model. }
\renewcommand{\namePaper}{NERO\xspace} 

Modern data-intensive applications demand high compute capabilities with strict power constraints. Unfortunately, such applications suffer from a significant waste of {both} execution cycles and energy in current computing systems due to the costly  
data movement between the compute units and the memory units. Weather prediction modeling is one such data-intensive application where we generate petabytes of data. Moreover, ongoing climate change calls for {fast and accurate} 
weather and climate {modeling}. However, {when solving large-scale {weather prediction}
simulations,} state-of-the-art CPU and GPU implementations suffer from limited performance and high energy consumption. {These implementations are {dominated by} complex irregular memory access patterns and low arithmetic intensity {that} pose fundamental challenge{s} to acceleration}. To overcome these challenges, in this chapter, we propose {and evaluate} the use of near-memory acceleration using a reconfigurable fabric with high-bandwidth memory (HBM). 

        
\section{Introduction} 
{Accurate weather prediction using detailed weather models is essential to {make} weather-dependent decisions in a timely manner.}
The Consortium for Small-Scale Modeling (COSMO)~\cite{doms1999nonhydrostatic} 
{built} one such weather model 
to meet the high-resolution forecasting requirements of weather services. The COSMO model is a non-hydrostatic atmospheric prediction model currently being used by a dozen nations for meteorological purposes and research applications. 

The central part of the COSMO model ({called \emph{dynamical core} or \emph{dycore}}) solves the Euler equations on a curvilinear grid and applies implicit discretization {(i.e., parameters are dependent on each other at the same time instance~\cite{bonaventura2000semi})} in the vertical dimension and 
explicit discretization {(i.e., 
{a solution is dependent on the previous system state}~\cite{bonaventura2000semi})} in the horizontal dimension. The use of different discretizations leads to three computational patterns~\cite{cosmo_knl}: {1)} horizontal stencils, {2)} tridiagonal solvers in the vertical dimension, and {3)} point-wise computation. These computational kernels are compound stencil kernels that operate on a three-dimensional grid~\cite{gysi2015modesto}.  
\emph{Vertical advection} ({\texttt{vadvc}}) and  \emph{horizontal diffusion} ({\texttt{hdiff}})  are such compound kernels found in the \emph{dycore} of the COSMO \gagannn{weather prediction} model. {These kernels} are representative {of} the data access {patterns and algorithmic} complexity of the entire COSMO model. {They} are similar to the kernels used in other weather and climate models~\cite{kehler2016high,neale2010description,doi:10.1175/WAF-D-17-0097.1}.  Their performance is dominated by memory-bound operations with unique irregular memory access patterns 
{and} low arithmetic intensity that often results in $<$10\% sustained floating-point performance on current CPU-based systems~\cite{chris}. 

\begin{figure}[h]
  \centering
  \includegraphics[bb=32 29 546 396,width=0.8\textwidth,trim={1cm 0.8cm 1.5cm 0.6cm},clip]{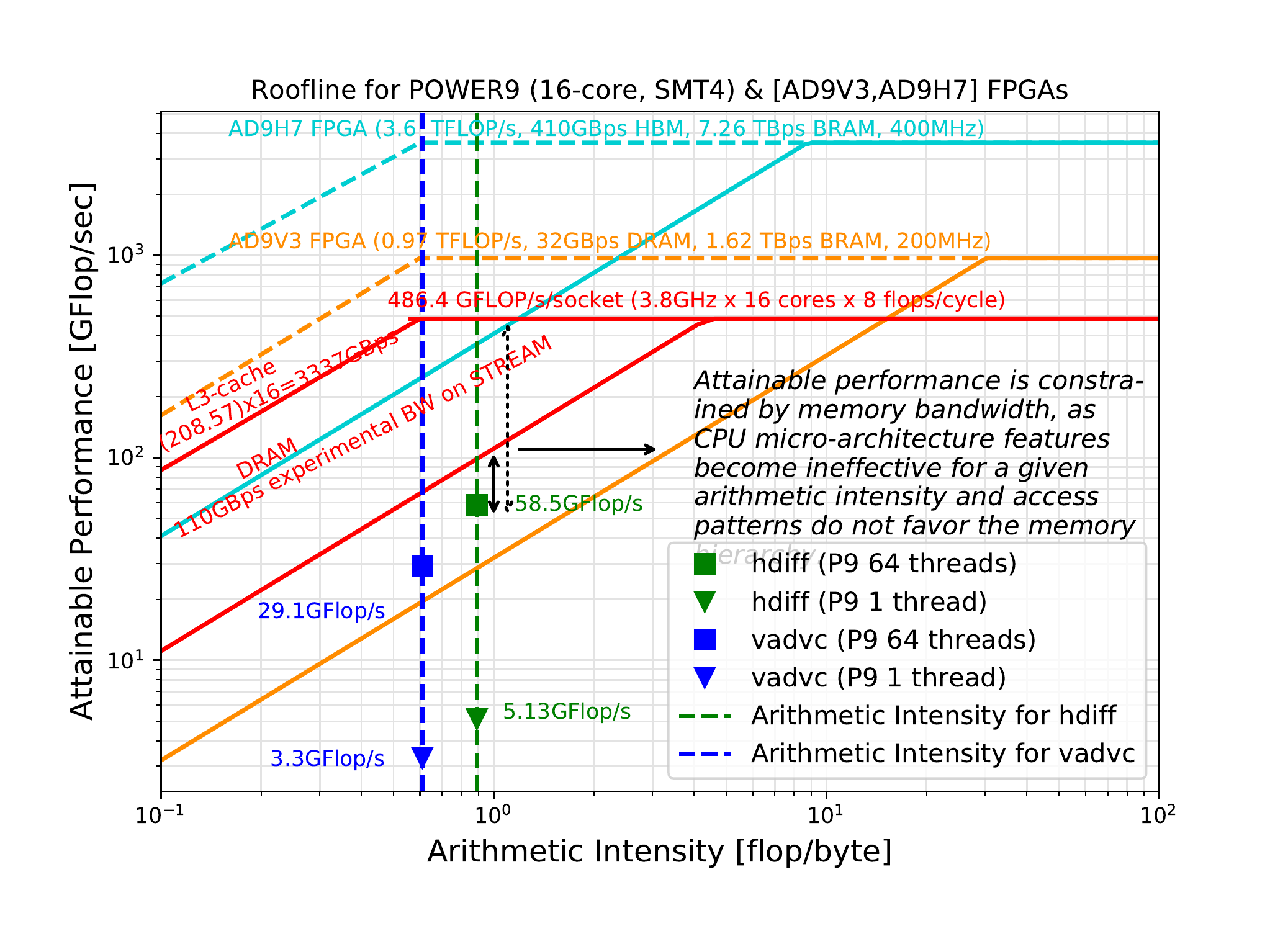}
    \caption{Roofline~\cite{williams2009roofline} for POWER9 (1-socket) showing vertical advection {(\texttt{vadvc})} and horizontal diffusion {(\texttt{hdiff})} kernels for single-thread and 64-thread 
    implementations. 
    {The plot shows also the rooflines} 
    of the FPGAs used in this chapter}
 \label{fig:nero/roofline}
\end{figure}

Figure~\ref{fig:nero/roofline} shows the roofline plot{~\cite{williams2009roofline}} for {an} IBM 16-core POWER9 CPU (IC922).\footnote{IBM and POWER9 are registered trademarks or common law marks of International Business Machines Corp., registered in many jurisdictions worldwide. Other product and service names might be trademarks of IBM or other companies.} After optimizing the \texttt{vadvc} and \texttt{hdiff} kernels for the POWER architecture by following the approach in~\cite{stencilOnPOWER8}, 
{they} achieve {29.1}~GFLOP/s and 58.5~GFLOP/s, {respectively}, {for 64 threads}. {Our roofline analysis 
{indicates} that these kernels are constrained by the host DRAM bandwidth.} 
{Their} low arithmetic intensity limits 
{their performance, which is one order of magnitude smaller than the peak performance,} and results in a fundamental memory bottleneck that 
standard CPU-based optimization techniques {cannot overcome}.

{\textbf{Our goal} is to overcome the memory} {bottleneck {of weather {prediction} kernels} by {exploiting near-memory {computation} capability on} FPGA accelerators with high-bandwidth memory (HBM)~\cite{6757501,hbm,lee2016smla} {that are attached} {to the host CPU}. 
Figure~\ref{fig:nero/roofline} shows the roofline model{s} of the two FPGA cards {(AD9V3~{\cite{ad9v3}} and AD9H7~{\cite{ad9h7}})} used in this chapter. 
FPGAs {can} {handle} 
irregular memory access patterns {efficiently} and 
offer significantly 
{higher} memory bandwidth {than the host CPU with} 
their on-chip URAMs {(UltraRAM)}, BRAMs {(block RAM)}, and {off-chip} HBM ({high-bandwidth memory} for the AD9H7 card).} 
However, taking full advantage of 
FPGAs for accelerating a workload is not a trivial task. To compensate {for} the higher clock frequency of the {baseline} CPUs, 
{our} FPGAs must exploit at least one order of magnitude more parallelism in a target workload. 
{This is challenging, as it requires sufficient FPGA programming skills to {map the workload and} optimize the design for the FPGA microarchitecture.}

As mentioned in Section~\ref{subsection:intro/data_centric}, {modern FPGA boards deploy} 
{new} cache-coherent interconnects, such as 
IBM Coherent Accelerator Processor Interface (CAPI)~\cite{openCAPI}, Cache Coherent Interconnect for Accelerators
(CCIX)~\cite{benton2017ccix}, and Compute Express Link (CXL)~\cite{cxlwhitepaper}, {which} allow tight integration of FPGAs with CPUs at 
high bidirectional bandwidth {({on} the order of tens of GB/s)}. However, memory-bound applications on FPGAs are limited by {the relatively} low DDR4 bandwidth (72 GB/s 
{for four independent dual-rank DIMM interfaces}{~\cite{VCU}}). 
{To overcome this limitation,} FPGA vendors have started offering devices {equipped} with HBM~\cite{xilinx_utlra,hbm_specs,intel_altera,lee2016smla} 
{with} a theoretical peak bandwidth of 410~GB/s. 
{HBM-equipped} FPGAs have the potential to \gagannn{reduce} the \gagannn{memory} bandwidth bottleneck, {but} a study of their advantages for 
{real-world memory-bound} applications 
is still~missing.\\


{We aim} to answer the following {research question}: \textbf{Can FPGA-based {accelerators} with HBM {mitigate} the performance bottleneck of {memory-bound} compound weather {prediction} kernels 
in an energy-efficient way?} As an answer to this question{,} we present \namePaper, a \underline{ne}ar-HBM accelerator for weathe\underline{r} predicti\underline{o}n. 
{We design and implement} \namePaper~
{on an FPGA with HBM to} optimize two 
kernels (vertical advection and horizontal diffusion), 
{which notably} represent 
{the} spectrum of {computational diversity} found in the COSMO {weather prediction} application. 
{We co-design a hardware-software framework and provide an optimized API to interface efficiently with the rest of the COSMO model, which runs on the CPU}.
Our {FPGA-based} solution {for \texttt{hdiff} and \texttt{vadvc}} leads to 
{performance improvements of $4.2\times$ and $8.3\times$ and}
{energy reductions} of $22\times$ and $29\times$, {respectively,} {with respect to optimized CPU implementations~\cite{stencilOnPOWER8}}. 

\section{Background} 
\label{sec:nero/background}
\juang{In this section, we first provide} 
an overview of the \juang{\texttt{vadvc} and \texttt{hdiff}} {compound stencils, 
\juang{which represent} a large fraction of the overall computational load of the COSMO nn{weather prediction} model}. 
Second, we introduce the CAPI SNAP {(Storage, Network, and Analytics Programming)} framework\footnote{https://github.com/open-power/snap} that we use to connect our \namePaper~accelerator to an IBM POWER9~system.

\subsection{{Representative} COSMO Stencils}
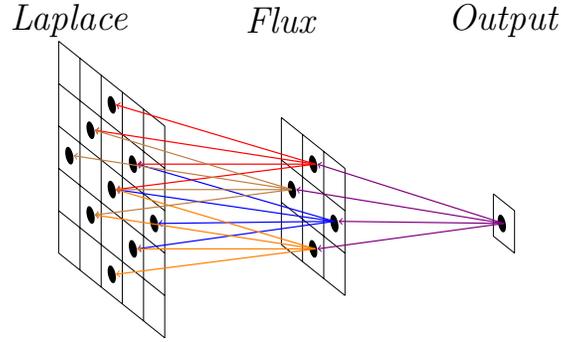
\begin{figure}[h]
\centering
    \resizebox{0.5\textwidth}{!}{

    \begin{tikzpicture}[rotate=90,transform shape]

    \gridThreeD{0}{8.25}{black};
    \node at (7.5,10.5)[ rotate=270] {\huge \textit{Laplace}};
      \dotesUpper{0}{8.25};
    \gridThreeDSecond{1}{4}{black};
    \node  at (7.5,5.5)[ rotate=270] {\huge \textit{Flux}};
     \lowerDots{3}{4};
     
    \gridThreeDThird{2}{0}{black};
     \node  at (7.5,0.25) [ rotate=270] {\huge \textit{Output}};
     \lowestDot{1.6}{-0.45}

      \draw [thick,->,blue] (2.75,4.25) -- (2.7,8.4);
       \draw [thick,->,blue] (2.75,4.25) -- (2.1,8.9);
       \draw [thick,->,blue] (2.75,4.25) -- (4.1,8.9);
       \draw [thick,->,blue] (2.75,4.25) -- (3.5,9.4);

       \draw [thick,->,orange] (2.1,4.7) -- (3.5,9.4);
       \draw [thick,->,orange] (2.1,4.7) -- (1.5,9.4);
       \draw [thick,->,orange] (2.1,4.7) -- (2.1,8.9);
       \draw [thick,->,orange] (2.1,4.7) -- (2.9,9.9);
     
       \draw [thick,->,red] (4.1,4.7) -- (3.5,9.4);
       \draw [thick,->,red] (4.1,4.7) -- (5.5,9.4);
       \draw [thick,->,red] (4.1,4.7) -- (4.1,8.9);
       \draw [thick,->,red] (4.1,4.7) -- (4.9,9.9);

           \draw [thick,->,brown] (3.5,5.25) -- (3.5,9.4);
           \draw [thick,->,brown] (3.5,5.25) -- (2.9,9.9);
           \draw [thick,->,brown] (3.5,5.25) -- (4.9,9.9);
           \draw [thick,->,brown] (3.5,5.25) -- (4.3,10.4);
         \draw [thick,->,violet] (2.7,0.25) -- (3.5,5.15);
         \draw [thick,->,violet] (2.7,0.25) -- (4.1,4.65);
         \draw [thick,->,violet] (2.7,0.25) -- (2.1,4.65);
         \draw [thick,->,violet] (2.7,0.25) -- (2.75,4.15);

    

\end{tikzpicture}
}

\caption{Horizontal diffusion compound kernel composition in a two dimensional plane
\label{fig:NERO/hdiffMemory}}
\vspace{-0.3cm}
\end{figure}

A stencil operation updates values in a structured multidimensional grid based on the values of a fixed local neighborhood of grid points. 
Vertical advection (\texttt{vadvc}) and horizontal diffusion (\texttt{hdiff}) from the COSMO model are two such compound \juang{stencil} kernels, which represent the typical code patterns found in the \emph{dycore} of COSMO.  
Algorithm~\ref{algo:nero/hdiffKernel} shows {the} pseudo-code for \texttt{vadvc} and \texttt{hdiff}  kernels. 
The horizontal diffusion kernel iterates over a 3D grid performing \textit{Laplacian} and \textit{flux} to calculate different grid points, as shown in Figure~\ref{fig:NERO/hdiffMemory}. Vertical advection has a higher degree of complexity since it uses the Thomas algorithm~\cite{thomas} to solve a tri-diagonal matrix of the velocity field along the vertical axis. Unlike the {conventional} stencil kernels, vertical advection has dependencies in the vertical direction, which leads to limited available parallelism.

\begin{algorithm}[h!]
\scriptsize
\setstretch{1}
\caption{Pseudo-code for vertical advection and horizontal diffusion {kernels}
used by the COSMO~\cite{doms1999nonhydrostatic} {weather prediction} model}
\label{algo:nero/hdiffKernel}
 \begin{algorithmic}[1]
    \Function{verticalAdvection}
    {$float*                ccol,
                float* dcol,
                float* wcon,
                float* ustage,\newline
                \phantom{\textbf{procedure} \texttt{VISIT\_SEQUENCE\_EVAL}(}
                float* upos,
                float* utens,
                float* utensstage$}
        \For{$c\gets2$ \textbf{to} $column-2$}
            \For{$r\gets2$ \textbf{to} row-2}
                 \Function{forwardSweep}{$float* ccol,
                float* dcol,
                float* wcon,float* ustage,\newline
                \phantom{\textbf{procedure} \texttt{VISIT\_SEQUENCE\_EVAL\quad\quad\quad}(}
                float* upos,
                float* utens,
                float* utensstage$}

                \For{$d\gets1$ \textbf{to} $depth$}
                       \State{\texttt{/* forward sweep calculation */}}
                \EndFor
                \EndFunction
                       
                \Function{backwardSweep}{$float* ccol,
                float* dcol,
                float* wcon,
                float* ustage,\newline
                \phantom{\textbf{procedure} \texttt{VISIT\_SEQUENCE\_EVAL\quad\quad\quad}(}
                float* upos,
                float* utens,
                float* utensstage$}     
                 
                \For{$d\gets depth-1$ \textbf{to}$ 1$}
                \State{\texttt{/* backward sweep calculation */}}
                \EndFor
                \EndFunction
            \EndFor
        \EndFor
      \EndFunction


\hrule

  \Function{horizontalDiffusion}{$float* src, float* dst$}    
       \For{$d\gets1$ \textbf{to} $depth$}
             \For{$c\gets2$ \textbf{to} $column-2$}
                    \For{$r\gets2$ \textbf{to} row-2}
                        \State{\texttt{/* {L}aplacian calculat{ion} */}}
                        \State{$lap_{CR}=laplaceCalculate(c,r)$}
                        \State{\texttt{/* row-laplacian */}}
                        \State{$lap_{CRm}=laplaceCalculate(c,r-1)$}      
                        \State{$lap_{CRp}=laplaceCalculate(c,r+1)$}
                        \State{\texttt{{/* column-laplacian */}}}
                        \State{$lap_{CmR}=laplaceCalculate(c-1,r)$}
                        \State{$lap_{CpR}=laplaceCalculate(c+1,r)$}
                        \State{\texttt{/* column-flux calculat{ion} */}}
                        \State{$flux_{C} = lap_{CpR} - lap_{CR}$}
                        \State{$flux_{Cm} = lap_{CR} - lap_{CmR}$}
                        \State{\texttt{/* row-flux calculation */}}
                        \State{$flux_{R} = lap_{CRp} - lap_{CR}$}
                        \State{$flux_{Rm} = lap_{CR} - lap_{CmR}$}
                        \State{\texttt{/* output calculat{ion} */}}
                        \State{$dest[d][c][r] = src[d][c][r] -
                            c1 * (flux_{CR}- flux_{CmR}) + (flux_{CR}- flux_{CRm})$}
                        
            \EndFor 
             \EndFor   
   \EndFor
      \EndFunction

\end{algorithmic}
\end{algorithm}

Such compound kernels are dominated by memory-bound operations with complex memory access patterns and low arithmetic intensity. This poses a fundamental challenge to acceleration. 
{CPU implementations of these kernels~\cite{stencilOnPOWER8} suffer from limited data locality and inefficient memory usage, as our roofline analysis in Figure~\ref{fig:nero/roofline} exposes}.

\subsection{CAPI SNAP Framework}
The OpenPOWER Foundation Accelerator Workgroup~\cite{open_power} created the CAPI SNAP framework, 
{an open-source environment for FPGA programming productivity}. 
CAPI SNAP {provides two key benefits}~\cite{wenzel2018getting}: (i) it enables an improved developer productivity for FPGA acceleration and eas{es the} use of CAPI's cache-coheren{ce mechanism}, and (ii) it places 
{FPGA-}accelerated compute engines, 
{also known as} FPGA \textit{actions}, closer to {relevant} data to achieve better performance. SNAP provides a simple API to invoke an accelerated \textit{action}, and also provides programming methods to {instantiate} customized accelerated \textit{actions} on the FPGA side. These accelerated \textit{actions} can be specified in C/C++ code that is then compiled to the FPGA target using the Xilinx Vivado High-Level Synthesis (\acrshort{hls}) tool~\cite{hls}.

\section{Design Methodology}
\label{sec:nero/design}
\subsection{NERO, A Near HBM Weather Prediction Accelerator}
The low arithmetic intensity of real-world weather {prediction} kernels limits the attainable performance on current multi-core systems. This sub-optimal performance is due to {the kernels'} complex memory {access} patterns and their inefficiency in exploiting a rigid cache hierarchy, {as quantified in {the} roofline plot in} Figure~\ref{fig:nero/roofline}. 
These kernels {cannot} fully utilize the available memory bandwidth, which leads to 
high 
{data movement overheads} in terms of latency and energy consumption. We address 
{these inefficiencies} by developing an architecture {that} combines fewer off-chip data {accesses} with higher throughput for the loaded data. {To this end}, our accelerator design {takes} a data-centric approach~\cite{mutlu2019,ghose2019processing,teserract,singh2019near,NAPEL,hsieh2016accelerating,tom,ahn2015pim,googleWorkloads,kim2018grim} {that exploits} near high-bandwidth memory acceleration.

\begin{figure*}[h]
\begin{subfigure}[t]{0.5\textwidth}
  \includegraphics[bb=9 17 525 278,width=1\linewidth,trim={0.5cm 0.5cm 0cm 0cm},clip]{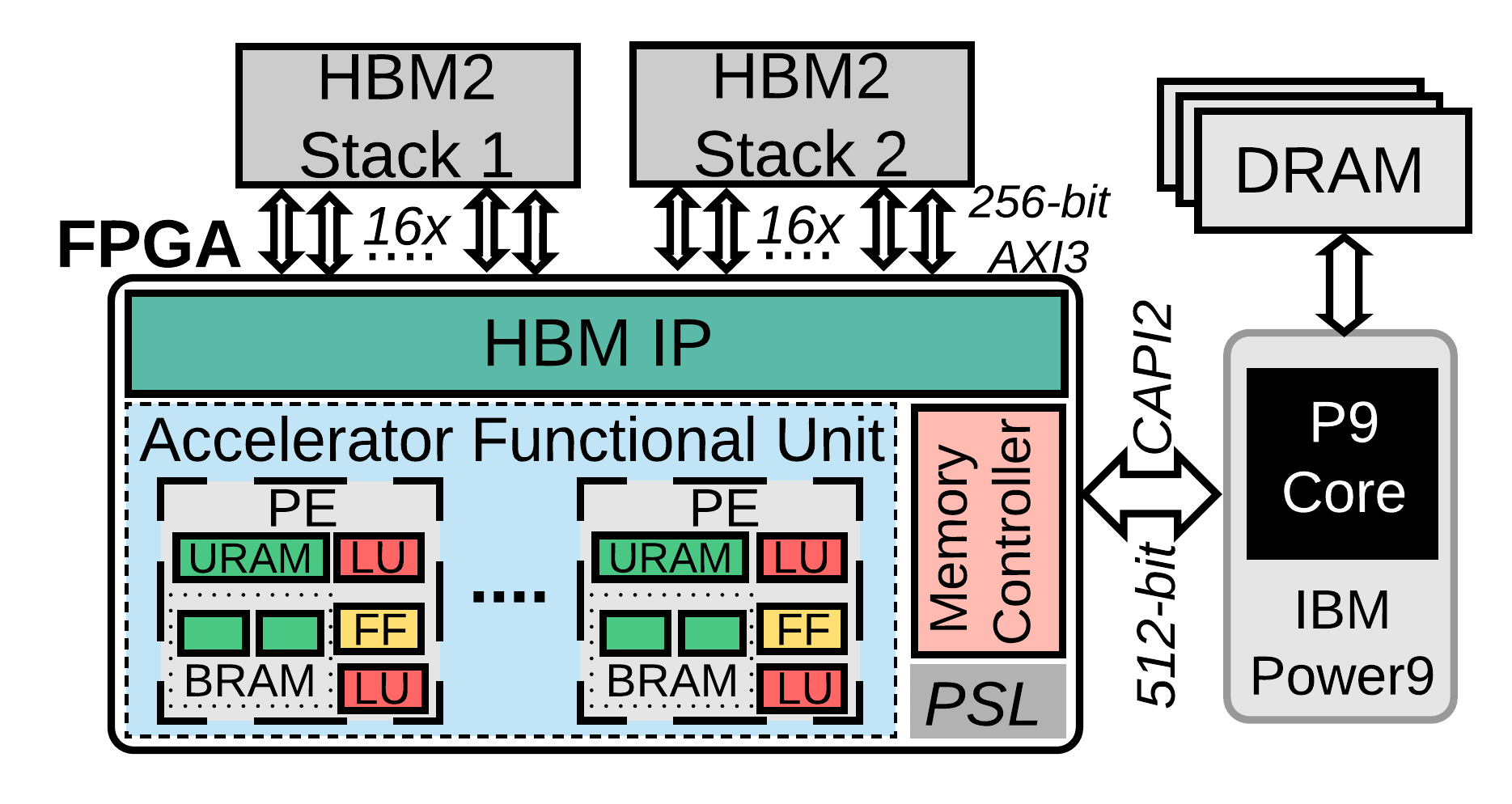}
   \caption{
   \label{fig:nero/system}}
\end{subfigure}%
\begin{subfigure}[t]{0.4\textwidth}
  \includegraphics[bb=0 0 623 531,width=0.95\linewidth,trim={0cm 0cm 0cm 0cm},clip]{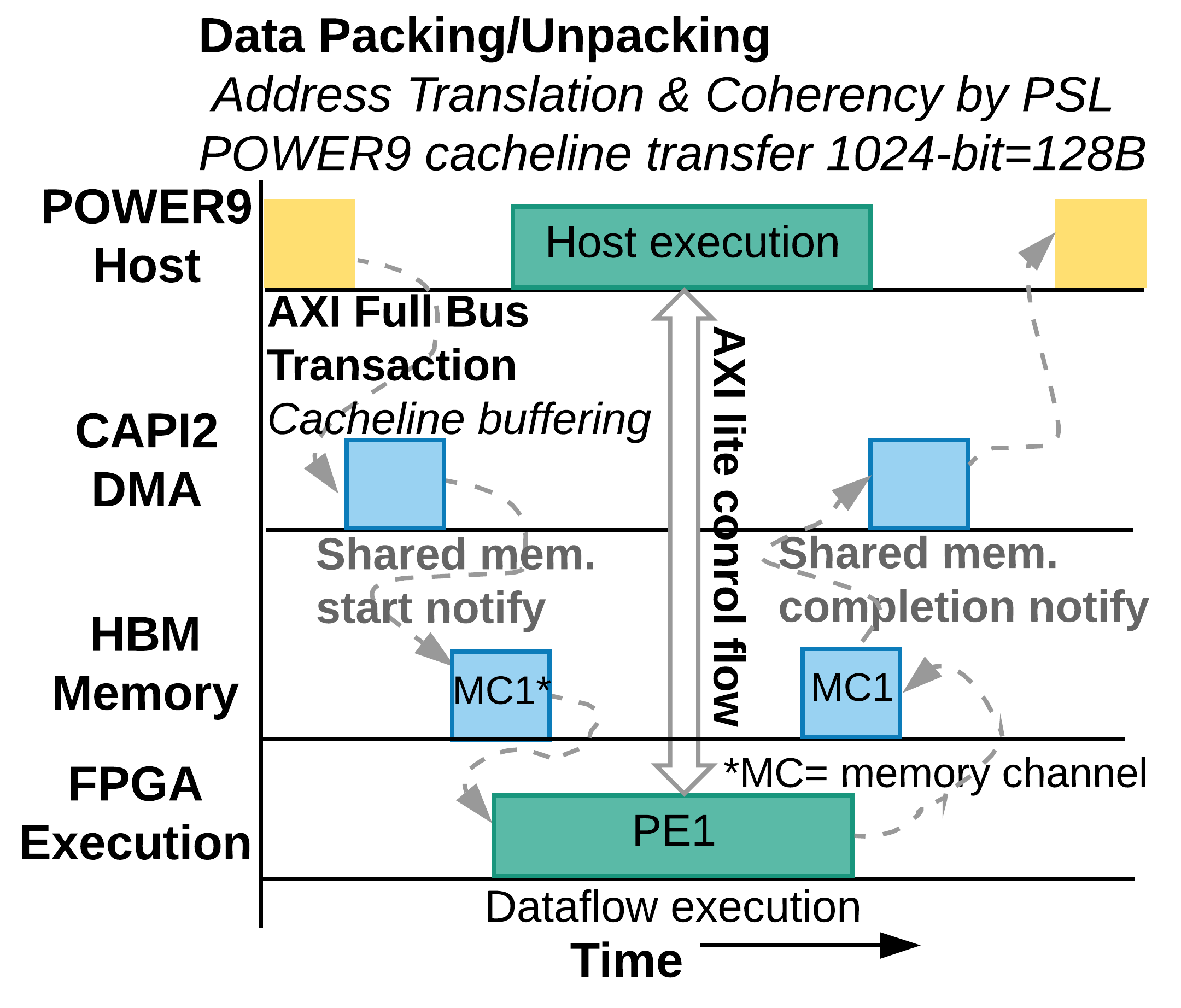}
   \caption{
  \label{fig:nero/execution}}
\end{subfigure}
\caption[(a) Heterogeneous platform with an IBM POWER9 system connected to an HBM-based FPGA board via~CAPI2. (b)  Execution timeline with data flow sequence from the  host  DRAM  to the onboard  FPGA  memory]{(a) Heterogeneous platform with an IBM POWER9 system connected to an HBM-based FPGA board via~CAPI2. (b)  Execution timeline with data flow sequence from the  host  DRAM  to the onboard  FPGA  memory\label{fig:complete_hbm_flow}}
\end{figure*}

\begin{figure}[h]
  \centering
  \includegraphics[bb=20 21 769 347,width=\linewidth,trim={0.3cm 0.2cm 0cm 0.2cm},clip]{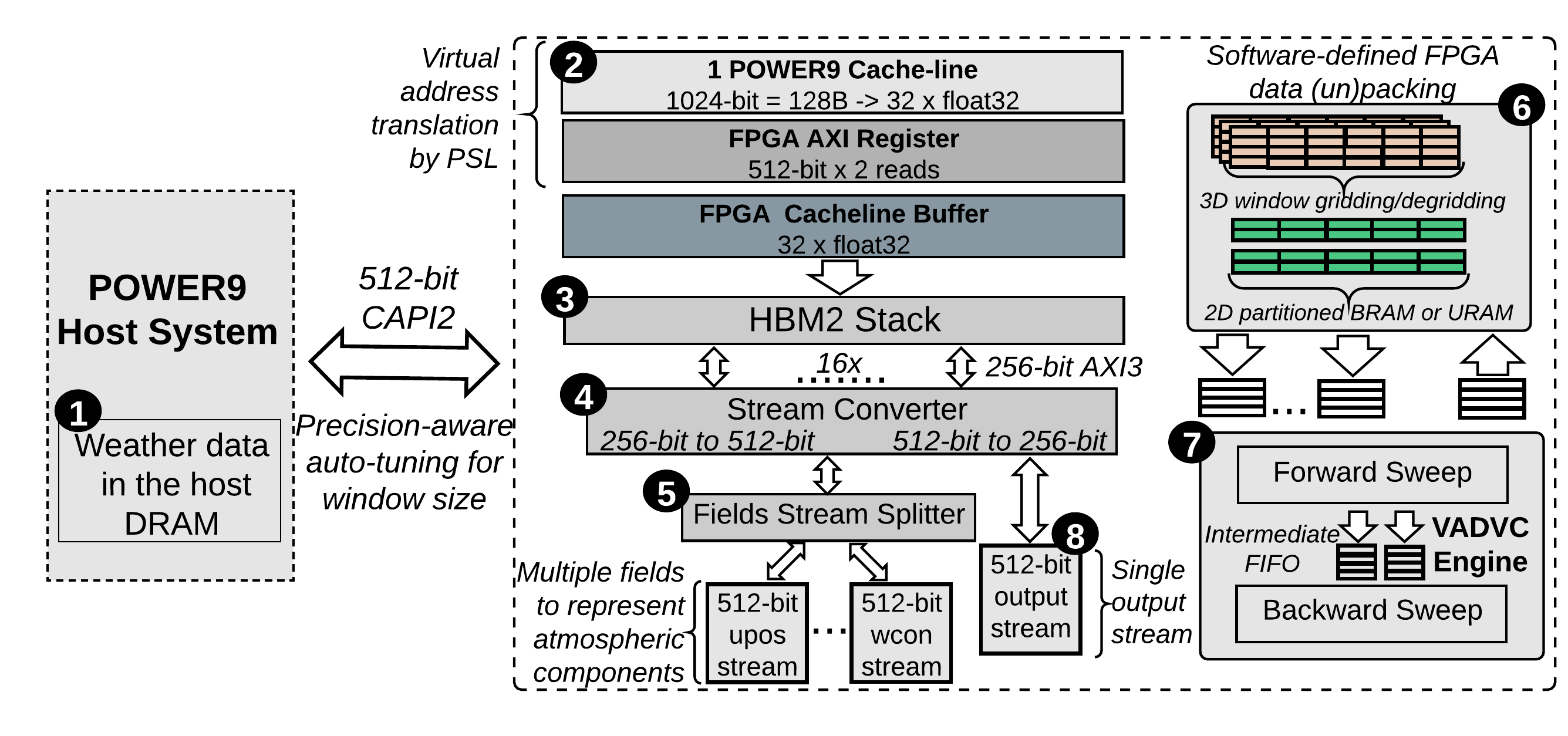}
  \caption{Architecture overview of~\namePaper~with data flow sequence from the host DRAM to the on-board FPGA memory via POWER9 cachelines. We {depict a} single {processing element (PE)} fetching data from a dedicated HBM port. The number of HBM ports scales linearly with the number of {PEs}. Heterogeneous partitioning of on-chip memory blocks {reduces} read {and} write latencies across the FPGA memory hierarchy
 \label{fig:nero/complete_flow}}
 \end{figure}

Figure~\ref{fig:nero/system} shows a high-level overview of our integrated system. An HBM-based FPGA is connected to a server system based on an IBM POWER9 processor using the Coherent Accelerator Processor Interface version 2 (CAPI2). 
The FPGA consists of two HBM stacks\footnote{We enable only a single stack based on our resource and power consumption analysis for the \texttt{vadvc} kernel.}, each with 16 \emph{pseudo-memory channels}~\cite{axi_hbm}. A channel is exposed to the FPGA as a 256-bit wide port, and in total, the FPGA has 32 such ports. The HBM IP provides 8 memory controllers (per stack) to handle the data transfer to and from the HBM memory ports. Our design consists of an \emph{accelerator functional unit} (\acrshort{afu}) that interacts with the host system through the power service layer (\acrshort{psl}), which is the CAPI endpoint on the FPGA. An AFU comprises of multiple \emph{processing elements} (\acrshort{pe}s) that perform compound stencil computation. Figure~\ref{fig:nero/complete_flow} shows the architecture overview of \namePaper. As vertical advection is the most complex kernel, we depict our architecture design flow for vertical advection. We use a similar design for the {horizontal diffusion} kernel.

The weather data, based on the atmospheric model resolution grid, is stored in the DRAM of the host system {(\circled{1} in Figure~\ref{fig:nero/complete_flow})}.  We employ {the} double buffering technique  between the CPU and the FPGA to hide the PCIe {(Peripheral Component Interconnect Express~\cite{pcie})} {transfer} {latency}.  {By configuring a buffer of 64 cache lines, between the AXI4 interface of CAPI2/PSL and the AFU, we can reach the theoretical peak bandwidth of CAPI2/PCIe (x8 PCIe Gen4 interface with a theoretical peak bandwidth 15.75 GB/s and a transfer rate of 16 GT/s).} We create a specialized memory hierarchy from the heterogeneous FPGA memories {(i.e., URAM, BRAM, and HBM)}. By using a greedy algorithm, we determine the best-suited hierarchy for our kernel. The memory controller (shown in Figure~\ref{fig:nero/system}) handles the data placement to the appropriate memory type {based on programmer's directives}.

{On the FPGA, following the initial buffering (\circled{2}), the {transferred} grid data is mapped onto the HBM memory (\circled{3}).  As {the} FPGA ha{s} limited resources, we propose a 3D window-based grid transfer from the host DRAM to the FPGA, facilitating a smaller, less power-hungry deployment. The window size represents the portion of the grid a processing element ({PE} in Figure~\ref{fig:nero/system}) would process. {Most FPGA developers manually optimize for the right window size. However, manual optimization is tedious because of the huge design space, and {it} requires expert guidance. 
Further, selecting an inappropriate window size lead{s} to sub-optimal results. 
{Our experiments (in Section~\ref{subsection:nero/evaluation}) show that}: (1) finding the {best} window size is critical in terms of {the} area vs. performance trade-off, and (2) the {best window} size depends on the datatype precision. Hence, instead of pruning the design space manually, we formulate 
{the search for} the {best} window size as a multi-objective auto-tuning problem taking into account the datatype precision. We make use of OpenTuner~\cite{opentuner}, 
{which} uses machine-learning techniques to guide the design-space~exploration.}}


Our design consists of multiple PEs (shown in Figure~\ref{fig:nero/system}) that exploit data-level parallelism in COSMO {weather prediction} kernels. {A dedicated HBM memory port is assigned to a specific PE}; therefore, we enable as many HBM ports as the number of PEs. This allows us to use the high {HBM} bandwidth {effectively} because each PE fetches from an independent port. 
In our design, we use a switch, which provides the capability to 
{bypass} the HBM, {when the} grid size {is small}, and map the data directly onto the FPGA's URAM {and BRAM}. The HBM port provides 256-bit data, {which} is half the size of {the} CAPI2 {bitwidth} (512-bit). Therefore, to match {the} CAPI2 bandwidth, we introduce a stream converter logic (\circled{4}) that converts {a} 256-bit HBM stream to {a} 512-bit stream (CAPI compatible) or vice versa.
From HBM, {a PE} reads a single stream of data that consists of all the fields\footnote{Fields represent atmospheric components like wind, pressure, velocity, etc. that are required for weather calculation.} {that} are needed for a specific COSMO kernel computation. {The PEs} use a fields stream splitter logic (\circled{5}) that splits a single HBM stream to multiple streams (512-bit each), one for each field.
  
{To optimize a PE{,} we apply various optimization strategies. First, we exploit the inherent parallelism in {a given} algorithm through hardware pipelining. {Second}, we partition on-chip memory to avoid the stalling of our pipelined design, since the on-chip BRAM/URAM has only two read/write ports. {Third}, all the tasks execute in a dataflow manner that enables task-level parallelism. \texttt{vadvc} is more { computationally complex} than \texttt{hdiff} because it involves forward and backward sweeps {with dependencies in {the} z-dimension}. While \texttt{hdiff} performs {only} {Laplacian} and flux calculations with dependencies in {the} x- and y-dimensions. 
Therefore, we demonstrate our design flow by means of {the} \texttt{vadvc} kernel (Figure~\ref{fig:nero/complete_flow}). 
Note {that} we show {only} a single port-based PE operation.  {However}, for multiple PEs, we enable multiple HBM ports.} 

We make use of memory reshaping techniques to  configure our memory space with multiple parallel BRAMs or URAMs~\cite{dioFPT}. {We form an intermediate memory hierarchy by decomposing (or slicing) 3D window data into a 2D grid. } This allows us to bridge the latency gap between the HBM memory and our accelerator. Moreover, it allows us to exploit the available FPGA resources efficiently. Unlike traditionally-fixed CPU memory hierarchies, which perform poorly with irregular access patterns and suffer from cache pollution effects, application-specific memory hierarchies {are} shown to improve {energy and latency} by tailoring the cache levels and cache sizes to {an} application's memory access patterns~\cite{jenga}.
  
The main computation pipeline (\circled{7}) {consists of a forward and a backward sweep logic}. The forward sweep {results} are stored in {an} intermediate buffer to allow for backward sweep calculation. {Upon} completion of {the} backward sweep, results are {placed in} an output buffer {that is followed by a degridding logic (\circled{6}). The degridding logic} converts the calculated results to a 512-bit wide output stream (\circled{8}). As there is only a single output stream (both in \texttt{vadvc} and \texttt{hdiff}), we do not need extra logic to merge the streams. The 512-bit {wide} stream goes through an HBM stream converter logic (\circled{4}) that converts the stream bitwidth to HBM port size (256-bit).  
  
{Figure~\ref{fig:nero/execution} shows the execution timeline from our host system to the FPGA board for a single PE. The host offloads the processing to an FPGA and 
{transfers} the required data {via DMA (direct memory access)} over {the} CAPI2 interface. The SNAP framework allows for parallel execution of the host and our FPGA PEs while exchanging control signals over the AXI lite interface{~\cite{axilite}}. On task completion, the AFU notif{ies} the host system via {the} AXI lite interface and 
{transfers} back the results {via DMA}. }

\subsection{\namePaper~Application Framework}
\begin{figure}[h]
  \centering
  \includegraphics[bb=8 21 882 664,width=0.6\linewidth,trim={0cm 0cm 0.1cm 0cm},clip]{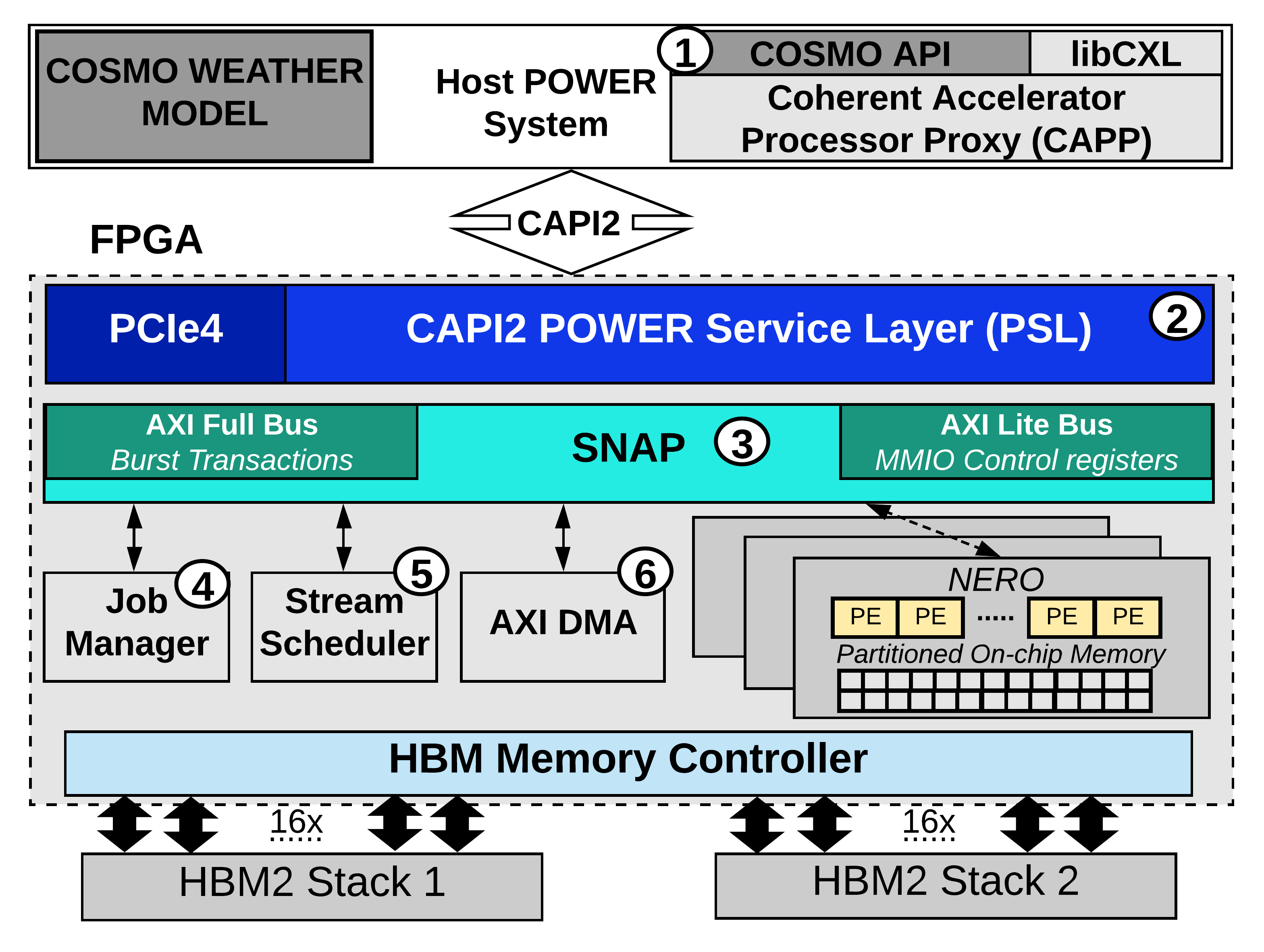}
  \caption{
  {\namePaper~application framework.} 
  We co-design our software and hardware using {the} SNAP framework. COSMO API allows the host to offload kernels to our FPGA platform
 \label{fig:nero/snap_api}}
 \end{figure}

Figure~\ref{fig:nero/snap_api} shows the \namePaper~application framework {to support our architecture}.  A software-defined COSMO API (\circledWhite{1}) handles offloading jobs to \namePaper~with an interrupt-based queuing mechanism.  This allows for minimal CPU usage (and, hence, power {usage}) during FPGA {operation}. \namePaper~employs an array of processing elements to compute COSMO kernels, such as vertical advection or horizontal diffusion. Additionally, we pipeline our PEs to exploit the available spatial parallelism. 
By accessing the host memory through 
{the CAPI2} cache-coherent link, \namePaper~acts as a peer to the CPU.  This is enabled {through} {the} Power-Service Layer (PSL) (\circledWhite{2}). { SNAP (\circledWhite{3}) allows for seamless {integration} of the COSMO API with our CAPI-based accelerator.} The job manager (\circledWhite{4}) dispatches jobs to streams, which are managed in the stream scheduler (\circledWhite{5}). The execution of a job is done by streams that determine which data is to be read from the host memory and sent to the PE array through DMA transfers (\circledWhite{6}). The pool of heterogeneous on-chip memory is used to store the input data from the main memory and the intermediate data generated by {each} PE.

\section{Results}
\label{sec:nero/results}

\subsection{System Integration}
We implemented our design on an Alpha-Data ADM-PCIE-9H7 card~\cite{ad9h7} featuring the Xilinx Virtex Ultrascale+ XCVU37P-FSVH2892-2-e~\cite{vu37p} and 8GiB HBM2 
{(i.e., two stacks of 4GiB each)}~\cite{hbm} with an IBM POWER9 as the host system. The POWER9 socket has 16 cores, each of which supports four-thread simultaneous multi-threading. We compare our HBM-based design to a conventional DDR4 DRAM~\cite{ad9v3} based design. 
{We perform the experiments for the DDR4-based design on} 
an Alpha-Data ADM-PCIE-9V3 card featuring the Xilinx Virtex Ultrascale+ XCVU3P-FFVC1517-2-i~\cite{vu37p}.

Table~\ref{tab:nero/systemparameters}
provides our system parameters.
We have co-designed our hardware and software interface around the SNAP framework while using the HLS design flow.

\begin{table}[h]
  \caption{System parameters and hardware configuration for the CPU and the FPGA board}
  \vspace{-0.4cm}
    \label{tab:nero/systemparameters}
      \begin{center}
      \footnotesize
\resizebox{0.75\linewidth}{!}{%
\begin{tabular}{|l@{\hspace{0.1\tabcolsep}}|p{7cm}|}

\hline
\textbf{Host CPU} 

 & 16-core IBM POWER9 AC922 \\&@3.2 GHz, 4\gagann{-way} SMT\\
 \hline
 \textbf{Cache-Hierarchy}&32 KiB L1-I/D, 256 KiB L2, 10 MiB L3 \\
 \hline
 \textbf{System Memory}&16x32GiB RDIMM DDR4 2666 MHz \\
\hline

 \begin{tabular}[c]{@{}l@{}}
 \textbf{HBM-based} \\\textbf{FPGA Board} \end{tabular}  &
  \begin{tabular}[c]{@{}l@{}}
 Alpha Data
ADM-PCIE-9H7\\
 Xilinx Virtex Ultrascale+ XCVU37P-2\\
  8GiB (HBM2) with PCIe Gen4 x8\\
 \end{tabular}\\
\hline
 \begin{tabular}[c]{@{}l@{}}
 \textbf{DDR4-based} \\\textbf{FPGA Board} \end{tabular}  & \begin{tabular}[c]{@{}l@{}}
 Alpha Data
ADM-PCIE-9V3\\
 Xilinx Virtex Ultrascale+ XCVU3P-2\\
 8GiB (DDR4) with PCIe Gen4 x8
  \end{tabular}\\



\hline

\end{tabular}
}
  \end{center}
 \end{table}

\subsection{Evaluation}
\label{subsection:nero/evaluation}
\begin{figure}[h]
  \centering
  \includegraphics[width=0.98\linewidth,trim={0.4cm 0.3cm 0.35cm 0.2cm},clip]{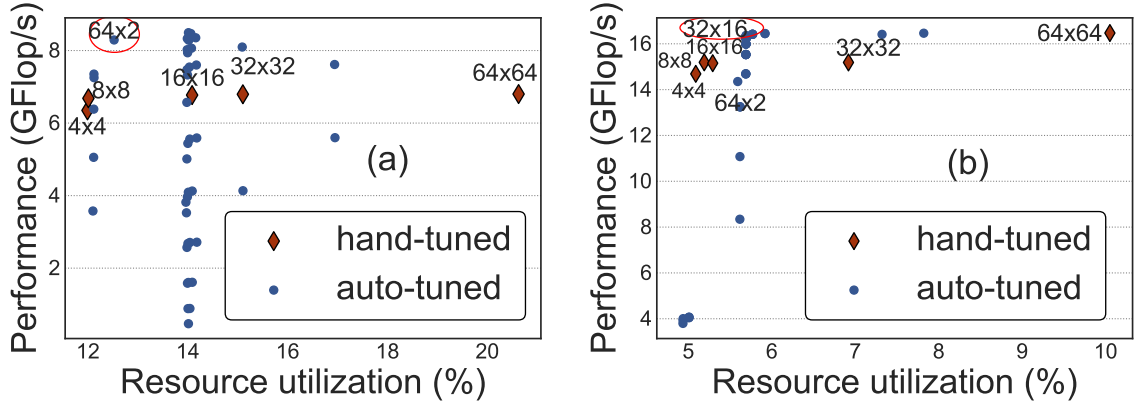}
      \caption{Performance {and FPGA resource utilization } {of} single \texttt{vadvc} PE{, as a function of tile-size,} using hand-tuning and auto-tuning for (a) single-precision (32-bit) and (b) half-precision (16-bit). We highlight the Pareto-optimal solution that we use for our \texttt{vadvc} accelerator \gagan{(with a red circle). Note that }the Pareto-optimal solution changes with precision
  \label{fig:nero/single_afu}}
 \end{figure}

We 
{run} our experiments using a $256\times256\times64$-point domain similar to the grid domain used by the COSMO {weather prediction} model. We employ an auto-tuning technique to determine a Pareto-optimal solution (in terms of performance and resource utilization) for our 3D window dimensions.
{The auto-tuning {with OpenTuner} exhaustively searches for every tile size in the x- and y-dimensions for \texttt{vadvc}.\footnote{\texttt{vadvc} has dependencies in {the} z-dimension; therefore, it cannot be parallelized in the z-dimension.} For \texttt{hdiff}, we consider sizes in all three dimensions. We define our auto-tuning as a multi-objective optimization with the goal {of maximizing} performance with  minimal resource utilization.} Section~\ref{sec:nero/design} provides further details on our design.
Figure~\ref{fig:nero/single_afu} shows hand-tuned and auto-tuned {performance and FPGA resource utilization} results for {\texttt{vadvc}}{, as a function of the chosen tile size. From the figure, we draw two observations.} 

{First, by using the auto-tuning approach and our careful FPGA microarchitecture design, we can get \gagan{P}areto-optimal results with a tile size of $64\times2\times64$ for single-precision {\texttt{vadvc}}\gagan{, which} gives us a peak performance of 8.49 GFLOP/s. For half-precision, we use a tile size of $32\times16\times64$ to achieve a peak performance of 16.5 GFLOP/s. We employ a similar strategy for {\texttt{hdiff}} to attain a single-precision performance of 30.3 GFLOP/s with a tile size of $16\times64\times8$ and a \gagan{half-precision} performance of 77.8 GFLOP/s \gagan{with} a tile size of $64\times8\times64$. 

Second, in FPGA acceleration, designers usually rely on expert judgement to find the appropriate tile-size and often adapt the design to {use} homogeneous tile sizes. However, as shown in Figure~\ref{fig:nero/single_afu}\gagan{,} such hand-tuned implementations lead to sub-optimal results in terms of either resource utilization~or~performance. 

{We conclude that} the {Pareto-optimal} tile size depends on the data precision used\gagan{:} a \gagan{good} tile-size for single-precision might lead to poor results when used with half-precision.

\gagan{Figure~\ref{fig:nero/perf} shows {single-precision} performance results for the (a) vertical advection and (b) horizontal diffusion kernels. For both kernels, we implement our design 
{on} an HBM- and {a} DDR4-based FPGA board. To compare the performance results, we  scale the number of PEs and analyze the change in execution time. For {the} DDR4-based design, we 
{can accommodate only} 4 PEs on {the} 9V3 board, while for the HBM-based design, we 
{can} fit 14 {PE}s before exhausting the on-board resources. {We draw four observations from the figure.}

 \begin{figure}[h]
  \centering
  \includegraphics[width=0.98\linewidth,trim={0.4cm 0.3cm 0.35cm 0.4cm},clip]{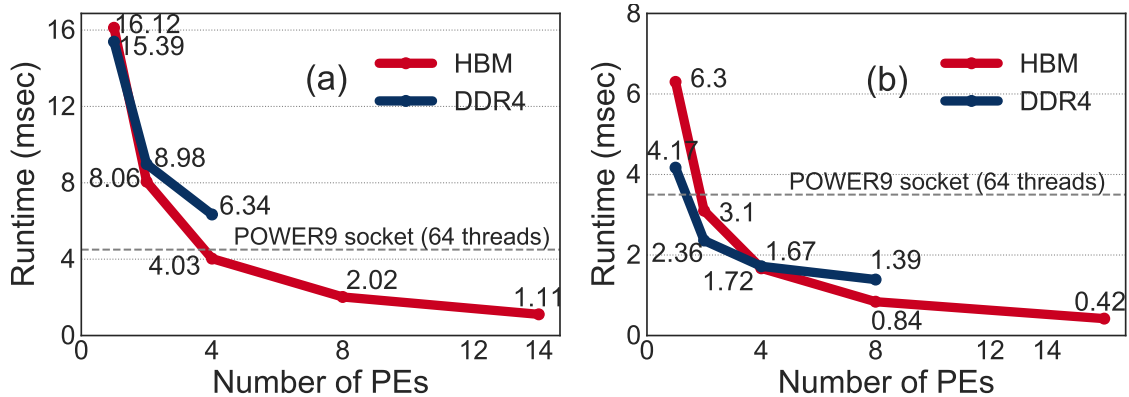}
   \caption{ {Single-precision} performance for (a) \texttt{vadvc} and (b) \texttt{hdiff}{, as a function of accelerator PE count on the HBM- and DDR4-based FPGA boards}. 
   \gagan{We also show {the}} single socket (64 threads) performance of \gagan{an} IBM POWER9 host system for both \texttt{vadvc} and \texttt{hdiff}
  \label{fig:nero/perf}}
 \end{figure}

First, our {full-blown} \gagan{HBM-based} {\texttt{vadvc}} and {\texttt{hdiff}} implementations} {provide} 157.1~GFLOP/s and 608.4~GFLOP/s \gagan{performance}, which are 4.2$\times$ and 8.3$\times$ higher than the performance of a complete~POWER9~socket. 
{For half-precision, if we use} 
the same amount of {PE}s as in single precision, {our accelerator} 
{reaches} a performance of 247.9~GFLOP/s for \texttt{vadvc} ($2.1\times$ {the} single-precision performance) and 1.2~TFLOP/s for \texttt{hdiff} ($2.5\times$ {the} single-precision performance). \gagan{{Our} DDR4-based design achieves 34.1~GFLOP/s and 145.8~GFLOP/s {for \texttt{vadvc} and {\texttt{hdiff}, respectively, which are}} 1.2$\times$ and 2.5$\times$ {the} performance {on the POWER9 CPU}.

Second, for a single PE, which 
{fetches} data from a single memory channel, {the} DDR4-based design provides higher performance {than the HBM-based design}. 
{This is because the DDR4-based FPGA has {a} larger bus width (512-bit) than an HBM port (256-bit). 
This {leads to} a lower transfer rate for an HBM port (0.8-2.1 GT/s\footnote{Gigatransfers per second.}) than for a DDR4 port (2.1-4.3 GT/s). 
One way to match the DDR4 bus width would be to have a single PE fetch data from multiple HBM ports in parallel. 
However, using more ports leads to higher power consumption ($\sim$1 Watt per HBM port).}

Third, as we increase the number of PEs, we observe a linear reduction in the execution time of {the} HBM-based design. 
This is because 
we can {evenly} divide the computation between multiple PEs{,} each \gagan{of which} 
{fetches data} from a separate HBM port. 

Fourth, {in the DDR4-based design, the use of only a single channel to feed multiple {PE}s leads to a congestion issue that causes a non-linear run-time reduction.} {As we {increase the number of accelerator PEs}, we observe that the} {PEs}} 
{compete for a single memory channel, which causes frequent stalls.} {This phenomenon leads to worse performance scaling characteristics for the DDR4-based design as compared to the HBM-based design.}


\subsection{{Energy} Analysis}
We compare the energy consumption of our accelerator to \gagan{a} {16-core} POWER9 host system. For the POWER9 system, we use the AMESTER\footnote{https://github.com/open-power/amester} tool to measure the active power\footnote{Active power denotes the difference between the total power of a complete socket (including CPU, memory, fans, I/O, etc.) when an application is running compared to when it is idle.} consumption. 
{We measure}~99.2~Watt\gagan{s} for \texttt{vadvc}, and ~97.9~Watt\gagan{s} for \texttt{hdiff} by monitoring {built-in} power sensors in 
{the POWER9}~system. 

{By executing these kernels on an HBM-based board, we {reduce the energy consumption by} $22\times$ for {\texttt{vadvc}} and $29\times$ for {\texttt{hdiff}} compared to 
{the} 16-core POWER9 system.} 
\gagan{Figure~\ref{fig:nero/energy_eff} shows the energy efficiency (GFLOPS per Watt) for {\texttt{vadvc}} and {\texttt{hdiff}} 
{on the HBM-} and DDR4-based designs.} 
We make three \gagan{major} observations from the figure. 

 \begin{figure}[h]
  \centering
  \includegraphics[width=0.98\linewidth,trim={0.4cm 0.3cm 0.35cm 0.4cm},clip]{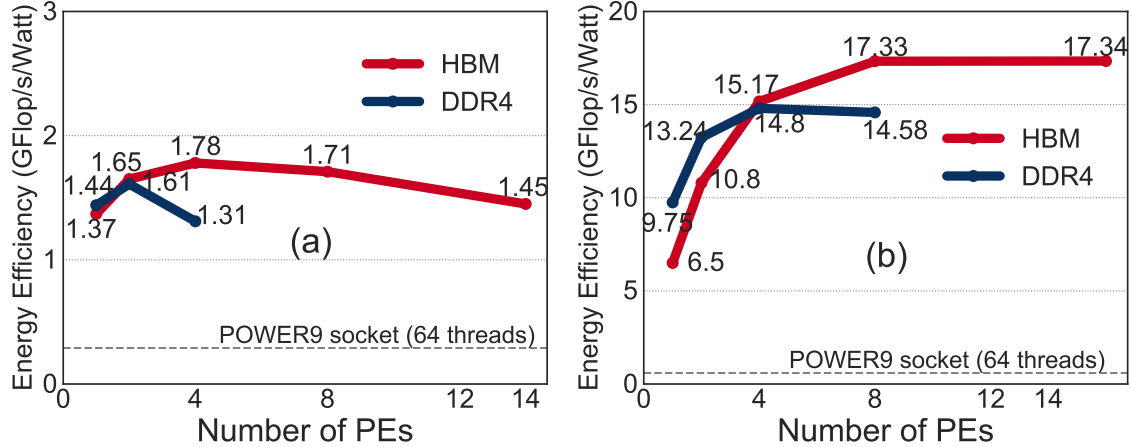}
     \caption{Energy efficiency for (a) \texttt{vadvc} and (b) \texttt{hdiff} on HBM- and DDR4-based FPGA boards. {We also show {the}} single socket (64 threads) energy efficiency of \gagan{an} IBM POWER9 host system for both \texttt{vadvc} and \texttt{hdiff}
  \label{fig:nero/energy_eff}}
 \end{figure}

First, with our 
{full-blown} \gagan{HBM-based} designs \gagan{(i.e., 14 PEs for \texttt{vadvc} and 16 PEs for \texttt{hdiff})}, we achieve 
energy efficiency {values} of 1.5~GFLOPS/Watt and 17.3~GFLOPS/Watt for {\texttt{vadvc}} and {\texttt{hdiff}}, respectively. 

Second, 
the DDR4-based \gagan{design} is more energy efficient {than the HBM-based design} {when the number of PEs is small}. 
{This observation is inline with our discussion about performance with small PE counts in Section~\ref{subsection:nero/evaluation}.} 
However, as we increase the number of {PE}s, the HBM\gagan{-based design provides better energy efficiency} for memory-bound kernels. \gagan{This is because} more data can be fetched {and processed} in \gagan{parallel} {via} multiple ports. 

Third, kernels like {\texttt{vadvc}}, with intricate memory access patterns, \gagan{are not able to} reach the peak {computational power} of FPGAs. {The large amount of control flow in {\texttt{vadvc}} leads to large resource consumption. Therefore, {when} increasing the PE count, we observe a high increase in power consumption with low energy efficiency. } 

{We conclude that} enabling many HBM ports might not always {be} beneficial in terms of energy consumption because each HBM port {consumes} $\sim$1 Watt of power consumption. However, \gagan{data-parallel} kernels like {\texttt{hdiff}} can achieve much higher performance in an energy efficient manner} {with more PEs and HBM~ports.} 

\subsection{{FPGA Resource Utilization}}
Table~\ref{tab:nero/utilization} shows the \gagan{resource} utilization \gagan{of \texttt{vadvc}} and \texttt{hdiff} on the AD9H7 board. {We draw two observations. First, there is a high BRAM consumption compared to other FPGA resources. This is because we implement input, field, and output signals as \texttt{hls::streams}. In {high-level synthesis}, by default, streams are implemented as FIFOs that make use of BRAM. Second,  \texttt{vadvc} has a much larger resource consumption  than \texttt{hdiff} because \texttt{vadvc} has higher computational complexity and requires {a larger} number of fields to \gagan{perform} the \gagan{compound} stencil calculation. \gagan{Note that for} \texttt{hdiff}, we can \gagan{accommodate} more {PE}s, but in this thesis, we  make use of only \gagan{a} single HBM stack. Therefore\gagan{,} we \gagan{use} 16 {PE}s \gagan{because} a single \gagan{HBM} stack offers 16 memory ports.}

\begin{table}[h]
  \caption{FPGA resource utilization {in} {our highest-performing HBM-based design{s for}} \texttt{vadvc} and \texttt{hdiff}}
  \vspace{-0.4cm}
    \label{tab:nero/utilization}
          \begin{center}
\resizebox{0.7\linewidth}{!}{%
\begin{tabular}{llllllc}
\toprule
\textbf{Algorithm} & \textbf{\acrshort{bram}} & \textbf{\acrshort{dsp}} & \textbf{\acrshort{ff}} & \textbf{\acrshort{lut}} & \textbf{\acrshort{uram}} \\ \hline
\texttt{vadvc}              & 81\%            & 39\%           & 37\%          & 55\%           & 53\%                         \\ 
\texttt{hdiff}              & 58\%            & 4\%            & 6\%           & 11\%           & 8\%                           \\ \bottomrule
\end{tabular}
}
  \end{center}
\vspace{-0.2cm}
\end{table}

 \section{Related Work}
\label{sec:nero/relatedWork}
\juang{To our knowledge, this is the first thesis to evaluate the benefits of using FPGAs equipped with \gagann{high-bandwidth memory (HBM) to accelerate} stencil computation.} \gagannn{We exploit near-memory capabilities of such FPGAs to accelerate important weather prediction kernels.}

\gagannn{Modern workloads exhibit limited locality and operate on large amounts of data, which causes frequent data movement between the memory subsystem and the processing units~\cite{ghose2019processing,mutlu2019,googleWorkloads,mutlu2019enabling}. This frequent data movement has a severe impact on overall system performance and energy efficiency. 
A way to alleviate this \emph{data movement bottleneck}~\cite{singh2019near, mutlu2019, ghose2019processing,mutlu2019enabling,googleWorkloads} is \emph{near-memory computing} (NMC), which consists of placing processing units closer to memory. 
NMC is enabled by new memory technologies, such as 3D-stacked memories~\cite{7477494,6757501,6025219,hbm,lee2016smla}, and also by cache-coherent interconnects~\cite{openCAPI, benton2017ccix, cxlwhitepaper},  which allow close integration of processing units and memory units. 
Depending on the applications \gagannn{and systems} of interest (e.g.,~\cite{nai2017graphpim,7056040,lee2018application,kang2013enabling,hashemi2016continuous,akin2015data,babarinsa2015jafar,lee2015bssync,chi2016prime,kim2016neurocube,asghari2016chameleon,boroumand2016lazypim,seshadri2015gather,liu2017concurrent,gao2015practical,morad2015gp,googleWorkloads,teserract,ahn2015pim,hsieh2016accelerating,hashemi2016accelerating}), prior works propose different types of near-memory processing units, such as general-purpose CPU cores~\cite{lee2018application,alian2018application,de2017mondrian,koo2017summarizer,7927081,teserract,nair2015active,6844483,googleWorkloads,boroumand2016lazypim,boroumand2019conda}, GPU cores~\cite{zhang2014top,7756764,tom,ghose2019demystifying}, reconfigurable units~\cite{7446059, jun2015bluedbm,istvan2017caribou,singh2019narmada}, or fixed-function units~\cite{ahn2015pim,hsieh2016accelerating,gu2016biscuit,nai2017graphpim,liu2018processing,kim2018grim,hashemi2016accelerating,hashemi2016continuous}}.

FPGA accelerators are promising to enhance overall system performance \gagann{with low power consumption.}
Past works~\cite{jun2015bluedbm, kara2017fpga, alser2019shouji, giefers2015accelerating,diamantopoulos2018ectalk,Lee:2017:EBG:3137765.3137776,alser2017,chai_icpe19,chang2017collaborative,jiang2020,alser2019sneakysnake} show that FPGAs can be employed {effectively} for a wide range of applications. The recent addition of HBM to \juang{FPGAs} 
\juang{presents an opportunity to exploit} 
high \juang{memory} bandwidth with \gagannn{the} low-power FPGA fabric. 
The potential of high-bandwidth memory~\cite{hbm,lee2016smla} has been explored in many-core processors~\cite{hbm_joins,ghose2019demystifying} and GPUs~\cite{hbm_gpu_data_intensive,ghose2019demystifying}. 
\juang{A recent work~\cite{wang2020} shows the potential of HBM for FPGAs with a memory benchmarking tool.}
\juang{\namePaper} is the first work to accelerate a real-world HPC weather prediction application using \juang{the} FPGA+HBM fabric. {Compared to a previous work~\cite{singh2019narmada} that \juang{optimizes only the} horizontal diffusion kernel \juang{on an FPGA with DDR4 memory}, our analysis reveals \juang{that the} vertical advection kernel has a much lower compute intensity with little to no regularity.
Therefore, this thesis \juang{accelerates} both kernels that together represent the algorithmic diversity of the entire COSMO \gagannn{weather prediction} model. Moreover, compared to 
\cite{singh2019narmada}, \juang{\namePaper~\gagannn{improves performance by}} $1.2\times$ on a DDR4\gagannn{-}based \gagann{board} and $37\times$ on an HBM-based board for horizontal diffusion by using a dataflow implementation with auto-tuning.}

\gagan{Enabling higher performance for stencil computations has been a subject of optimizations 
across the \juang{whole} computing stack~\cite{sano2014multi,7582502,chi2018soda,de2018designing,christen2011patus,  datta2009optimization, meng2011performance, henretty2011data, strzodka2010cache, tang2011pochoir,gonzalez1997speculative,armejach2018stencil,gysi2015modesto}}. \juang{Szustak~\etal} accelerate the MPDATA advection scheme on multi-core CPU~\cite{szustak2013using} and computational fluid dynamics kernels on FPGA~\cite{mpdata}. Bianco~\etal~\cite{bianco2013gpu} \juan{optimize the COSMO \gagannn{weather prediction} model} for GPUs while Thaler~\etal~\cite{cosmo_knl} \juan{port} COSMO to a many-core system.  Wahib~\etal~\cite{wahib2014scalable} \juang{develop} an analytical performance model for choosing an optimal GPU-based execution strategy for various scientific applications, including COSMO. 
Gysi~\etal~\cite{gysi2015modesto} \juang{provide} guidelines for optimizing stencil kernels for CPU--GPU systems.


 \section{Conclusion}
\label{sec:conclusion}
\juang{We introduce} \namePaper, the first design and implementation on a reconfigurable fabric \juang{with \gagann{high-bandwidth memory} (HBM)} to accelerate representative weather prediction kernels, i.e., vertical advection (\texttt{vadvc}) and horizontal diffusion (\texttt{hdiff}), from a real-world weather prediction application. 
These kernels are compound stencils that are found in various weather prediction applications, including the COSMO  model. \gagannn{We show that c}ompound kernels do not perform well on conventional architectures due to their complex data access patterns and low data reusability, which \gagann{make} them memory-bounded. 
\juang{Therefore, they greatly benefit from our near-memory computing solution \gagannn{that} takes advantage of the high \gagann{data transfer} bandwidth of HBM.}


\juang{\namePaper's implementations of \texttt{vadvc} and \texttt{hdiff} outperform the optimized software implementations on a 16-core POWER9 with \gagann{4-way multithreading} by $4.2\times$ and $8.3\times$, with $22\times$ and $29\times$ less energy consumption, respectively. 
We conclude that hardware acceleration on \gagann{an} FPGA+HBM fabric is a promising solution for compound stencil\gagann{s} present in weather prediction applications. \gagann{We hope that o}ur \gagann{reconfigurable near\gagannn{-}memory} accelerator inspire\gagann{s} developers of different high-performance computing applications that suffer from \gagann{the} memory bottleneck.
}

 \chapter{Low Precision Processing for High Order Stencil Computations}
\chapternote{The content of this chapter was published as \emph{``Low Precision  Processing  for  High Order  Stencil Computations''} in Springer LNCS 2019.}
\label{chapter:low_precision_stencil}
Modern scientific workloads have demonstrated the inefficiency of using high-precision formats. Moving to a lower bit format or even to a different number system can provide tremendous gains in terms of performance and energy efficiency. This chapter explores the applicability of different number formats and searches for the appropriate bit-width for three-dimensional stencil kernels, which are among the most widely used scientific workloads. Further, we demonstrate the achievable performance of these kernels on state-of-the-art hardware that includes a host CPU connected to an FPGA. An FPGA provides us with the capability to implement arbitrary fixed-point precision. Thus, this chapter fills the gap between current hardware capabilities and future systems for stencil-based scientific applications.


%
%
\vspace{-0.7cm}

\section{Introduction}
Stencil computation is essential for numerical simulations of finite difference methods~(FDM)~\cite{ozicsik2017finite} and is applied in iterative solvers of linear equation systems. We use stencil computation in a wide range of applications, including computational fluid dynamics~\cite{huynh2014high}, image processing~\cite{hermosilla2008non}, weather prediction modeling~\cite{doms1999nonhydrostatic}, etc. A stencil operation~\cite{gysi2015modesto} defines a computation sequence where elements in a multidimensional grid are updated using data values from a subset of its neighbors based on a fixed pattern.

Stencil computation is applied to data structures that are generally much larger than the available system's cache capacity~\cite{datta2008stencil}. Stencils are cache-unfriendly because the amount of data reuse within a stencil iteration is limited to the number of points in a stencil. Due to the cache-unfriendly, complex data
access patterns, and low operational intensity~\cite{stencilOnPOWER8,singh2020nero}, stencil compute kernels do not perform well on traditional CPU or GPU-based systems. 

High-performance implementations of stencils on modern processors operate using a single-precision or a double-precision floating-point data type. The floating-point format is the most widely supported datatypes by our current hardware devices. Using this data type in real-world applications, which use large grid sizes, puts enormous stress on the memory subsystem.  Therefore, storing data in the memory using a smaller number of bits can decrease the memory footprint and provide reductions in latency and energy consumption. The industry trend~\cite{finnerty2017reduce} shows a clear shift away from using floating-point representation. For example, applications like
neural networks can use an 8-bit fixed-point format or lower precision without significant loss in accuracy~\cite{iwata1989artificial}. Hence, in this chapter, we examine the use of different number systems -- fixed-point, posits, floating-point -- and analyze the precision tolerance of three-dimensional stencil kernels, one of the most widely used kernels in real-world applications.

\section{Background}
\label{sec:low_precision/background}
This section provides details on the stencil kernels used and discusses the relevance of the precision analysis.
\vspace{-0.2cm}
\subsection{Stencil Benchmark}
Stencil computation updates a multi-dimensional grid based on a specific computation pattern. A stencil kernel's performance on the current multicore system depends heavily on the data mapping of the grid. For instance, suppose a 3D grid in (\textit{row}, \textit{column}, \textit{depth}). When the grid is stored by \textit{row}, accessing data elements in the other dimensions typically results in cache eviction. This issue is because, for real-world applications, the problem size is too large to fit in the processor cache.
\begin{figure}[htp]

\begin{subfigure}[t]{0.5\textwidth}
\centering
\includegraphics[bb=54 138 703 460,width=\linewidth,trim={0.8cm 4.9cm 0.4cm 2cm},clip]{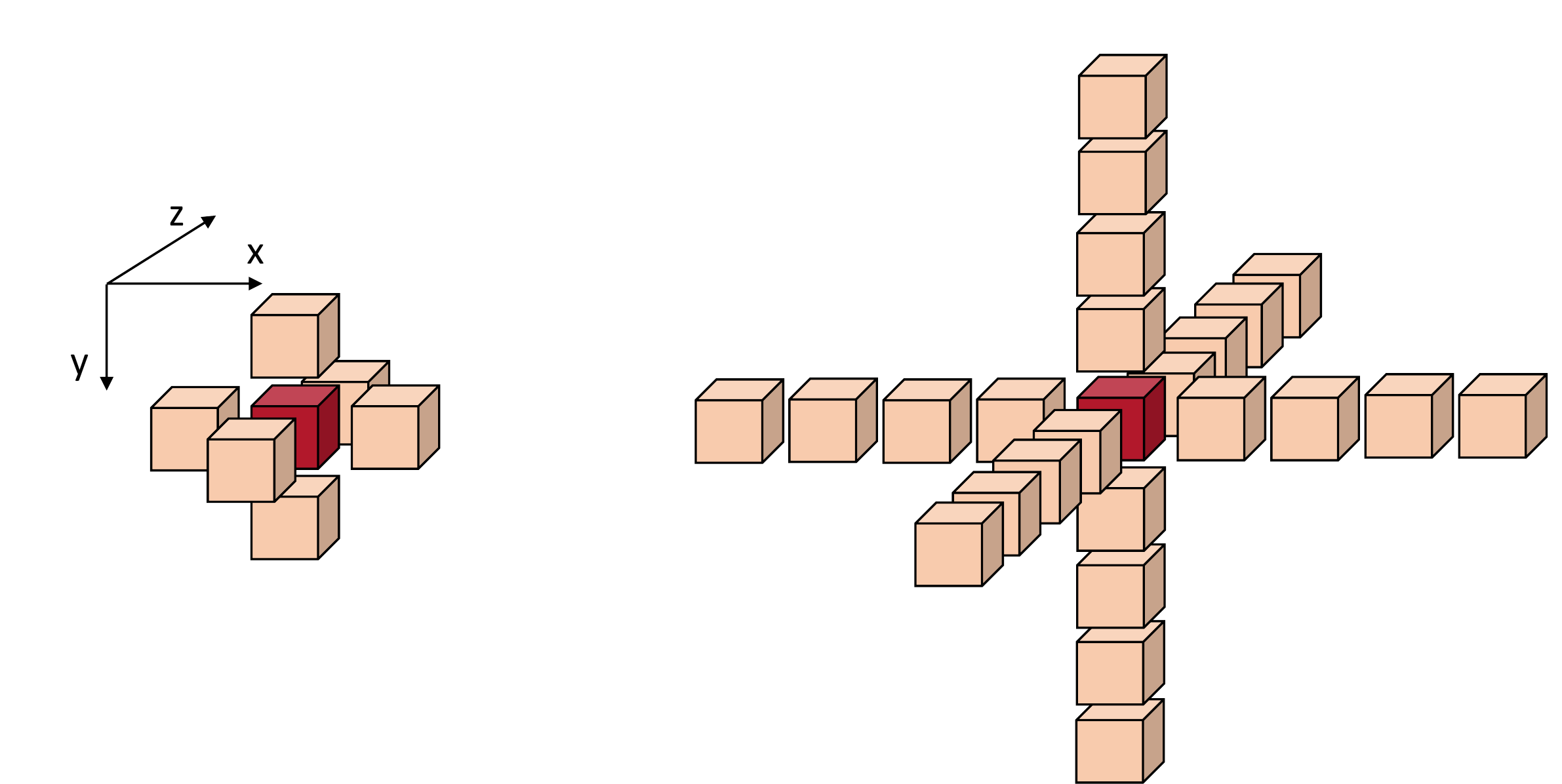}
 \caption{
\label{fig:low_precision/stencil7&25}}
\end{subfigure}
  \begin{subfigure}[t]{0.5\linewidth}
\centering
\resizebox{0.8\textwidth}{!}{
    \begin{tikzpicture}[rotate=90,transform shape]

    \gridThreeD{0}{8.25}{black};
    \node at (7.5,10.25)[ rotate=270] {\huge \textit{Laplace}};
      \dotesUpper{0}{8.25};
    \gridThreeDSecond{1}{4}{black};
    \node  at (7.5,5.5)[ rotate=270] {\huge \textit{Flux}};
     \lowerDots{3}{4};
     
    \gridThreeDThird{2}{0}{black};
     \node  at (7.5,0.25) [ rotate=270] {\huge \textit{Output}};
     \lowestDot{1.6}{-0.45}

      \draw [thick,->,blue] (2.75,4.25) -- (2.7,8.4);
       \draw [thick,->,blue] (2.75,4.25) -- (2.1,8.9);
       \draw [thick,->,blue] (2.75,4.25) -- (4.1,8.9);
       \draw [thick,->,blue] (2.75,4.25) -- (3.5,9.4);

       \draw [thick,->,orange] (2.1,4.7) -- (3.5,9.4);
       \draw [thick,->,orange] (2.1,4.7) -- (1.5,9.4);
       \draw [thick,->,orange] (2.1,4.7) -- (2.1,8.9);
       \draw [thick,->,orange] (2.1,4.7) -- (2.9,9.9);
     
       \draw [thick,->,red] (4.1,4.7) -- (3.5,9.4);
       \draw [thick,->,red] (4.1,4.7) -- (5.5,9.4);
       \draw [thick,->,red] (4.1,4.7) -- (4.1,8.9);
       \draw [thick,->,red] (4.1,4.7) -- (4.9,9.9);

           \draw [thick,->,brown] (3.5,5.25) -- (3.5,9.4);
           \draw [thick,->,brown] (3.5,5.25) -- (2.9,9.9);
           \draw [thick,->,brown] (3.5,5.25) -- (4.9,9.9);
           \draw [thick,->,brown] (3.5,5.25) -- (4.3,10.4);
         \draw [thick,->,violet] (2.7,0.25) -- (3.5,5.15);
         \draw [thick,->,violet] (2.7,0.25) -- (4.1,4.65);
         \draw [thick,->,violet] (2.7,0.25) -- (2.1,4.65);
         \draw [thick,->,violet] (2.7,0.25) -- (2.75,4.15);

    

\end{tikzpicture}
}
 \caption{
 \label{fig:low_precision/stencilHdiff}}
\end{subfigure}
\vspace{-0.1cm}
\caption{(a) 7-point stencil and 25-point elementary stencils (b) Compound horizontal diffusion stencil that is used by the COSMO weather prediction model 
\label{fig:low_precision/kernels}}
\vspace{-0.25cm}
\end{figure}
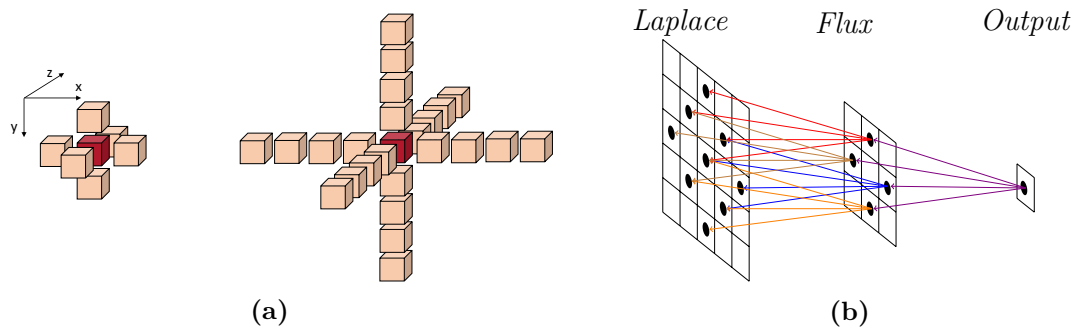
This chapter focuses on both a \sevpoint and \tfivepoint 3D elementary stencil and a compound horizontal diffusion (\hdiff) stencil, shown in Figure~\ref{fig:low_precision/kernels}. These kernels access a three-dimensional grid and have complex access patterns.  The 3D \sevpoint and \tfivepoint (see Figure~\ref{fig:low_precision/stencil7&25}) stencils commonly arise
from the finite difference method for solving partial differential equations~\cite{stencilOnPOWER8}. 
 The \sevpoint stencil performs
eight FLOPS per grid point, while the \tfivepoint stencil performs twenty-seven FLOPS per grid point (without any common subexpression elimination). Thus, the arithmetic intensity,
the ratio of FLOPS performed for each byte of memory traffic, is much higher
for the \tfivepoint stencil than the \sevpoint stencil.  

As discussed in Chapter~\ref{chapter:nero}, stencil patterns in a real-world weather prediction application consist of a collection of stencils that performs a sequence of element-wise computations. Horizontal diffusion kernel is an example of one such kernel that executes each
stencil using a separate loop nest. It iterates over a 3D grid that performs \textit{laplacian} and \textit{flux}, as depicted in Figure~\ref{fig:low_precision/stencilHdiff}, as well as calculations for different grid points. Such compound kernels have intricate memory access patterns because they apply a series of elementary stencil operations. Although such implementations may be straightforward to write, they are not efficient in terms of data
locality, memory usage, or parallelism.






\vspace{-0.2cm}
\subsection{Precision Optimization}
IEEE-754 floating-point representation has become the universal standard in modern computing systems. Floating-point numbers have a mantissa and exponent component with an additional bit to represent the sign of a number. In terms of computing resources, this floating-point arithmetic requires complex circuitry leading to high latency and power consumption~\cite{finnerty2017reduce}. 

The use of low-precision arithmetic with a minimal loss in the accuracy has been proposed as a promising alternative to the commonly used floating-point arithmetic for emerging workloads, e.g.,
machine learning and graph processing. From the system perspective, there are two main benefits of moving to a
lower precision. First, the hardware resources
for a given silicon area may enable higher operations per second
(OPS) at a lower precision as these operations require less hardware
area, and thus power. Note, this also necessitates efficient memory traffic management. Secondly, many operations are memory bandwidth bound~\cite{singh2019near,NAPEL}, and reducing precision would allow for better cache usage and reduction of bandwidth bottlenecks.
Thus, data can be moved faster through the memory hierarchy to maximize the utilization of computing resources.
\section{Methodology}
\label{sec:low_precision/methodology}
\begin{figure}
\centering
\includegraphics[bb=2 21 848 194,width=\linewidth,trim={0cm 0.7cm 0cm 0cm},clip]{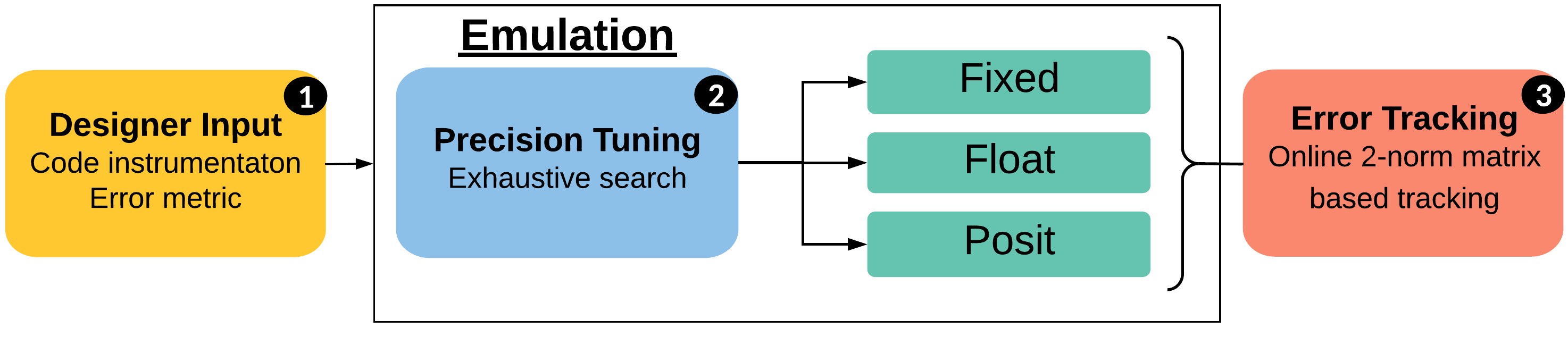}
\vspace{-0.3cm}
\caption{Overview of application precision exploration. The designer inputs the code with an appropriate precision template. Exhaustive precision exploration is performed for different number systems that include fixed-point arithmetic, floating-point arithmetic, and posit arithmetic. While exploring, error tracking is performed using the 2-norm matrix approach
\label{fig:low_precision/methodology}}
\end{figure}

The following section provides detail on our methodology to explore precision for different number systems, as depicted in Figure~\ref{fig:low_precision/methodology}. In the first phase (\circled{1}), we analyze and instrument a part of an application for which the precision exploration needs to be performed. In the next phase (\circled{2}), we execute an exhaustive search to find the appropriate precision based on the number system used. In this chapter, we make use of fixed-point, floating-point, and posit number systems. During the exhaustive design space exploration, continuous error tracking (\circled{3}) is performed to measure the extent of accuracy deviation compared to the IEEE floating-point arithmetic format.

\vspace{-0.5cm}

 \paragraph{Accuracy:} In our experiments, for precision tuning, we consider the induced 2-norm of a matrix~\cite{normError} as our measure of the accuracy. A matrix norm is a vector norm in a vector space. The induced 2-norm of an \textit{m}$\times$\textit{n}  matrix \textit{A} is the supremum of the ratio between the 2-norm of a vector \textit{Ax} and the 2-norm of \textit{x}, where \textit{x} is an  \textit{n}-dimensional vector. We calculate the relative norm or mean relative error (MRE)  $\epsilon_i$ to indicate how close the predicted value $A_{i}^\prime$ is to the actual value $A_{i}$. MRE provides an unbiased estimate of the error variance between two matrices.
\vspace{-0.3cm}
\begin{align}
 \epsilon_i= \frac{||A_{i}^\prime -A_{i}||_{2}}{||A_{i}||_{2}}
\label{eq:low_precision/2}
\end{align}

\subsection{Evaluated Arbitrary Precision}
\begin{figure}
\centering
\includegraphics[bb=1 3 707 241,width=\linewidth]{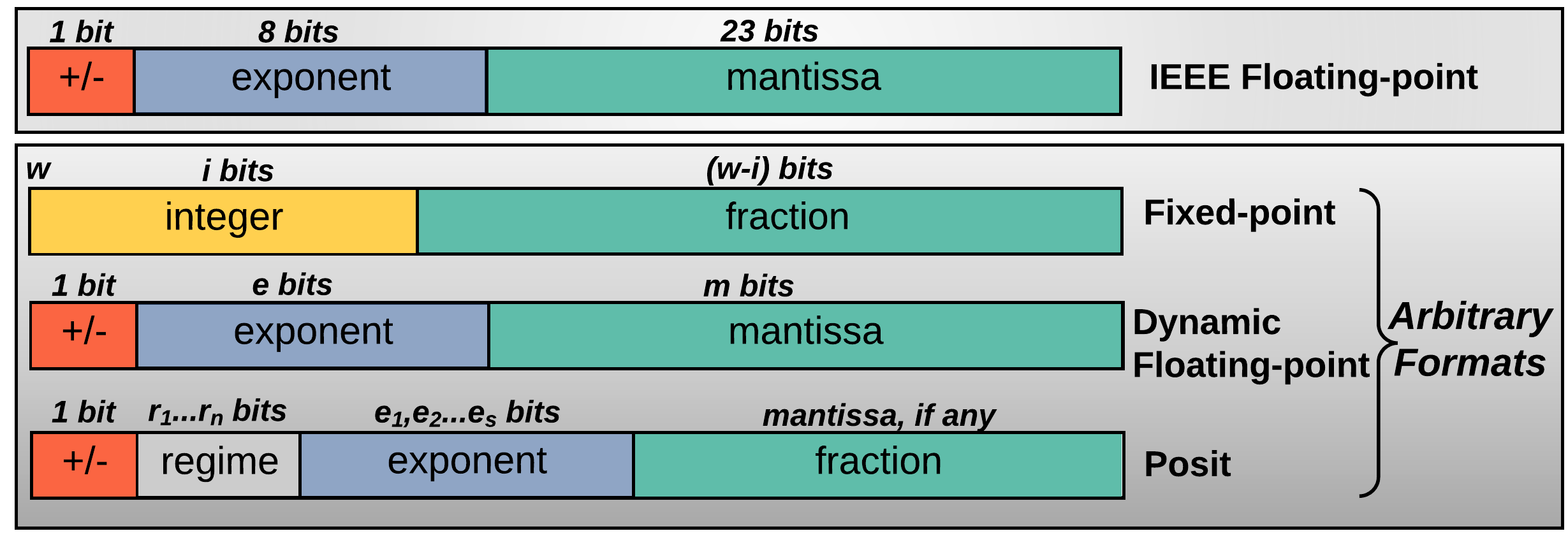}
\caption{Arithmetic types used with widths indicated above each field. IEEE single precision floating-point number is 32-bit where a positive sign bit is represented by a 0 and a negative by 1.  Fixed-point has fixed integer and fraction bits where w (total bits) could be any multiple of 2, based on the bitwidth of the data path. Dynamic floating-point arithmetic uses arbitrary exponent and mantissa bits. A posit number~\cite{gustafson2017beating} is similar to floating-point with additional bits for the regime part. It has $e_s$ exponent bits, but depending upon the data this could be omitted (same is valid for mantissa bits)
\label{fig:low_precision/arithemacticTypes}}
\end{figure}

As an alternative to the currently used IEEE single and double-precision floating-point representation, we explore the precision tolerance of 3D stencil kernels using the following number formats (see Figure~\ref{fig:low_precision/arithemacticTypes}):

1) Fixed-Point Arithmetic: A fixed-point consists of an integer and a fraction part where total width could be any multiple of 2, based on the bit-width of the data path. Compared to the floating-point format, fixed-point numbers simplify the logic by fixing the radix point. 

In an FPGA, the fixed-point format offers a more resource-efficient alternative to the floating-point implementation. This efficiency is because floating-point support often uses more than 100$\times$ as many gates compared to fixed-point support~\cite{finnerty2017reduce}.

2) Dynamic Floating-Point Arithmetic: By lowering the precision of a floating-point format, we could retain the advantages of floating-point arithmetic (e.g., higher dynamic range) with a lower bit-width. Dynamic floating-point arithmetic uses an arbitrary number of bits for the exponent and significand (or mantissa) parts of a floating-point number.

3) Posit Arithmetic: Posit\cite{gustafson2017beating} borrows most of the components from the IEEE 754 floating-point scheme, such as the exponent and fraction (or mantissa) fields. However, posit has an additional \textit{regime} bit introduced to create a tapered accuracy, which lets small exponents have more accuracy. One
could choose to either represent a large number by assigning more bits to the exponent
field or opt for more decimal precision by having more fraction bits. 

Figure~\ref{fig:low_precision/arithemacticTypes} shows the different datatypes explored in this chapter. While analyzing these types, there are several things to take into account. Firstly, posit can provide the highest dynamic range compared to the other number systems, and fixed-point offers the lowest~\cite{carmichael2018deep}. Additionally, floating-point numbers are susceptible to rounding errors and could lead to an overflow or underflow~\cite{gustafson2017beating}. We determine the precision bit-width through bit accurate simulations for different bit-width configurations. While changing bitwidth, we analyze the trend of the relative error.
\vspace{-0.2cm}

\section{Evaluation}
\label{sec:low_precision/evaluation}
We use IBM\textsuperscript{\textregistered} POWER9 as the host system comprising of 16 cores, each of which supports four-thread simultaneous multi-threading. Table~\ref{tab:low_precision/systemparameters}
provides complete details of our system parameters. To provide a full-scale analysis of stencil optimization techniques, we set the grid size of all stencil kernels as $1280\times1080\times960$, much larger than the on-chip cache capacity of POWER9, with input data distribution as a Gaussian function. The problem size dictates which input dataset would reside in the cache; hence is an important parameter while measuring the system performance.  Note: in Chapter~\ref{chapter:nero}, we use a grid domain used by the COSMO weather
prediction model.

  \vspace{0.2cm}
\begin{table}[h]
  \caption{System parameters and hardware configuration for the CPU and the FPGA board}
  \vspace{-0.4cm}
    \label{tab:low_precision/systemparameters}
   
      \begin{center}
      \footnotesize
\resizebox{0.75\linewidth}{!}{%
\begin{tabular}{|l@{\hspace{0.1\tabcolsep}}|p{7cm}|}

\hline
\textbf{Host CPU} 

 & 16-core IBM POWER9 AC922 \\&@3.2 GHz, 4\gagann{-way} SMT\\
 \hline
 \textbf{Cache-Hierarchy}&32 KiB L1-I/D, 256 KiB L2, 10 MiB L3 \\
 \hline
 \textbf{System Memory}&16x32GiB RDIMM DDR4 2666 MHz \\
\hline
 \begin{tabular}[c]{@{}l@{}}
 \textbf{DDR4-based} \\\textbf{FPGA Board} \end{tabular}  & \begin{tabular}[c]{@{}l@{}}
 Alpha Data
ADM-PCIE-9V3\\
 Xilinx Virtex Ultrascale+ XCVU3P-2\\
 8GiB (DDR4) with PCIe Gen4 x8
  \end{tabular}\\



\hline

\end{tabular}
}
  \end{center}
 \end{table}

For precision tuning of the fixed-point number system, we use the Xilinx fixed-point library from the Vivado 2018.2 tool~\cite{xilinx_hls}. We use the C++ template-based \emph{FloatX} (Float eXtended) library~\footnote{https://github.com/oprecomp/FloatX} to explore arbitrary precision for floating-point arithmetic. Software-based posit implementation is available as part of the ongoing efforts to develop an ecosystem for posit evaluation\footnote{https://github.com/stillwater-sc/universal}. All three libraries are provided as a C++ header format, which allows us to replace the data types in the source code of the application and study the effect of low precision using the same software toolchain as that of the application itself.
We develop a highly optimized FPGA accelerator for all the kernels to make a performance comparison between floating-point and fixed-point number systems. We implement these designs on an Alpha-Data ADM-PCIE-9V3~\cite{ad9v3} card featuring the Xilinx Virtex Ultrascale+ XCVU3P-FFVC1517-2-i device.

\subsection{Emulated Precision Tuning}
\begin{figure}[h]
 \centering
 \includegraphics[bb=10 12 725 264,width=\linewidth]{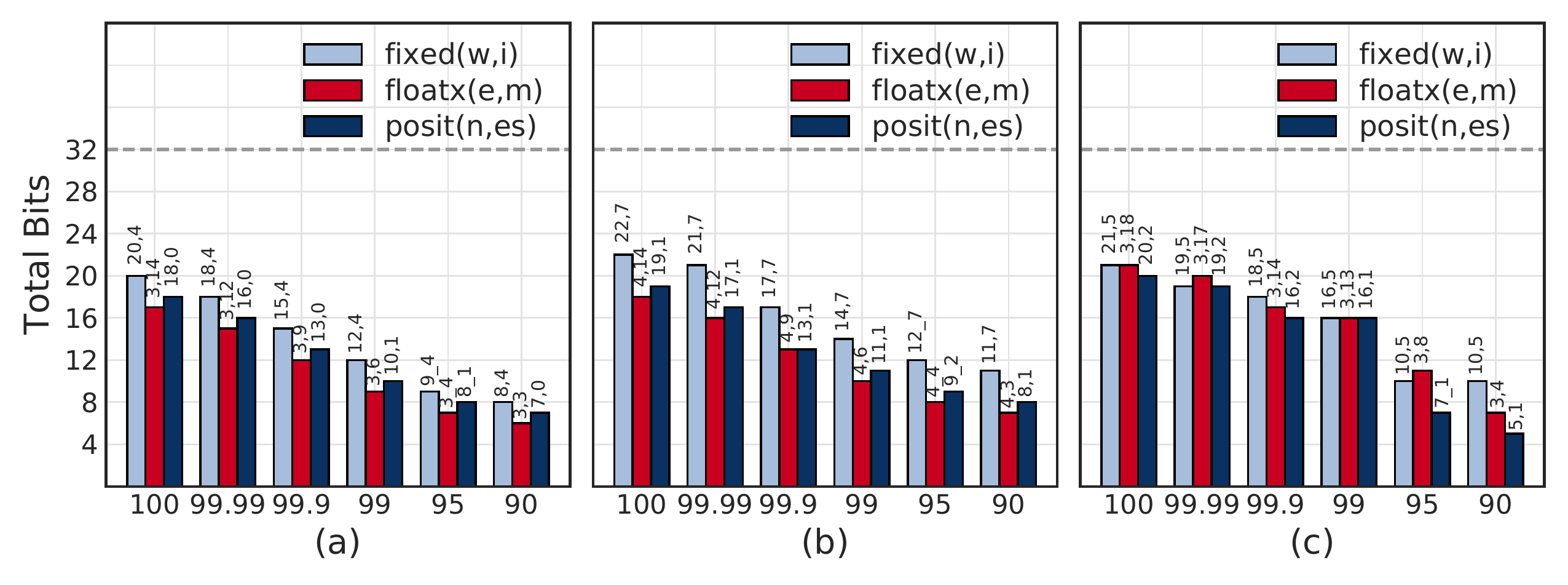}
 \caption{Total bits vs accuracy (percentage) for (a) \sevpoint, (b) \tfivepoint, and (c) horizontal diffusion compared to single-precision IEEE floating-point representation. Notation fixed (\textit{w},\textit{i}) defines a fixed number with total \textit{w} bits including \textit{i} integer bits. With floatx, \textit{e} refers to the exponent bits and \textit{m} defines the mantissa. In the case of the posit number system, \textit{n} is the total number of bits with \textit{es} bits for the exponent part
 \label{fig:low_precision/precisionResults}}
 \end{figure}

The tuning process analyzes multiple configurations for each of the arithmetic types considered. The tuner re-executes the program for each
configuration and computes the error on its output values to
provide a measure of the resultant accuracy.
Figure~\ref{fig:low_precision/precisionResults} shows the precision results for the considered workloads for three different number systems. The accuracy is compared to the most ubiquitously used IEEE single-precision floating number system. 

For all the kernels, we can achieve full accuracy with much lower bits. Moreover, as the error tolerance increases, we could use a lower number of total bits. Based on this, we make three observations. First, in the case of a \sevpoint and \tfivepoint stencil, we could reduce bits by more than $50\%$ for the considered three data types, with a precision loss of only $1\%$. Second, elementary 3D stencil kernels (\sevpoint and \tfivepoint) could not exploit the high dynamic range offered by posit. Therefore, with a lower bit-width floating-point arithmetic, we could achieve better results. Third, the weather compound kernel comparatively needs a higher dynamic range; therefore, with $0.1\%$ tolerance in the accuracy, we could cut the number of bits to half compared to the IEEE floating-point and move to a posit of (16,2). This observation motivates the use of posit number format can be useful in the domain of weather prediction modeling.

 \subsection{Case Study for Current Multi-Core Systems and Arbitrary Precision Supported Hardware}
We perform a case study to measure the capabilities of current state-of-the-art hardware platforms.
 We tune the considered stencil kernels both for IBM POWER9 CPU and for a high-end FPGA platform. Our FPGA is coherently attached to our host CPU through the CAPI2 link. For the FPGA and the POWER9 node, we use the AMESTER\footnote{https://github.com/open-power/amester} tool to measure the active power consumption.
 
 Our current FPGA devices only support floating-point and arbitrary fixed-point arithmetic. Therefore, we compared hardware implementations across the stencil benchmarks for floating-point single and half precision with fixed-point datatype for the bit width that gave similar accuracy to the floating-point. Note, as current state-of-the-art hardware devices do not support the posit data type, we did not include it in our hardware comparison because the emulation of posit data type would be expensive in an FPGA and would lead to unfair comparisons with other data types. In Appendix~\ref{chapter:precise_fpga}, we develop PreciseFPGA, an automated framework to obtain an application-aware optimal fixed-point configuration without exhaustively searching the entire design space.

\begin{figure}[h]
\centering
\begin{subfigure}[h]{.6\textwidth}
  \centering
  \includegraphics[bb=1 7 847 346,width=\linewidth]{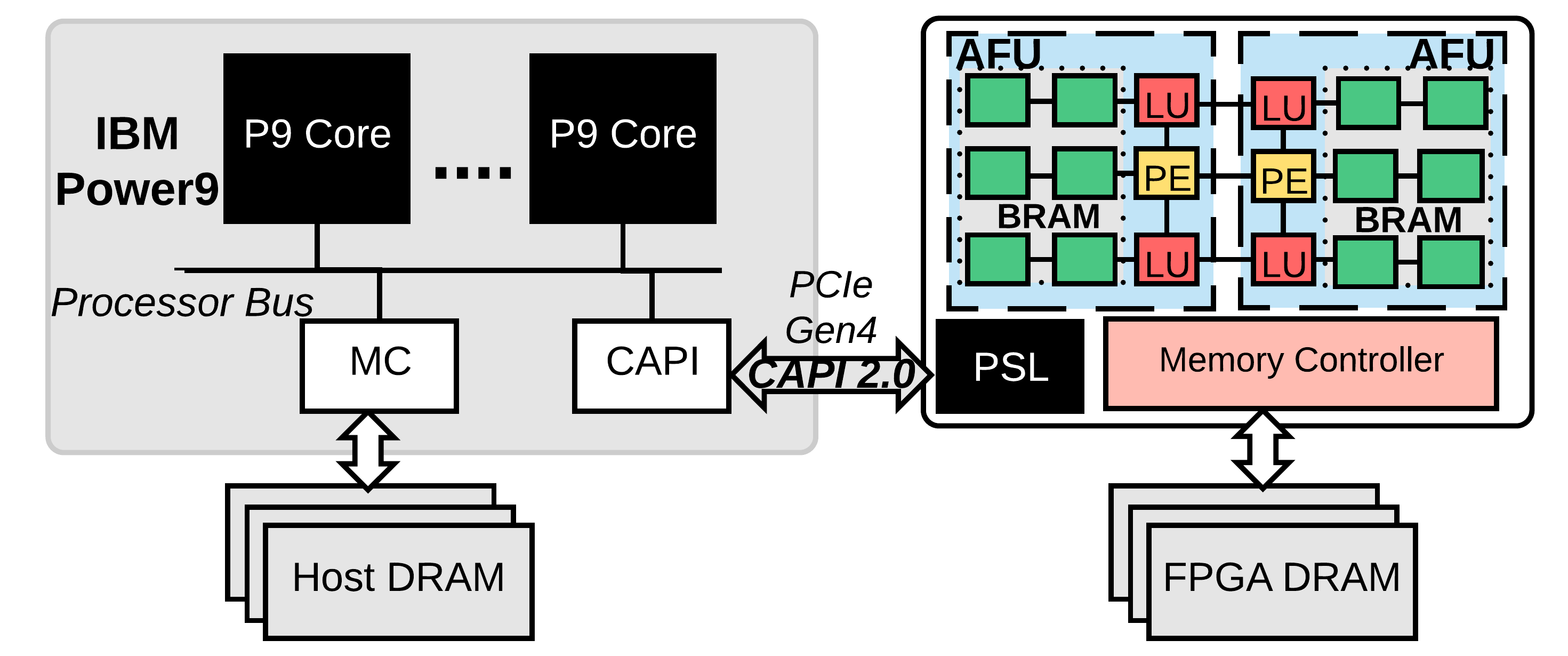}
  \caption{
  \label{fig:low_precision/system}}
\end{subfigure}%
\begin{subfigure}[h]{.4\textwidth}
  \centering
  \includegraphics[bb=5 4 339 192,width=\linewidth]{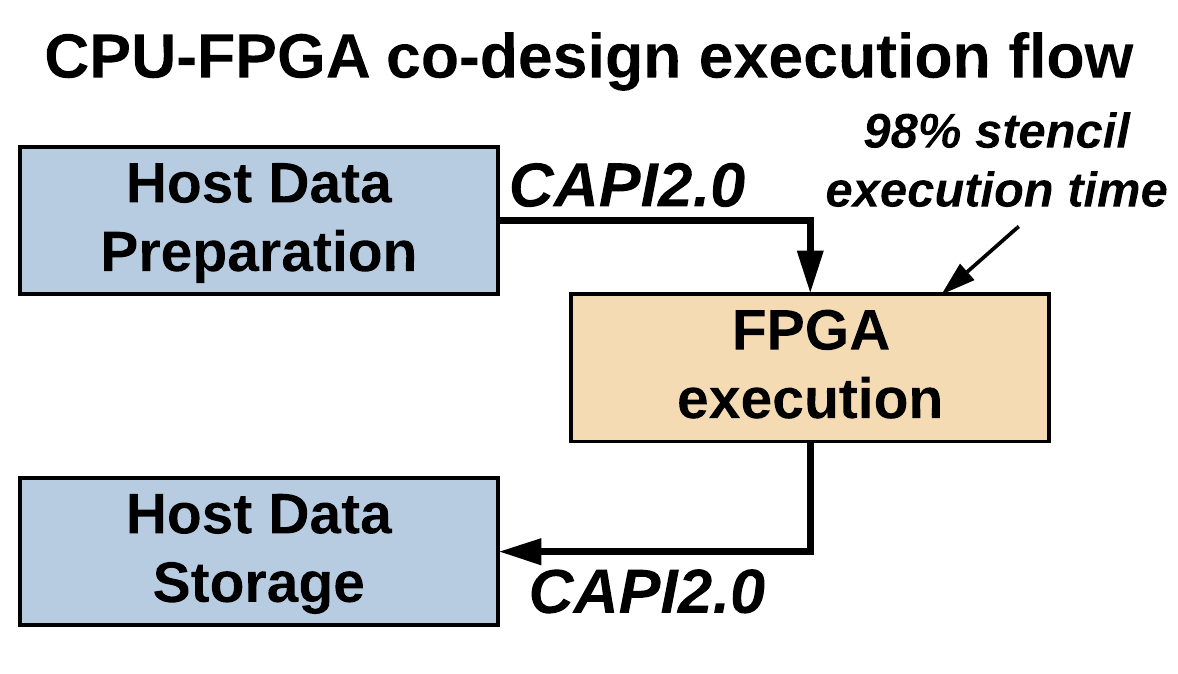}
    \caption{
  \label{fig:low_precision/execution}}
\end{subfigure}
 \vspace{-0.1cm}
\caption{(a) CAPI 2-based accelerator platform with IBM\textsuperscript{\textregistered} POWER9 (b) FPGA is acting as a peer to the  CPU by accessing the main memory through a high-performance CAPI2 link, enabled by PSL. Data flow sequence from the  Host  DRAM  to the onboard  FPGA  memory. A software-defined API handles offloading jobs to accelerators with an interrupt-based queuing mechanism that allows minimal CPU usage (thus, power) during FPGA use}
\end{figure}

 In Figure~\ref{fig:low_precision/system}, we show a high-level overview of our integrated system. The FPGA is connected to a server system, based on the IBM\textsuperscript{\textregistered} POWER9 processor, using IBM\textsuperscript{\textregistered} coherent accelerator processor interface 2.0 (CAPI~2.0). The FPGA implementation consists of accelerator function units (AFU) that interact with the power service layer (PSL), which is the CAPI endpoint on the FPGA. The co-designed execution flow is shown in Figure~\ref{fig:low_precision/execution}. We provide the experimental results of tuning stencil kernels for current CPU and FPGA-based systems. 
 
\begin{figure}[h]
 \centering
 \includegraphics[width=1.1\linewidth,trim={1.2cm 0.6cm 0.4cm 0.4cm},clip]{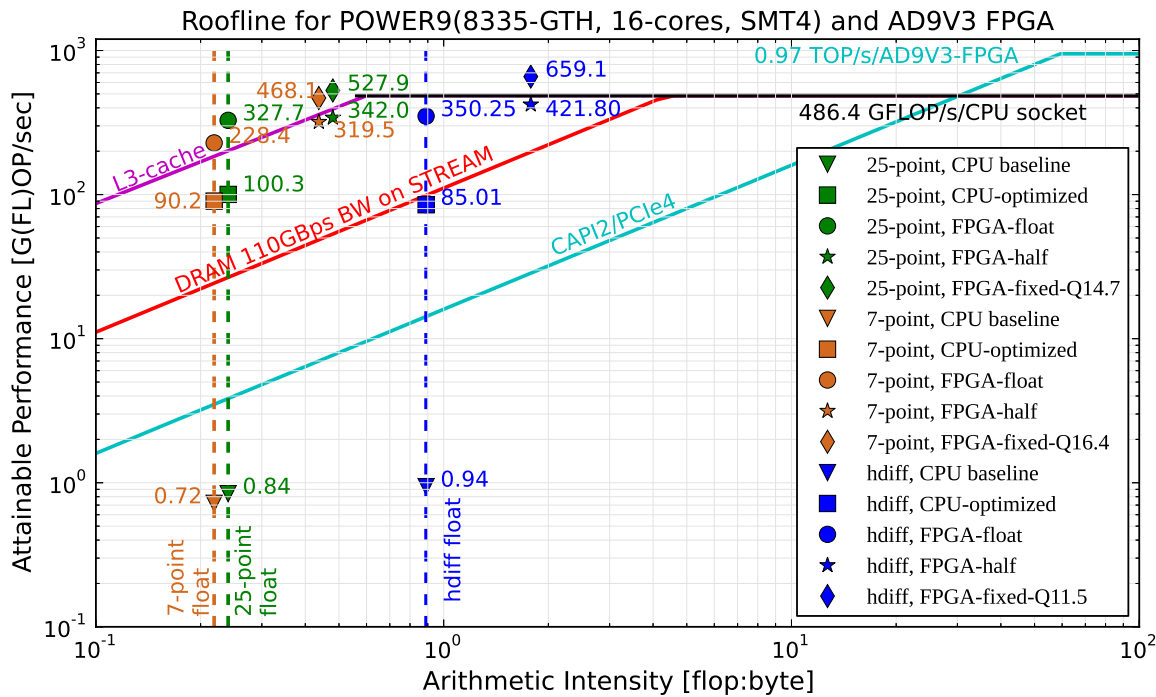}
 \caption{Roofline~\cite{williams2009roofline} for POWER9 (1-socket) showing elementary stencil (\sevpoint and \tfivepoint) and horizontal diffusion (\texttt{hdiff}) kernels for single-thread baseline and 64-thread fully-optimized 
    implementations. 
    The plot shows also the roofline 
    of the FPGA with attained performance of our examined stencils using different precision data types
 \label{fig:low_precision/roofline}}
 \end{figure}

 Figure~\ref{fig:low_precision/roofline} shows the roofline of the three stencil kernels  (\sevpoint, \tfivepoint, and \hdiff)  that we use in this study. By mapping both, arithmetic intensity of all examined stencils and peak attainable GFLOP/sec (GOP/sec for fixed-point) on the roofline of our heterogeneous system (CPU+FPGA), we make the following three observations. First, we observe that compiler and tiling optimizations~\cite{stencilOnPOWER8} lead to 125.2$\times$ 119.4$\times$ and 90.4$\times$ speedup compared to non-optimized CPU implementations for \sevpoint, \tfivepoint, and \hdiff, respectively. The memory bandwidth constrains the performance of elementary stencils (\sevpoint, \tfivepoint) since the stencil data for the 3D grid cannot be mapped to contiguous memory location leading to limited cache locality. Although \hdiff has a higher arithmetic intensity, its access patterns are more complex because it applies a series of elementary stencil operations with different stencil patterns. 
 
Second, we observe that the floating-point FPGA implementations increase the additional speedup to 2.5$\times$, 3.3$\times$, and 4.1$\times$ compared to the CPU-optimized implementation for \sevpoint, \tfivepoint, and \hdiff, respectively. By effectively using the FPGA's on-chip memory, the FPGA-based implementations are not constrained by the DRAM memory bandwidth. However, the CAPI2/PCIe4 link offers an order of magnitude less bandwidth than that of the host CPU's DRAM. Since our platform offers memory-coherent access of FPGA to the system memory, we build a pipelined execution, where communication time for transferring data from host to FPGA memory is masked with the actual FPGA processing~\cite{diamantopoulos2018ectalk}. This technique allows us to exploit FPGA processing capabilities completely. 

Third, we have measured additional gains by replacing a single-precision floating-point data type with a lower precision data type.  Specifically, in the roofline of Figure~\ref{fig:low_precision/roofline}, we plot the performance of three stencils using half and fixed-point data types. The specific bit-width for the integer and fractional part of the fixed point was selected at 99\% accuracy, i.e., Q14.7 for \tfivepoint, Q16.4 for \sevpoint, and Q11.5 for \hdiff. Arithmetic intensity is improved for both half and fixed data types since the bytes fetched from memory are half that of the single-precision floating-point (i.e., 2 bytes instead of 4 bytes). Since fixed-point implementations use fewer resources on an FPGA than float and half, we were able to add more accelerators on the same FPGA device, allowing us to measure 468.1, 527.9, and 659.1 GOPs/sec for \sevpoint, \tfivepoint, and \hdiff, respectively. These numbers are very close to the theoretical peak performance of   0.97 TOPs/s offered by our FPGA device 
\footnote{While the three stencils comprise different access patterns and acceleration kernels, the primary operations, i.e., vectorized multiply-accumulate computation (MAC), which define the FPGA micro-architecture, remain the same. Using vectorized 
MAC, we have calculate 0.97 TOPs/s theoretical top performance for stencils for our AD9V3 FPGA.}.



\begin{figure}[h]
 \centering
 \includegraphics[bb=14 12 509 191,width=\linewidth]{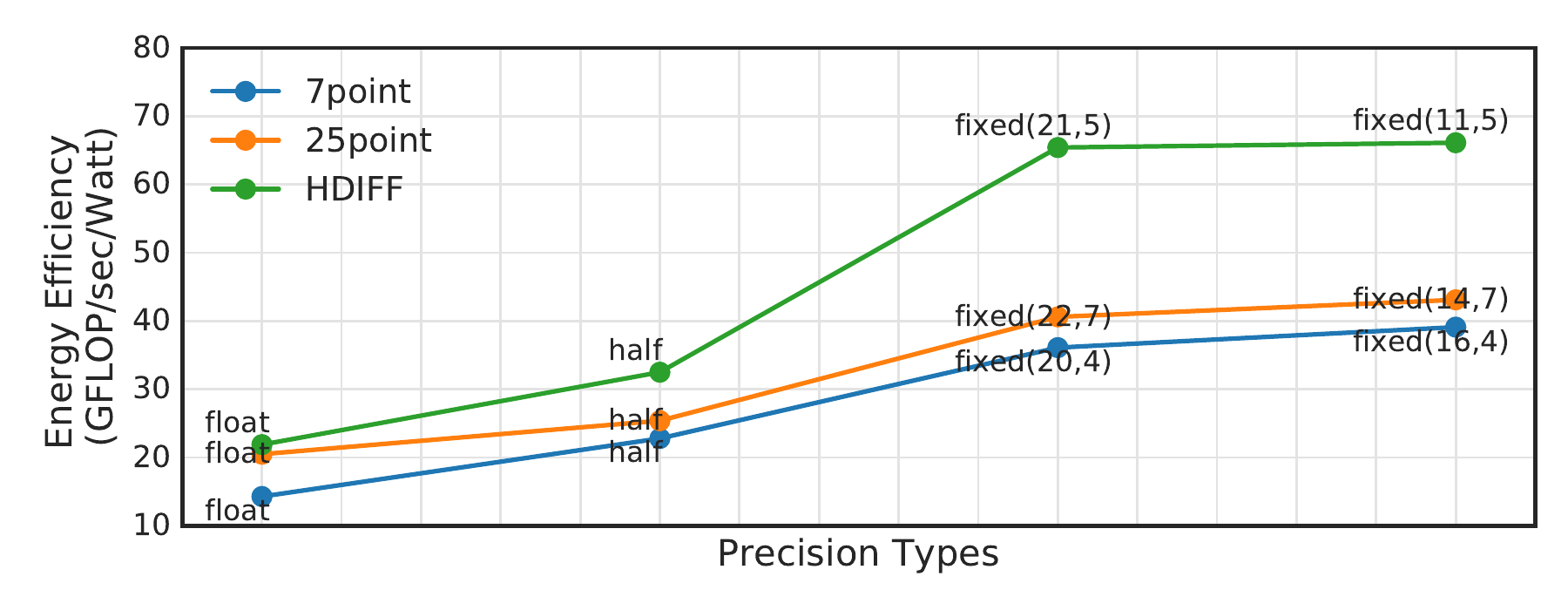}
 \vspace{-0.8cm}
 \caption{Evaluated design points for different stencil kernels. The plot shows energy efficiency (GFLOPS/Watt) with different precision implementations on an Alpha-Data ADM-PCIE-9V3~\cite{ad9v3} card featuring the Xilinx Virtex Ultrascale+ XCVU3P-FFVC1517
 \label{fig:low_precision/energyEff}}
 \end{figure}

Table~\ref{tab:low_precision/fpgaResources} shows the resource utilization for our examined stencil kernels on an FPGA using different precision data types. In all the scenarios, going from single to half-precision increases the performance with a corresponding reduction in the number of resources. Further, moving to fixed-point arithmetic representation increases the performance due to a decrease in the number of bytes loaded at the cost of LUT utilization. However, the utilization of other FPGA resources is reduced. Figure~\ref{fig:low_precision/energyEff} shows the achieved energy efficiency with different precision data types. As the number of bits reduces, we see an increase in energy efficiency for all considered kernels. Designs implemented in fixed-point will always be more efficient than their equivalent in floating-point alternative because fixed-point implementations consume fewer resources and less power (see Table~\ref{tab:low_precision/fpgaResources}). As these stencil kernels do not require the high dynamic range achievable with
floating-point, moving to fixed-point implementations could provide better energy efficiency. In the case of \hdiff, we see a huge increase in energy efficiency on moving to a lower precision. This increase is because \hdiff is a compound kernel; therefore, each elementary stencil's energy improvement with lower precision leads to much higher cumulative gains.
\begin{table}[h]
\renewcommand{\arraystretch}{0.95}
\setlength{\tabcolsep}{8pt}
\centering
   \caption{FPGA resource utilization and performance for the examined stencil kernels on FPGA testbeds, with different precisions}
    \label{tab:low_precision/fpgaResources}
\resizebox{\textwidth}{!}{
\begin{tabular}{p{1.6cm}Hp{2cm}ccccccc}
\toprule
\multirow{2}{*}{\textbf{Kernel}} & \multirow{2}{*}{Data Size} & \multirow{2}{*}{\textbf{Precision}}&\multirow{2}{*}{\textbf{Accuracy} \textbf{(\%)}}  & \multicolumn{4}{c}{\textbf{Utilization} (\%)} & \multirow{2}{*}{\begin{tabular}[c]{@{}l@{}}\textbf{Performance}\\ \textbf{(GLOP/s)}\end{tabular}} & \multirow{2}{*}{\begin{tabular}[c]{@{}l@{}}\textbf{Energy} \\ \textbf{(mJ)}\end{tabular}} \\ \cline{5-8}
&                            &                          &  & {BRAM}    & {DSP}   & {FF}   & {LUT}    &                                                                                 &                                                    \\ \hline
\sevpoint                  & $1280\times1080\times960  $              & float         &100             & 38      & 35     & 18   & 29     & 228.4                                                                           & 4617.2                                                                     \\
\sevpoint                  & $1280\times1080\times960    $           & half &99.95                      & 25      & 24     & 15   & 28     & 319.5                                                                           & 2887.6                                                                     \\
\sevpoint                  &$1280\times1080\times960$                 & fixed (20,4)    &100           & 16      & 12     & 49   & 95    & 467.6                                                                           & 1832.3                                                                     \\
\sevpoint                  & $1280\times1080\times960 $               & fixed (16,4)      &99.96         & 12      & 12     & 47   & 92.5   & 468.1                                                                          & 1689.4                                                                     \\
\tfivepoint                 & $1280\times1080\times960$              & float    &100                  & 42      & 62    & 36   & 44     & 327.7                                                                           & 1608.7                                                                     \\
\tfivepoint                 & $1280\times1080\times960$            & half     & 99.06                  & 32      & 43   & 32   & 43     & 342.1                                                                         & 1541.5                                                                     \\
\tfivepoint                 & $1280\times1080\times960$              & fixed (22,7)      & 100         & 29       & 21    & 56   & 95    & 527.9                                                                         & 1510.3                                                                     \\
\tfivepoint                 & $1280\times1080\times960 $               & fixed (14,7)       &99.05        & 19       & 21    & 55   & 91     & 528.9                                                                         & 1497.9                                                                     \\
\hdiff                   & $1280\times1080\times960$                & float    &100                  & 52      & 89    & 65   & 61     & 350.3                                                                       & 3010.5                                                                      \\
\hdiff                   &$1280\times1080\times960  $                 & half  &98.02                     & 44      & 84    & 35   & 57     & 421.8                                                                         & 2031.1                                                                      \\
\hdiff                   &$1280\times1080\times960   $                & fixed (21,5)    &100           & 24      & 45     & 77  & 76     & 653.9                                                                        & 1007.9                                                                       \\
\hdiff                   & $1280\times1080\times960   $                   & fixed (11,5)   &97.92            & 14       & 35     & 69   & 71     & 659.1                                                                   & 997.9                        \\\bottomrule                                                     
\end{tabular}}
\vspace{-0.4cm}
\end{table}

\section{Related Work}
\label{sec:low_precision/relatedWork}
Floating-point representation is the most widely supported data type by current hardware devices. Recently, in many application domains, there has been a significant amount of research to explore error resilience across the complete stack of computer architecture from application to device physics. \fc{A large body of literature~\cite{iwata1989artificial,menard2002methodology,de2017understanding,higham2019simulating,esmaeilzadeh2012neural,menard2002automatic,ha2021precision} has analyzed the benefits of using lower precision fixed-point computation compared to floating-point for different application domains.} With the emergence of the posit number system~\cite{gustafson2017beating}, research into lower precision with these alternate number systems is regaining attention. In neural-networks, Langroudi~\etal~\cite{langroudi2018deep} demonstrate a minimum accuracy degradation by using a 7-bit posit format. In another study,  Kl{\"o}wer~\etal~\cite{klower2019posits} show the applicability of posit in weather modeling. However, they used a different weather prediction model than COSMO, which uses different numerical solvers for predicting weather.

High-performance implementations of stencils on modern processors usually use the IEEE single-precision or double-precision floating-point data types. There have been various efforts to improve these kernels for different architectures using various software-based optimization techniques. Datta~\etal~\cite{datta2008stencil} optimized the 2D and 3D stencil for multicore architectures using several hardware adherent optimizations. Similarly, Nguyen~\etal~\cite{nguyen20103} worked on algorithm optimization for CPU and GPU-based systems. Gysi~\etal~\cite{gysi2015modesto} provided guidelines for optimizing complex kernels for CPU--GPU systems using analytic models. \ou{ Gan~\etal~\cite{gan2013accelerating} use a mixed-precision approach for 13-point shallow water equation (SWE) stencils. The authors use fixed-point representation for variables that require a lower precision, while floating point precision for other variables.}  However, to the best of our knowledge, this thesis is the first to study the precision tolerance for scientific 3D stencil kernels, including a compound weather prediction stencil kernel, for a wide range of number systems, i.e., fixed-point arithmetic, floating-point arithmetic, and posit arithmetic.

\section{Conclusion}
\label{sec:low_precision/conclusion}
Stencils are one of the most widely used computational kernels across various real-world applications. This chapter analyzed the precision tolerance for different 3D stencil kernels using fixed-point, floating-point, and posit number systems. We demonstrated by exhaustive precision exploration that these kernels have a margin to move to a lower bit-width with minimal loss of accuracy using different number formats.

Further, in a case study, we measured the performance of these kernels on a state-of-the-art multi-core platform and designed lower bit-width-based accelerators for all considered 3D stencil kernels on an FPGA platform. FPGA is the only device that gives us the capability to implement arbitrary fixed-point precision data types. Hence, we leveraged this capability to show the advantages of accelerating these kernels with lower precision compared to the ubiquitous IEEE floating-point format.  In the future, we can use this analysis technique in an integrated design-flow to build efficient systems for stencil-based applications. Another future direction would be to study the effects of low precision processing not only for streaming applications, e.g., stencil and convolution, where computation is done locally but also for iterative applications where errors accumulate.


\chapter[\texorpdfstring{NAPEL: Near-Memory Computing Application Performance Prediction via Ensemble Learning}{NAPEL: Near-Memory Computing Application Performance Prediction via Ensemble Learning}
]{\chaptermark{header} NAPEL: Near-Memory Computing Application Performance Prediction via Ensemble Learning}
\chaptermark{NAPEL}
\label{chapter:napel}
\renewcommand{\namePaper}{NAPEL\xspace} 

\chapternote{The content of this chapter was published as \emph{``NAPEL: Near-Memory Computing Application Performance Prediction via Ensemble Learning''} in DAC 2019.}

To harness the power of near-memory computing architectures, having a high-level performance model can offer fast turnaround times in early design stages. This chapter proposes an NMC model that combines technology parameters and application specific characteristics to evaluate the performance of a workload. A statistical design of experiment (DoE) methodology is first employed to select a set of representative design points for simulation. These sampled points well represent the whole space of possible input configurations. Then, ensemble learning is leveraged to develop a model that can predict performance metrics for new unseen applications on these novel architectures. Further, this model can act as a classifier to predict whether or not should an application be offloaded to NMC cores.  

\section{Introduction}
\vspace{-0.1cm}

As discussed in Chapter~\ref{chapter:background}, past works~\cite{hsieh2016accelerating,Azarkhish:2016:DEP:2963802.2963805,tom, boroumand2016lazypim,googleWorkloads,kim2018grim, ahn2015pim,teserract,boroumand2019conda} show that NMC architectures can be employed {effectively} for a wide range of applications, {including graph processing, databases, neural networks, bioinformatics}. However, a common challenge all such past works {face} is how to evaluate the performance and energy consumption of the NMC {architectures} for different workloads systematically and {accurately} in a reasonable amount of time~\cite{singh2018review,mutlu2019}. In the early design stage, system architects use simulation techniques (e.g., \cite{7063219,tom, boroumand2016lazypim, Azarkhish:2016:DEP:2963802.2963805,teserract}) for architectural performance and energy evaluation.
However, this approach is extremely slow, because a single simulation for a real-world application with a representative dataset typically takes hours or even days. Specifically, the speed of a cycle-accurate simulator is in the range of a few thousand instructions per second~\cite{sanchez2013zsim}, which is {orders of magnitude} slower than native execution. 

\textbf{Our goal} is to enable fast early-stage design space exploration of NMC architectures without having to rely on time-consuming simulations. 
To this end, we propose the \textit{\underline{N}MC \underline{A}pplication performance and energy \underline{P}rediction framework using \underline{E}nsemble machine \underline{L}earning} (NAPEL). 
The key idea is to use ensemble learning to build a model that, once trained for a fraction of programs on a number of {architecture configurations}, can predict the performance and energy consumption of \emph{different} applications {on the same NMC architecture}. 
The ensemble learning mechanism we use is random forest (RF)~\cite{breiman2001random}. 
NAPEL can make performance and energy predictions for an {average} application on a specific architecture {$220\times$ faster} than using simulation.
Previous ML-based approaches~\cite{6827943,calotoiu2013using,wu2011scalaextrap} {perform} extrapolations to predict, for example, the performance {of a known application} for a bigger dataset.
In contrast, NAPEL can make predictions for \emph{previously-unseen} applications, after being trained with data from applications that are different from the applications {that} we want to predict. 

NAPEL still needs to run simulations to gather training data that is required to construct {its} predictive model. 
As discussed above, running simulations is very time-consuming if we apply a brute-force approach to run all the application-input configurations needed for training data. 
To alleviate this problem, we use a technique called \emph{design of experiments} (DoE)~\cite{natrella2010nist} to extract representative data with a {small} number of experimental runs {(between 11 and 31 for the evaluated applications)}. 
Specifically, we employ a DoE variant called \emph{central composite design} (CCD), which allows us to explore the interactions and nonlinear effects between the application input parameters and the output response {(i.e., performance and energy consumption)}. 







\section{NAPEL}
\label{sec:napel/methodology}
\vspace{-0.1cm}
NAPEL is a {performance and energy estimation} framework that targets the early stages of {NMC} system design. In this section, we describe the main components of the framework. First, we give an overview of NAPEL training and prediction (Section~\ref{sec:napel/overview}). Second, we describe the target NMC architecture we consider in this chapter (Section~\ref{sec:napel/nmcarch}). Third, we explain the code-instrumentation process for the applications used to generate training datasets and for the applications under performance and energy prediction (Section~\ref{sec:napel/code}). Fourth, we describe the two most important components of NAPEL training: the design of {experiments} methodology (Section~\ref{subsec:napel/doe}) and the ensemble machine learning (ML) technique (Section~\ref{sec:napel/rf}).

\vspace{-0.2cm}

\subsection{Overview}
\label{sec:napel/overview}

\vspace{-0.1cm}

\begin{figure}[h]
\centering
 \includegraphics[bb=17 0 1395 705,width=1.0\linewidth]{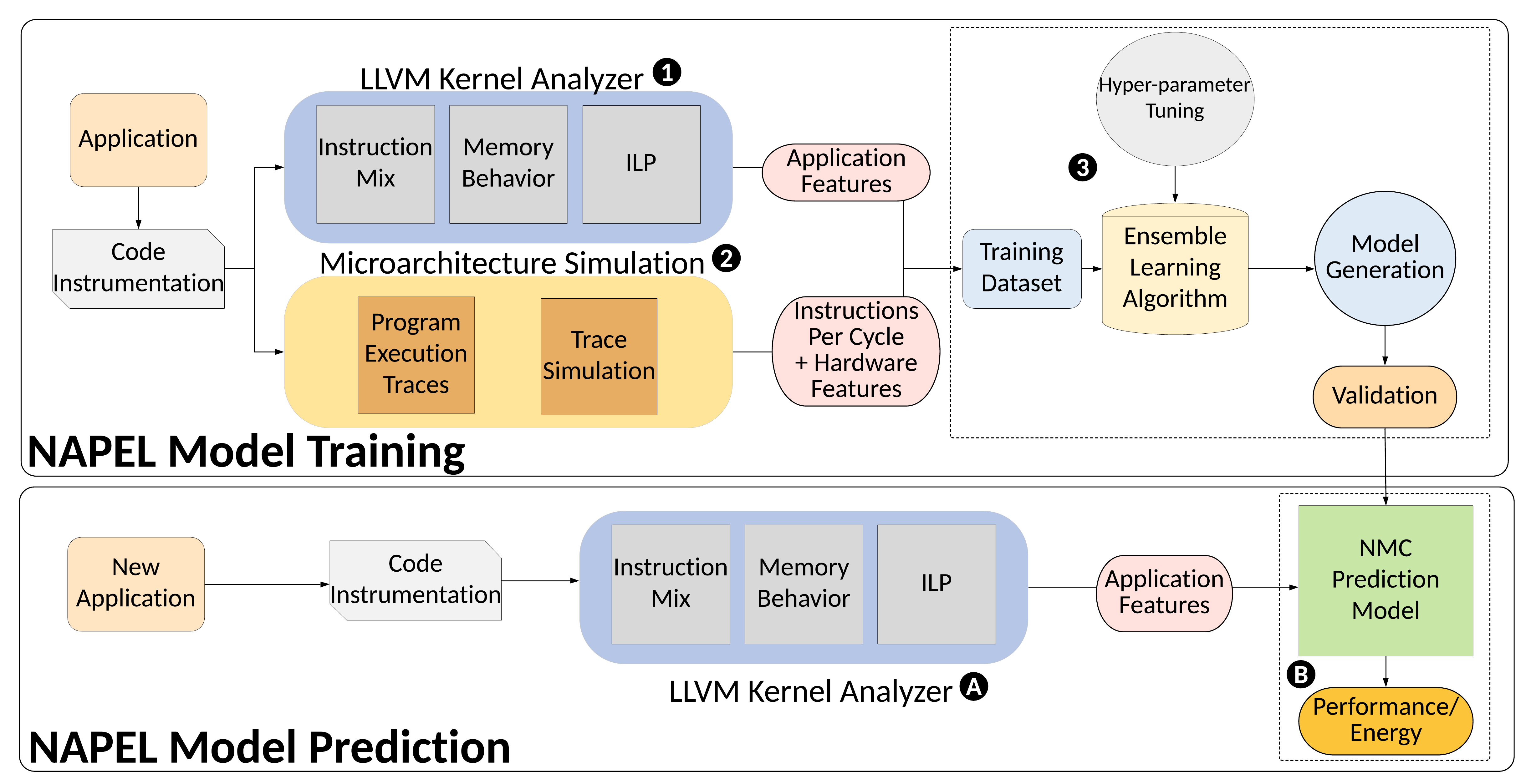}
\vspace{-0.8cm}
\caption{Overview of NAPEL training and prediction
\label{fig:napel/model}}
\vspace{-0.3cm}
\end{figure}

NAPEL is based on ensemble learning. {Thus,} it needs to be trained before {it can} predict performance and energy consumption.
Figure~\ref{fig:napel/model} depicts the {key} components of NAPEL training and prediction.


\noindent \textbf{Model Training.} NAPEL training consists of three phases. 
The first phase (\circled{1} {in Figure~\ref{fig:napel/model}}) is an LLVM-based~\cite{llvm} kernel {analysis} phase (Section~\ref{sec:napel/code}), which extracts architecture-independent workload characteristics. 
First, we instrument applications or parts of them that we use to gather data for model training. We consider the instrumented codes for execution on NMC compute units with a specific architecture configuration. 
Second, we characterize the instrumented codes in a microarchitecture-independent manner by using a specialized plugin of the LLVM compiler framework~\cite{Anghel2016}. This type of  characterization excludes any hardware dependence and captures the inherent {characteristics} of workloads.

In the second phase~\circled{2}, microarchitectural simulations are performed to gather architectural responses for training. 
For the simulations, we use central composite design (\acrshort{ccd})~\cite{mariani2017predicting}, a technique for {the design of experiments (DoE) method}~\cite{montgomery2017design}. With CCD, we can minimize the number of simulation experiments to gather training data for NAPEL while ensuring {good} quality of the training data (Section~\ref{subsec:napel/doe}). 
The generated simulator responses along with application properties from the first phase and the {microarchitectural} parameters form the input to our ML algorithm.

In the third phase~\circled{3}, we train our ML algorithm (Section~\ref{sec:napel/rf}). During this phase, we perform additional tuning of our ML algorithm's hyper-parameters. Hyper-parameters are sets of ML algorithm variables that can be tuned to optimize the accuracy of the prediction model. 
{We validate the prediction model against performance and energy simulation results from the second phase.} 
Once trained, the framework can predict the performance and energy of a {previously-unseen} application on a specific NMC architecture.

\vspace{0.1cm}
\noindent \textbf{Model Prediction.} NAPEL prediction has only two phases. 
The first phase (\circled{A} {in Figure~\ref{fig:napel/model}}) is the same LLVM-based kernel analysis phase as in NAPEL training. {This phase extracts} architecture-independent features of the workload for which NAPEL will predict the performance and energy consumption. 
The second phase~\circled{B} {performs} the prediction by using the trained model. We feed the model with the architecture-independent workload features and {the model provides} the performance and energy estimations.

\vspace{-0.1cm}

\subsection{NMC Architecture}
\label{sec:napel/nmcarch}
\vspace{-0.1cm}

\begin{figure}[h]
\centering
\includegraphics[bb=24 168 808 467,width=0.70\linewidth]{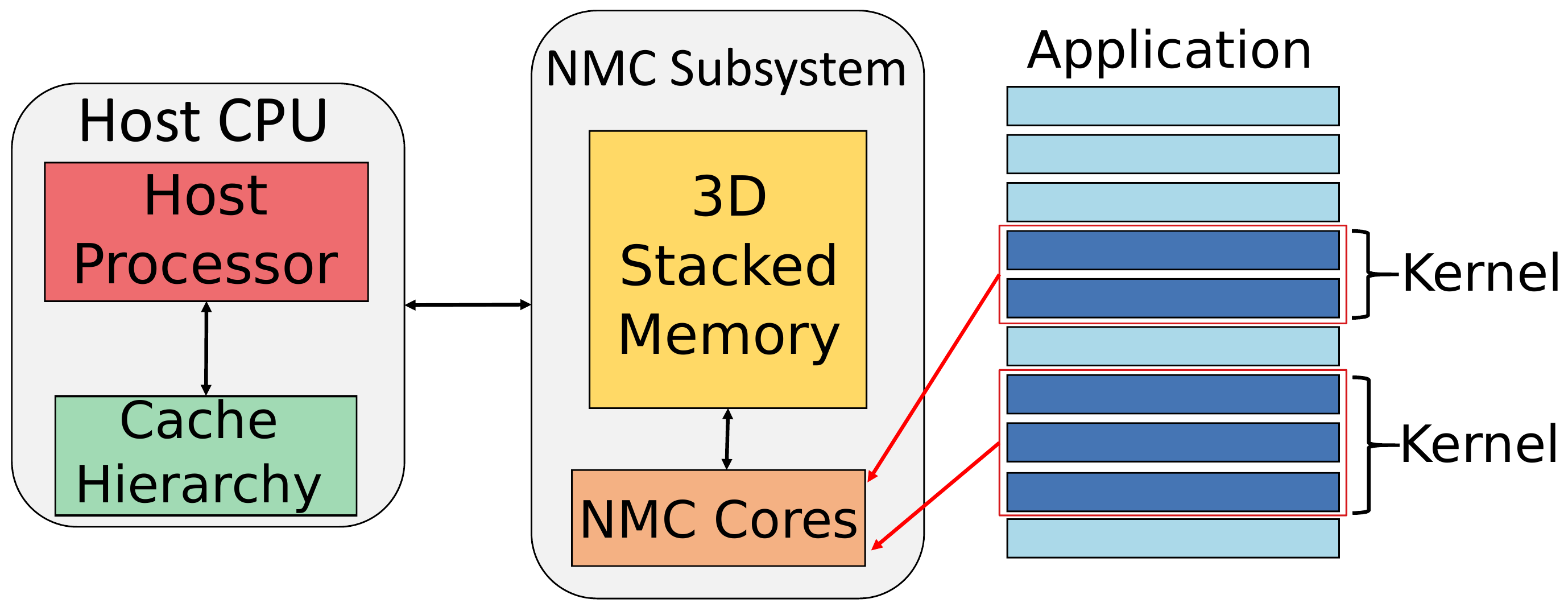}
\vspace{-0.4cm}
\caption{Overview of a system with NMC capability. On the right, an abstract view of application code with kernels {that are} offloaded to NMC
\label{fig:napel/systemNMC}}
\vspace{-0.2cm}
\end{figure}

Figure~\ref{fig:napel/systemNMC} depicts the reference computing platform that we consider in this chapter. It contains a host processor and an external memory equipped with NMC compute units. 

\vspace{-0.3cm}

The NMC \juan{subsystem} consists of a 3D-stacked memory~\cite{6757501,gao2015practical, teserract, lee2016smla} with processing elements (PEs) embedded in its logic layer. 
The memory is divided into several vertical DRAM partitions, called vaults, each with its own DRAM controller in the logic layer. 
In this chapter, we model NMC PEs as in-order, single-issue cores {with a private cache} as proposed in previous work~\cite{teserract,gao2015practical}, taking into account the limited thermal and area budget in the logic layer. 
NAPEL can be extended to support other types of general-purpose cores and accelerators by selecting the appropriate {architectural} features (see Table~\ref{tab:napel/application}) for training the {NAPEL model}.

\vspace{-0.2cm}

\subsection{Code Instrumentation and Analysis}
\label{sec:napel/code}
\vspace{-0.1cm}

In the first phase of NAPEL training and prediction, the programmer annotates the region of the source code, {called \emph{kernel} (\textit{k}),} \juang{which} is a candidate for offloading to NMC (i.e., execution on NMC processing elements). Then, that specific region is converted into an LLVM intermediate-representation (IR), {which} provides the basis for performing hardware-independent kernel analysis. 
{Hardware-independent} profiling enables us to generate {an} application profile \textit{p} independently of the {NMC design}. 

The {application} profile \textit{p(k, d)} is obtained by executing the instrumented application kernel \textit{k} while processing a dataset \textit{d}. \textit{p(k, d)} is a vector where each parameter is a statistic about an application feature \textit{f}. 
Table~\ref{tab:napel/application} lists the main application features {we extract} {by using the LLVM-based PISA analysis tool~\cite{Anghel2016}}. 
{We select} these features to analyze the memory access behavior of an application (data reuse distance, memory traffic, memory footprint, etc.), which is key for {assessing} the suitability of NMC for the application. 
Ultimately, the application profile \textit{p} has 395 features, which includes all the sub-features {of each metric we consider}. {Such a} large number of features enables complex relationships to be identified between the analyzed {application and its performance and energy consumption} on the underlying {NMC} architecture~\cite{mariani2017predicting}.

\vspace{-0.2cm}

{\footnotesize

\begin{table}[h]
\centering
\caption{Main application and architectural features}
\label{tab:napel/application}
\resizebox{0.85\linewidth}{!}{%
\begin{tabular}{p{0.4\linewidth} p{0.62\linewidth}}
\toprule
\textbf{Application Feature} & \textbf{Description} \\ 
\midrule
Instruction Mix & Fraction of instruction types (integer, floating point, memory read, memory write, etc.) \\
\acrshort{ilp}                      & Instruction-level parallelism on an ideal machine. \\
Data/Instruction reuse distance   & For a given distance $\delta$, probability of reusing one data element/instruction (in a certain memory location) before accessing $\delta$ other unique data elements/instructions (in different memory locations). \\
Memory traffic                      & Percentage of memory reads/writes that need to access the {main memory}, assuming a {cache} of size equal to the maximum reuse distance. \\
Register traffic    & Average number of registers {per} instruction. \\
Memory footprint    & Total memory size used by the application. \\ 
\bottomrule

\textbf{NMC Arch. Features }   & \textbf{Description} \\ 
\midrule
Core type & In-order\\
\#PEs & Total number of near-memory processing units\\
Core frequency & Operating frequency of the core\\
Cache line size & Total size of a cache line (bytes) \\
\#cache-lines & Number of cache lines\\
DRAM layers & Number of stacked DRAM layers\\
Size of DRAM& Total size of memory (bytes) \\
Cache access fraction & Cache hit ratio \\
DRAM access fraction & Cache miss ratio \\

\bottomrule
\end{tabular}
}

\end{table}
}


\vspace{-0.1cm}

\subsection{Central Composite Design}
\label{subsec:napel/doe}
\vspace{-0.1cm}
In the second phase of NAPEL training, we use the design of experiments (DoE) {method}~\cite{montgomery2017design} as a way to minimize the number of experiments to train NAPEL without {sacrificing} the amount and quality of the information gathered by the experiments. 
DoE {is a set of} statistical techniques meant to {locate} a {small set of points in a parameter space with the goal of representing the whole parameter space}. 
The traditional brute-force approach to collecting training data is time-consuming: the sheer number of experiments renders detailed simulations intractable. Thus, the DoE strategy to gather a training dataset is a critical component of our model. 

\begin{SCfigure}[][h]
  \centering
  \caption{Central composite DoE for two parameters (\textit{x},~\textit{y}). For example, for \emph{atax} (\textit{x},~\textit{y}) are (\textit{dimension},~\textit{threads})}
  \hspace{3cm}
\includegraphics[bb=21 136 715 725,width=3.2cm,trim={0cm 0cm 0cm 0cm},clip]{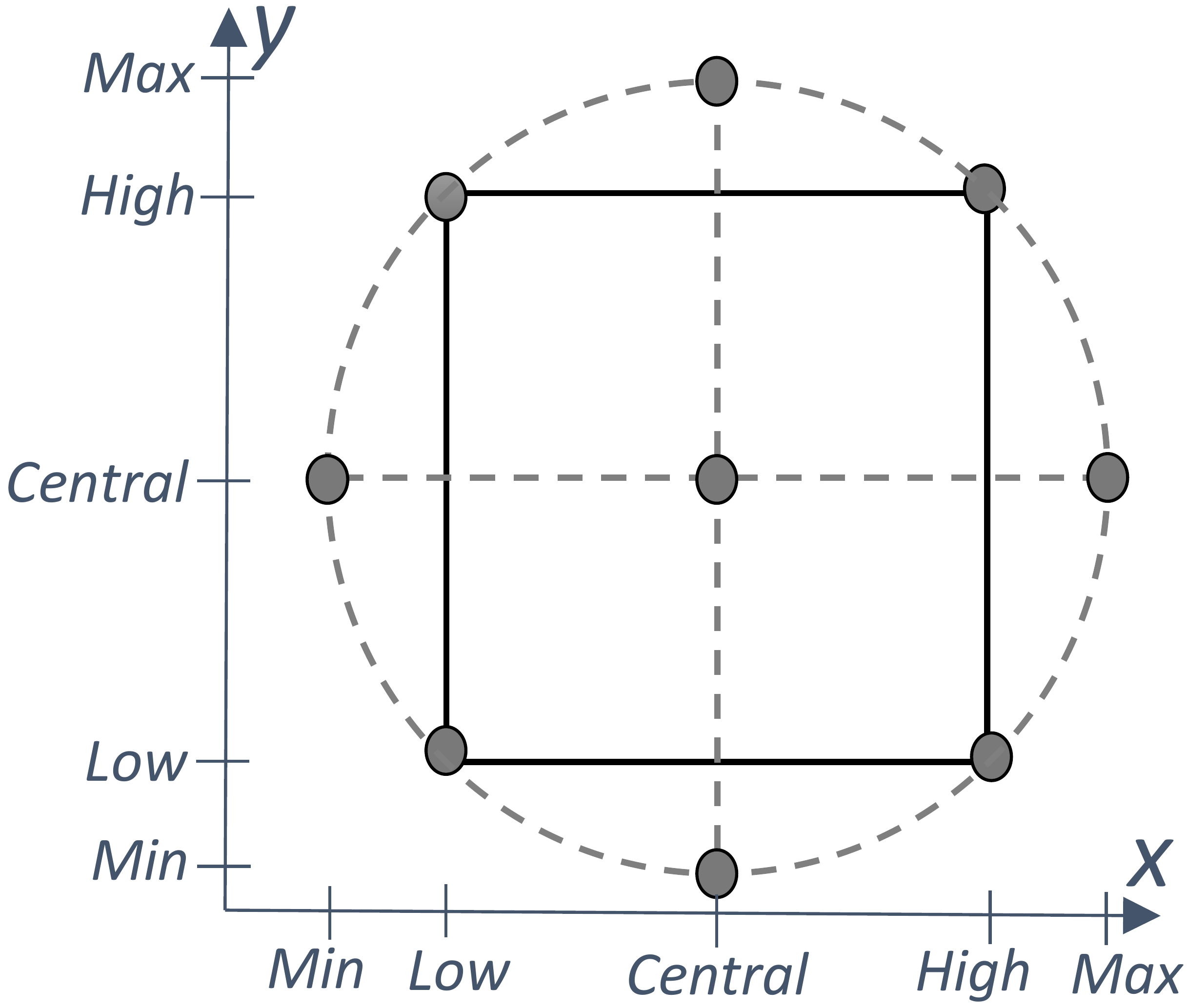}
\label{fig:napel/doe}
\vspace{-0.4cm}
\end{SCfigure}

We apply the Box--Wilson central composite design (CCD)~\cite{mariani2017predicting}, the goal of which is to minimize the uncertainty of a nonlinear polynomial model that accounts for parameter interactions. 
While applying CCD, we treat the application input dataset \textit{d} as a parameter vector (e.g., dataset size,~number of threads, etc.) and {each input} configuration as a point in {a multidimensional} parameter space. 
For example, application \textit{atax} from the PolyBench benchmark suite~\cite{pouchet2012polybench} has two significant {parameters} (\textit{dimension},~\textit{threads}) (see Table~\ref{tab:napel/results}). 
In CCD, each input parameter in the vector \textit{d} can {have} one of five levels: {\textit{minimum},~\textit{low},~\textit{central},~\textit{high},~\textit{maximum}}. 
{First, we select these levels for each parameter.
For example, for \textit{atax}, the levels of \textit{dimension} are (500,~1250,~1500,~2000,~2300). 
Second, we place in the parameter space a point for each parameter combination (i.e., input configuration) with \textit{low} and \textit{high} levels (the corners of the solid-line square in Figure~\ref{fig:napel/doe}). 
In the case of \textit{atax}, the points (\textit{dimension}, \textit{threads}) are (1250,~8), (1250,~32), (2000,~8), (2000,~32). 
Third, we draw a multidimensional sphere (represented as a circle in Figure~\ref{fig:napel/doe}) that circumscribes the initial square. 
This sphere generalizes the DoE to capture the nonlinearity in the system. 
Fourth, we obtain additional points on the sphere by combining the \textit{central} level of each parameter with the \textit{maximum} and \textit{minimum} levels of the other parameters. For \textit{atax}, these points (\textit{dimension}, \textit{threads}) are (1500,~4), (1500,~64), (500,~16), (2300,~16). 
Fifth, we include the \textit{central} configuration, {which} is (1500,~16) for \textit{atax}.}

\vspace{-0.2cm}


We run these {DoE-selected application-input} configurations on different architectural configurations to collect the training dataset. {Table~\ref{tab:napel/results} lists the parameter levels for the evaluated applications. We include a \textit{test} configuration, which we use in Section~\ref{sec:napel/usecase}.}

\begin{table}[h!]


\caption{Evaluated applications and their DoE parameters (``DoE param.''). For each DoE parameter, we show its five levels (\textit{minimum},~\textit{low},~\textit{central},~\textit{high},~\textit{maximum}) and \textit{test} input}
\label{tab:napel/results}
\vspace{-0.4cm}
\begin{center}
\resizebox{0.8\linewidth}{!}{%
\begin{tabular}{p{0.8cm}p{4cm}p{1.9cm}|p{0.7cm}p{0.7cm}p{1cm}p{0.7cm}p{0.7cm}p{0.3cm}}
\toprule
\multicolumn{3}{c|}{\textbf{Application}}  & \multicolumn{6}{c}{\textbf{DoE Parameter Levels}}    \\ 

 Name& Description& DoE Param. & Min& Low& Central & High& Max& Test \\ \hline

atax    & Matrix Transpose \newline and Vector Mult.   & Dimensions    \newline Threads               &500\newline4&1250 \newline8&1500\newline16&2000\newline32&2300\newline64  &8000 \newline 32 \\

bfs      & Breadth-first \newline Search                        &  Nodes\newline Weights    \newline Threads    \newline Iterations          &400k \newline1 \newline1 \newline30 &   800k \newline2\newline9\newline40&   900k \newline4 \newline16 \newline65  &1.2m \newline25   \newline32 \newline70 &1.4m\newline49 \newline64 \newline80  &1.0m \newline 4\newline32\newline95 \\

bp    & Back-propagation                & Layer Size  \newline Seed   \newline Threads    \newline Iterations                                   &800k \newline 2 \newline 4 \newline 1&1m \newline 4 \newline 8 \newline 3&2m \newline 5 \newline 16\newline 9 &3.5m  \newline10\newline 32\newline 16 &4m \newline 12 \newline 64 \newline 25 &1.1m \newline 5\newline 32\newline 9\\

chol   & Cholesky \newline Decomposition                  & Dimensions             \newline Threads  \newline Iterations     &  64 \newline 4\newline 10&  384 \newline 8\newline 20&128 \newline 16\newline 30&320 \newline 32\newline 50&512 \newline 64\newline 80 &2000 \newline 32 \newline 60  \\

gemv &Vector Multiply \newline and Matrix\newline Addition  & Dimensions        \newline Threads      \newline Iterations             &    500  \newline4  \newline 50      &750\newline8\newline 60&1250\newline16\newline 80&2000\newline32\newline 100&2250\newline64   \newline 150 &8000 \newline 32  \newline 60  \\

gesu  & Scalar, Vector, and \newline Matrix Mult.  & Dimensions       \newline Threads     \newline Iterations              &          500  \newline4\newline10        &750\newline8\newline20&1250\newline16\newline40&2000\newline32\newline50&2250\newline64\newline60 &8000 \newline 32 \newline50  \\

gram   & Gram-Schmidt \newline Process                    & Dimension$_i$ \newline Dimension$_j$    \newline Threads                                     &  64 \newline64 \newline 4&  384 \newline384 \newline 8&128 \newline128 \newline 16&320 \newline320 \newline 32&512 \newline512 \newline 64        &2000 \newline 2000 \newline32  \\

kme     & K-Means \newline Clustering          & Data Size     \newline Clusters  \newline Threads      \newline Iterations                             & 100k \newline3\newline1\newline 10&300k \newline 5\newline9\newline 20&700k \newline 6\newline1\newline 30&    900k \newline 7\newline32\newline 40&  1.2m  \newline8\newline64\newline 50 & 819k  \newline 5\newline32\newline 30\\

lu     & LU Decomposition                          & Dimensions      \newline Threads     \newline Iterations            &196 \newline 4\newline 98 &256 \newline 8\newline 128&320 \newline 16\newline 256&420 \newline 32 \newline 420 &512 \newline 64 \newline 512  &2000 \newline 32  \newline 2000   \\

mvt     & Matrix Vector \newline Product     &Dimensions          \newline Threads     \newline Iterations        &     500  \newline4       \newline10 &750\newline8 \newline20&1250\newline16 \newline30&2000\newline32 \newline50&2250\newline64 \newline60 &2000 \newline 32   \newline40 \\

syrk     & Symmetric Rank-k \newline Operations              & Dimension$_i$     \newline Dimension$_j$ \newline  Threads                               & 64 \newline64 \newline4&128 \newline 128 \newline8&320 \newline 320 \newline16&    512 \newline 512 \newline 32&  640  \newline640  \newline 64      &2000 \newline 2000 \newline32     \\

trmm      & Triangular Matrix \newline Multiply       & Dimension$_i$\newline Dimension$_j$              \newline Threads           &196 \newline196 \newline 4 &256 \newline256 \newline 8&320 \newline320 \newline 16&420 \newline 420 \newline32 &512 \newline512 \newline 64 &2000 \newline 2000 \newline32   \\ 
\bottomrule

\end{tabular}
}
\end{center}


\end{table}

\vspace{-0.2cm}

\subsection{Ensemble Machine Learning}
\label{sec:napel/rf}
\vspace{-0.1cm}

The third phase of NAPEL training is the training of the ML algorithm. 
As we retrieve hundreds of application features from the application analysis, we make use of the random forest (\acrshort{rf})~\cite{breiman2001random} algorithm, which embeds automatic procedures to screen many input features. 
RF is an ensemble ML algorithm, {which, starting from a root node, constructs a tree and iteratively grows the tree by associating it with a splitting value for an input variable to generate two child nodes}. 
Each node is associated with a prediction of the target metric equal to the mean observed value in the training dataset for the input subspace the node represents. 
This input subspace is randomly {sampled} from the entire training dataset.

We employ RF to capture the intricacies of {new} NMC architectures by predicting instructions per cycle (\acrshort{ipc}) when {executing} an application {near} memory. Formally, we predict IPC$(p,~a)\sim$ IPC$(k,~d,~a)$, where \textit{p} is the hardware-independent application profile representation of kernel \textit{k} when processing input dataset \textit{d} on an architecture configuration \textit{a}. \juan{The input data gathered to} train our RF model \juan{has} three parts: (1) a hardware-independent application profile \textit{p(k,~d)}, (2) an architectural design configuration \textit{a}, and (3) {responses corresponding to each pair $(p,~a)$}. To gather the architectural responses, kernel \textit{k} belonging to training set T with input dataset \textit{d} is executed on an architectural simulator, simulating an architecture configuration \textit{a}. This produces IPC\textit{(k,~d,~a)} for that configuration and is used as a \textit{label} while training our RF algorithm. 

We improve NAPEL training by tuning the algorithm's hyper-parameters~\cite{scikit-learn}. Hyper-parameter tuning can provide better performance estimates for some applications. 
{First, we perform as many iterations of the cross-validation process as hyper-parameter combinations}. 
{Second, we compare all the generated models by evaluating them on the testing set, and select the best one}. 

After training our RF algorithm, we can predict the IPC of a kernel that is \emph{not} in the training set. 
The predicted IPC can be used for performance evaluation of a kernel on an NMC system. The execution time $\Pi_{\textrm{NMC}}$ of the kernel offloaded to NMC can be calculated as $\Pi_{\textrm{NMC}}=\frac{I_{\textrm{offload}}}{IPC.f_{\textrm{core}}}$, 
where $f_{\textrm{core}}$ is the frequency of the NMC processing cores and $I_{\textrm{offload}}$ is the total number of offloaded instructions. Similarly, we build another model for energy prediction where we use energy consumption as a \textit{label} when we train our RF algorithm.

\vspace{-0.3cm}

\section{Experimental Results}
\label{sec:napel/results}

\subsection{Experimental Setup}

We consider different workloads from the PolyBench~\cite{pouchet2012polybench} and Rodinia~\cite{5306797} benchmark suites {that} cover a {wide} range of domains, such as image processing, machine learning, graph processing, radio astronomy.
First, we instrument the region of code that is considered for offloading to NMC processing elements. 
Second, {we apply} CCD to these workloads to select a {small} set of application input configurations that represent the space of possible input configurations. Third, we carry out the LLVM-based~\cite{llvm} microarchitecture-independent characterization to extract application metrics (Table~\ref{tab:napel/application}) by using the PISA analysis tool~\cite{Anghel2016}. 

We evaluate host performance on a real IBM POWER9 system~\cite{power9} and NMC performance on a state-of-the-art simulator, Ramulator~\cite{7063219}. We extend Ramulator with a 3D-stacked memory model to simulate the NMC processing elements~\cite{ramulator-pim-repo}.
Table~\ref{tab:napel/systemparameters} summarizes the system details used for the host system and {the NMC} system. 
We collect dynamic execution traces of the instrumented code with a Pin tool. {We feed} the acquired traces to Ramulator. 
We {use} the simulation results {as training data for} our RF algorithm. 
Once trained, {we use NAPEL to predict the performance and energy consumption of previously-unseen applications}.

\vspace{-0.2cm}

{\footnotesize
\begin{table}[h]
  \caption{System parameters and configuration}
    \label{tab:napel/systemparameters}
\vspace{-0.4cm}
\begin{center}
\resizebox{0.78\linewidth}{!}{%
\begin{tabular}{lp{8.6cm}}

\toprule
\textbf{Host CPU System}\\ \hline 

Configuration & IBM\textsuperscript{\textregistered} POWER9 AC922 @2.3 GHz, 16 cores (4-way SMT), 32 KiB L1 cache, 256 KiB L2 cache, 10 MiB L3 cache, 16x32GiB RDIMM DDR4 2666 MHz \\
\hline

 \textbf{NMC System} \\

\hline
Cores  &32$\times$ single issue, in-order execution @ 1.25 GHz\\
L1-I/D& 2-way, cache size = 2 cache lines, 64B per cache line \\
DRAM Module & 32 vaults, 8 stacked-layers, 256B row buffer; 4GB total size; closed-row policy \\
Off-chip Link & 16-bit full duplex high-speed serializer/deserializer (SerDes) I/O link @ 15 Gbps~\cite{7477494}\\


\bottomrule

\end{tabular}
}
\end{center}

\end{table}
}


\subsection{{Model} Training and Prediction Time}
\vspace{-0.1cm}

Table~\ref{tab:napel/timing} shows the time for performing training simulations (see ``DoE run (mins)'') {with the selected DoE configurations (``\#DoE conf.'') to gather training data. The table also includes the time for training and tuning (``Train+Tune (mins)'') and the prediction time (``Pred. (mins)'') for each application}. Once {the model is trained}, the DoE simulation time is amortized {every time we predict performance and energy consumption for a}  previously unseen application. Thus, quick exploration and {large} prediction time savings compared to simulation are possible for a previously unseen application.

\vspace{-0.2cm}

\begin{table}[h]

\caption{Number of DoE configurations (``\#DoE conf'') for gathering training data (``DoE run (mins)''), NAPEL training time (``Train+Tune (mins)''), including tuning, and NAPEL prediction time (``Pred. (mins)'').}
\label{tab:napel/timing}
\resizebox{0.965\linewidth}{!}{%
\begin{tabular}{p{0.1\linewidth}|llll}
\toprule
\multicolumn{1}{c|}{\textbf{Application}}  & \multicolumn{4}{c}{\textbf{Training/Prediction Time}}  \\ 

Name &\#DoE conf.&DoE run (mins)&Train+Tune (mins) &Pred. (mins)\\ \hline

atax   & 11& 522 &34.9 &0.49  \\

bfs      &   31&1084 & 34.2&0.48 \\

bp   & 31&1073 &43.8 &0.47\\

chol  & 19&741 &34.9&0.49   \\

gemv&19&741  &24.4 & 0.51  \\

gesu  &19& 731 &36.1 &0.51   \\

gram  &19& 773  &36.5 & 0.52  \\

kme    & 31&742    &36.9&0.55\\

lu    &19& 633 &37.9 & 0.51  \\

mvt   &  19&955 &38.0 &0.54     \\

syrk    & 19&928 &35.7 &0.51    \\

trmm    &19& 898 &37.6 & 0.48  \\ 
\bottomrule

\end{tabular}
}
\vspace{-0.4cm}

\end{table}

\begin{SCfigure}[][h]
\hspace{-0.3cm}
\includegraphics[bb=16 14 296 162,width=0.6\linewidth]{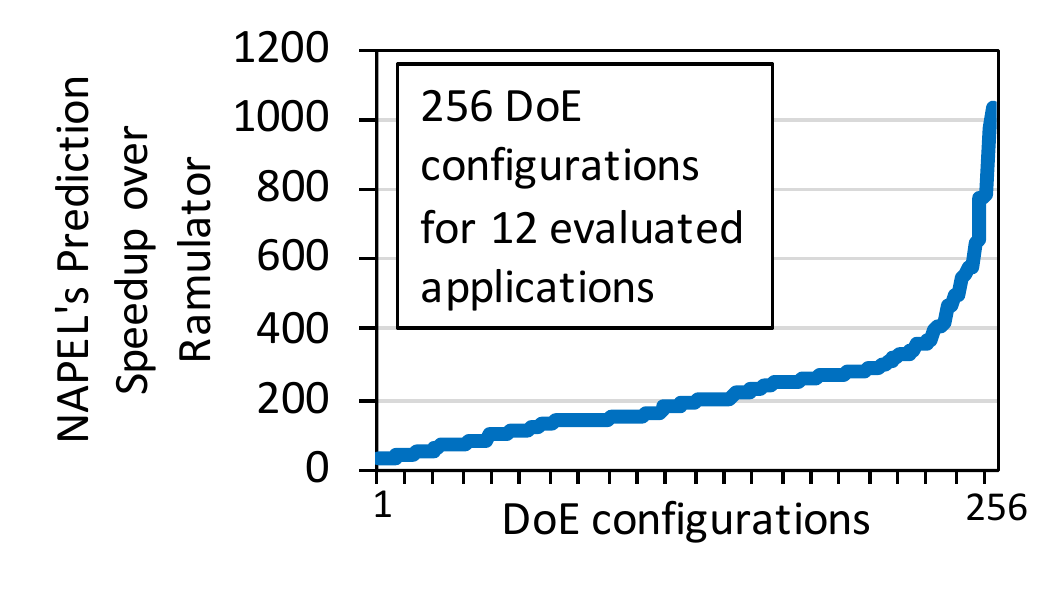}
\caption{NAPEL's prediction speedup (in increasing order) over Ramulator for 256 DoE configurations.}
\label{fig:napel/pred}
\vspace{-0.6cm}
\end{SCfigure}

For all the evaluated applications, we compare the prediction time using trained NAPEL models with the prediction time using {Ramulator} simulations. 
{Figure~\ref{fig:napel/pred} shows NAPEL's prediction speedup over Ramulator for 256 DoE configurations for all the evaluated workloads.}
We observe that NAPEL is, on average, {$220\times$ (min.~$33\times$, max.~$1039\times$) faster than simulation.} 

\vspace{-0.3cm}

\subsection{Accuracy Analysis}
\vspace{-0.1cm}

We analyze the accuracy of NAPEL for previously unseen applications by {performing} cross-validation~\cite{scikit-learn}. 
To evaluate the prediction accuracy for a particular application, our training data comprises all the collected data (using an LLVM kernel analyzer and a microarchitecture simulator) for all applications \emph{except} the application for which the prediction will be \juang{made}. 
We repeat the same process to gather test prediction results for all applications, yet every time we test for a particular application, we do \emph{not} include it in the training set. 
Therefore, when predicting performance and energy consumption of an application {on NMC}, {we do \emph{not} use any data related to that application}. 
This makes the prediction more difficult because the ML algorithm has no knowledge of {the application to be predicted}. 
Thus, the test set differs from the training set as much as applications differ from each other. 
We evaluate the accuracy of the proposed model in terms of relative error $\epsilon_i$ to indicate how close the predicted value $y_{i}^\prime$ is to the actual value $y_{i}$. 
We calculate the mean relative error (MRE) for each application with Equation~\ref{eq:napel/2}.

\vspace{-0.3cm}

\begin{align}
 MRE= \frac{1}{N} \displaystyle \sum_{i=1}^{N} \epsilon_i=\frac{1}{N} \displaystyle \sum_{i=1}^{N}\frac{|y_{i}^\prime -y_{i}|}{y_{i}}
\label{eq:napel/2}
\end{align}

\vspace{-0.1cm}

\begin{figure}[h]
\centering
\includegraphics[bb=240 99 724 438,width=0.90\linewidth,trim={8cm 3cm 8cm 3cm},clip]{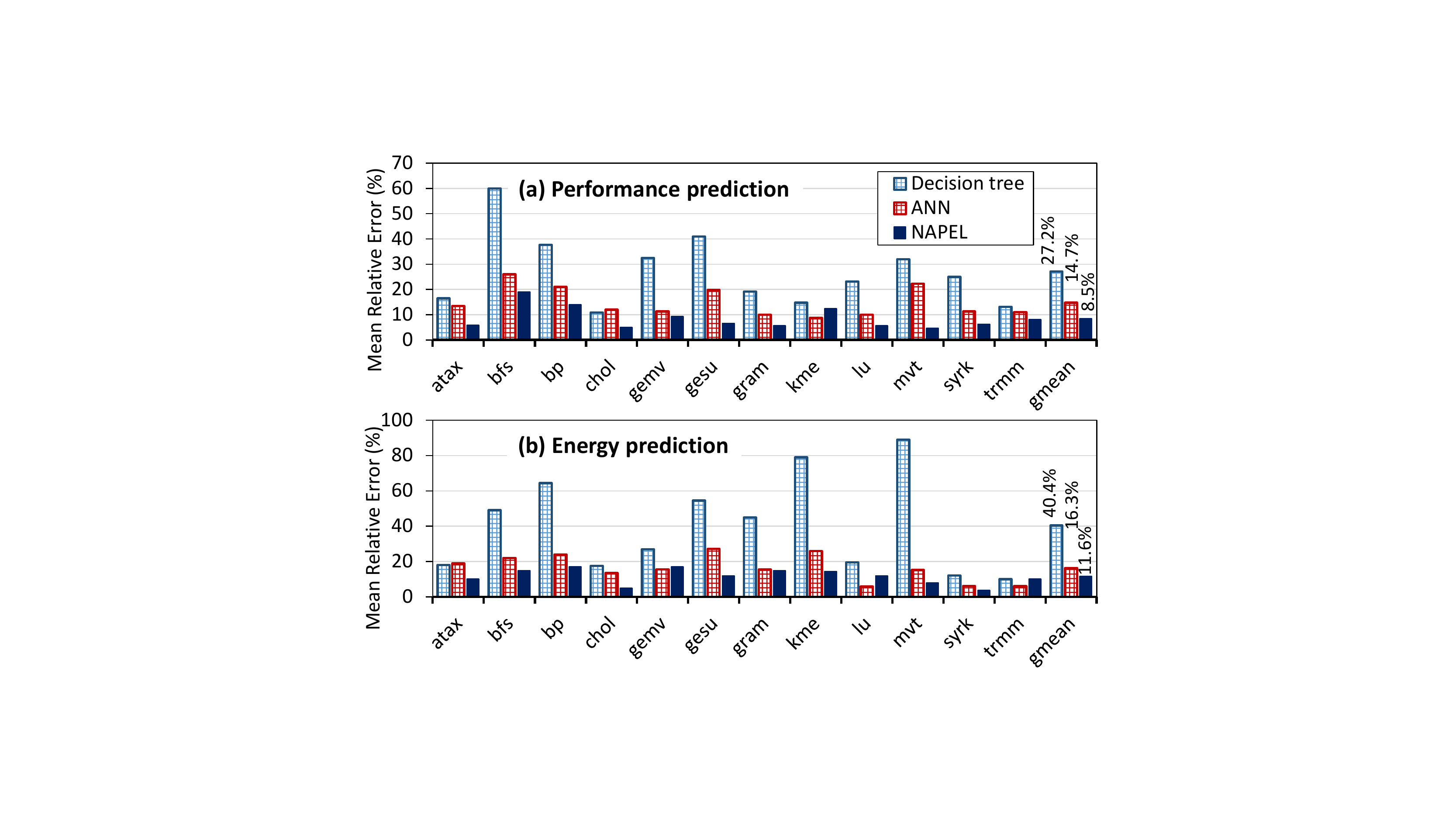}
\vspace{-0.45cm}
\caption{Mean relative error for performance (a) and energy (b) predictions using NAPEL vs. other methods.}
\label{fig:napel/results}
\vspace{-0.25cm}
\end{figure}

Figure~\ref{fig:napel/results} shows {NAPEL's} MRE for the workloads in Table~\ref{tab:napel/results}. {NAPEL's average} MRE is 8.5\% for performance predictions and 11.6\%  for energy-consumption predictions. 
The highest error is for \textit{bfs}, \textit{bp}, and \textit{kmeans} because these applications exhibit quite different characteristics compared to the other {evaluated} applications. 
In Figure~\ref{fig:napel/results}, we also compare {NAPEL} with two other ML algorithms that can be used \juang{to predict} performance and energy consumption: an artificial neural network (ANN) based on Ipek~\etal~\cite{ipek2006efficiently} and a linear decision tree used by Guo~\etal~\cite{guo2013microarchitectural}. 
We make the following three observations. 
First, NAPEL is {1.7$\times$ (1.4$\times$) and 3.2$\times$ (3.5$\times$) more accurate in terms of performance (energy) prediction than the ANN and the linear decision tree, respectively}.
Second, the {linear} decision tree is very inaccurate, as {shown by its} high MRE. Decision trees are suitable {mainly} for linear regression, {so they cannot capture the nonlinearity present in NMC performance and energy}.
Third, ANN is more accurate than the decision tree, but it is {less accurate than} NAPEL for almost all workloads. 
{ANN requires a much larger training dataset to reach NAPEL's accuracy. 
When running these experiments, we also observe that the ANN takes more training time than NAPEL with hyper-parameter tuning (up to 5$\times$).} 
\vspace{-0.2cm}

\vspace{-0.2cm}

\subsection{Use Case: NMC-Suitability Analysis}
\label{sec:napel/usecase}
\vspace{-0.1cm}

In this section, we use NAPEL to perform an NMC-suitability analysis, i.e., to assess the potential benefit of offloading a workload to NMC. 
This analysis compares the energy-delay product (\acrshort{edp}) of executing a workload on the NMC units, which we obtain from NAPEL's predicted NMC performance and energy consumption, to the {measured EDP} of executing the workload on a host processor.
We use \juang{EDP} as our major metric of reference in this analysis because both energy and performance are critical criteria for evaluating NMC suitability.

\begin{figure}[h]
\centering
\includegraphics[bb=15 9 439 162,width=0.85\linewidth]{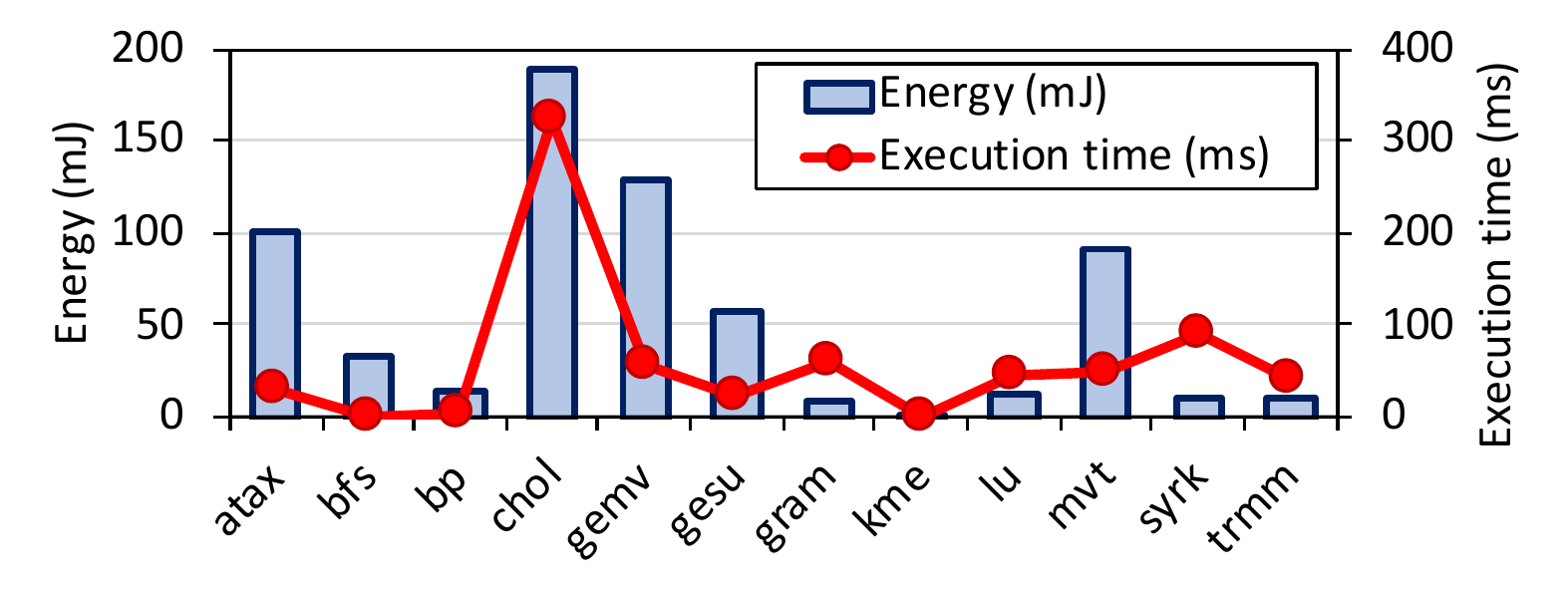}
\vspace{-0.5cm}
\caption{Execution time and energy on an IBM POWER9.
\label{fig:napel/ibmPower}}
\vspace{-0.3cm}
\end{figure}

In order to obtain EDP results for the host system, we use a POWER9 system with 16 cores each supporting four-thread simultaneous multi-threading. 
We measure power consumption by monitoring built-in power sensors on our host system via the AMESTER\footnote{https://github.com/open-power/amester} tool. 
Figure~\ref{fig:napel/ibmPower} shows the execution time and energy consumption of each workload on the POWER9. 
For the EDP results on the NMC system, we use NAPEL with tuned hyper-parameters. 


\begin{figure}[h]
\centering
\includegraphics[bb=5 5 391 139,width=0.85\linewidth]{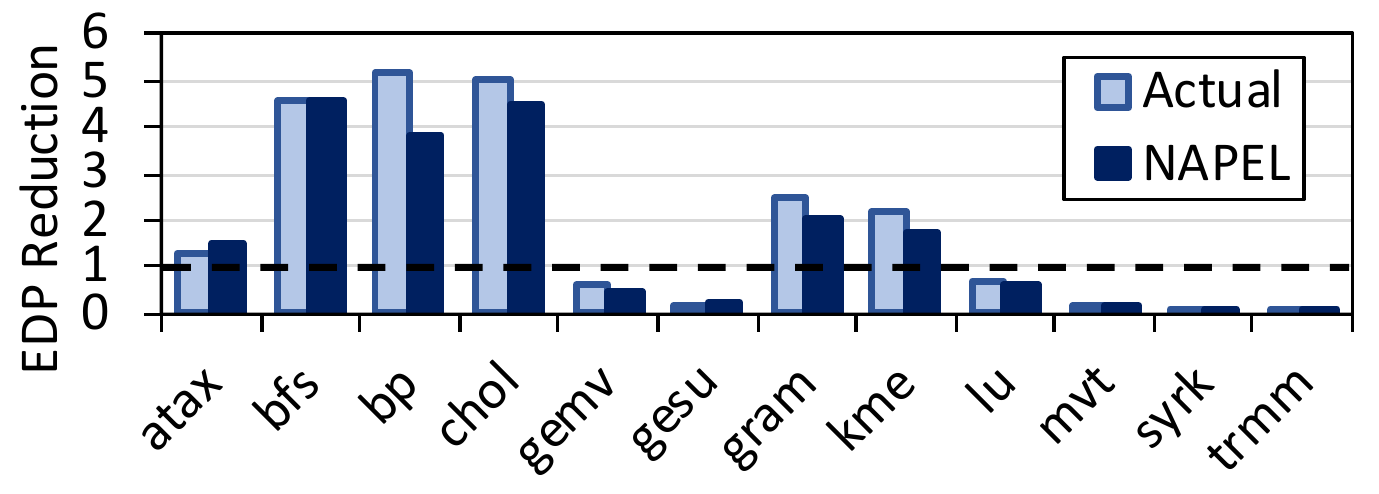}
\vspace{-0.4cm}
\caption{Estimated EDP reduction of offloading to NMC units versus \juanggg{execution} on the baseline host CPU. ``Actual'' denotes the estimation with Ramulator. ``NAPEL'' \juanggg{denotes} NAPEL's prediction results.
\label{fig:napel/edp}}
\vspace{-0.3cm}
\end{figure}
Figure~\ref{fig:napel/edp} shows estimated EDP reduction when executing each application on the NMC system compared to {executing} the same application on the host system using the \textit{test} dataset (see Table~\ref{tab:napel/results}). 
For each application, we show two bars: (1) NAPEL's estimated EDP reduction, and (2) the estimated EDP reduction obtained by simulating the application using the cycle-accurate Ramulator~\cite{7063219} (``Actual''). 
We make five observations. 
First, NAPEL {estimates the same workloads to be NMC suitable} as Ramulator does (i.e., workloads with EDP reduction greater than 1). 
Second, the MRE of NAPEL's EDP prediction is between 1.3\% and 26.3\% (14.1\% on average). 
Third, \textit{gemver}, \textit{gesummv}, \textit{lu}, \textit{mvt}, \textit{syrk}, and \textit{trmm} are not suitable for NMC, since {their} EDP reduction is less than 1. These applications have enough data locality to leverage the host cache hierarchy. 
Fourth, \textit{bfs}, \textit{bp}, \textit{cholesky}, \textit{gramschmidt}, and \textit{kmeans} are good \juang{fits} for NMC. These applications are memory intensive and have irregular memory access patterns, so the host execution suffers from expensive offchip data movement.
Fifth, \textit{atax} benefits from the host cache hierarchy when performing vector multiplication, which has high data locality. However, it also performs matrix transposition, which is memory intensive. For {\textit{atax}-like} workloads, the introduction of a small cache or scratchpad memory in the NMC compute units (larger than the 128B L1 cache in Table~\ref{tab:napel/systemparameters}) can be beneficial, such that the data locality of \juanggg{of the application} can still be exploited.

\vspace{-0.4cm}

\section{Related Work}
\label{sec:napel/relatedWork}
\vspace{-0.1cm}

The lack of evaluation tools is a critical challenge \juan{to} the adoption of near-memory computing (NMC)~\cite{mutlu2019,singh2018review}.
The importance of architecture simulators is widely acknowledged. However, simulators are generally very slow as they may take hours to simulate even \juang{a} single configuration~\cite{guo2013microarchitectural,sanchez2013zsim}.

Recent works propose ML-based performance prediction methods for faster early-stage design space exploration of different architectures. 
Table~\ref{tab:napel/related} lists recent works (including NAPEL) that use different prediction techniques for several architectures. 

{\scriptsize
\begin{table}[h]
\caption{Related works in different domains}
    \label{tab:napel/related}
\centering
\resizebox{0.8\linewidth}{!}{%
\begin{tabularx}{\columnwidth}{p{4cm}p{4.0cm}p{2.4cm}X}
\toprule
\textbf{Name}         & \textbf{Approach}          & \textbf{Architecture} & \textbf{DoE}                                \\ \midrule
Joseph \etal~\cite{1598116} & Linear Regression & CPU          & D-optimal Design                       \\
Ipek \etal~\cite{ipek2006efficiently}         & ANN               & CPU          & Variance Based Sampling             \\

Wu \etal~\cite{7056063}     & ANN               & GPU          & None                     \\
Guo \etal~\cite{guo2013microarchitectural}   & Model Tree        & CPU          & None                     \\

Mariani \etal~\cite{mariani2017predicting} & Random Forest, \newline Genetic Algorithm & HPC & D-optimal Design, CCD\\

SemiBoost~\cite{li2018processor}    & ANN               & CPU          & Latin Hypercube Sampling               \\\bottomrule
\textbf{NAPEL}     & \textbf{Random Forest}     & \textbf{NMC}     & \textbf{CCD}    \\ \bottomrule
\end{tabularx}
}
\end{table}
}

Joseph~\etal~\cite{1598116} and Guo~\etal~\cite{guo2013microarchitectural} use linear regression models to predict CPU performance. Linear models cannot accurately capture nonlinearity between application and processor responses, as shown in Figure~\ref{fig:napel/results}. 
Wu~\etal~\cite{7056063} use an ANN for GPU performance prediction without applying the DoE technique. 
Unlike NAPEL, \juangg{this work uses} traditional, time-consuming brute-force techniques to collect the training dataset. 
In the HPC domain, Mariani~\etal~\cite{mariani2017predicting} predict the performance of applications on cloud architectures using random forest and genetic algorithms, which are trained using DoE techniques. 
Ipek~\etal~\cite{ipek2006efficiently} use an ANN with variance-based sampling for CPU performance prediction. 
Likewise, Li~\etal~\cite{li2018processor} use an ANN with Latin hypercube sampling for design-space exploration of multicore CPUs. 
To our knowledge, NAPEL is the first performance and energy-prediction framework for NMC architectures that uses {machine-learning} models. 
NAPEL can make accurate predictions for previously unseen applications {on NMC architectures.}

\section{Conclusion}
\label{sec:napel/conclusion}
\vspace{-0.1cm}

{We introduce} NAPEL, the first high-level \juan{machine learning}-based prediction framework for fast and accurate {early-stage} performance and energy-consumption estimation {on NMC architectures}.
NAPEL avoids time-consuming simulations to predict the performance and energy consumption of previously unseen applications on \juan{various} NMC architecture configurations. 
{To achieve this}, NAPEL relies on random forest, an ensemble learning technique, to build {its} prediction models. 

NAPEL is {$220\times$ faster} than a state-of-the-art NMC simulator, with an accuracy loss in performance (energy) prediction \juan{of only} 8.5\% (11.6\%) compared to the simulator. 
Compared to an artificial neural network, NAPEL is 1.7$\times$ (1.4$\times$) more accurate in performance (energy) prediction. 
NAPEL can accurately perform fast design-space exploration for different applications and NMC \juan{architectures}.
{We hope the NAPEL approach enables faster development of NMC systems and inspires the development of other alternatives to simulation for NMC performance and energy estimation.}

 \renewcommand{\namePaper}{LEAPER\xspace} 
\chapter[\texorpdfstring{\namePaper: Modeling FPGA-Based Systems via Few-Shot Learning}{\namePaper: Modeling Cloud FPGA-based Systems via Few-Shot Learning}
]{\chaptermark{header} \namePaper: Modeling FPGA-Based Systems via Few-Shot Learning}
\chaptermark{\namePaper}
\label{chapter:leaper}

\chapternote{A part of this chapter is published as \emph{``Modeling FPGA-Based Systems via Few-Shot Learning''} in FPGA 2021.
}

Machine-learning-based models have recently gained traction as a way to overcome the slow downstream implementation process of FPGAs by building models that provide fast and accurate performance predictions. However, these models suffer from two main limitations: (1) a model trained for a specific environment cannot predict for a new, unknown environment; (2) training requires large amounts of data (features extracted from FPGA synthesis and implementation reports), which is cost-inefficient because of the time-consuming FPGA design cycle. In a cloud system, where getting access to platforms is typically costly, error-prone, and sometimes infeasible, collecting enough data is even more difficult. FPGA-based cloud environments are usually $2\times$ more expensive than CPU-only cloud environments. Therefore, before deploying a cloud instance, a user cares whether the attained performance while using an FPGA would justify the incurred cost (both in terms of designing an accelerator and deploying in the cloud). To overcome these limitations, in this chapter,  we propose \namePaper, a \textit{transfer learning}-based approach for FPGA-based systems that adapts an existing ML-based model to a new, {unknown} environment. 

\section{Introduction}

The need for energy-efficiency from {edge} to cloud computing has boosted the widespread adoption~of~FPGAs. 
{In cloud computing~\cite{arefmicrosoft,choi2019depth,aws,alibaba,pourhabibi2020optimus,byma2014fpgas}, 
FPGA's flexibility is not just about being able to make use of reconfigurable hardware for a diverse set of workloads~\cite{10.5555/3294624}. {Its flexibility can also be attributed to the}  {cloud} deployment model that spans from on-premises clusters to compute, storage, and networking capacity in public clouds and even out to the edge where AI and analytics are being increasingly deployed because of latency and data movement issues~\cite{mutlu2019}.} 


An FPGA is highly configurable as its circuitry can be tailored to perform any task~\cite{umuroglu2017finn,umuroglu2020logicnets,kara2018columnml, dai2017foregraph}. 
The large configuration space of FPGA and {the} complex interactions among configuration options lead many 
developers to explore individual optimization options in an ad-hoc manner. 
Moreover, FPGAs have infamously low productivity due to the time-consuming FPGA {implementation} process~\cite{o2018predictive}.
 A common challenge that past works {have faced} is how to evaluate the performance of an FPGA implementation in a reasonable amount~of~time~\cite{o2018hlspredict}.  Thus, the development of efficient FPGA accelerators has required tremendous engineering effort due to the complexity of the FPGA configuration space and difficulties in evaluating performance.
To overcome this problem, researchers have {recently} employed machine learning (ML)-based models~\cite{o2018hlspredict,wang2020lutnet,xppe_asp_dac,dai2018fast} to {estimate}  the performance of an FPGA-based system quickly. {These models are} based, in turn, on traditional ML {approaches}. 

Traditional ML models have four fundamental issues {that can reduce the usability for assessing FPGA performance, especially in a cloud environment}. First, they are trained for specific workloads, fixed hardware, and/or a set of inputs. {Therefore,} when presented with a different feature-space distribution {because of a new workload or hardware}, {an ML model must} 
be retrained from scratch. Otherwise, {the model} will perform poorly because the trained model does not have a notion of the new, {unknown} environment.\footnote{{In this chapter, we consider an application or hardware platform as an environment.}} Therefore, traditional ML-based models have limited \textit{re-usability}.

Second, learning-based approaches, { such as neural networks}, require a considerable {number} of samples to construct a useful prediction model. {Collecting such a large number of samples is} often slow and time-consuming {due to the very long FPGA implementation cycle}. 
Similarly, in {settings} such as cloud computing, where getting access to platforms is typically \textit{costly}, \textit{error-prone}, and sometimes \textit{infeasible}, the data collection process {is even more difficult}. 

{Third, traditional machine learning with limited samples {is} prone to serious \emph{overfitting} problems {(i.e., when a model matches too closely to the training data)}~\cite{dai2007boosting}, limiting model generalization.}
Fourth, it is impossible to {construct} one model for all different scenarios as the interpretation of data changes over time. Thus, ML models are prone to \emph{concept drift}, where the accuracy of an ML-based model could degrade due to a change in the statistical properties of a target variable {(e.g., using a different dataset for an application than the one used during the training of an ML-model)}~\cite{model_drift}.\\

\textbf{Our research aims to answer the following question: for an FPGA-based system, can we leverage an ML-based performance model trained on a low-end local system to predict the performance in a new, unknown, high-end FPGA-based system?}

{{As an} answer to this question,} we present \namePaper\footnote{We call our mechanism \namePaper~because it allows us to hop or ``leap'' between machine learning models.}, a transfer learning-based model that predicts the performance of a new, unknown high-end FPGA-based system. We train \namePaper~ on a low-end local system. 
\namePaper~ uses predictive modeling to train an ML-based model 
and statistical techniques to collect representative training data set efficiently. 
\namePaper uses \emph{transfer learning}~\cite{pan2009survey} to leverage a trained \textit{base model} (for a low-end system) and adapt it efficiently to an unknown \emph{{target}} environment {(i.e., a high-end system)} with a few samples from the target environment as possible. 

\begin{figure}[h]
  \centering
  \includegraphics[bb=4 14 1147 485,width=1\linewidth,trim={0.2cm 0.9cm 0.4cm 0.3cm},clip]{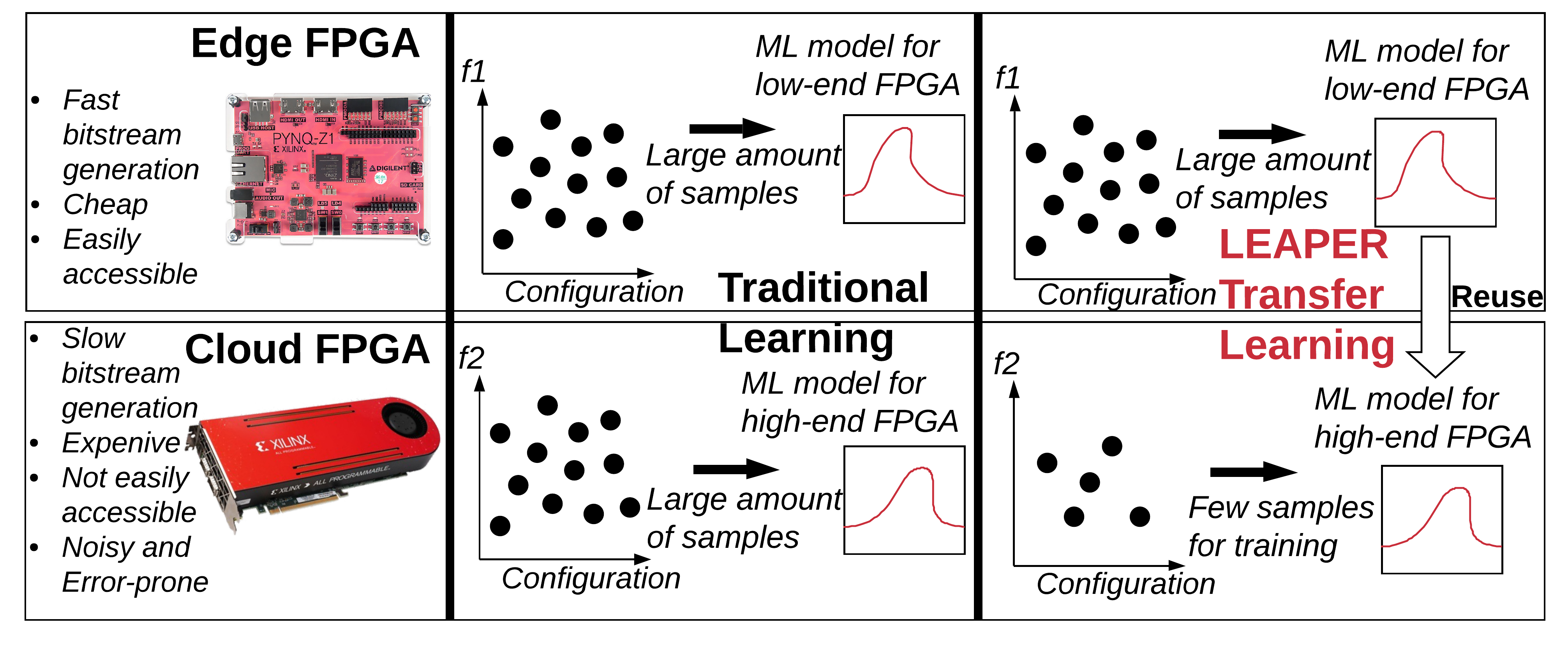}
     \vspace{-0.3cm}
  \caption{Comparison of traditional learning approach and LEAPER. {Traditional learning methods are costly because they build models only for a specific environment. } LEAPER allows transfer of models from a low-cost edge FPGA, where data collection is easier, to a high-cost cloud FPGA environment to build cheaper and faster models.}
  \label{fig:transfer_intro}
 \end{figure}

{ Figure~\ref{fig:transfer_intro} demonstrates the traditional approach of building models and the LEAPER transfer learning-based approach. Using the traditional approach, we would 
{need} to create two separate prediction models, {one} for the {low-end} edge {environment} and {another for the high-end} cloud environment, each {one} requiring a large number of samples. In contrast, \namePaper~provides the ability to reuse the prediction model  built on an {\emph{inexpensive}} edge FPGA for {performance prediction on a target} cloud environment~\footnote{Note: LEAPER is not limited to a cloud environment.} using only a few samples {from the target environment}. {This allows {developers} to avoid the slow downstream implementation process of FPGAs 
{by generating} cheaper and faster performance models} using transfer learning.  }
This paradigm is also referred to as \textit{few-shot learning}~\cite{fsl}. The idea behind {few-shot learning} is that, similar to humans, algorithms can learn from past experiences and transfer the knowledge to accomplish previously-{unknown} tasks more efficiently. The transferred models usually have high generality~\cite{fsl} and can overcome {concept drift}~\cite{concept_drift}.

\section{\namePaper}
\label{sec:LEAPER_SIGMETRICS/methodology}

\namePaper~is a performance 
{and} resource estimation 
{approach} to \textit{transfer} ML-based models across different FPGA-based platforms. First, we give an overview of \namePaper~(Section~\ref{subsec:LEAPER_SIGMETRICS/overview}). Second, we describe the two 
 components of \namePaper~that are used to generate training datasets: (1) FPGA-based accelerator configuration options and application features used for training our base model (Section~\ref{subsec:LEAPER_SIGMETRICS/fpga_deploy}), and (2) the \textit{design of experiments} (DoE)~\cite{montgomery2017design} methodology (Section~\ref{subsec:LEAPER_SIGMETRICS/lhs}). Third, we briefly describe the {base} model training (Section~\ref{subsec:LEAPER_SIGMETRICS/base_model}).  Fourth, we explain the 
 \juan{key} component of \namePaper:
 the transfer learning technique (Section~\ref{subsec:LEAPER_SIGMETRICS/ensemble_transfer}). 

\subsection{Overview}
\label{subsec:LEAPER_SIGMETRICS/overview}
\begin{figure*}[h]
  \centering
  \includegraphics[bb=44 60 2325 1016,width=0.9\linewidth,trim={0.8cm 0.8cm 0.4cm 1cm},clip]{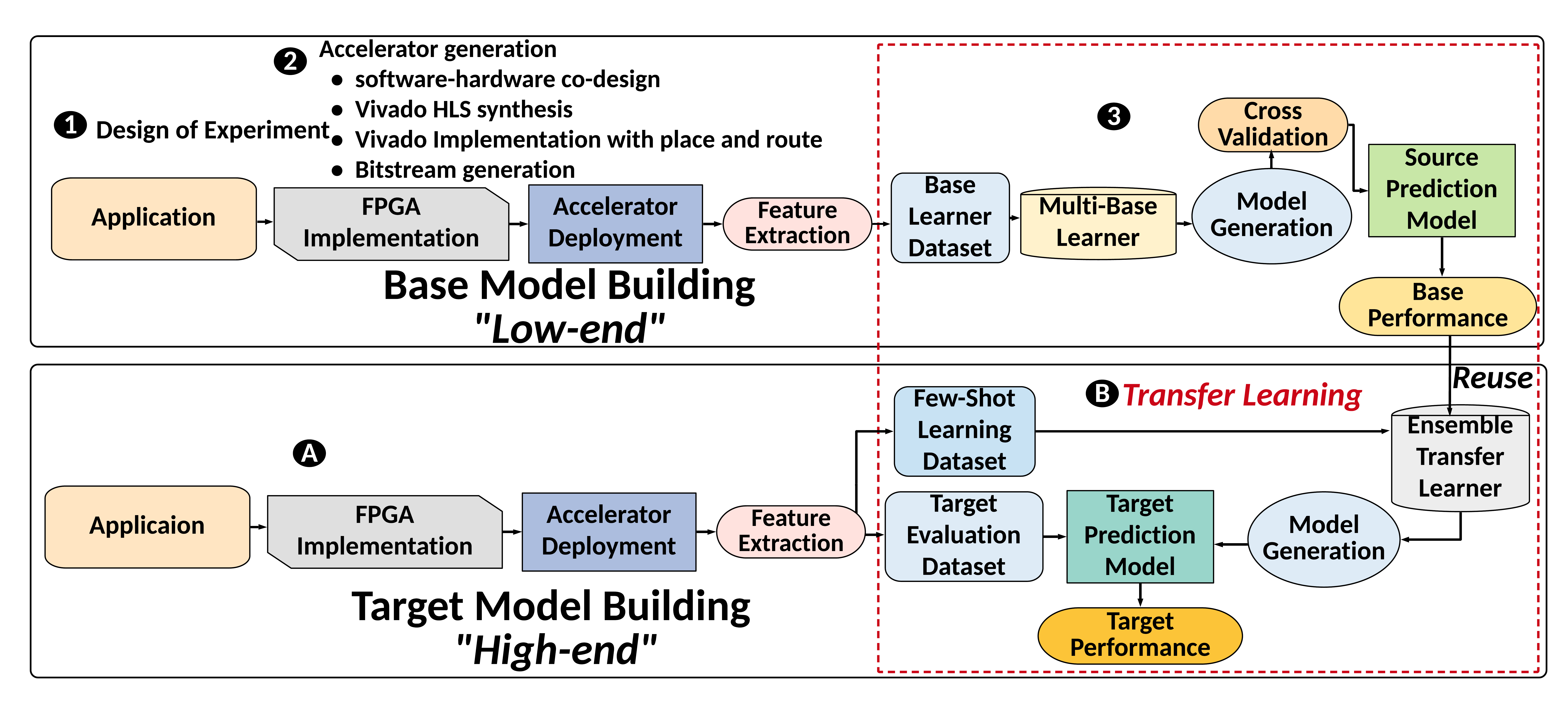}
     \vspace{-0.1cm}
  \caption{Overview of 
  {\namePaper}. Base Model Building: {\namePaper~builds} ML models that predict 
performance 
and resource usage for an application {on an FPGA}. Target Model Building: with \textit{few-shot learning}, \namePaper~adapts base models to a new, unknown environment, 
{from} which only a few labeled samples are 
{needed}.
}
  \label{fig:LEAPER_SIGMETRICS/transfer_overiew}
 \end{figure*}

Figure~\ref{fig:LEAPER_SIGMETRICS/transfer_overiew} depicts the key components of \namePaper. \juan{The upper part of the figure describes the construction of the base model, while the lower part shows the phases for the target model.}

\noindent \textbf{Base Model Building.} \namePaper~base model building consists of three phases. In the first phase (\circled{1} in Figure~\ref{fig:LEAPER_SIGMETRICS/transfer_overiew}), 
{we employ} 
\textit{Latin hypercube sampling} (LHS)~\cite{li2018processor} 
to select a small set of input configurations that well represent the entire space of input configurations ($c_{lhs}$) to build a highly accurate \textit{base learner}. We use LHS to minimize the number of experiments needed to gather training data for \namePaper~while ensuring {good} quality training data (Section~\ref{subsec:LEAPER_SIGMETRICS/lhs}). In the second phase~\circled{2}, FPGA implementations are made with a software-hardware co-design process. Once the FPGA design has been implemented, the \juan{resulting FPGA-based} accelerator is deployed \juan{in a system} with a host CPU. 
{Then,} we 
\juan{run} {the} $c_{lhs}$ configurations on 
\juan{this FPGA-based system} to gather responses for training our base model. The generated responses, along with \juan{the} applied configuration options {(\textit{ref}. Table~~\ref{tab:LEAPER_SIGMETRICS/pragma})}, form the input to our base ML algorithm. 
In the third phase~\circled{3}, we train our ML algorithm (Section~\ref{subsec:LEAPER_SIGMETRICS/base_model}) using ensemble learning~\cite{opitz1999popular}. 
We divide our dataset into 10 equal subsamples during cross-validation, of which 1 set is used for validation.  
Once trained, the framework can predict the performance and resource usage \juan{on the base system (with a low-end FPGA)} of 
{previously-unseen} configurations, 
\juan{which} are not part \juan{of the} $c_{lhs}$ \juan{configurations used during training.} 

\noindent \textbf{Target Model Building.} {To transfer the base model, which we built in the previous stage, to a target cloud environment, we introduce the \emph{few-shot learning} target model stage.} In the first phase (\circled{A} {in Figure~\ref{fig:LEAPER_SIGMETRICS/transfer_overiew}}) of {the} target model building, {we repeat the accelerator generation step to get a few samples.} 
We perform this step to create our \textit{few-shot} transfer learning dataset ($c_{tl}$), 
{ which is used to adapt the base model to the target cloud environment.}
In the final phase \circled{B}, we train our transfer learners (see Section~\ref{subsec:LEAPER_SIGMETRICS/ensemble_transfer}) to leverage the base model to perform predictions for a new, {unknown} target environment (new application or hardware).  
\subsection{FPGA Configuration Options and Application Features}
\label{subsec:LEAPER_SIGMETRICS/fpga_deploy}
The ML feature vector used for training an ML model is composed of FPGA configuration options (Table~\ref{tab:LEAPER_SIGMETRICS/pragma}) and application features (Table~\ref{tab:LEAPER_SIGMETRICS/application_features}).
Table~\ref{tab:LEAPER_SIGMETRICS/pragma} describes commonly used HLS pragmas that belong to {our} FPGA configuration {options} for both the base and the target environment and constitute part of our ML feature vector. \juan{We select these HLS pragmas because they are used to optimize and tune the performance of FPGA implementations~\cite{xilinx_hls}.}
\begin{table}[h]
\centering
  \caption{{The FPGA configuration options used to train our ML-models.}}
    \label{tab:LEAPER_SIGMETRICS/pragma}
\footnotesize   
        \resizebox{\linewidth}{!}{%
\begin{tabular}{p{0.25\linewidth} p{0.62\linewidth}}
\toprule
\textbf{Configuration} &  \textbf{Description}  \\ \midrule
Pipelining (PL)       &   Enabled/Disabled \\
Partitioning (PR)   &  Block/Cyclic/Complete (Factor: $2^n, 1\leq n \leq 6$)) \\

Inlining (I)        &   Enabled/Disabled function inlining \\
Dataflow (D)        &  Task level pipelining \\
Read burst (R)     &   Read data burst from the host \\
Write burst (W)    &   Write data burst from the host \\
Unrolling  (U)     &   Unrolling factor (Factor: $2^n, 1\leq n \leq 6$) \\
FPGA Frequency (F)     &  Four-different frequency levels for the FPGA logic   \\
\bottomrule
\end{tabular}
}
\end{table}

Both \textit{loop pipelining} \juan{(PL)} and \textit{loop unrolling} \juan{(U)} can improve application performance significantly. 
To enable simultaneous memory accesses, \textit{array partitioning} \juan{(PR)} \juan{divides arrays} into smaller memory units of arbitrary dimensions 
to map \juan{them} 
to 
\juan{different memory banks.} This optimization produces considerable speedups but consumes more resources. \textit{Inlining} \juan{(I)} ensures that a function is instantiated as dedicated hardware. \textit{Dataflow} \juan{(D)} allows parallel execution of tasks, which is similar to multi-threading in CPUs. 
In addition, \textit{burst read} \juan{(R)} and \textit{write}  \juan{(W)} access to/from the host guarantee that the accelerator is not stalled for data. Moreover, \textit{FPGA frequency} \juan{(F)} affects \juan{not only} performance 
\juan{but also} resource consumption. For instance, to meet the FPGA timing requirements, the FPGA tool tries to insert registers between the flip-flops, \juan{which increases the resource consumption}.

In total, our configuration options \juan{for a particular application} consist of \juan{up to} 4,608 configurations. 
The actual configuration space of an application depends 
on the specific application characteristics (see Table~\ref{tab:LEAPER_SIGMETRICS/app_details}). 
\juan{For example, we include} loop unrolling \juan{in the configuration space when an application contains loops that can be unrolled.} 

\juan{For each application kernel \textit{k} processing a dataset \textit{d}, we obtain an application profile \textit{p(k, d)}.} 
\textit{p(k, d)} is a vector where each parameter is a statistic about an application feature. 
Table~\ref{tab:LEAPER_SIGMETRICS/application_features} lists the main application features {we extract} {by using the LLVM-based PISA analysis tool~\cite{Anghel2016}}. 
{We select} these features to analyze the behavior of an application (data reuse distance, memory traffic, memory footprint, etc.). 
Ultimately, the application profile \textit{p} has 395 features, which includes all the sub-features {of each metric we consider}.



\begin{table}[h]
\centering
\caption{Main application features extracted from LLVM.}
\label{tab:LEAPER_SIGMETRICS/application_features}
\footnotesize
\begin{tabular}{p{0.24\linewidth} p{0.62\linewidth}}
\toprule
\textbf{Application Feature} & \textbf{Description} \\ 
\midrule
Instruction Mix & Fraction of instruction types (integer, floating point, memory read, memory write, etc.) \\
ILP                      & Instruction-level parallelism on an ideal machine. \\
Data/Instruction reuse distance   & For a given distance $\delta$, probability of reusing one data element/instruction (in a certain memory location) before accessing $\delta$ other unique data elements/instructions (in different memory locations). \\
Register traffic    & Average number of registers {per} instruction. \\
Memory footprint    & Total memory size used by the application. \\ 
\bottomrule

\end{tabular}
\end{table}


\pagebreak
\subsection{Latin Hypercube Statistical Sampling}
\label{subsec:LEAPER_SIGMETRICS/lhs}
{Running experiments to collect training data for all available optimization options can be an extremely time-consuming process. For example,} 
the configuration options ($\mathbb{CO}$) of 
{only} eight parameters {with two possible values each} 
{entails} $2^8=256$ different configuration inputs. 
If we spend 6 hours (e.g., FPGA downstream implementation process on an ADM-PCIE-KU3 {FPGA board}~\cite{adku3}) to collect one training point, it would take {us} 
$\sim$64 days to collect data for all configurations 
for just \textit{one} application on a \textit{single} platform. 
{This ``brute-force'' approach to collecting training data is too time-consuming: the sheer number of experiments renders \juan{a} detailed implementation intractable}.

To create a 
{cost-effective} model, we use the \juan{\emph{design of experiments} (DoE) methodology}~\cite{montgomery2017design} 
to minimize the number of experiments needed 
{for training data collection} without sacrificing the amount and quality of the information gathered 
\juan{from} the experiments. 
DoE is a set of statistical techniques meant to locate a small set of points in  parameter space to represent the entire parameter space.
{In particular, we make use of \juan{a type of DoE called}  \textit{Latin hypercube sampling} (LHS)~\cite{li2018processor} because it allows each of the critical parameters to be represented in a fully stratified manner (i.e., dividing the configuration space into subgroup before further sampling), which provides a better coverage~\cite{fang2002centered}.}

\begin{SCfigure}[][h]
\vspace{0.8cm}
\centering
    \resizebox{0.3\textwidth}{!}{

    \begin{tikzpicture}

\draw[step=0.25cm,gray,very thin] (0,0) grid (2,2);

\draw[thick,->] (0,0) -- (2.4,0);
\draw[thick,->] (0,0) -- (0,2.2);
\filldraw [color=gray!60, fill=gray!5,very thick] (2,2) circle (2pt);
\filldraw [color=gray!60, fill=gray!5,very thick] (0.25,1.5) circle (2pt);
\filldraw [color=gray!60, fill=gray!5,very thick] (0.75,0) circle (2pt);
\filldraw [color=gray!60, fill=gray!5,very thick] (1,1.25) circle (2pt);
\filldraw [color=gray!60, fill=gray!5,very thick] (1.75,0.25) circle (2pt);
\filldraw [color=gray!60, fill=gray!5,very thick] (1.5,0.75) circle (2pt);
\filldraw [color=gray!60, fill=gray!5,very thick] (1.25,1.75) circle (2pt);
\filldraw [color=gray!60, fill=gray!5,very thick] (0,1) circle (2pt);
\filldraw [color=gray!60, fill=gray!5,very thick] (0.5,0.5) circle (2pt);

\node at (0,2.4)[ rotate=0] {\small \textit{x1}};
\node at (2.6,0)[ rotate=0] {\small \textit{x2}};
\end{tikzpicture}
}
\vspace{-1cm}
\caption{LHS with 2 parameters where the input space is divided into equal intervals and 9 non-overlapping sample points are chosen.
\label{fig:LEAPER_SIGMETRICS/LHS}}
\end{SCfigure}
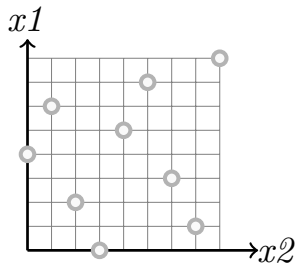

LHS divides each parameter range into \textit{k} intervals and takes only one sample from each interval with equal probability, which is more efficient than a random approach and more cost-effective than a ``brute-force'' approach. 
To apply 
{LHS, we} 
choose \textit{m} sample points, each from a specific interval,  which together we refer to as \textit{c$_{lhs}$}. 
{Thus,} LHS guarantees effective space-filling, i.e.,  LHS spreads out points with the aim of encouraging a diversity of data~\cite{li2018processor}. 
Figure~\ref{fig:LEAPER_SIGMETRICS/LHS} illustrates LHS with two parameters \juan{\textit{x1} and \textit{x2}}, 
\juan{which create an} input space \juan{that} is divided into equal-area intervals. 

\subsection{Base Model {Building}}
\label{subsec:LEAPER_SIGMETRICS/base_model}

The third phase of \namePaper~is the base learner training phase. Formally, { in a learning task, $\mathcal{X}$ represents feature space with label $\mathcal{Y}$, where a machine learning model is responsible for estimating a function $f: \mathcal{X} \to \mathcal{Y}$.} 
{\namePaper~predicts the execution time (resource consumption) $\mathcal{Y}$ for a tuple} $(p,~k,~c)$ {that belongs to the ML feature space $\mathcal{X}$}, where \textit{p} is an {FPGA-based} {configuration option (see Section~\ref{subsec:LEAPER_SIGMETRICS/fpga_deploy})} 
{that runs an} application characteristics \textit{k}, with 
{an optimization configuration vector \textit{c}}. 

We use two base learners.  Our base learners are non-linear algorithms that {can} capture the intricacies of FPGA architectures by predicting the execution time 
or resource consumption. 
Our first algorithm is the \emph{random forest} (RF)~\cite{breiman2001random}. RF consists of an ensemble of learners where it aggregates the predictions of the weak learners to select the best prediction. This technique is called \textit{bagging}. We use RF to avoid a complex feature-selection scheme 
\juan{since RF} embeds automatic procedures \juan{that are able} to screen many input features~\cite{mariani2017predicting}, \juan{like the ones we selected in the previous section}. 
Starting from a root node, RF constructs a tree and iteratively grows the tree by associating 
\juan{a node} with a splitting value for an input 
\juan{feature} to generate two child nodes. 
Each node is associated with a prediction of the target metric, equal to the observed \juan{mean} value in the training dataset for the input subspace that the node represents. 
Our second learner is \emph{gradient boosting}~\cite{gradientboosting}, which consists of an ensemble of learners. Gradient boosting aims to \textit{boost} \juan{the accuracy of a weak learner}  
by using other learners to correct its predictions. {Bagging reduces model variance and boosting decreases errors~\cite{kotsiantis2004combining}. {Therefore, we use random forest and gradient boosting together to increase the predictive power of our final base model.}}


{The training dataset for our base model} has two parts: (1) an optimization configuration vector \textit{c}, whose representation remains invariant across different environments, and (2) {the} {responses corresponding to each 
{tuple} $(p,~k,~c)$}. 
To gather the architectural responses, {we run each} application \textit{k} belonging to {the} training set $\mathbb{T}$ with {an} input dataset \textit{d} 
on an FPGA-based platform \textit{p}, {deploying} a configuration \textit{c}. 
{This way, we obtain the execution time for the tuple \textit{(p,~k,~d)}, which we can use} 
as a \textit{label} ($\mathcal{Y}$) for training our base learner {for performance prediction}. 
{We build a similar model} to predict resource consumption, where we use {the} resource consumption ($\eta_{\{BRAM, FF, LUT,  DSP\}}$) {of the tuple \textit{(p,~k,~d)}} as a \textit{label} when we train our base learner {for resource consumption}. 
After training our base learners, 
we can predict the {execution} time (resource usage) ($\hat{f_s}:\mathcal{X}_s \to \mathcal{Y}_s$) of 
{tuples \textit{(p,~k,~d)}} that are \emph{not} in the training set. We use 10-fold cross-validation to validate our base learner's performance, {whereby the data is divided into ten validation sets.} 
\subsection{{{Cloud} Model Building via Transfer Learner}}
\label{subsec:LEAPER_SIGMETRICS/ensemble_transfer}
{The real strength of 
LEAPER 
comes from its ability to transfer trained FPGA models.}
{\namePaper~defines} a target environment $\tau_{t}$ as an environment for which we wish to build a prediction model {$\hat{f}_t$ where, however, data collection is expensive}, and 
a source environment $\tau_{s}$ as an environment 
for which we can \textit{cheaply} collect many samples to build an ML model ${f}_s$. {In our case, $\tau_{s}$ is a low-cost edge FPGA, while $\tau_{t}$ is a high-cost cloud FPGA. } {\namePaper~then transfers the ML model for $\tau_{s}$ to $\tau_{t}$.}

Algorithm~\ref{algo:LEAPER_SIGMETRICS/transfer} presents \namePaper's transfer learning approach. 
\juan{\namePaper~ trains transfer learners (TL) that} 
transform the source performance and utilization model $f_{s}$ to \juan{the target model} $f_{t}$ by using {a few} sample observations from {both} the source and the target environments, which we refer to as \textit{c$_{tl}$}. This helps us to avoid measuring all c$_{lhs}$ from a cost-prohibitive target cloud environment. We select \textit{c$_{tl}$} from \textit{c$_{lhs}$} by applying {a} probability-based sampling {technique} called \textit{reservoir sampling}~\cite{vitter1985random}. 
\juan{Reservoir sampling} 
assigns an equal probability of being selected to every element of 
\juan{a} population \juan{(i.e., \textit{c$_{lhs}$})}. 
\juan{By using the selected} \textit{c$_{tl}$}, we generate a transfer model $\hat{h}_t$. Finally, to build $\hat{f}_t$ from ${f}_s$ we use $\hat{h}_t$ that performs a non-linear transformation of the predictions of ${f}_s$. We use non-linear transfer learners because, based on our analysis (\juan{Section}~\ref{subsection:LEAPER_SIGMETRICS/relatedness_analysis}), non-linear models can capture the nonlinearity present in the FPGA performance and configuration options.

\begin{figure}[h]
  \centering
\begin{minipage}{\linewidth}
\centering
{
\footnotesize
\begin{algorithm}[H]
 \caption{\namePaper's~transfer learning}
 \label{algo:LEAPER_SIGMETRICS/transfer}
  \begin{algorithmic}[1]
   \State \textbf{Input:}{(1) Base learner ($f_s$) i.e., trained low-end edge FPGA prediction model,\\  \hskip3em (2) Sub-sampled {\emph{few-shot learning} dataset $c_{tl}\subset c_{lhs}$ from the\\\hskip3embase and the target model}}   
  \State \textbf{Output:}{ Target {cloud FPGA} model $\hat{f}_t:\mathcal{X}_t \to \mathcal{Y}_t$}
\State \textbf{Initialization:} {Maximum number of iterations M}
\While{$M\not=0$}
    \State Normalize the feature vector
    \State Train ensemble transfer learners (TL) with $c_{tl}$
   \State Find the candidate TL ($\hat{f}_t$):
   \State \hskip1em  $\hat{h}_t:\mathcal{X}_{tl} \to \mathcal{Y}_{tl}$ that minimizes the error over the $c_{lhs}-c_{tl}$ \label{algo:LEAPER_SIGMETRICS/line:findTL}
    \State Compute the mean relative error: 
   \State \hskip1em $\epsilon_{mre}=\frac{1}{c_{lhs}-c_{tl}} \displaystyle\sum_{i=1}^{c_{lhs}-c_{tl}} \frac{|y_{t}^{acc} -y_{t}^{pred}|}{y_{t}^{acc}}$
   
    \State {Use identified $\hat{h}_t$ to transform predictions of $f_s$:}
   \State \hskip1em$\hat{f}_t$=$\hat{h}_t(f_s)$ \hskip7em where $f_s: \mathcal{X}_s \to \mathcal{Y}_s$
    \State$M \leftarrow M - 1$
\EndWhile
\textbf{return}  $\hat{f}_t$
\end{algorithmic}
\end{algorithm}
}
\end{minipage}
\end{figure}

{In transfer learning, a weak relationship between the base and the target environment can decrease the predictive power for the target environment model. This degradation is referred to as a \textit{negative transfer}~\cite{jamshidi2017transfer}. To avoid this,} we use an ensemble 
model trained on the transfer set (i.e., {the} \textit{few-shot learning} dataset in Figure~\ref{fig:LEAPER_SIGMETRICS/transfer_overiew}) as our transfer learners (TLs). Our first TL is based on TrAdaBoost~\cite{dai2007boosting}, a boosting algorithm, which is a learning framework that fuses many weak learners into one strong predictor by adjusting the weights of training instances. {The motivation behind such an approach is that by fusing many weak learners boosting can improve the overall predictions
in areas where the previously grown learners did not perform well.} We use Gaussian process regression~\cite{Gaussian} as our second TL. It is a Bayesian non-parametric algorithm that calculates the probability distribution over all the appropriate functions that fit the data.  To transfer a trained model, we train TrAdaBoost and Gaussian progression, which are our candidate TLs. We choose the one that has minimum transfer error (see Line~\ref{algo:LEAPER_SIGMETRICS/line:findTL}).



\section{{Evaluation}}
\label{sec:LEAPER_SIGMETRICS/evaluation}

{We evaluate \namePaper~using six} 
 benchmarks (see~Table~\ref{tab:LEAPER_SIGMETRICS/app_details}), which are hand-tuned for FPGA execution 
covering 
{several application} domains, i.e., \textbf{(1) image processing}:  histogram {calculation (\textit{hist})~\cite{gomez2017chai}, and canny edge detection (\textit{cedd})~\cite{gomez2017chai}}; \textbf{(2) machine learning}: binary long short term memory (\textit{blstm})~\cite{diamantopoulos2018ectalk}, digit recognition (\textit{digit})~\cite{zhou2018rosetta}; 
\textbf{(3) databases}: relational operation (\textit{select})~\cite{ds-gomezluna-2015}; 
and \textbf{(4) {data reorganization}}: stream compaction (\textit{sc})~\cite{gomez2017chai}. These kernels are specified in C/C++ code that is compiled to the FPGA target. 

\subsection{Hardware Platform and Tools}
With high adoption of FPGAs in the cloud, various emerging CPU-FPGA platforms with competing cache-coherent interconnect standards are being developed, such as  
 the IBM Coherent Accelerator Processor Interface (CAPI)~\cite{openCAPI}, the Cache Coherent Interconnect for Accelerators
(CCIX)~\cite{benton2017ccix}, the Ultra Path Interconnect (UPI)~\cite{UPI}, and the Compute Express Link (CXL)~\cite{cxlwhitepaper}.  

{The benefits of employing such cache-coherent interconnect links for attaching FPGAs to CPUs, as opposed to the traditional DMA-like communication protocols (e.g., PCIe), are not only the ultra lower-latency and the higher bandwidth of the communication, but most importantly, the ability of the accelerator to access the entire memory space of the CPU coherently, without consuming excessive CPU cycles. Traditionally, the host processor has a shared memory space across its cores with coherent caches. Attached devices such as FPGAs, GPUs, network and storage controllers are memory-mapped because of which they use a DMA to transfer data between local and system memory across an interconnect such as PCIe. The attached devices can not see the entire system memory but only a part of it. Communication between the host processor and attached devices requires an inefficient software stack, including user-space software, drivers, and kernel-space modules, in comparison to the communication scheme between CPU cores using shared memory. Especially when DRAM memory bandwidth becomes a constraint, requiring extra memory-to-memory copies to move data from one address space to another is cumbersome~\cite{Fang2020}. This is the driving force of the industry to push for coherency and shared memory across CPU cores and attached devices, like FPGAs. This way, the accelerators act as peers to the processor cores. 

Based on this upcoming trend, in this thesis, we adopt cache-coherent FPGA accelerators, both for the low-end and the high-end systems. Specifically, we select a \textit{low-end} edge PYNQ-Z1~\cite{zynq} as the source platform to build base learners. We use the Accelerator Coherency Port (ACP) port~\cite{ACP} for attaching accelerators to the ARM Cortex A9 CPU of PYNQ-Z1. In addition, we select a CAPI-based system as the target platform } that provides the most mature coherent accelerator-based ecosystem with a production-ready cloud offering through Nimbix Cloud~\cite{nimbix}. 

We make use of CAPI in a coarse-grained way since we offload the entire application to the FPGA. In this case, CAPI ensures that the FPGA accelerators access the entire CPU memory with the minimum number of memory copies between the host and the FPGA, e.g., avoiding the intermediate buffer copies that a traditional PCIe-based DMA invokes~\cite{10.1145/2897937.2897972}. However, depending on the application, the CAPI protocol can be employed in finer-grained algorithm-hardware co-design, like \textit{ExtraV} \cite{Lee:2017:EBG:3137765.3137776}, where the authors aggressively utilize the fine-grained communication capability of CAPI to boost graph analytics performance.  Table~\ref{tab:LEAPER_SIGMETRICS/systemparameters} summarizes the system details of the source and our {on-premise research cloud} environment as our target platform.

For accelerator implementation and deployment, we leverage CAPI-compatible tools offered by Xilinx. In particular, we use the {Xilinx} SDSoC~\cite{sdsoc} design tool for implementing the low-end system
$\tau_{s}$ and the Vivado HLS~\cite{hls} with IBM CAPI-SNAP framework\footnote{https://github.com/open-power/snap} for the high-end system $\tau_{t}$. The SNAP framework provides seamless integration of an accelerator~\cite{10.1007/978-3-030-34356-9_25} and allows to exchange of control signals between the host and the FPGA processing elements over the
AXI lite interface~\cite{axilite}. On task completion, the processing element notifies
the host system via the AXI lite interface and transfers back
the results via CAPI-supported DMA transactions. 

As derived from the indicative prices listed at the right column of Table~\ref{tab:LEAPER_SIGMETRICS/systemparameters}, the total cost of ownership (TCO) of a high-end system can be more than 100x of that of the low-end one; thus, it can be prohibitive for a bare-metal deployment for many users. Complementary, moving a workload to a cloud FPGA instance should offer such a speedup that it compensates for the extra design time, effort, and cost of this decision by the end-user. LEAPER helps a user to rapidly quantify such a decision by experimenting with low-cost and broadly available FPGAs, like the PYNQ-Z1.

\begin{table}[h]
  \caption{{  Evaluated applications; description including their major kernels and the input dataset. For major kernels, we mention the optimization space where $\times$ represents the optimization being applied to multiple loops or elements. see Table~\ref{tab:LEAPER_SIGMETRICS/pragma} for description of the optimization options.}}
    \label{tab:LEAPER_SIGMETRICS/app_details}
        \renewcommand{\arraystretch}{1}
\setlength{\tabcolsep}{14pt}
  \resizebox{1\linewidth}{!}{%
\begin{tabular}{l@{\hspace{0.9\tabcolsep}}l@{\hspace{0.9\tabcolsep}}l@{\hspace{0.9\tabcolsep}}l@{\hspace{0.9\tabcolsep}}l}
\toprule
  \textbf{Application}                  & \textbf{Domain} & \textbf{Major Kernels}     & \textbf{Dataset}    & \textbf{Optimization space}                                                                     \\ \hline
    blstm~\cite{diamantopoulos2018ectalk} &  \begin{tabular}[c]{@{}l@{}} Machine \\learning\end{tabular}     & \begin{tabular}[c]{@{}l@{}}Hidden lay. fw\\ Hidden lay. back \\ Output layer\end{tabular} & Fraktur OCR~\cite{yousefi2015binarization} & 
    \begin{tabular}[c]{@{}l@{}}2$\times$PL, 3$\times$PR(2,4) I, 2$\times$U \\ 2$\times$PL, I, 2$\times$U  \\ PL, I, U+D, R, W, F\end{tabular}

\\\hline 
    
     cedd~\cite{gomez2017chai}             & \begin{tabular}[c]{@{}l@{}}  Image \\  proc. \end{tabular}   & \begin{tabular}[c]{@{}l@{}}
     Gaussian filter\\
     Sobel filter\\
    Suppress. filter\\
    Hysteresis filter\\
    \end{tabular}            & \begin{tabular}[c]{@{}l@{}}   Frame-354$\times$626 \\ 1000 frames               \\
    \end{tabular}  &
    \begin{tabular}[c]{@{}l@{}}  
    PL, PR(2,4), I, U\\
    PL, PR(2,4), I, U\\
    PL, PR(2,4), I, U\\
    PL, I, U+D, R, W, F\\

     \end{tabular} 
    \\
\hline 
    digit~\cite{zhou2018rosetta}             & \begin{tabular}[c]{@{}l@{}}  Machine \\learning \end{tabular}    & \begin{tabular}[c]{@{}l@{}}Hamming dist.\\
    KNN voting\end{tabular}          & \begin{tabular}[c]{@{}l@{}} MNIST\\18000 train \\
    2000 test  \end{tabular}        &       \begin{tabular}[c]{@{}l@{}}       
     2$\times$PL, 3$\times$PR(2,4), I, 4$\times$U\\
I+R,W,F\\

\end{tabular}
\\\hline

    hist~\cite{gomez2017chai}                      &\begin{tabular}[c]{@{}l@{}}   Image\\ proc.  \end{tabular}   & Histogram avg.     &    \begin{tabular}[c]{@{}l@{}}   Input-1536$\times$1024      \\
    Bins-256\\
        \end{tabular} & \begin{tabular}[c]{@{}l@{}} PL, PR(all), I, D,\\ R, W, U, F  \end{tabular} \\
\hline

   select~\cite{gomez2017chai}         & \begin{tabular}[c]{@{}l@{}}   Data-\\ base  \end{tabular}         & Selection                                                                         &     1048576 inputs        & \begin{tabular}[c]{@{}l@{}}  PL, I, D, R, W, F       \end{tabular}     
   \\\hline 
   
    sc~\cite{gomez2017chai}                & \begin{tabular}[c]{@{}l@{}}   Data\\ reorg.  \end{tabular}  & \begin{tabular}[c]{@{}l@{}}Count\\ Compact\end{tabular}       &                  1048576 inputs    & \begin{tabular}[c]{@{}l@{}}PL, I, D, R, W, U, F       \end{tabular}          \\ \bottomrule
\end{tabular}
}
\end{table}


 \begin{figure*}[t]
  \begin{subfigure}{0.495\textwidth}
   \includegraphics[width=\textwidth,trim={0cm 0cm 0cm 0cm},clip]{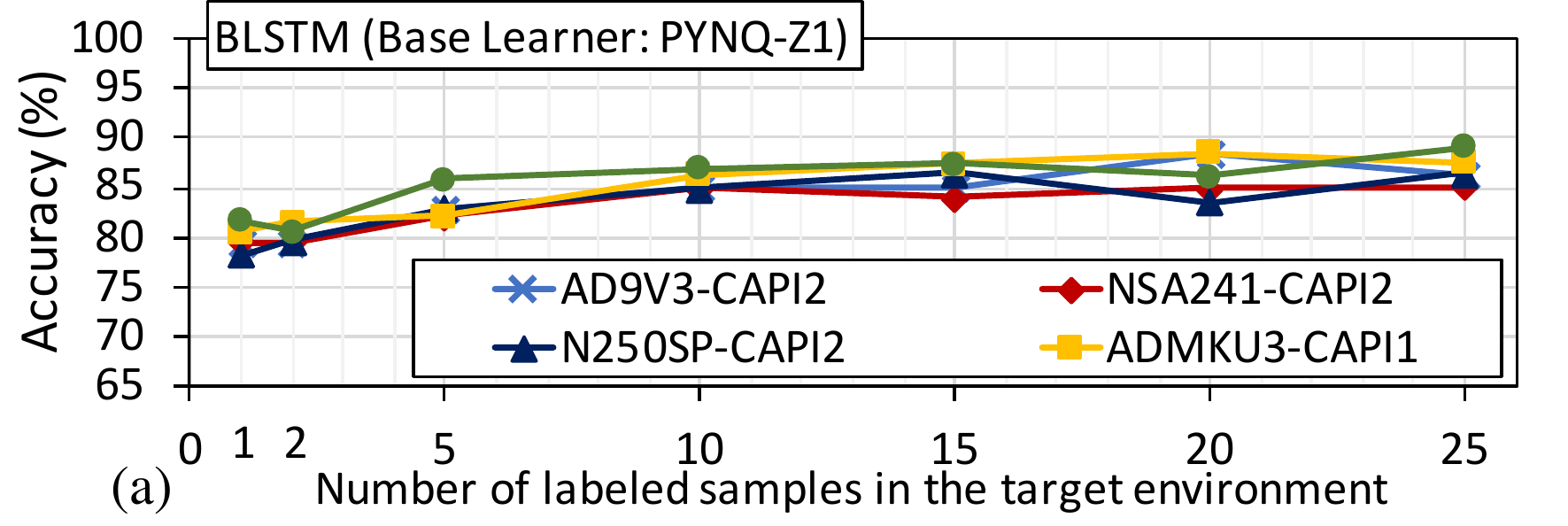}
    \label{fig:LEAPER_SIGMETRICS/board_blstm}
  \end{subfigure}
  \begin{subfigure}{0.495\textwidth}
  \includegraphics[width=\textwidth,trim={0cm 0cm 0cm 0cm},clip]{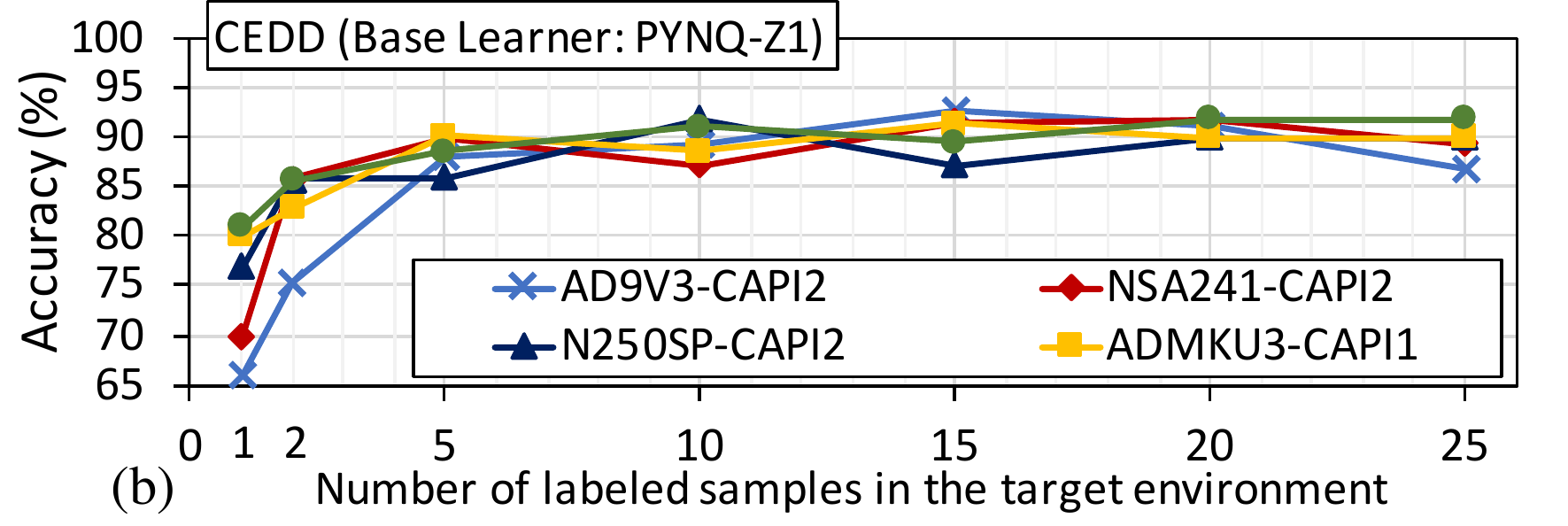}
    \label{fig:LEAPER_SIGMETRICS/board_cedd}
  \end{subfigure}
   \begin{subfigure}{0.495\textwidth}
\includegraphics[width=\textwidth,trim={0cm 0cm 0cm 0cm},clip]{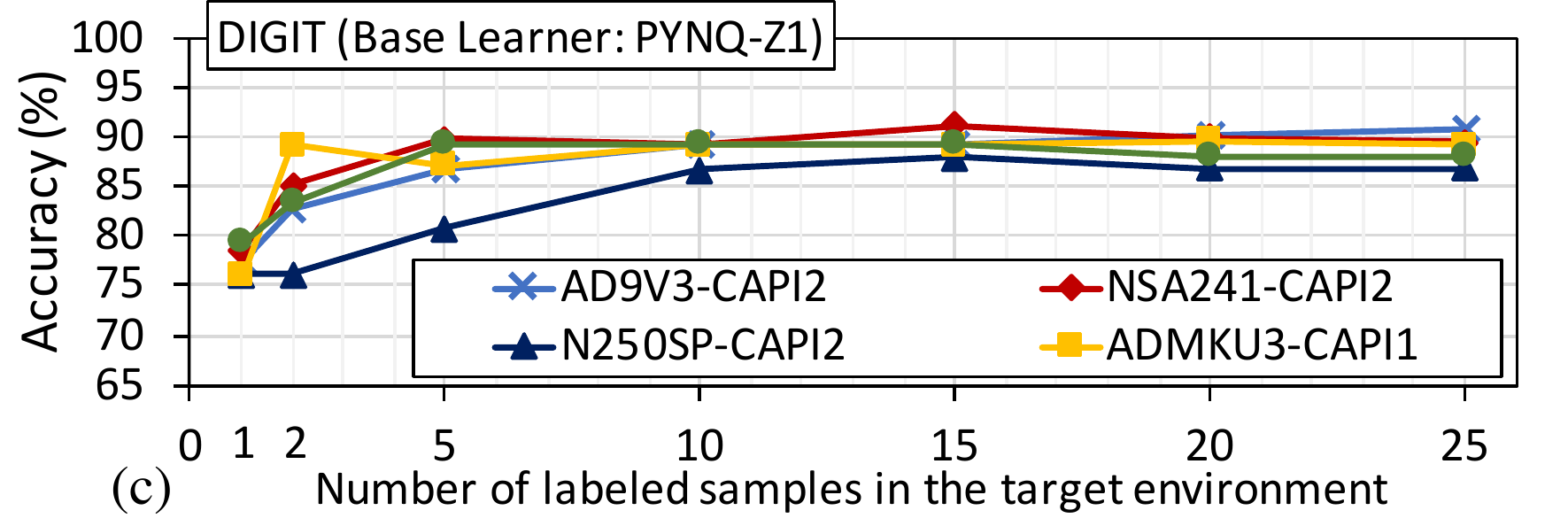}    
    \label{fig:LEAPER_SIGMETRICS/board_digit}
  \end{subfigure}
   \begin{subfigure}{0.495\textwidth}
    \includegraphics[width=\textwidth,trim={0cm 0cm 0cm 0cm},clip]{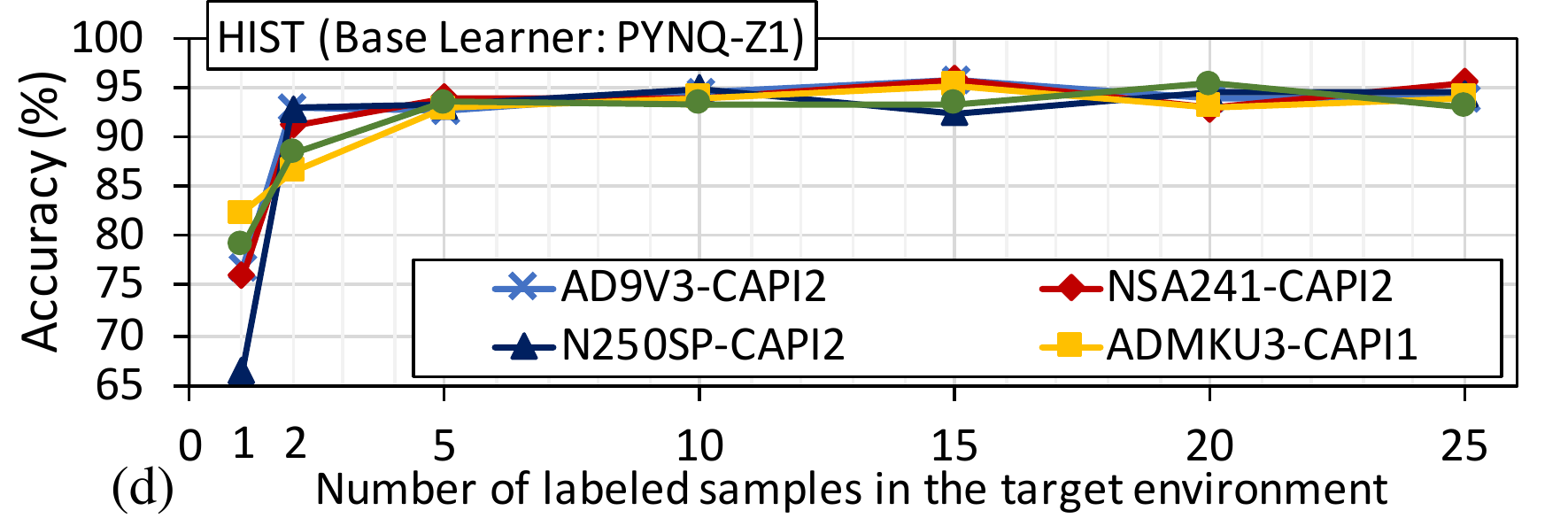}
    \label{fig:LEAPER_SIGMETRICS/board_hist}
  \end{subfigure}
     \begin{subfigure}{0.495\textwidth}
    \includegraphics[width=\textwidth,trim={0cm 0cm 0cm 0cm},clip]{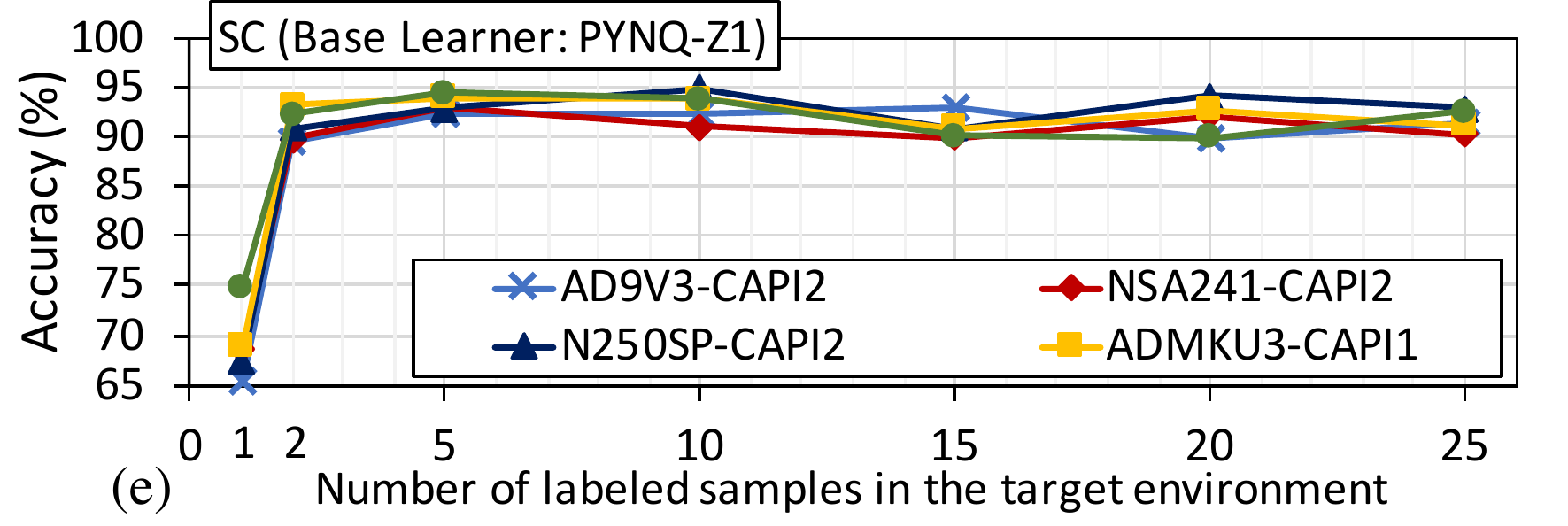}
    \label{fig:LEAPER_SIGMETRICS/board_sc}
  \end{subfigure}
     \begin{subfigure}{0.495\textwidth}
    \includegraphics[width=\textwidth,trim={0cm 0cm 0cm 0cm},clip]{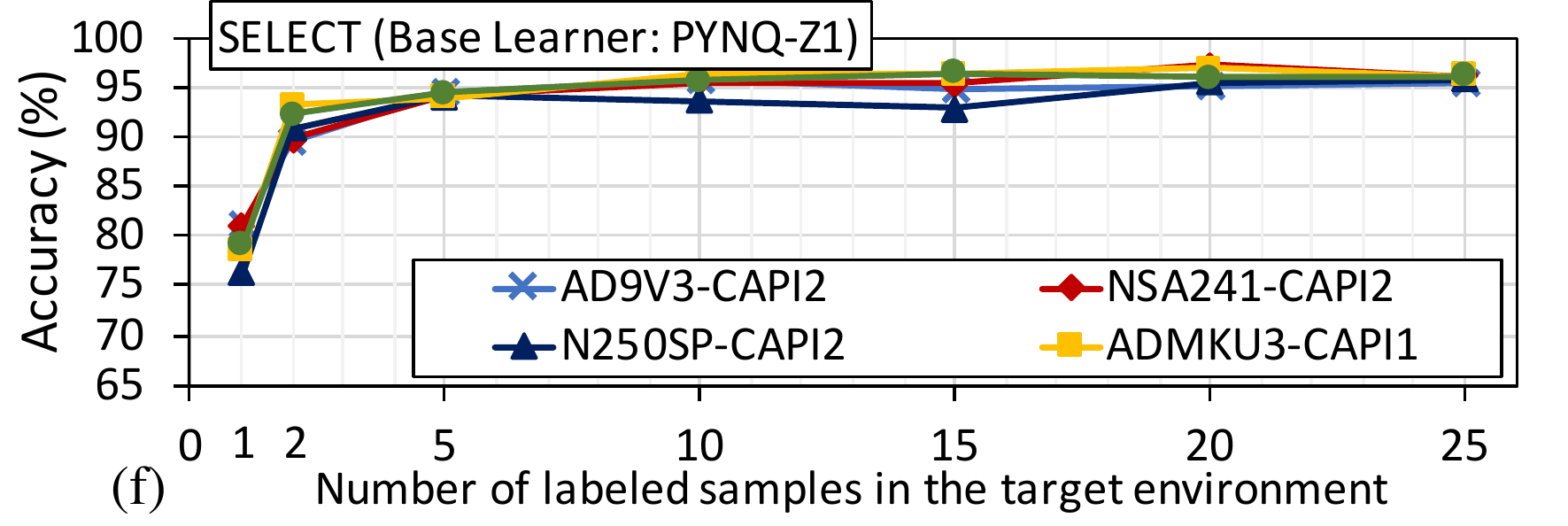}
    \label{fig:LEAPER_SIGMETRICS/board_select}
  \end{subfigure}
    \vspace{-0.3cm}
  \caption{\namePaper's accuracy for 
  {transferring} base models across CAPI-enabled {cloud} FPGA-based systems. {The legends indicate the target platforms}. The base model was trained on a low-end PYNQ-Z1 board and, for each application, we \textit{transfer} this model to different high-end cloud FPGA-based 
  platforms using different samples {(horizontal axis)} from the target platform. Once trained using \textit{few shot}, the transferred model makes predictions for all other configurations in the target platform.}
   \label{fig:LEAPER_SIGMETRICS/acros_board}
\end{figure*}

 \begin{figure*}[t]
  \begin{subfigure}{0.495\textwidth}
   \includegraphics[width=\textwidth,trim={0cm 0cm 0cm 0cm},clip]{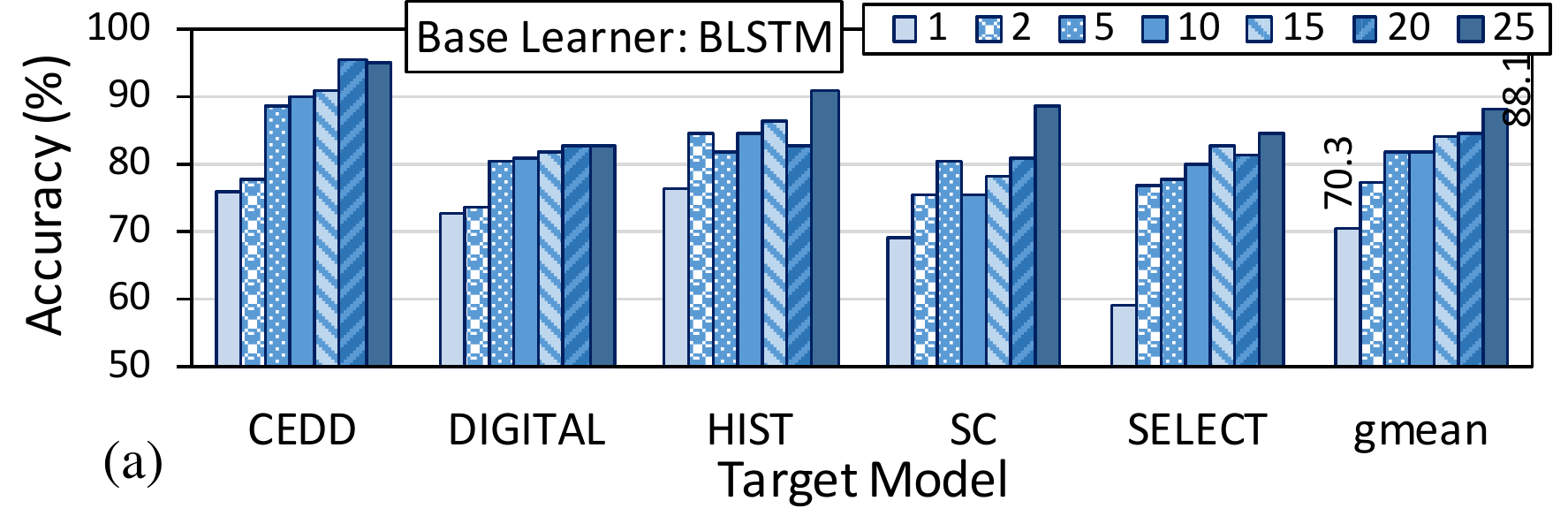}
    \label{fig:LEAPER_SIGMETRICS/app_blstm}
  \end{subfigure}
  \begin{subfigure}{0.495\textwidth}
  \includegraphics[width=\textwidth,trim={0cm 0cm 0cm 0cm},clip]{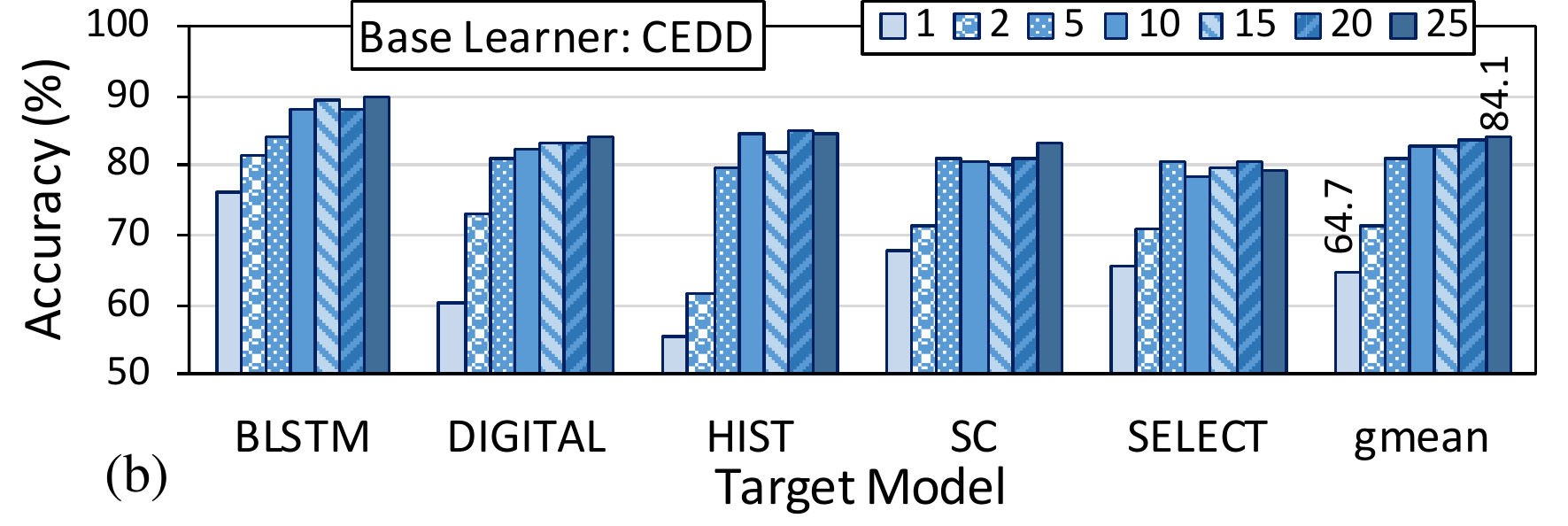}
    \label{fig:LEAPER_SIGMETRICS/app_cedd}
  \end{subfigure}
   \begin{subfigure}{0.495\textwidth}
\includegraphics[width=\textwidth,trim={0cm 0cm 0cm 0cm},clip]{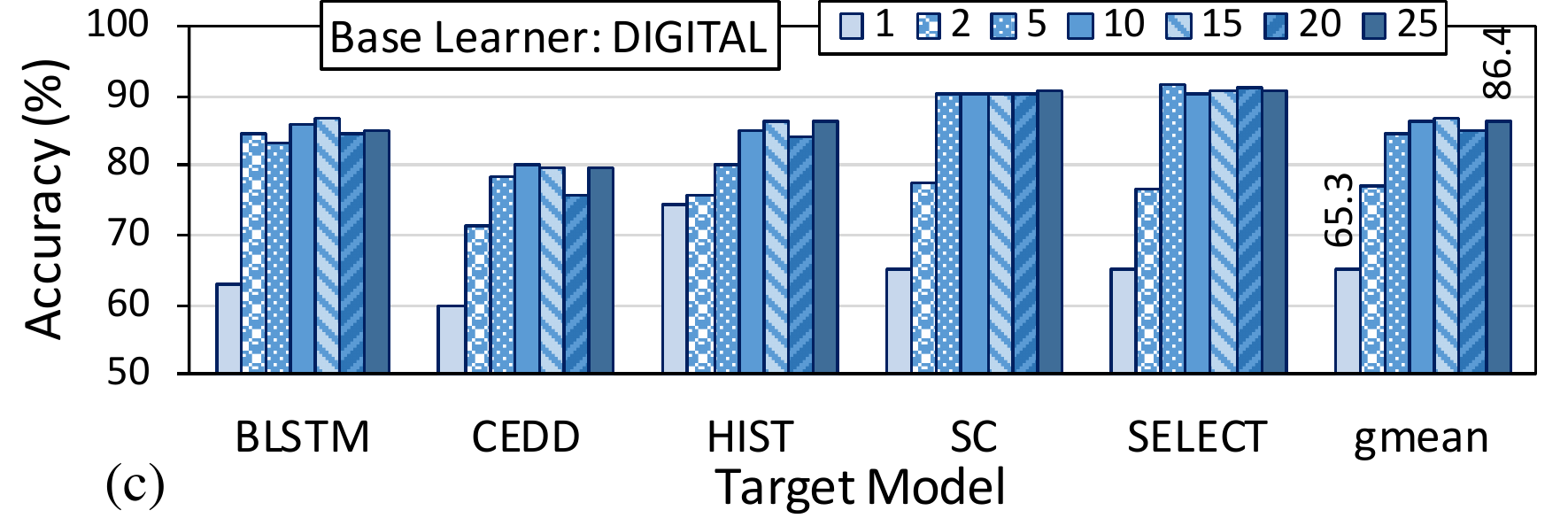}    
    \label{fig:LEAPER_SIGMETRICS/app_digit}
  \end{subfigure}
   \begin{subfigure}{0.495\textwidth}
    \includegraphics[width=\textwidth,trim={0cm 0cm 0cm 0cm},clip]{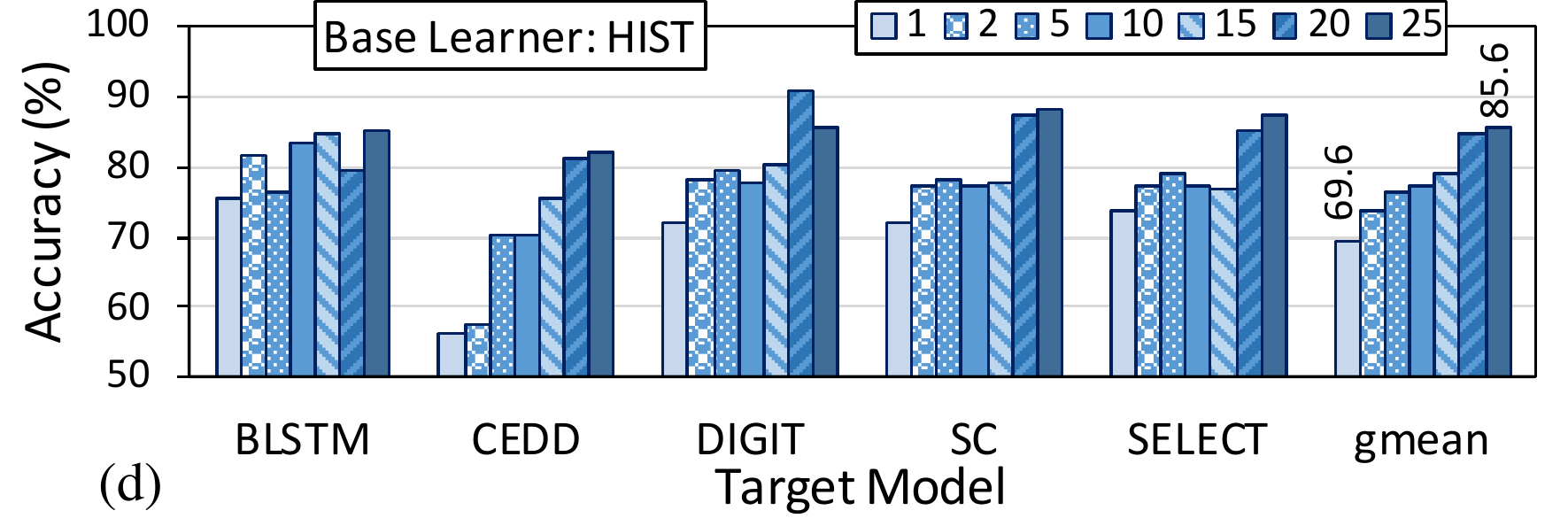}
    \label{fig:LEAPER_SIGMETRICS/app_hist}
  \end{subfigure}
     \begin{subfigure}{0.495\textwidth}
    \includegraphics[width=\textwidth,trim={0cm 0cm 0cm 0cm},clip]{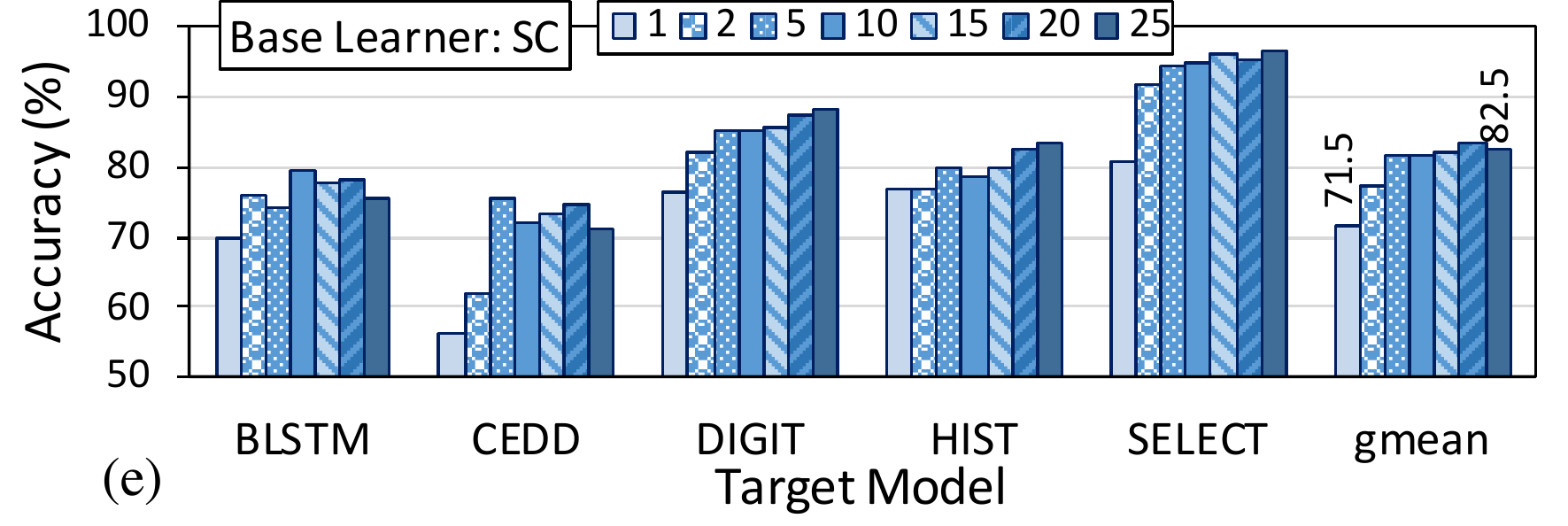}
    \label{fig:LEAPER_SIGMETRICS/app_sc}
  \end{subfigure}
     \begin{subfigure}{0.495\textwidth}
     \includegraphics[width=\textwidth,trim={0cm 0cm 0cm 0cm},clip]{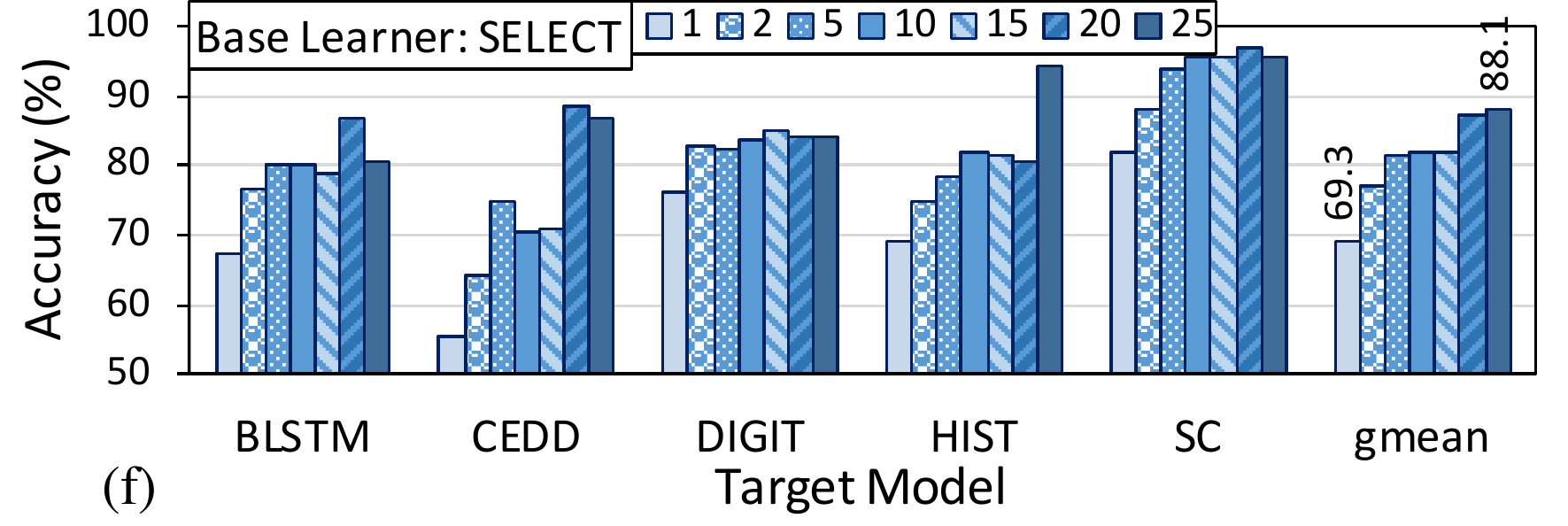}
    \label{fig:LEAPER_SIGMETRICS/app_select}
  \end{subfigure}
\vspace{-0.3cm}
  \caption{\namePaper's accuracy for 
  {transferring} base models across various applications. {The legends indicate the number of samples}. Each plot represents a different application as a base learner. We \textit{transfer} these base learners, trained on the PYNQ-Z1 platform, by using invariant configuration features to build a target application model.}
   \label{fig:LEAPER_SIGMETRICS/across_app}
\end{figure*}
\subsection{{Target Model }Accuracy Analysis}
\namePaper~is used to transfer a trained model using \textit{few-shot learning} to the $\tau_{t}$'s {optimization} space. We then analyze the accuracy of the newly-built target model to predict the performance and resource utilization of all the other configurations in 
$\tau_{t}$. 
We evaluate the accuracy of the 
{transferred} model in terms of the relative error $\epsilon_i$ to indicate the proximity of the predicted value $y_{i}^\prime$ to the actual value $y_{i}$ across \textit{N} test samples. The mean relative error (MRE) is calculated with Equation~\ref{eq:LEAPER_SIGMETRICS/2}.
\vspace{-0.08cm}
\begin{align}
 MRE= \frac{1}{N} \displaystyle \sum_{i=1}^{N} \epsilon_i=\frac{1}{N} \displaystyle \sum_{i=1}^{N}\frac{|y_{i}^\prime -y_{i}|}{y_{i}}.
\label{eq:LEAPER_SIGMETRICS/2}
\end{align}
\vspace{-0.05cm}

\noindent {\textbf{Performance Model Transfer.}}
Figure~\ref{fig:LEAPER_SIGMETRICS/acros_board} shows \namePaper's accuracy for transferring \textit{across different {cloud} platforms}. We make the following four observations.

First, as we increase the number of labeled samples, the target model accuracy increases. However, the accuracy saturates and,  with 5-10 \textit{shots}, we can achieve an accuracy as high as 80 to 90\%. 

Second, compared to applications with multiple complex kernels (\textit{blstm}, \textit{cedd}, \textit{digit}), simpler kernels (\textit{hist}, \textit{sc}, \textit{select}) can be {more} easily transferred using fewer samples. Applications with multiple kernels have a larger optimization space. The large optimization space leads to more complex interactions that have compounding effects with other optimization options because we are 
{modeling for} multiple kernels rather than just a single~kernel. 
Additionally, simple kernels such as \textit{sc} and \textit{select} have been implemented using \textit{hls stream} interfaces. Here rather than storing intermediate data in local FPGA memories, we read streams of data, and hence certain complex optimizations (like array partitioning) cannot be applied. This leads to a change in the feature space of different environments.

Third, less severe changes are more amenable to transfer as the source, and target models are more closely related, e.g., 
{transferring} to CAPI1 {(PCIe Gen3 with $\sim$~3.3 GB/s bandwidth)} 
{from low-end PYNQ with PCIe Gen2 $\sim$~1.2 GB/s bandwidth }
{entails} a smaller increment in bandwidth 
than moving to CAPI2, which offers R/W bandwidth of $\sim$~12.3 GB/s. 

Fourth, 
 change in {the} technology node from one FPGA to another has a lower impact than changing the external bandwidth to a new {interconnect} standard on the {transferring} process.

Figure~\ref{fig:LEAPER_SIGMETRICS/across_app} shows \namePaper's accuracy for transferring ML models \emph{across different applications}. We make the following three observations.
First, we make a similar observation in the case of transferring models across different platforms: as we increase the number of training samples, the mean relative error for most applications decreases. 
{Second, we notice} that the most considerable improvement 
{in accuracy occurs} when our sample size $(c_{tl})$ 
{is} between 2 to 10. In most cases, {the accuracy} saturates after 20 samples. 
{Third}, in some cases, we see a decrease in accuracy {when increasing the number of samples}. This result could be attributed to training with a small amount of data, which can sometimes lead to overfitting~\cite{dai2007boosting}. 

We 
{further explain} our results by measuring the divergence of performance distributions to quantify the statistical distance between the source and target models, see Section~\ref{subsection:LEAPER_SIGMETRICS/relatedness_analysis}. We also compare the average accuracy of predicting performance across boards and applications using different TLs (see Table~\ref{tab:LEAPER_SIGMETRICS/compare}).
\\
%

\begin{table}[t]
\centering
  \caption{System parameters and configuration.}
    \label{tab:LEAPER_SIGMETRICS/systemparameters}
    \renewcommand{\arraystretch}{1}
\setlength{\tabcolsep}{14pt}
    \resizebox{\linewidth}{!}{
\begin{tabular}{@{}lllllr@{}}
\toprule
\multicolumn{5}{l}{\textbf{Low-end base system}} & \textbf{Indicative price}                                                                                                                                                                               \\ \midrule

\textbf{Embedded Board}    & 
\multicolumn{4}{l}{ PYNQ-Z1 ZYNQ~\cite{zynq} XC7Z020-1CLG400C } & \$199\footnote{https://store.digilentinc.com/pynq-z1-python-productivity-for-zynq-7000-arm-fpga-soc/, Accessed 12 Jan. 2021} \\

&\multicolumn{4}{l}{ARM Cortex-A9 processor  @650MHz, dual-core} \\ 

\toprule
\toprule
\multicolumn{5}{l}{\textbf{{On-prem cloud} target system with OpenStack\cite{openstack} and KVM Hypervisor}}                                                                                                                                                                               \\ \midrule
\textbf{Host Configuration}    & 
\multicolumn{4}{l}{IBM\textsuperscript{\textregistered} POWER9 AC922 @2.3 GHz, 16 cores} & \$55000-\$75000\footnote{https://www.microway.com/product/ibm-power-systems-ac922/, Accessed 12 Jan. 2021} \\ 
&\multicolumn{4}{l}{4-way SMT, 32 KiB L1 cache,256 KiB L2 cache,} \\ 
&\multicolumn{4}{l}{10 MiB L3 cache, 16x32GiB RDIMM DDR4 2666 MHz \footnote{.}} \\ 

\midrule
\textbf{FPGA Description} & \multicolumn{4}{l}{}                                                                                                                                 &                             \\ 

 \textbf{Board} &\textbf{FPGA Family}        &\textbf{Device}   &\textbf{Interface}   \\    
ADM-PCIE-8K5~\cite{ad8k5}   &Kintex UltraScale &XCKU115-2   & CAPI-1  & & N/A \\
ADM-PCIE-KU3~\cite{adku3}    &Kintex UltraScale &XCKU060-2  & CAPI-1  & & N/A \\
Semptian NSA241~\cite{nsa241}& Virtex UltraScale\tnote{+} &XCVU9P-2  & CAPI-2  & & N/A  \\
ADM-PCIE-9V3~\cite{ad9v3}   &Virtex UltraScale\tnote{+} & XCVU3P-2  & CAPI-2   & & N/A  \\
N250SP~\cite{250SP}&Kintex UltraScale\tnote{+} &KU15P-2  &CAPI-2  & & N/A  \\      
\bottomrule
\end{tabular}
}
\begin{flushleft}
\scriptsize \textsuperscript{6} https://store.digilentinc.com/pynq-z1-python-productivity-for-zynq-7000-arm-fpga-soc/, Accessed 12 Jan. 2021.\\
\scriptsize \textsuperscript{7} https://www.microway.com/product/ibm-power-systems-ac922/, Accessed 12 Jan. 2021.\\
N/A\footnote{.}: Not available indicative price from an online store, but in the region of \$2500-\$5000 for our purchased on-prem cards.

\end{flushleft}
\end{table}

\begin{figure}[h]
\begin{subfigure}[b]{\textwidth}
\centering
   \includegraphics[width=0.6\linewidth]{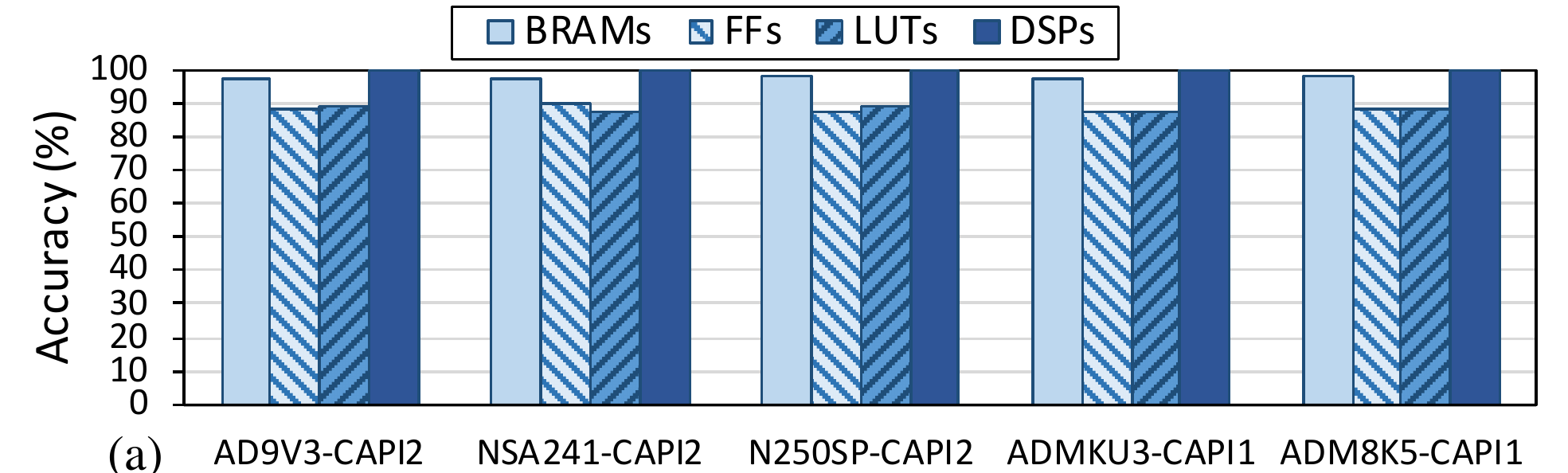}
         \phantomcaption
    \label{fig:LEAPER_SIGMETRICS/area_board} 
    \vspace{0.1cm}
\end{subfigure}
\begin{subfigure}[b]{\textwidth}
\centering
   \includegraphics[width=0.6\linewidth]{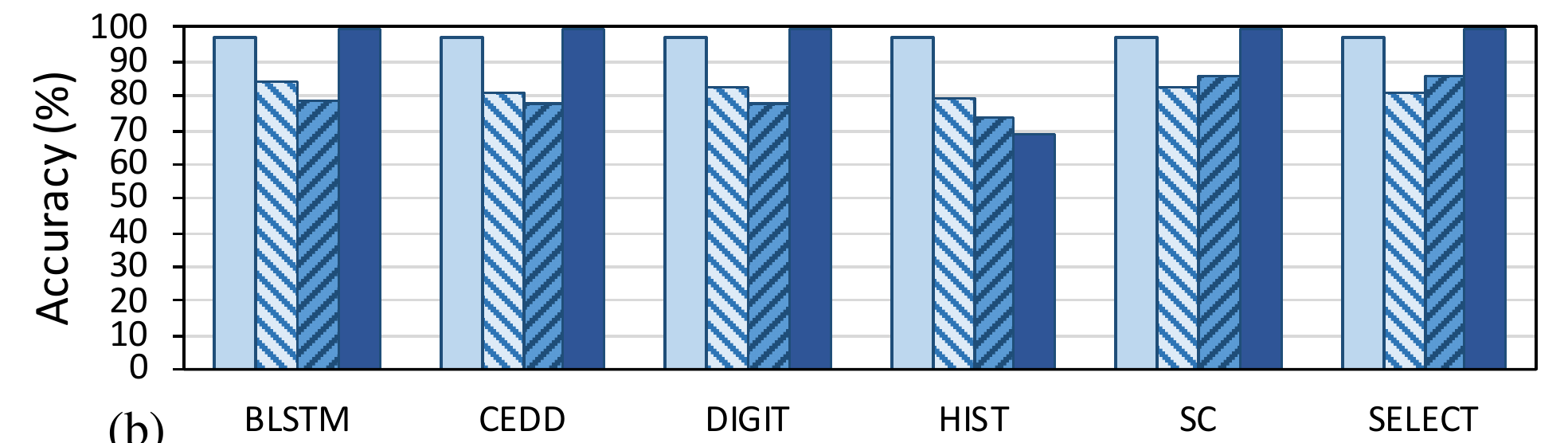}
      \phantomcaption
    \label{fig:LEAPER_SIGMETRICS/area_app}

\end{subfigure}
\vspace{-0.3cm}
\caption[Two numerical solutions]{\namePaper's average accuracy for transferring FPGA resource usage models through \textit{5-shot} using (a) a base learner trained on a low-end PYNQ-Z1 to different high-end target FPGA boards (horizontal axis), and  (b) different applications as base learners (horizontal axis) to all the target applications, on low-end PYNQ-Z1 board.
}
\label{fig:LEAPER_SIGMETRICS/area_transfer} 
\end{figure}

\noindent \textbf{Resource Model Transfer.} By using \namePaper, 
we can also train a resource consumption model on a \textit{low-end} source environment and transfer it to a high-end cloud target environment.  Figure~\ref{fig:LEAPER_SIGMETRICS/area_transfer} shows the accuracy of a target model trained by \textit{5-shot} {transfer} learning for predicting 
{a} resource utilization vector $\eta_{\{BRAM, FF, LUT, DSP\}}$. Note:~the reported accuracy is for the transferred model, i.e., using a base model (low-end FPGA) to predict a target model (high-end {cloud} FPGA) after few-shot learning. 

In Figure~\ref{fig:LEAPER_SIGMETRICS/area_board}, the horizontal axis depicts the target platform, while the base learner is trained on a PYNQ-Z1 board. In the case of an application model transfer (Figure~\ref{fig:LEAPER_SIGMETRICS/area_app}), the platform remains unchanged (PYNQ-Z1), while the base learner application changes (horizontal axis).
We make 
{three} observations. First, the resource model shows low error rates for predicting BRAM and DSP. This is attributed to the fact that the technological configuration of these resources remains relatively unchanged across platforms (e.g., BRAM is implemented as 18 Kbits in both the source and target platforms). Second, flip-flops and look-up-tables have 
comparably higher error rates 
because the configuration of CLB\footnote{A configurable logic block (CLB) is the fundamental component of an FPGA, made up of look-up-tables (LUTs) and flip-flops (FF).} slices varies with 
{the} transistor technology and FPGA family. {Third}, Figure~\ref{fig:LEAPER_SIGMETRICS/area_app} shows the mean accuracy for transferring different application-based models on a fixed FPGA board. 
{We observe} relatively low accuracy for DSP consumption while transferring a base model trained on \textit{hist}. This low accuracy is because the \textit{hist} FPGA implementation does not make use of DSP units. However, all other applications utilize DSPs. Therefore, the ML model trained on \textit{hist} is not able to perform well in other \textit{transferred} environments.

\begin{table}[t]
\centering
  \caption{Average accuracy (\%) comparison of \namePaper~ with decision tree (DT) and adaBoost (ADA) as TL for~\textit{5-shot}~transfer.}
    \label{tab:LEAPER_SIGMETRICS/compare}
    {\footnotesize
    \renewcommand{\arraystretch}{1}
\setlength{\tabcolsep}{14pt}
    \resizebox{0.65\linewidth}{!}{%
\begin{tabular}{@{}llll@{}}
\hline
\textbf{Environment}               & \textbf{DT}   & \textbf{ADA}   & \textbf{\namePaper} \\ \midrule
Across Board       & 77.7 & 83.2 & 89.8  \\
Across Application & 70.6 & 73.5 & 81.2   \\ \hline
 \vspace{-0.3cm}
\end{tabular}
}}
\end{table}
\subsection{Target {Cloud FPGA }Model Building Cost}
\label{subsection:LEAPER_SIGMETRICS/timing}
Table~\ref{tab:LEAPER_SIGMETRICS/timing} shows the time for collecting (see ``DoE run 
{(hours)}'') {the 50 sampled DoE configurations ($c_{lhs}$) {that we use} to gather training data. {Please note that while the process of synthesis and P\&R of the high-end system's FPGA, which is needed to obtain the maximum operating clock frequency and the resources' utilization, can be carried out offline, most of the cloud providers are offering VMs with all the appropriate software, IPs and licenses needed to generate an FPGA image ready to be deployed at their cloud infrastructure (e.g., the Vivado AMI of AWS~\cite{AMI}). This justifies the argument of the cost of the cloud environment for DoE runs. Specifically, we use a Linux image with FPGA and SoC development tools, IPs, and licenses to generate bitstreams for the selected Xilinx devices of Table~\ref{tab:LEAPER_SIGMETRICS/systemparameters} on the Nimbix cloud.}

Table~\ref{tab:LEAPER_SIGMETRICS/timing} also includes the {execution} time 
on the ADM-PCIE-KU3 {cloud} platform (``Exec (msec)'') and the transfer time (``Transfer (msec)'') for each model}. By using transfer learning, the DoE runtime is amortized and, by using a few labeled samples $c_{tl}$ (``5-shot (hours)'') from the target platform, we can transfer a previously trained model and make predictions for all the other configurations for the target platform. 
{As a result}, quick exploration and {significant} time savings {(at least 10.2$\times$) are possible when transferring a model (i.e., ``5-shot (hours)'' + ``Transfer (msec)'') as compared to building a new model from scratch (i.e., ``DoE run (hours)'') + ``Exec (msec)'')}. Note, LHS reduces training samples from 500+ to 50, while  \textit{5-shot} transfer learning further reduces this space to 5, so we achieve $100\times$ effective~speedup compared to a traditional ``brute-force'' approach.

\begin{table}[h]
\caption{DoE time for gathering sampled data points for a single CPU-FPGA platform (``DoE run (hours)''), DoE execution time on the deployed platform (``Exec (ms)''), Estimated Cost on a cloud platform (``Est. Cost (\$)''), time for gathering 5 labeled samples (``5-shot (hours)''), \namePaper~time including the transfer time (``Transfer (msec)''), ``Speedup'' over building a new model from scratch using just the DoE data (still 
{more} cost-efficient than traditional ``brute-force'' training).}
\label{tab:LEAPER_SIGMETRICS/timing}
\centering
\resizebox{\linewidth}{!}{%
 \begin{tabular}{l|lll|lll|l}
\hline
\multicolumn{1}{c}{\textbf{Application}}  & \multicolumn{7}{c}{\textbf{Transfer Time}}  \\ 

Name &
 \begin{tabular}{l}    
 DoE run \\(hours)
\end{tabular}&
 \begin{tabular}{l}      
Exec \\(msec)

\end{tabular}&
 \begin{tabular}{l}    
Est.Cost$^{7}$\\(\$)

\end{tabular}&
 \begin{tabular}{l}    
5-shot \\(hours)

\end{tabular}
& \begin{tabular}{l} 
Transfer \\(msec)

\end{tabular}&
 \begin{tabular}{l}  
Est.Cost$^{7}$\\(\$)

\end{tabular}
&
 \begin{tabular}{l} 
Speedup\\

\end{tabular} \\
\hline
blstm   &  135  &  1245 & 168.7 & 13 & 55.6 & 16.2 & 10.4  \\

cedd    &  124  & 2217 & 155.0 & 12 & 26.5 & 15.0 & 10.3 \\

digit   & 122   & 873 & 152.5 & 12 & 58.8 & 14.9 & 10.2\\

hist    & 97  &  1104  &  121.2 & 9 & 17.1   & 11.3 & 10.8   \\

sc      & 104  & 4018 &  130.0 & 10 & 27.9 & 12.4 & 10.4 \\

select  &  106   & 3978 & 132.5 & 10 & 27.6 &  12.5 & 10.6 \\

\hline
\end{tabular}
}

\begin{flushleft}
{\tiny{$^{7}$The cost is estimated based on an enterprise online cost estimator~\cite{nimbix-calc}, using a public-cloud system with configuration  akin to  
our on-prem system. Specifically, we have selected an \textit{n2}~(8-core, 64GB RAM VM - 1.25\$/h) for bitstream generation (x86) and 
an \textit{np8f1}  instance (160-thread POWER8, 1TB RAM, ADM-PCIE-KU3 with CAPI-1 - 3\$/h) for deployment.} \\
}
\end{flushleft}

\end{table}







Table~\ref{tab:LEAPER_SIGMETRICS/low_high_end} mentions the performance and resource utilization for our considered applications both on a low-end system and a high-end cloud system. We use \namePaper to obtain the performance and resource utilization for the high-end cloud configuration.
\begin{table}[h]
  \caption{{Execution time and resource utilization for low-end base configuration (PYNQ-Z1) and high-end cloud configuration, a Nimbix \textit{np8f1} instance (POWER8, ADM-PCIE-KU3 with CAPI-1).}}
 \label{tab:LEAPER_SIGMETRICS/low_high_end}
    \renewcommand{\arraystretch}{1}
\setlength{\tabcolsep}{14pt}
  \resizebox{\linewidth}{!}{%
\begin{tabular}{l|l|l|l|l|l|l}
\toprule
\textbf{Application} & \textbf{Config.} & \textbf{Exec (msec)} & \textbf{BRAM} & \textbf{DSP} & \textbf{FF} & \textbf{LUT} \\ \hline
\multirow{ 2}{*}{blstm}         &      low-end              &   4200              &    80\%       &    15\%          &     24\%           &   47\%           \\
                                &high-end  &   1245              &    62\%       &    8\%          &     12\%           &   21\%           \\

\multirow{ 2}{*}{cedd}          &    low-end                &    10254              &     83\%          &    37\%          &   95\%          &     97\%         \\
                                &   high-end &    2217             &   56\%                &      3\%          &   75\%            &              94\%         \\

\multirow{ 2}{*}{digit}         &    low-end                &   2458             &     94\%              &       33\%         &     79\%          &            85\%     \\
                                & high-end  &       873          &     84\%              &       12\%         &     24\%          &            75\%              \\

\multirow{ 2}{*}{hist}          &    low-end                &    6173              &      94\%         &     0\%         &     11\%       &         37\%     \\
                                &  high-end  &   1104              &     67\%             &    0\%           &    5\%          &             30\%           \\

\multirow{ 2}{*}{sc}            &   low-end                 &   19306              &    82\%       &    0.4\%          &     12\%           &   25\%           \\
                                &high-end  &   4018              &    91\%       &    0.1\%          &     12\%           &   23\%     \\
 
\multirow{ 2}{*}{select}        &       low-end              &   18306              &    82\%       &    0.4\%          &     12\%           &   25\%           \\
                                &high-end  &   3918              &    91\%       &    0.1\%          &     12\%           &   23\%     \\

   \bottomrule
\end{tabular}
}
\end{table}

\subsection{Base Model Accuracy Analysis}
\begin{figure}[h]
  \centering
  \includegraphics[width=0.8\linewidth,trim={0.2cm 0.1cm 0.1cm 0.1cm},clip]{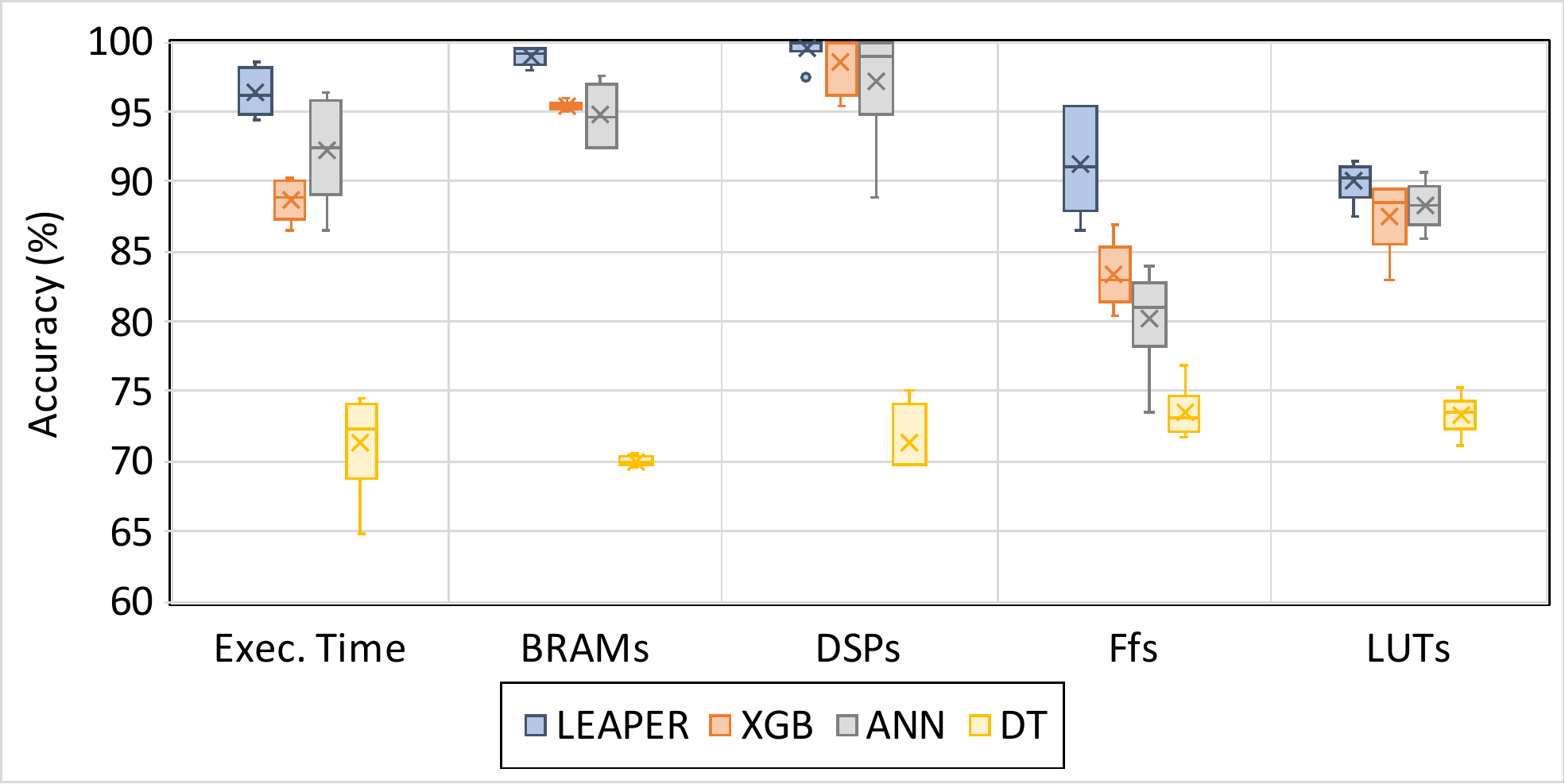}
  
  \caption{Mean accuracy for performance and resource utilization predictions using \namePaper's base model~vs. other methods: XGB (XGBoost), an artificial neural network (ANN), and a decision tree (DT).   }
  \label{fig:LEAPER_SIGMETRICS/base_mode}
 \end{figure}

We also evaluate the accuracy of our base model. The base model (Section~\ref{subsec:LEAPER_SIGMETRICS/base_model}) is trained on our low-end edge PYNQ board using $c_{lhs}$ configurations sampled using LHS. The base can predict performance (resource utilization) outside the base learner dataset (DoE configurations) $c_{lhs}$ (see Figure~\ref{fig:LEAPER_SIGMETRICS/transfer_overiew}). To assess our base model, we use 30 previously unseen configurations that are not part of $c_{lhs}$ on the base system, and we evaluate the mean relative error for all 30 unseen configurations on all six applications.  Figure~\ref{fig:LEAPER_SIGMETRICS/base_mode} shows \namePaper's base model accuracy results. 

We compare \namePaper~to three other ML algorithms
that are also trained using $c_{lhs}$ configurations to predict performance and resource consumption: XG-Boost (XGB) based on  Dai~\etal~\cite{dai2018fast},
an artificial neural network (ANN) used by Makrani~\etal~\cite{xppe_asp_dac} and a traditional
decision tree (DT). We make three observations. First, on average, \namePaper~is more accurate in terms of performance (resource utilization) prediction than the other ML models. Second, XGBoost is more data-efficient than ANNs~\cite{dai2018fast} and can perform better than decision trees~\cite{NAPEL}. Third, ANN is more accurate than the decision tree, but it is less accurate
than \namePaper. This is because ANN requires a much larger
training dataset to reach LEAPER’s accuracy~\cite{li2018processor}.

\subsection{{Why Does \namePaper~Work?}}
 \label{subsection:LEAPER_SIGMETRICS/relatedness_analysis}
{To explain our transfer learning results}, we analyze the degree of \textit{relatedness} between the source and target environments. 
We perform a correlation analysis followed by a divergence analysis of the performance distributions of the environments.

{Using the correlation analysis,} we make the following four observations. First, we see a high correlation of 0.76 to 0.97 {between the source and target  execution time for different target hardware platforms}, which indicates that the target model's performance behavior can easily be learned using the source environment.  
As the linear correlation is not 1 for all platforms, the use of a nonlinear transfer model is substantiated. Second, as we increase the external bandwidth {of the target platform (i.e., CAPI1 to CAPI2)}, 
the correlation becomes lower because the hardware change is much more \textit{severe}, coming from a low-end FPGA with limited external bandwidth. 
Third, compared to using a single transfer learner, an ensemble of learners 
can perform more accurate and robust predictions (see Figure~\ref{fig:LEAPER_SIGMETRICS/acros_board}, Figure~\ref{fig:LEAPER_SIGMETRICS/across_app}, and Figure~\ref{fig:LEAPER_SIGMETRICS/area_transfer}). 
Fourth, the correlation between applications on a single platform is lower (0.45 to 0.9) because of {the} varying application characteristics {and optimization space.}

We measure the relatedness of the performance distributions of the source ($P(\tau_{s})$) and the target applications  ($P(\tau_{t})$) because application-based environments exhibit a low linear correlation. We employ the Jensen-Shannon Divergence (JSD)~\cite{jsd} (see Table~\ref{tab:LEAPER_SIGMETRICS/jsd}) to quantify the statistical distance between $P(\tau_{s})$ and $P(\tau_{t})$. The lower the values of JSD, the more similar the target environment is to the source (i.e., if $D_{JSD}(P(\tau_{t}) || P(\tau_{s}))=0$ implies the distributions are identical, and 1 indicates unrelated distributions). This analysis confirms the trend observed from transferring application models (Figure~\ref{fig:LEAPER_SIGMETRICS/across_app}), i.e., the more closely related the source and target applications, 
the fewer 
samples are required to train our nonlinear transfer learners. The measured distance between the tasks is proportional to the error of the target~task. {As can be seen from Table~\ref{tab:LEAPER_SIGMETRICS/jsd}  for many applications, transfer learning to build accurate models is feasible since their JSD is limited. } {Similar trends are observed for the resource~utilization~model.}

    \begin{table}[h]
\centering
  \caption{Jensen-Shannon Divergence (JSD)~\cite{jsd} between 
  performance distributions of {different applications}. JSD measures statistical distance between two probability distributions.}
    \label{tab:LEAPER_SIGMETRICS/jsd}
      {\footnotesize
    \renewcommand{\arraystretch}{1}
\resizebox{0.65\linewidth}{!}{%
\begin{tabular}{@{}clllllll@{}}
\toprule   
\multirow{10}{*}{\rotatebox{90}{\textbf{Target Model}}} & \multicolumn{7}{c}{\textbf{Base Learner}}\\

                              & \multicolumn{1}{l}{}       & blstm & cedd & digit & hist & sc   & select \\ 
                              \cmidrule{2-8} 
                              & \multicolumn{1}{l|}{blstm}  & 0.00     & 0.24 & 0.34  & 0.25 & 0.31 & 0.30   \\
                              & \multicolumn{1}{l|}{cedd}   & 0.24  & 0.00     & 0.49  & 0.54 & 0.41 & 0.40   \\
                              & \multicolumn{1}{l|}{digit}  & 0.34  & 0.49 & 0.00      & 0.25 & 0.21 & 0.21   \\
                              & \multicolumn{1}{l|}{hist}   & 0.25  & 0.54 & 0.25  & 0.00     & 0.25 & 0.24   \\
                              & \multicolumn{1}{l|}{sc}     & 0.30  & 0.40 & 0.21  & 0.24 & 0.00     & 0.05   \\
                              & \multicolumn{1}{l|}{select} & 0.30  & 0.41 & 0.21  & 0.25 & 0.05 & 0.00       \\ \bottomrule
\end{tabular}
}}
\end{table}

\subsection{Discussion and Limitations}
 \label{subsection:LEAPER_SIGMETRICS/limit}
 \paragraph*{\textbf{Transfer to a new platform and application simultaneously.}}
In supervised learning, transferring both to a new platform and application at the same time would lead to sub-optimal results (as observed in~\cite{xapp}). This is because we would perform two-levels of the transfer. Moreover, the ML model needs to have some notion of the target environment. Therefore, we explicitly exclude this option in this chapter.

\paragraph*{\textbf{FPGA resource-constrained environments.}}
{During partial reconfiguration~\cite{beckhoff2012go} or in a multi-tenant environment~\cite{pourhabibi2020optimus}, we are often constrained by limited resources~\cite{ting2020dynamic}. In such scenarios, resource management is more efficient to be controlled by a middleware layer~\cite{vaishnav2020fos}. We do not assume such as a middleware. Therefore, our analysis targets bare-metal systems. In the future, we aim to extend our work to such scenarios as well.}
{
\paragraph*{\textbf{\namePaper~generality to other platforms.}}
\namePaper, in essence, is a framework for building and then transferring models from a small edge platform to any new, unknown FPGA-based environment. We demonstrated our approach using the cloud as our target environment because cloud systems often use expensive, high-end FPGAs, e.g., Amazon AWS~F1 cloud~\cite{aws}, Alibaba Elastic cloud with FPGAs~\cite{alibaba}, etc. Thus, we can achieve tangible gains in terms of cost, efficiency, and performance.
}

\paragraph*{\textbf{Effect of FPGA resource saturation.}}

An FPGA gives us the flexibility to map operations to different resources. For example, we can map a multiplication operation either to a CLB or a DSP slice. The deciding factor is the operand width (as shown in Appendix~\ref{chapter:precise_fpga}). If the operand width is smaller than the DSP slice width, the operation is mapped to a CLB else to a DSP unit. An ML-model can be trained to learn such relations. However, we avoid it in our current chapter.

\section{Related Work}
\noindent\textbf{Transfer Learning.} Recently, transfer learning~\cite{chen2010experience,pan2009survey,baumann2018classifying} has gained traction to decrease the cost of learning by transferring knowledge. Valov~\etal~\cite{valov2017transferring} investigated the transfer of application models across different CPU-based environments using linear transformations. Jamshidi~\etal~\cite{jamshidi2017transfer} demonstrated the applicability of using nonlinear models to transfer CPU-based  performance models. {The works above influenced the design of~\namePaper}. In contrast, we focus on FPGA-based systems, where we tailor the hardware circuitry to an application by leveraging the large high-dimensional optimization
space and has very low productivity due to the time-consuming downstream implementation process. Moreover, we use an ensemble of transfer learners that transfers accurate models to a target environment via \textit{few-shot learning}. \\

\noindent\textbf{ML-based FPGA Modeling.} Recent works propose ML-based methods to overcome the issue of low productivity with FPGAs. O'Neal~\etal~\cite{o2018hlspredict} use CPU performance counters to train several ML-based models to predict FPGA performance and power consumption. Makrani~\etal~\cite{xppe_asp_dac} trained an ANN to predict application speedup across different FPGAs. Dai~\etal~\cite{dai2018fast} use ML to predict post-implementation resource utilization from pre-implementation results. However, these solutions become largely impractical once the platform, the application, or even the size of the workload changes. \namePaper~proposes to reuse previously built models on a low-end source environment to accelerate the learning of ML models on a high-end target environment through transfer learning. Unlike \namePaper, past works~\cite{tulikaML} apply traditional, time-consuming brute-force techniques to collect training dataset. These techniques quickly become intractable when the number of optimization parameters increases due to the curse of dimensionality~\cite{bellman1966dynamic}.\\

\noindent\textbf{Analytical FPGA Modeling.} {Analytic models abstract low-level system details and
provide quick performance estimates at the cost of accuracy. These approaches~\cite{linAnalyzer,zhao2017comba,hls_scope} analyze dataflow graphs and apply mathematical equations to approximate resource usage or performance after the HLS pre-implementation phase. These approaches enable {quick early-stage design studies, however, analytical models} cannot model the intricacies of the complete FPGA implementation process~\cite{dai2018fast}. Therefore, these approaches provide crude estimates of the actual performance. Moreover, these models require expert knowledge to form mathematical equations.  In contrast,~\namePaper~does not require expert knowledge to construct equations. \namePaper~learns from the data, taking into account the complete downstream implementation process, and provides the capability to transfer models from an edge-FPGA to a high-end cloud FPGA environment.}

 LEAPER is the first method to deal with transfer learning on FPGAs, which are infamous for low productivity due to the time-consuming mapping process. \namePaper~not only allows us to predict performance for different configurations (Figure~\ref{fig:LEAPER_SIGMETRICS/transfer_overiew} ``Base model building'' stage) but also provides the ability to transfer the models (Figure~\ref{fig:LEAPER_SIGMETRICS/transfer_overiew} ``Target model building - Few-shot learning'' stage). \namePaper~is orthogonal to previous approaches (ML and non-ML) as it also provides the ability to ``transfer'' models in milli-seconds (\textit{ref}. Table~\ref{tab:LEAPER_SIGMETRICS/timing}).   Additionally, in this chapter, we leverage the \textit{design of experiment} techniques to reduce the training overhead dramatically and still build accurate models. 

\section{Conclusion}

We introduce \namePaper, {the first} \textit{transfer learning}-based approach for FPGA-based systems.  \namePaper combines statistical techniques and transfer learning to minimize the ML training data collection overhead.  It overcomes the inefficiency of traditional ML-based methods by accurately \textit{transferring} an existing ML model built on an {inexpensive}, low-end FPGA platform to a new, {unknown}, high-end environment.

 
 \dd{The experimental results} show that we can develop cheaper (with \textit{5-shot}), faster (up to $10\times$), and highly accurate (on average 85\%) models to predict performance and resource consumption in a new, {unknown} target {cloud} environment. 
 We believe that \namePaper~would open up new avenues for research on FPGA-based systems from edge to cloud computing, and hopefully, inspires the development of other alternatives to traditional ML-based models.










\renewcommand{\namePaper}{Sibyl\xspace} 
\chapter[\texorpdfstring{\namePaper: Adaptive and Extensible  Data Placement in Hybrid Storage Systems Using Online Reinforcement Learning}{\namePaper: Adaptive and Extensible  Data Placement in Hybrid Storage Systems Using Online Reinforcement Learning}
]{\chaptermark{header} \namePaper: Adaptive and Extensible  Data Placement in Hybrid Storage Systems Using Online Reinforcement Learning}
\chaptermark{\namePaper}
\label{chapter:sibyl}

\go{Hybrid storage systems (HSS) use multiple different storage devices to \gonn{provide} high and scalable storage capacity at high performance. Data placement across different devices is critical to maximize the benefits \gonn{of} such a hybrid system. Recent research proposes various techniques that aim to accurately identify performance-critical data to \gca{place} it in a ``best-fit'' storage device. Unfortunately, most of these techniques are 
\rakeshisca{rigid}, which (1) limits their \gon{adaptivity} to perform well \gon{for} a wide range of \gon{workloads} and storage device configurations, and (2) makes it difficult for designers to extend these techniques \gon{to} different storage system configuration\gon{s} (e.g., \gon{with \gonn{a} different number \gonn{or different types} of} storage devices) than the configuration \rbc{they are} designed for.  }
\go{Our goal} is to design a new data placement technique for hybrid storage \gon{systems} that \gon{overcomes these issues and provides: (1) \emph{adaptivity}\gonn{,} by}  
\emph{continuously learn\gon{ing}} from and \go{adapt\gon{ing} to} the \gon{workload} and the storage device characteristics, and (2) \emph{easy \gon{extensibility}} to  a wide range of \go{\gon{workloads} and} HSS configurations.

We introduce \emph{\namePaper}, \gon{the first technique that uses} 
reinforcement learning \gca{for data placement in hybrid storage systems}. \namePaper observes different features \gon{of} the \gon{running workload}  \gon{as well as the}  storage devices to make system-aware data placement decisions. For every decision \gon{it makes}, \namePaper receives a reward from the system that it uses to evaluate the long-term \gon{performance} impact of its decision and continuously optimizes its data placement policy online. 

We implement \namePaper on  \emph{real} \gonn{systems with various} \go{HSS configurations, including dual- and tri-hybrid storage systems}, and extensively compare it against \gon{four previously proposed data placement techniques (both heuristic- and machine learning-based) } over a wide range of \go{\gon{workloads}}.
Our results show that  
 \namePaper~\gon{ provides \gonzz{21.6\%/19.9\%} performance improvement in a performance-oriented\gonn{/cost-oriented} HSS configuration 
 compared to the best previous data placement technique.} 
Our evaluation using an HSS configuration with three different storage devices shows that \namePaper outperforms \gonn{the state-of-the-art} \gonzz{data placement} 
policy by 23.9\%-48.2\%\gonn{,} while significantly reducing the system architect's burden 
in designing a data placement mechanism that can simultaneously incorporate three storage devices.
We show that \namePaper  achieves 80\% of the performance of an {oracle} policy that has \gon{complete} knowledge~of~future access patterns while incurring \gon{a \gonn{very modest} storage}  overhead of only \gonzz{124.4} KiB. 

\section{Introduction}
\label{sec:Sibyl/introduction}
Hybrid storage systems (HSS) take advantage of both \gon{fast-yet-small} storage devices and large-yet-slow storage devices to deliver high storage capacity at low latency~\cite{meza2013case,bailey2013exploring,smullen2010accelerating,lu2012pram,tarihi2015hybrid,xiao2016hs,wang2017larger,lu2016design,luo2015design,srinivasan2010flashcache,reinsel2013breaking,lee2014mining,felter2011reliability,bu2012optimization,canim2010ssd,bisson2007reducing,saxena2012flashtier,krish2016efficient,zhao2016towards,lin2011hot,chen2015duplication,niu2018hybrid, oh2015enabling,liu2013molar,tai2015sla,huang2016improving,kgil2006flashcache,kgil2008improving,oh2012caching,yang2013hec,ou2014edm,appuswamy2013cache,cheng2015amc,chai2015wec,dai2015etd,ye2015regional,chang2015profit,saxena2014design,li2014enabling,zong2014faststor,do2011turbocharging,lee2015effective,baek2016fully,liu2010raf,liang2016elastic,yadgar2011management,zhang2012multi,klonatos2011azor}. The key challenge in designing a high-performance and cost-effective hybrid storage system is to accurately identify the performance-critical\gon{ity} of  application \gon{data} and place \gon{data} in the ``best-fit'' storage device~\cite{niu2018hybrid}. 


{Past works~\cite{matsui2017design,sun2013high,heuristics_hyrbid_hystor_sc_2011,vasilakis2020hybrid2,lv2013probabilistic,li2014greendm,guerra2011cost,elnably2012efficient,heuristics_usenix_2014,doudali2019kleio,ren2019archivist,cheng2019optimizing,raghavan2014tiera,salkhordeh2015operating,hui2012hash,xue2014storage,zhang2010automated,zhao2010fdtm,shi2013optimal,wu2012data,ma2014providing,iliadis2015exaplan,wu2009managing,wu2010exploiting,park2011hot}  propose many different data placement techniques to improve the performance of an HSS. We identify \gon{two major} shortcomings \gon{of} prior proposals that significantly limit their performance: \gonn{lack of (1) adaptivity \gonzz{to workload changes and the storage device characteristics,} and (2) extensibility.}

\head{(\gon{1a}) Lack of adaptivity \gca{to workload changes}} \gon{To guide data placement, past techniques consider only a limited number of \gon{workload} characteristics} ~\cite{matsui2017design,sun2013high,heuristics_hyrbid_hystor_sc_2011,vasilakis2020hybrid2,lv2013probabilistic,li2014greendm,guerra2011cost,elnably2012efficient,heuristics_usenix_2014,lv2013hotness,montgomery2014extent}.
Designers statically tune the {parameters values} for all considered \gon{workloads} at design time based on
\rcam{empirical analysis and designer} experience, and expect those {statically-fixed values} to be equally effective \rnlast{for} a wide range of \rcam{dynamic }\gon{workload \rcam{demands}} and system configurations seen in the real world. As a result, such data placement techniques \gon{cannot easily} adapt 
\rakeshisca{to} a wide range of \gonn{dynamic} \gon{workload} demands and significant\gon{ly} \gon{underperform}  when compared to an oracle technique that has \gon{complete} knowledge of future \gon{storage} access patterns (up to $41.1\%$ lower performance, \gonn{ref.} Section \ref{sec:Sibyl/motivation_limitations}).


\head{(\gon{1b}) \gon{Lack of adaptivity to changes in device types and configurations}} Most prior \gon{HSS} data placement techniques \gonn{(e.g.,~\cite{matsui2017design,sun2013high,heuristics_hyrbid_hystor_sc_2011,vasilakis2020hybrid2,lv2013probabilistic,li2014greendm,guerra2011cost,elnably2012efficient,heuristics_usenix_2014,doudali2019kleio,ren2019archivist})}
\rakeshisca{do not adapt well to changes in the}
underlying storage device characteristics (e.g., \gonn{changes in the level of} asymmetry in the \mbox{read/write} latenc\gonzz{ies},  \gon{or the number and types of storage devices}
). As a result, \rcam{existing techniques} cannot effectively take  into account the cost of data \gca{movement} between  storage devices while making data placement decisions. This lack of 
\rcam{adaptivity} leads to highly inefficient data placement policies, especially in HSSs with significantly-different device access latencies \gon{than what prior techniques were designed for} (as show\rcam{n} in Section \ref{sec:Sibyl/motivation_limitations}). 

\head{(2) Lack of extensibility} A large number of prior \gon{data placement} techniques \rcam{(e.g., ~\cite{matsui2017design,sun2013high,heuristics_hyrbid_hystor_sc_2011,lv2013probabilistic,li2014greendm,guerra2011cost,elnably2012efficient,heuristics_usenix_2014})} are typically designed for an HSS that consists of only two storage devices. As modern HSSs \gon{already} incorporat\gon{e} more than two types of storage devices\rakeshisca{~\cite{ren2019archivist, matsui2017tri, matsui2017design}}, system architects need to put significant effort into extending prior techniques to accommodate more than two devices. We observe that a \rcam{state-of-the-art} heuristic-based data placement technique optimized for an HSS with two storage devices~\cite{matsui2017tri} often leads to \gf{suboptimal} performance in an HSS with three storage devices (up to \gonzz{48.2}\% lower performance, \gonn{ref.} Section \ref{subsec:Sibyl/trihybrid}).  

\textbf{Our goal} is to develop a new, efficient, and high-performance data placement mechanism for hybrid storage systems that \gonn{provides} (1) \gon{\emph{adaptivity}}\gonn{,}  
by \emph{continuously learning} from and {adapting to} the \gon{workload and  storage device characteristics,} and  (2) \emph{\gon{easy extensibility}} to a wide range of \gon{\gon{workloads} and HSS} configurations.
}

\textbf{Key ideas.} To this end, we propose \namePaper, 
a reinforcement \linebreak learning-based data placement technique for hybrid storage systems.\footnote{In Greek mythology, \namePaper is an oracle who makes accurate prophecies\rcam{~\cite{wiki:Sibyl}}.} Reinforcement learning (RL)~\cite{sutton_2018} is a goal-oriented decision-making process in which an autonomous agent learns to \gon{take} optimal actions that maximize a reward function by interacting with an environment.    {The key idea of \namePaper is to design the data placement module in hybrid storage systems as a reinforcement learning agent that \emph{autonomously learns} and adapts to the best-fit data placement policy for the \gon{running} \gon{workload} and \gon{the current} hybrid storage \gonn{system} configuration.} 
For every storage \gon{page} access,  \namePaper observes
different features from the running \gon{workload} and the underlying storage system (e.g., access \gonz{count} of the current request, \gon{remaining capacity in the fast storage, etc.}). \rbc{It uses the features as  \textbf{\emph{state}} information to \gon{take} a data placement \textbf{\emph{action}} (i.e., which device to place the page into). \gont{For every action,} \namePaper receives 
a \gon{delayed} \textbf{\emph{reward}}} from the system in terms of per-\gup{request} latency. 
\gca{This} reward encapsulates the internal device characteristics  of \juanggg{an} HSS (such as read/write latenc\rcamfix{ies}, latency of garbage collection, \rcamfix{queuing delays,} \gont{error handling latencies,} and write buffer state). \namePaper uses this reward to estimate
the long-term impact of its \gon{action (i.e., data placement} decision\gon{)} on the overall application performance \gup{and continuously optimizes its data placement policy online} \gon{to maximize the long-term benefit \gonn{(i.e., reward)} of its actions}.

\rbc{\textbf{Benefits.}} 
\rbc{Formulating the data placement module as an RL agent} \rbc{enables a human designer} to specify only \emph{what} performance objective \rbc{the} data placement module should target, rather than designing and implementing a new data placement \gon{policy} that requires \gca{explicit specification of} \emph{how}  to achieve the performance objective. The use of RL not only \gon{enables} the data placement module to \emph{autonomously} learn the ``best-fit'' data placement strategy for a wide range of \gon{workloads} and hybrid storage \gon{system} configurations but also significantly reduces the burden of a human designer \gon{in finding a good data placement policy}. 

\rbc{\textbf{Challenges.}} While RL provides a promising alternative to existing data placement techniques, we identify two main  challenges in applying RL to data placement in an HSS.

\head{(1) Problem formulation}  The RL agent's effectiveness depends on how the data placement problem is cast as a reinforcement learning-based task. Two key issues arise  \gonn{when} formulating \gon{HSS} data placement as an RL problem\rnlast{:}  (1) \gonn{taking into account} the \gonn{latency} asymmetry \gonn{within and across} storage devices, and (2) deciding which actions to reward and penalize (\gca{also known as the} \emph{credit assignment problem}~\gca{\cite{minsky1961steps}}). 
First, we need to make the agent aware of the asymmetry in read and write latencies of \gon{each} storage \gon{device} and \gonx{the} differences in latencies across \gon{multiple} storage devices. Real-world storage devices could have \gon{dynamic} latency variations due to their complex hardware and software components (e.g., internal  caching, garbage collection, \gon{error handling, multi-level cell reading,} etc.)~\cite{cai2015data,grupp2009characterizing, jung2012nandflashsim, cai2014neighbor, cai2017error, cui2017dlv, Cai2018, park2021reducing}. 
Second, \gonn{if the fast storage is running \gonzz{out} of free space, there might be evictions \gup{in the background} \gon{from} the fast storage to the slow storage.

\gon{As a result, when} we reward the agent, \gon{not only there is a \gon{variable} \gonn{and delayed} reward,
~but} it is \gon{also} hard to properly \gon{assign} \emph{credit} or \emph{blame} \gon{to} different decisions.

\head{(2) Implementation overhead} A \gon{workload} could have \gon{hundreds of} thousands \rcam{ of} pages \gon{of \gonn{storage} data}, making it challenging to efficiently handle the large data footprint with a low design overhead \gon{for the learning agent.} 

To address the first challenge, we use two \gon{main} techniques. First, we design a \textit{reward} structure in terms of \emph{request latency}, \gon{which} allows \namePaper~to learn the \gon{workload and storage} device characteristics 
\nh{when} \gonn{continuously and frequently} interacting with a hybrid storage system. We  \gca{add} a negative penalty \gca{to the reward structure} in case of eviction, \gonn{which helps  \gca{with handling} the credit assignment problem and \gont{encourages the} agent \gont{to} place only performance-critical pages in the fast storage.} {Second, we perform thorough hyper-parameter tuning \gon{to find parameter values that work well for a wide variety of workloads.}} 
To address the second challenge, we use two main techniques. First, we divide states into a small number of bins that reduce \gon{the} state space, which directly \gon{affects}  the implementation overhead. {Second, instead of adopting a traditional table-based RL approach  (e.g., \cite{ipek2008self}) to store \gonn{the} agent's state-action information (collected by interacting with \gon{an HSS})\gonn{, which} can \gonn{easily} introduce significant performance overhead\gonn{s in the presence of}  a large \gont{state/action} space}, we use a {simple} feed-forward neural network~\cite{zell1994simulation} with only two hidden layers of 20 and 30 nodes, respectively. 


\textbf{Key results.}} \gup{We  evaluate \namePaper~\gf{using \gon{two different dual-HSS} configurations and two different tri-HSS configurations.} We use fourteen diverse \gon{storage} traces from \rakeshisca{Microsoft Research Cambridge (MSRC)} \cite{MSR} collected on real enterprise servers. \gon{We  evaluate \namePaper on workloads from  FileBench~\cite{tarasov2016filebench} on which it has never been trained}. \gon{We compare \namePaper to four state-of-the-art data placement techniques.} We demonstrate four key results. \gon{ First, \namePaper~\gonn{ provides \gonzz{21.6\%/19.9\%} performance improvement in a performance-oriented\gonn{/ cost-oriented} HSS configuration 
 compared to the best previous data placement technique.}} 
Second,  \namePaper outperforms the best-performing supervised learning-based technique  \gon{ on workloads it has never been trained on} by 46.1\% and 54.6\%, on average, } \gon{in} performance-oriented and cost-oriented HSS \gca{configurations, respectively}. 
Third,  \namePaper provides 23.9\%-48.2\% higher performance in tri-hybrid storage systems than a \gonn{state-of-the-art} heuristic-based \gon{data placement} technique \gonn{demonstrating that} \namePaper \gon{~is} \gonn{easily} extensible and alleviates the designer's burden in \gon{finding} sophisticated data placement mechanisms \gon{for new and complex HSS configurations}. Fourth, \namePaper's performance benefits come with a low \gagan{storage} implementation overhead of only \gonzz{124.4} KiB. 

\vspace{0.2em}
\noindent This work makes the following  \textbf{major contributions}:
\begin{itemize}[leftmargin=*, noitemsep, topsep=0pt]
 \item  We show on real \gonn{hybrid storage systems (HSSs)}  that prior state-of-the-art \gont{HSS data placement mechanisms} fall 
  short of the oracle placement due to: lack of  \gon{(1) adapt\gonn{i}vity to workload changes and \gonn{storage device characteristics}, and (2) extensibility. } 
  \item We propose \namePaper
  , a new self-\gon{optimizing}  mechanism 
  \gon{that uses} reinforcement learning \gon{to make data placement decisions in hybrid storage systems}. \namePaper{} 
  \go{dynamically}
  \emph{learns}\gont{,} \gonn{ using both multiple workload features and system-level feedback information\gont{, how} to continuously adapt its policy to improve \rnlast{its} long-term performance \rnlast{for} a workload.}
    
    \item We conduct an in-depth evaluation of \namePaper on \gon{real \gonn{systems with various} HSS configurations\gont{,}} showing that it outperforms \gonn{four} state-of-the-art 
    techniques over a wide variety of applications with a low implementation overhead.
    \item We provide an \rbc{in-depth} explanation of \namePaper's actions that show that \namePaper performs dynamic data placement decisions by learning \gonn{ changes in the level of asymmetry in the read/write latenc\gonzz{ies} and the number and types of storage devices.} 
    \item \gca{ We \gon{freely} open-source \namePaper to aid 
 {future research \gon{in} data placement for storage systems~\cite{sibylLink}}.} 
    
    \end{itemize}
 
\section{Background\label{sec:Sibyl/background}}
\vspace{-0.1cm}
\subsection{Hybrid Storage Systems (HSSs)\label{subsec:Sibyl/bg_hybrid}}

\begin{figure}[h]
  \centering
   \includegraphics[width=0.6\linewidth]{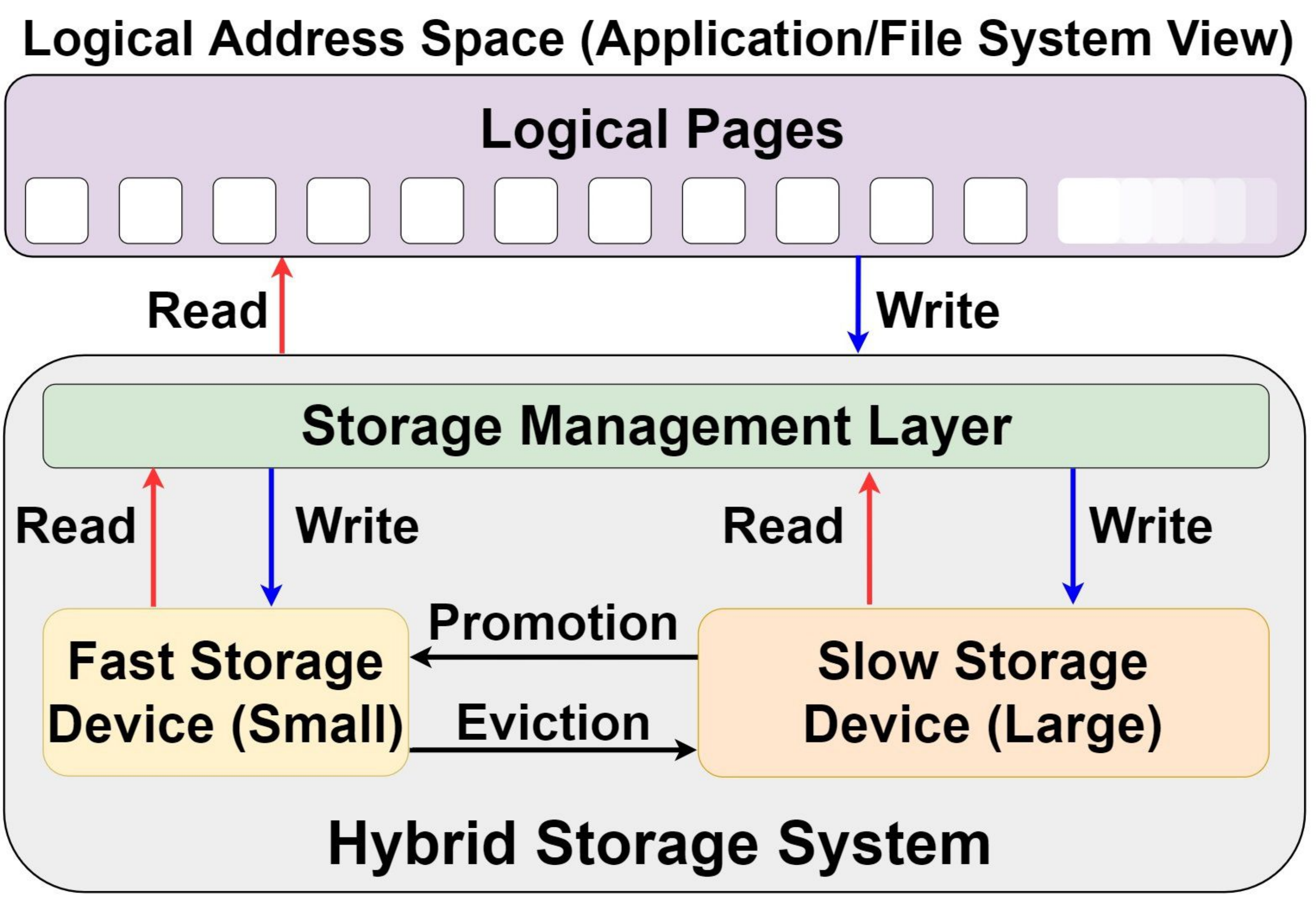}
  \caption{Overview of a hybrid storage system} 
  \label{fig:Sibyl/hybrid}
  \end{figure}

\fig{\ref{fig:Sibyl/hybrid}} depicts a typical HSS \gca{consisting} of a fast-yet-small \gon{storage} device \rcam{(e.g., \cite{inteloptane,samsung2017znand})} and a large-yet-slow \gon{storage} device \rcam{(e.g., \cite{intels4510, intelqlc, seagate, adatasu630})}. \gon{Traditional hybrid storage systems~\cite{heuristics_hyrbid_hystor_sc_2011,feng2014hdstore,micheloni2018hybrid} were designed with a smaller NAND flash-based SSD and a larger HDD. 
Nowadays, hybrid storage systems come with emerging {NVM} devices \rcam{(e.g.,~\cite{kim2014evaluating, tsuchida201064mb, kawahara20128, choi201220nm})} coupled with slower {high-density} NAND flash devices~\cite{matsui2017design, oh2013hybrid, lee2010high, okamoto2015application}. } The storage management layer can be implemented either as system software running on the host system or as \gca{the} firmware of a hybrid storage device (e.g., flash translation layer (FTL) in flash-based SSDs \rcam{\cite{cai2017error,gal2005algorithms}}), depending on the configuration of the HSS. In this chapter, we \gon{demonstrate and implement our ideas in \gonzz{the} storage management layer \gonzz{of} the operating system (OS)\gonzz{,} but they can be easily implemented in firmware as well.} The storage management layer {in the OS} orchestrates host I/O requests across heterogeneous {devices}, which are connected via an NVM Express (NVMe)~\cite{nvme} or SATA~\cite{sata} interface. {The storage management layer} provides the operating system with a unified logical \rcam{address} space (like the multiple device driver (md) kernel module in Linux~\cite{linux}).\rcam{~As illustrated in \fig{\ref{fig:Sibyl/hybrid}}, the unified logical address space is divided into a number of logical pages (e.g., 4 KiB pages). The pages in the logical address space are assigned to an application. The storage management layer translates \rcamfix{a} read/write performed on \rcamfix{a} logical page into a read/write operation on a target storage device based on the data placement policy. In addition, the storage management layer manages data migration between the storage devices in an HSS. When data currently stored in the slow storage device is moved to the fast storage device, it is called \emph{promotion}. Promotion is usually performed when \rcamfix{a page} in the slow storage device is accessed frequently. Data is moved from the fast storage device to the slow storage device during an \emph{eviction}. Eviction typically occurs when the data in the fast storage device is infrequently accessed or when the fast storage device becomes full.}

\gon{The performance of {a} hybrid storage system highly depend{s} on the ability of the storage management layer (\fig{\ref{fig:Sibyl/hybrid}}) {to \gonn{effectively} manage diverse devices and}
workloads~\cite{matsui2017design, ren2019archivist}.
This \gonn{diversity} presents a {challenge} 
for system {architects} 
\gonn{when they} design an intelligent data placement policy. 
{A desirable policy has} to effectively utilize the low latency characteristics of the fast device {while making} optimal use of its small capacity 
and \gon{should provide easy extensibility to a wide range of workloads and HSS configurations. }}

\section{Motivation}\label{sec:Sibyl/motivation_limitations}


To assess the effectiveness of existing \gon{HSS} data placement techniques under diverse workloads and hybrid storage configurations, we evaluate {state-of-the-art } heuristic-based (\emph{\cde}~\cite{matsui2017design} and \emph{\hps}~\cite{meswani2015heterogeneous}) and supervised learning-based  (\emph{\arcc}~\cite{ren2019archivist}) techniques. We also implement an RNN-based data placement technique (\kleio), adapted from hybrid main memory~\cite{doudali2019kleio}. To evaluate the effect of underlying storage device technologies, we use {three  different storage devices}: high-end {(\textsf{H})}~\cite{inteloptane}, middle-end {(\textsf{M})}~\cite{intels4510}, and low-end {(\textsf{L})}~\cite{seagate}, configured into two different hybrid storage configurations: a performance-oriented HSS (\hmssd) and a cost-oriented HSS (\hlssd).  
Table~\ref{tab:Sibyl/devices} provides details of our system and devices. We restrict the fast storage capacity to 10\% of the working set size of a workload, which ensures eviction of data from  fast storage to slow storage when  fast storage capacity is full.

{\cde}~\cite{matsui2017design} \gon{allocates} hot or random write requests in the faster storage, {whereas} cold and sequential write requests are evicted to the slower device. 
\hps~\cite{meswani2015heterogeneous} uses the access \gonz{count} \gca{of pages} to periodically migrate cold pages to the slower storage device. 
\arcc~\cite{ren2019archivist} uses a neural network classifier to predict the target device for data placement. 
\emph{\kleio}, adapted from~\cite{doudali2019kleio}, is a supervised learning-based mechanism {that} exploits recurrent neural networks (RNN) to \gca{predict the hotness of a page} \gon{and place hot pages in fast storage}.
We compare the above policies with three extreme \gon{baselines}:
(1) \slow{}, {where all data reside\rcamfix{s in} \gon{the} slow \gon{storage \rcamfix{device} (i.e., there is no fast storage \rcamfix{device})}}, (2) \fast, {where all data resides in \gon{the}  fast \gon{storage} \rcamfix{device}}, {and (3) \textsf{Oracle}}
\cite{meswani2015heterogeneous}, which exploits \gon{complete} {knowledge of} future I/O-access patterns  
{to perform data placement and to }\js{select victim data blocks for eviction from} the fast device. 

We identify \gon{two major}  shortcomings \gon{of the \gonn{state-of-the-art} baseline data placement techniques}: \gonzz{lack of (1) adaptivity to workload changes and the storage device characteristics, and (2) extensibility.}

\begin{figure}[h]
    \centering
 \includegraphics[width=1\linewidth,trim={0.2cm 0cm 0.2cm 0cm},clip]{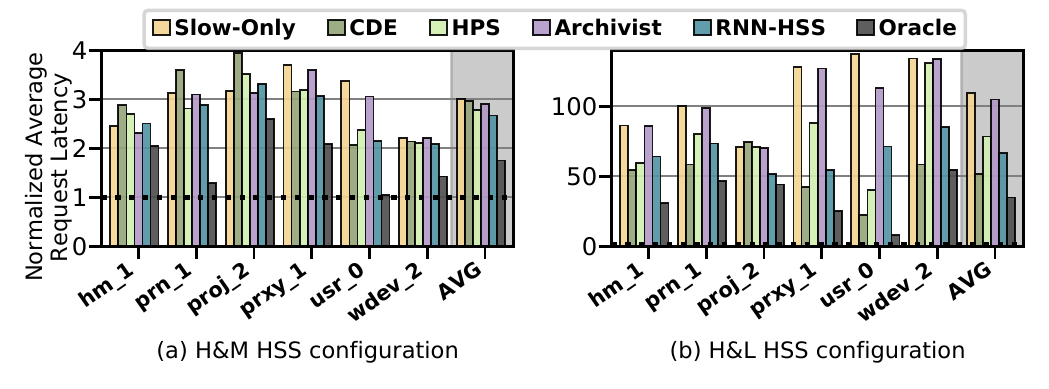}
 \vspace{-0.55cm}
    \caption{Average request latency normalized to \fast~policy} 
    \vspace{-0.1cm}
    \label{fig:Sibyl/motivation_iops}
\end{figure}

\head{(1a) Lack of adaptivity \gca{to workload changes}} 
\fig{\ref{fig:Sibyl/motivation_iops}} shows \gonzz{the} \gon{average request} latency of \gon{all} policies, \gon{normalized to \fast,} under two different hybrid storage configurations. \gon{We make the following \gonn{three} observations.}
First, \gon{all the baseline}  techniques are only effective under specific workloads, showing significantly \gonn{lower} performance \gonn{than} \textsf{Oracle} in most workloads.  \cde, \hps, \arcc, and \kleio~achieve comparable performance to \textsf{Oracle} {for specific workloads} \gagan{(e.g.,  \textsf{hm\_1} for \hps~in \textsf{H\&M},  \textsf{usr\_0} for \cde~in \textsf{H\&L}, \gon{\textsf{hm\_1} for \arcc~in \textsf{H\&M}, and \kleio in \textsf{proj\_2} for \cde~in \textsf{H\&L} }). Second, the baselines show a {large} \gon{average} performance \gon{loss} of 41.1\% (32.6\%), 37.2\% (55.5\%), \gonn{39.7\% (66.7\%), and 34.4\% (47.6\%)} \gon{compared} to \textsf{Oracle}'s performance, under the \textsf{H\&M} (\textsf{H\&L}) hybrid {storage} configuration, respectively.} 
\gonn{Third, in \hmssd, the baseline techniques provide a performance improvement of only 1.4\%, 7.4\%, 3.5\%, and 11.3\% compared to \slow.  }

\begin{SCfigure}[][h]
  \centering
 \includegraphics[width=0.7\linewidth,trim={0cm 0.02cm 0.2cm 0cm},clip]{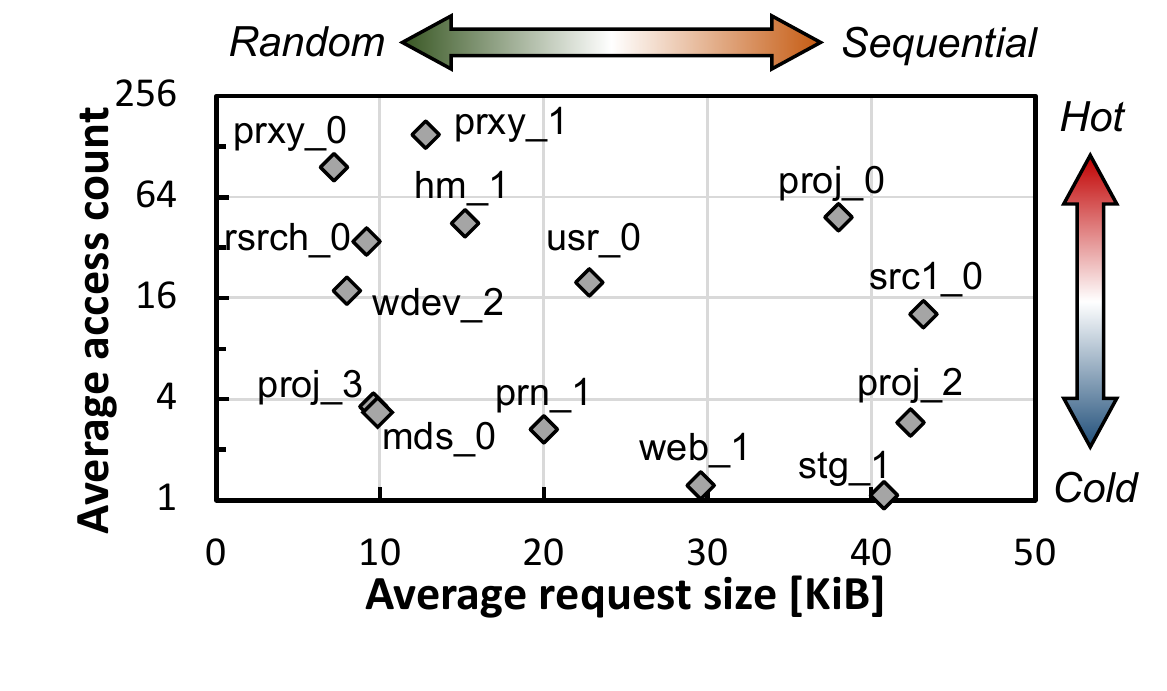}
 \vspace{-75pt}
 \caption{\gonx{Randomness and hotness} characteristics of real-world MSRC \gonx{workloads}~\cite{MSR}}
\label{fig:Sibyl/apps}
\vspace{1.2cm}
\end{SCfigure}

{

\gonn{We conclude that \gon{all four baselines} consider only a limited number of \gon{workload} characteristics to construct a data placement technique, which leads to a significant performance gap compared to the \oracle policy. Thus, there is no single  policy that} 
\juangg{works} well for all \gon{the workloads}. 

\gon{To further analyze the characteristics  of our evaluated workloads, 
we} plot the \gca{average} hotness (y-axis) and randomness (x-axis) in \fig{\ref{fig:Sibyl/apps}}. 
\gagan{We provide details on these workloads in Table~\ref{tab:Sibyl/workload}.} \gonn{  In these workload traces,} each \gon{storage} request is labeled with a timestamp that indicates the time \gonzz{when} the request was issued from the processor core. Therefore, the time interval between two consecutive I/O requests represents the time the core has spent computing.  {We quantify a workload’s hotness (or coldness) using the average access 
\rcamfix{count}, \gca{which is} the average \rcamfix{of the access counts of all pages} in a workload; the higher (lower) the average access \rcamfix{count}, the hotter (colder) the workload. We quantify a workload’s randomness using the average request size in the workload; the higher (lower) the average request size, the more sequential (random) the workload. } 
\gon{From \fig{\ref{fig:Sibyl/apps}}\rcamfix{,} we make \rcamfix{the}  following two observations. First, }the average hotness and randomness {vary widely between \gon{workloads}}. \gon{Second, we observe that each of our evaluated \gon{workload}\rcamfix{s}  exhibits highly dynamic behavior throughout its execution. For example, in Figure~\mbox{\ref{fig:Sibyl/dynamic}}, we show the execution \gonn{timeline} of \texttt{rsrch\_0}, \gonn{which \gonx{depicts} the accessed address\gonx{es} and request sizes.} 
\gon{We conclude that} an efficient policy needs to incorporate continuous adaptation to highly dynamic changes in workload behavior.}}  


{
\begin{SCfigure}[][h]
  \centering
  \hspace{0.2cm}
 \includegraphics[width=0.6\linewidth,trim={0cm 0cm 0.2cm 0cm},clip]{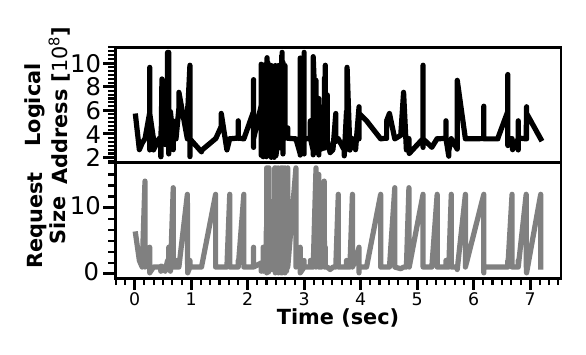}
\vspace{-45pt}
  \caption{\gor{\gonz{Timeline of} accessed logical addresses and request sizes during \gont{the} execution of \texttt{rsrch\_0} workload}}
\label{fig:Sibyl/dynamic}
\vspace{0.6cm} 
\end{SCfigure}
}

\vspace{0.1cm}
\head{\gon{(1b) Lack of adaptivity to changes in device types and configurations}} 
\juangg{\gon{There are a wide variety and number of storage devices 
~\cite{samsung2017znand, seagate, inteloptane, cheong2018flash,micron3dxpoint, hady2017platform, intelp4610, intelqlc, intels4510, adatasu630, kim2014evaluating, tsuchida201064mb, kawahara20128, choi201220nm, matsui2017design, oh2013hybrid, lee2010high, okamoto2015application} that can be used  to configure \gca{an HSS}.} The \gca{underlying storage} technology used in \gca{an HSS}  significantly influences \gca{the} effectiveness \gca{of a data placement policy}.} \gon{We demonstrate this with an example  observation from \fig{\ref{fig:Sibyl/motivation_iops}}}. In the \textsf{H\&M} configuration (\fig{\ref{fig:Sibyl/motivation_iops}(a)}), we \gon{observe} that  for certain workloads (\textsf{hm\_1} and \textsf{prn\_1}), both \cde  and \hps provide rather 
\juangg{low} performance even compared to \slow. Similarly, \arcc and  \kleio provide lower performance for \textsf{hm\_1} and \textsf{proj\_2} in \hmssd compared to \slow.   
While in the \textsf{H\&L} configuration (\fig{\ref{fig:Sibyl/motivation_iops}(b)}), we  observe that \cde, \hps, \arcc, and \kleio~
\juangg{result in} lower latency than \slow for the respective workloads. 
\gonn{Thus, we conclude that both heuristic-based and learning-based data placement policies lead to poor performance due to their inability to holistically take into account the device characteristics. The high diversity in \gagan{device} characteristics makes it very difficult for a system architect to design a generic  data-placement technique that is suitable for all \acrshort{hss} configurations. }

\head{(2) Lack of extensibility} \gon{As} modern \gon{HSSs already incorporate more than two types of storage  devices~\cite{ren2019archivist, matsui2017tri, matsui2017design, meza2013case}, system architects need  to put significant effort into extending prior \gonn{data placement} techniques to
accommodate more than two devices.} \rakeshisca{In Section \ref{subsec:Sibyl/trihybrid}, we evaluate the effectiveness of a state-of-the-art heuristic-based policy~\cite{matsui2017tri} \gon{for different tri-HSS configurations, \gonn{comprising} of three different storage devices.} This policy is based on the \cde~\cite{matsui2017design} policy that divides pages into hot, cold, and frozen data and allocates these pages to \textsf{H}, \textsf{M}, and \textsf{L} devices, respectively. \gon{ A system architect needs to statically define the hotness values and explicitly handle the eviction and promotion between the three devices during  design-time.} 
Through  our evaluation in Section \ref{subsec:Sibyl/trihybrid}, we conclude that such a heuristic-based policy (1) lacks extensibility, \rcam{thereby increasing}  
 the system architect's effort, and (2) leads to lower performance when compared to an RL-based solution (up to 48.2\% lower).} 

 \gon{Our} empirical study  shows \js{that \textbf{the \gonn{state-of-the-art} {heuristic- and learning-based} data placement techniques are rigid and far from 
optimal,}} which strongly motivates us to develop a new data placement technique \gonn{to achieve significantly higher performance than existing policies}.
The new technique should provide \mbox{(1) \emph{adaptivity}} to better capture the features \gonn{and dynamic changes} in I/O-access patterns and storage device characteristics, 
and \mbox{(2) \emph{easy}} \emph{extensibility} to a wide range of \gon{workloads} and HSS configurations. \gonn{Our goal is to develop such a technique using reinforcement learning.}  

\vspace{-0.2cm}
\section{{Reinforcement Learning}}
\vspace{-0.1cm}
\subsection{{Background}}
\label{background:rl}
\begin{figure}[h]
  \centering
    \includegraphics[width=0.7\linewidth,trim={2.2cm 7.6cm 3cm 1.4cm},clip]{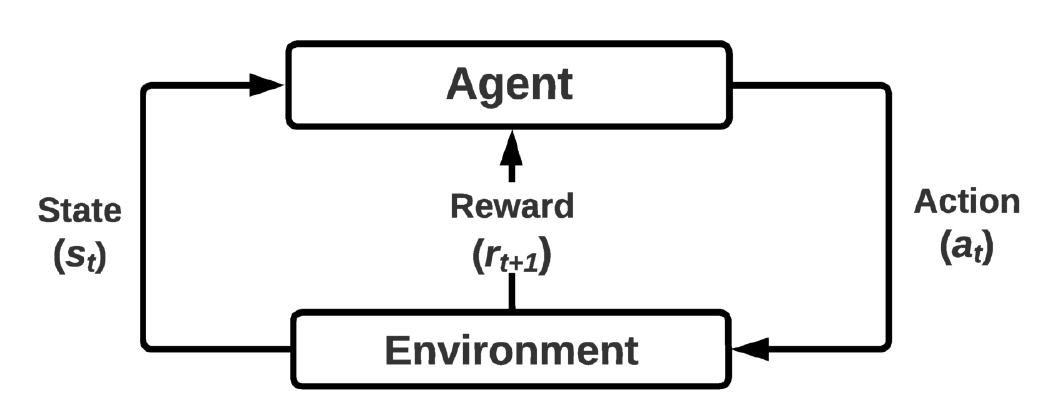}
  \caption{Main components of \gon{general} RL}
\label{fig:rl_basic}
 \end{figure}

Reinforcement learning (RL)~\cite{sutton_2018} is a class of machine learning (ML) algorithms that involve an \emph{agent} learning to achieve an objective by interacting with its \emph{environment}{, as shown in \fig{\ref{fig:rl_basic}}}. The agent starts from an initial representation of its environment in the form of an initial state\footnote{{State is a representation of an environment using different features.} } $s_0$ $\in$ $S$, where $S$ is the set of all  possible states. 
Then, at each \emph{time step} $t$, the agent performs an \emph{action} $a_t$ $\in$ $A$ in state $s_t$ ($A$ represents the set of possible actions) and moves to the next state $s_{t+1}$. The agent receives a numerical reward $r_{t+1}$ $\in$ $R$, \gon{which could be \textit{immediate} or \textit{delayed} in time},  for action $a_t$ that \gca{changes} the environment state from $s_t$ to $s_{t+1}$. The sequence of states and actions starting from an initial state to the final state is called an \emph{episode}. 
The agent \gca{\gonx{makes} decisions and receives corresponding rewards} 
while trying to maximize the \textit{accumulated} reward, \gont{\rcamfix{as opposed to} \gonz{maximizing} the reward for \gonz{only} \emph{each} action}. In this way, the agent can optimize for the long-term impact of its decisions. 

\gon{
The policy $\pi$ governs an agent's action in a state.  The agent’s goal is to find the optimal policy that maximizes the cumulative reward\footnote{The total cumulative reward is also known as the \emph{return}~\cite{sutton_2018}.} collected from the environment
over time.  
The agent finds an optimal policy $\pi^{*}$ by calculating the optimal action-value function \gont{($Q^{*}$), also known as the \textbf{Q-value}} of the state-action pair, where $Q(S,A)$ represents the expected cumulative reward by taking an action A in a given state S.  

\gonz{Traditional RL methods (\rbc{e.g.,}~\cite{watkins1989learning,qlearning_ML_1992,rummery1994line,ipek2008self,pythia}) use a tabular approach with a lookup table to store the \rbc{Q-values associated with each} state-action pair. 
These approaches can lead to high storage and computation overhead for environments with a large number of states and actions. 
 To overcome this issue, \rbc{\emph{value function approximation} was proposed.~\cite{silver2016mastering,silver2017mastering,mnih2013playing,liang2016deep}}. Value function approximation replaces \gonzz{the} lookup table with \gonzz{a} supervised-learning model~\cite{sutton1999policy,baird1995residual,liang2016deep,mnih2013playing,silver2016mastering,silver2017mastering}, which provides the capability to generalize over a large number of state-action pairs with a low storage and computation overhead. } }

\vspace{0.1cm}
\vspace{-0.2cm}
\subsection{Why Is RL  a Good Fit for Data Placement in Hybrid Storage Systems?}

We choose RL \gon{for data placement in HSS} due to the following advantages
compared to heuristic-based (e.g.,~\cite{matsui2017design,
meswani2015heterogeneous}) and supervised learning-based (e.g.,~\cite{ren2019archivist}) techniques.

\head{(1) \gon{Adaptivity}} As discussed in Section \ref{sec:Sibyl/introduction} and Section \ref{sec:Sibyl/motivation_limitations}, a data placement technique should have the ability to adapt to changing workload demands and underlying device characteristics. This \gon{adaptivity}  requirement of data placement makes RL a good fit \gca{to model} data placement. 
The RL agent works autonomously \gon{in an HSS} using the provided \gca{state} features and reward to \gon{\gonx{make} data placement decisions} without any human intervention.

\head{(2) Online learning} Unlike an \emph{offline} learning-based approach, an RL agent uses an \emph{online} learning approach. Online learning allows an RL agent to  continuously adapt  its decision-making policy using  system-level feedback and specialize to \juangg{the} current workload and system configuration. 

\gont{
\head{(3) Extensibility} RL provides the ability to easily extend a mechanism with \gonzz{a} \gonff{small} effort required to implement the extension. As shown in Section \ref{subsec:Sibyl/trihybrid}, unlike heuristic-based mechanisms,  RL can be easily extended to different \gonff{types and} number of storage devices\gonff{. Such extensibility} reduces the system architect's burden in designing sophisticated data placement mechanisms.} 

\gonff{\head{(4) Design Ease} \gonz{With RL, the designer of the HSS} does not need to specify \gonz{a data placement} policy. They need to specify \emph{what} to optimize (via reward function) but not \emph{how} to optimize it. }

\gon{
\head{(5) Implementa\gont{tion Ease}} RL provides ease of implementation \gonff{that requires a} \gonz{small}  computation overhead. 
{As shown} in Section \ref{sec:Sibyl/results}, \textit{function approximation}-based~RL~techniques can generalize over all the possible state-action pairs by using a simple feed-forward neural network to provide high performance at low implementation overhead {\gont{(compared to} sophisticated RNN-based mechanisms}).}

\section{\namePaper: RL Formulation}
\label{sec:Sibyl/rl_formulation}

\begin{figure}[h]
  \centering
 \includegraphics[width=0.85\linewidth,trim={1.12cm 3.5cm 1.8cm 5.5cm},clip]{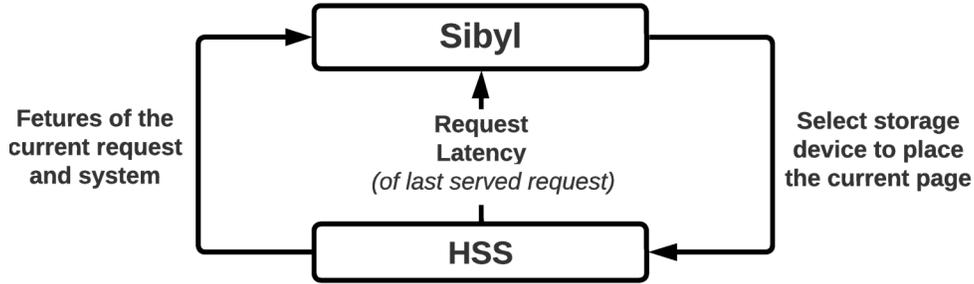}
  \caption{Formulating data placement as an RL problem}
\label{fig:rl_formulation}
 \end{figure}

\fig{\ref{fig:rl_formulation}} shows our formulation of  data placement as an RL problem. We design \namePaper as an RL agent that learns to perform accurate and system-aware data placement decisions by interacting with the hybrid storage system. With every storage request, \namePaper observes \gup{multiple \gon{workload}} \gont{and system-level} features \gca{as \gonzz{a} \emph{state}} to make a data placement decision. After every \emph{action}, \namePaper receives a \emph{reward} \gca{in terms of \gont{the served} request latency} that takes into account the data placement decision and internal storage system state. \namePaper's goal is to find an optimal data placement policy that 
\gont{maximizes} overall performance for \gon{the running} workload and \rcam{the current} system configuration.
\gon{To reach its performance goal, \namePaper needs to minimize the \gont{average request} latency \gon{of the running workload} \gont{by maximizing} the use of the fast storage device while avoiding the eviction penalty \gont{due to non-performance critical pages}. }

\head{Reward} 
\gon{After every data placement decision} at time-step\footnote{\gor{In HSS, a time-step is defined as a new storage request.
}}  $t$, {\namePaper gets a reward 
from the environment} at time-step $t+1$ that acts as a feedback to \namePaper's previous action. 
{To achieve \namePaper's performance goal, we craft \gon{the} reward \gon{function} \textit{R} as follows:
\begin{equation}
R =\begin{cases}
       {\scalemath{1}{ \frac{1}{L_t} }}& 
        \begin{gathered}
            \textit{if no eviction \gon{of a page from the}} \\[-\jot]
            \textit{\gon{fast storage to the slow storage}}
        \end{gathered}\\
        max(0,\frac{1}{L_t}-\textit{$R_p$}) & \text{\textit{in case of eviction} }
    \end{cases}
\end{equation}
{where $L_t$ and \textit{$R_p$} represent \rcamfix{the} \gonff{last} \gont{served} \gca{request latency}  and eviction penalty, respectively. \gont{If the fast storage is running \gonff{out} of free space, there might be evictions in the background from the fast
storage to the slow storage. Therefore, we add an eviction penalty (\textit{$R_p$}) to  guide \namePaper to  place only performance\gonzz{-}critical pages in the fast storage. 
} 
\gon{We empirically select \gont{\textit{$R_p$} to be equal to} 0.001$\times$$L_e$ ($L_e$ is the time spent in evicting pages from the fast storage to the slow storage)}\rcamfix{, which}  prevents the agent from aggressively \gonff{placing} all requests \gonff{into} the fast storage device.}}

\gont{$L_t$} \gonff{(request latency)}  is the time taken to service \gont{the last} read or write  I/O request \gont{from the OS}. Request latency can faithfully capture the status of the hybrid storage system, as it significantly varies depending on the request type, device type, and the internal state \gont{and \gonz{characteristics}} of the device (e.g., such as read/write latenc\rcamfix{ies}, \gonzz{the} latency of garbage collection, \rcamfix{queuing delays, \gonz{and error handling latencies})}.
\rcamfix{Intuitively, if} \gont{$L_t$} is low (high), i.e., if the agent serves a storage request from the fast (slow) device, the agent receives a high (low) reward. However, \gonff{if there is an eviction,} we \rcam{penalize} the agent \gonff{so as} \gup{ to \rcamfix{encourage} the agent \rcamfix{to} place only performance-critical pages in the fast storage device.} 
   We need the \rcamfix{eviction} penalty \gonff{to be} 
   large enough to \rcamfix{discourage} the agent \rcamfix{from evicting} and small enough not to deviate the learned policy too much on a \gonff{placement} decision \gonz{that leads} to higher latency.  

\head{State}
At each time-step $t$, the state features for a particular \mbox{read/write} request  are collected in an \emph{observation} vector. 
  \gon{We  perform feature selection~\cite{kira1992featureselection} to determine the best state features to include in \namePaper's \gont{observation} vector.} 

We use a limited number of features due to two reasons. First, a limited feature set allows us to reduce the implementation overhead of our mechanism \gont{(\gonff{see} \ref{sec:Sibyl/overhead_analysis})}. Second, we empirically observe that our RL agent is more sensitive to the reward structure than to the \gca{number} of features in the \gonff{observation vector}.  Specifically, using the request latency as a reward provides indirect feedback on the internal timing \rcam{characteristics} \emph{and} the current state \gon{(e.g., queu\rcamfix{e}ing delays, buffer dependencies, effects of garbage collection, read/write latenc\gonzz{ies},  write buffer state, \gonzz{and} \rcamfix{error handling latencies})} of the hybrid storage system. Our observation aligns with a recent study~\cite{silver2021reward} that argues that the reward is the most critical component \gca{in RL to find an optimal decision-making policy}.

In our implementation of \namePaper{}, the observation vector is a 6-dimensional tuple:
\vspace{-0.1cm}
\begin{equation}
\gonx{
O_t = (size_t, type_t, intr_t, cnt_t, cap_t, curr_t).}
\end{equation}

\noindent
Table~\ref{tab:Sibyl/state} 
{lists our \js{six}} selected features. \gon{We \rcamfix{quantize the representation of each state} into a small number of bins \gonff{to} reduce \gonff{the} \gont{storage overhead of state representation}.} These features can be captured in the block layer of the storage system \gup{and stored in a separate metadata table (\ref{sec:Sibyl/overhead_analysis})}.  
$size_t$ represents the size \gon{of the current request} in terms of the number of pages associated with \gon{it}. It {indicates whether the incoming request} is sequential or random. 
{$type_t$ \gon{(request type)} differentiates between read and write requests, \gon{important} for data placement decisions since storage devices have asymmetric read and write latencies.} 
$intr_t$ (access interval)  and $cnt_t$ (access \gonz{count}) represent the temporal and spatial reuse \gon{characteristics} of \gont{the currently requested page}\gca{, respectively}.  \gup{Access interval  
\rcamfix{is defined as} the number of page accesses between two  references to the same page. 
\gf{ Access \gonz{count} 
\rcamfix{is defined as} \gonff{the} total \gon{number of} accesses to \gon{the page}.} \gonz{These metrics} provide
insight into \gonzz{the} \gonz{dynamic} \gon{behavior of the currently requested page.}
\footnote{We did not use the reuse distance as a locality metric due to its high computation overhead during online profiling~\cite{zhong2009program}.}
} 
$cap_t$ is a global counter \gonz{that tracks} the remaining capacity in the fast storage \gon{device, which }is an important feature since our agent's goal is to maximize the use of the limited fast storage capacity \gca{while avoiding evictions} \gon{from the fast storage device}.  \gup{By including this feature, the agent can learn}
to \gonff{avoid the eviction penalty (i.e., learn to } restrain {itself} from \gonff{placing in fast storage} non-performance critical pages that would lead to evictions). 
 \gon{$curr_t$ \gont{is the} current placement of the requested page. Since every data placement decision affects the decision for future requests, $curr_t$ guides \namePaper to perform past-aware decisions.}


{
\begin{table}[h]
 \caption{State features \gon{used by} \namePaper}
    \label{tab:Sibyl/state}
\centering
\small
 \renewcommand{\arraystretch}{0.9}
\setlength{\tabcolsep}{2pt}
  \resizebox{1\linewidth}{!}{%
\begin{tabular}{l||l|c|c}
\hline
\textbf{Feature}             & \textbf{Description} & \textbf{\# of bins} & 
\textbf{Encoding (bits)}\\ 
\hline
\gonx{$size_t$}                      &  Size of the \gont{requested page (in pages)} &  8  & 8 \\
\gonx{$type_t$}                      &  Type of the \gont{current request (read/write) }  &2 & 4  \\
\gonx{$intr_t$}                      &  \gh{Access} interval \gon{of the requested page}&64 & 8 \\ 
\gonx{$cnt_t$}                      & Access \gonz{count} of the \gon{requested} page &64 &8   \\
\gonx{$cap_t$}   
&  Remaining capacity in the fast storage device  &8&8    \\
\gonx{$curr_t$}                      &   Current {placement} \gon{of the requested page} \gonff{(fast/slow)} &2 & 4\\
\hline
\end{tabular}
}
\end{table}

}

\head{Action} 
{At each time-step $t$, in a given state, \namePaper selects an action \gup{($a_t$ in Figure~\ref{fig:rl_formulation})} from \gon{all} possible actions. \gca{In a hybrid storage system with two devices,  possible actions are: } placing data in \gon{(1) the fast storage device}} 
or (2) the slow storage device. \gont{This is easily extensible to $N$ storage devices, where $N\ge3$.} 

\section{\namePaper: Design}
\label{sec:Sibyl/mechanism}
 \begin{figure*}[t]
  \centering
  \includegraphics[width=\linewidth,trim={0cm 1.4cm 0cm 0.2cm},clip]{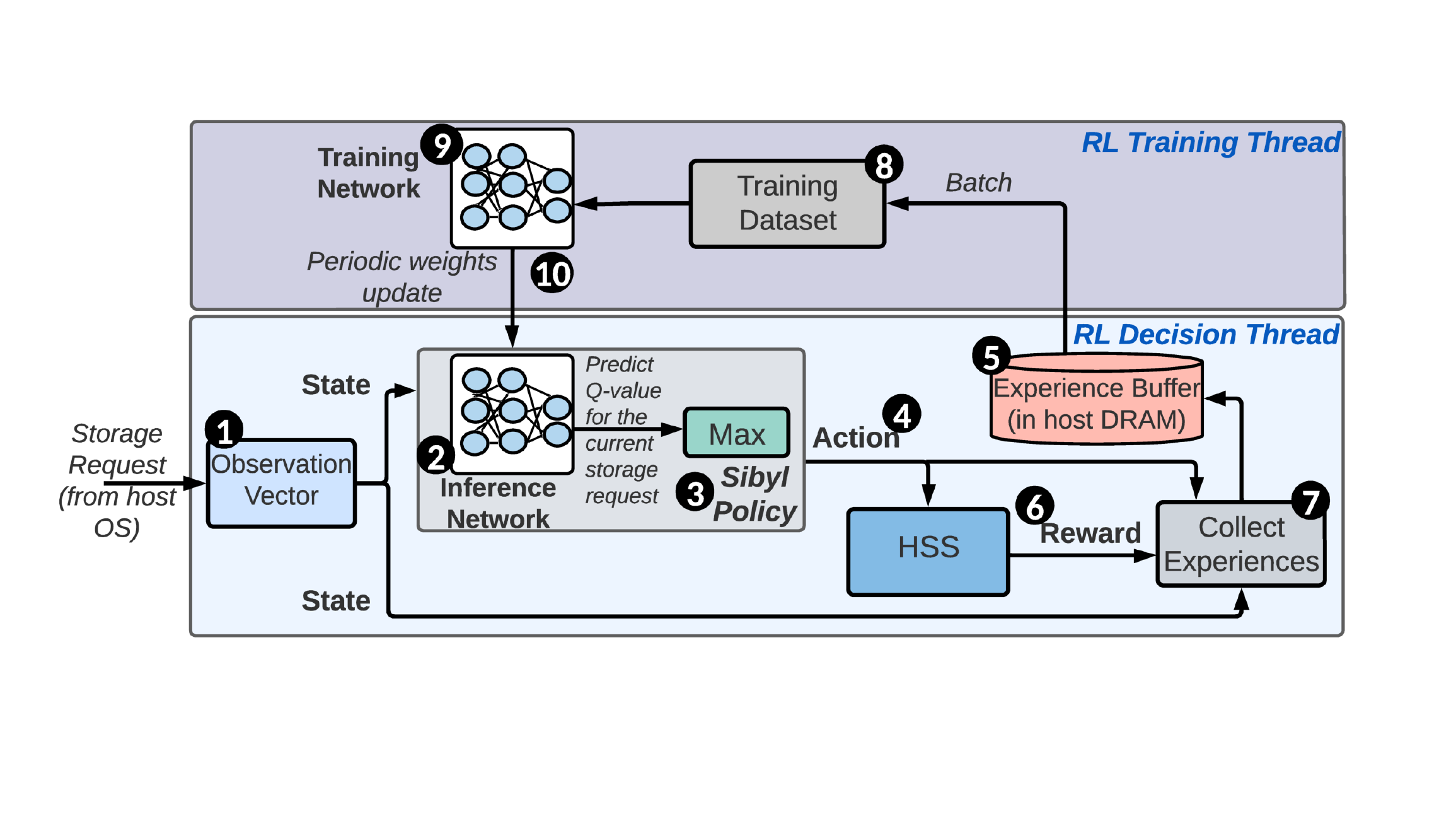}
   \vspace{-1.7cm}
   \caption{Overview of \namePaper 
  \label{fig:Sibyl/curator}}
\end{figure*}

\begin{figure}[t]
  \centering
  \includegraphics[width=0.5\linewidth,trim={0cm 1.4cm 0cm 0.2cm},clip]{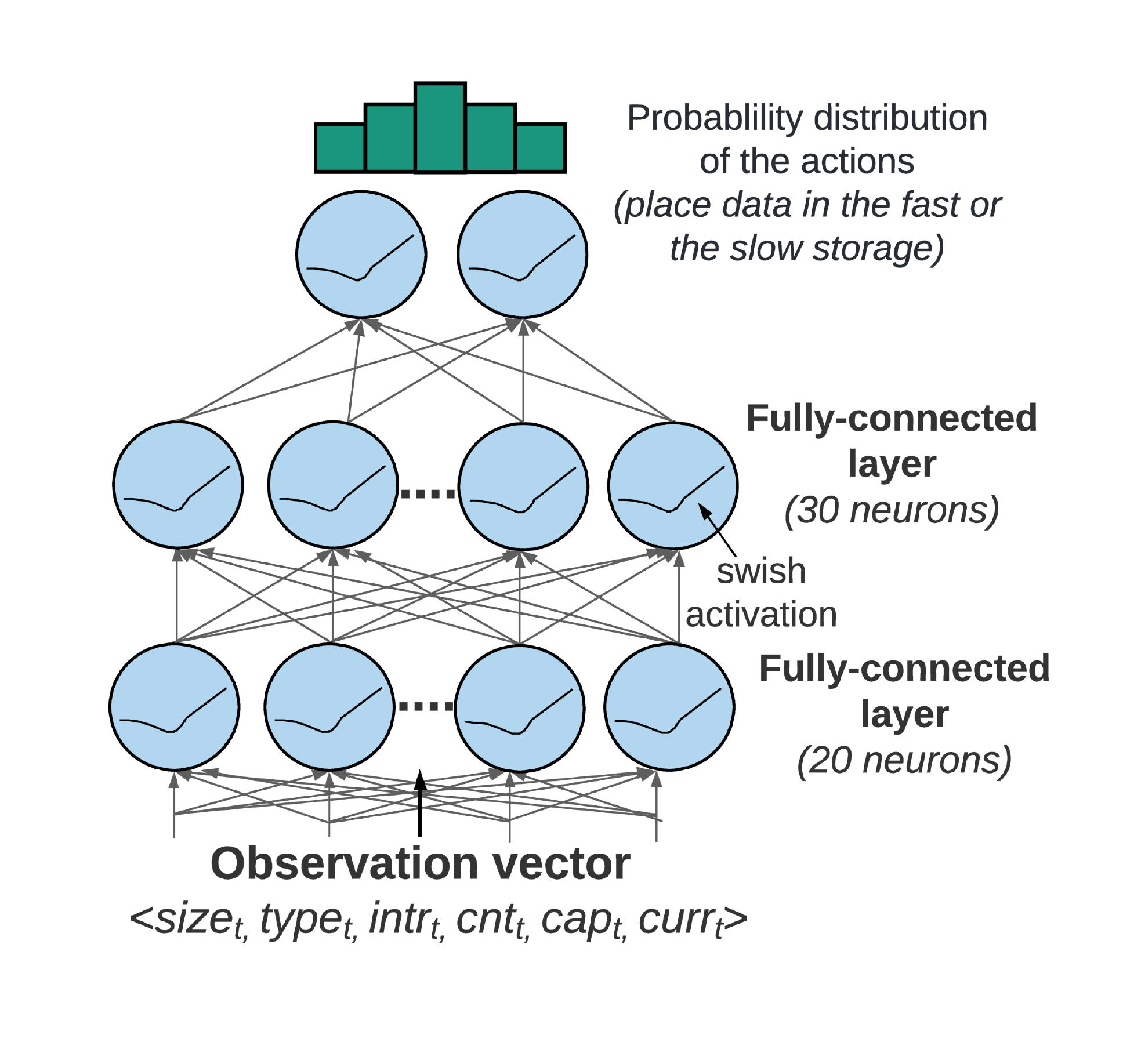}
   \vspace{-0.5cm}
   \caption{\gon{\gonz{Training}} network design using as input the \gonx{state} features from Table~\ref{tab:Sibyl/state}. \gon{The inference network \gonz{is identical except it is used only for inference}}
  \label{fig:Sibyl/model}}
\end{figure}

 \rbc{We implement \namePaper in \gon{the storage management layer of the host system's \gonz{operating system}}}. 
 Figure~\ref{fig:Sibyl/curator} shows a high-level overview of \namePaper. \gonz{Sibyl} is \rbc{composed of} two parts, \gonz{each implemented as a separate thread},  that run in parallel 
{\rbc{(1)} the \emph{RL decision thread}, \gont{where Sibyl decides the data placement \gonx{\circled{4}} of the current storage request  while collecting  \gonz{information \circled{7} about its decisions \gonx{\circled{4}} and their effects} \gonx{\circled{6}} in an \emph{experience buffer} \circled{5}, and} 
\rbc{(2)} the \emph{RL training thread}}, \gonff{where} \namePaper uses the collected \textit{experiences}\footnote{Experience is a representation of a transition from one time step to another, in terms of $\langle State, Action, Reward, Next State \rangle$.} \gonx{\circled{8}}  to \gca{ update its decision-making policy online \gonx{\circled{9}}.} \gont{Sibyl continuously learns from its past decisions \gonz{and their impact}. Our two-threaded implementation avoids that the learning (i.e., training) interrupts \gonz{or delays} data placement decisions for incoming requests. To enable the parallel execution of the two threads, we duplicate the neural network \gonz{that is} used to \gonz{make} data placement \gonz{decisions}. While one network \gonz{(called the \emph{inference network} \circled{2})} is deployed \gonx{(i.e., makes decisions)} the second network \gonz{(called the} \gonz{training} network \circled{9}), is  trained in the background. \gonx{The} inference network is used \emph{only} for inference, while \gonx{the}
training network is used \emph{only} for training.} \gonz{Therefore, \gonx{\namePaper does} \textit{not} perform a separate training step for the inference network and} \gonx{instead} periodically cop\gonx{ies} the  \gonz{\gonz{training}} network weights to the \gonz{inference} network \circleds{10}.

\gont{For every new storage request to the HSS, Sibyl uses the state information \circled{1} to \gonz{make} a data placement decision \gonx{\circled{4}}. 
The inference network  predicts the Q-value for each available action given the state information. Sibyl policy \circled{3} selects the action with the maximum Q-value \gonx{or, with a low probability, a random action for exploration} and performs the data placement. 
}

\vspace{-0.2cm}
 \subsection{\namePaper Data Placement Algorithm}
\label{subsec:Sibyl/mechanism_rl_algo}

Algorithm~\ref{algo:rl_algo} describes \gon{how} \namePaper~\gon{performs} {data placement}  \grakesh{for} an HSS. \gon{Initially, the experience buffer is allocated to hold $e_{EB}$ entries \gonz{(line~\ref{algo:intial_eb})}, and the \gonz{training} and the inference network weights are initialized to \gonz{random values } \gonz{(lines~\ref{algo:intial_exp} and \ref{algo:intial_inf})}.} \gon{When a storage request is received \gonz{(line~\ref{algo:intial_req})}}, Sibyl policy (\circled{3}  Figure~\ref{fig:Sibyl/curator}) either (1) randomly selects an action  with $\epsilon$ probability (lines~\ref{algo:exploration1}-\ref{algo:exploration2}) to perform exploration in an HSS environment,  or \gon{(2) selects \gonx{the action that maximizes the Q-value, based on information stored in the inference network}} 
(lines~\ref{algo:exploitation1}-\ref{algo:exploitation2}). After performing the \gon{selected} action \gont{(line~\ref{algo:act})}, \namePaper collects its reward\gonx{, whose value depends on whether an eviction is needed from fast storage} (lines~\ref{algo:reward_if}-\ref{algo:reward_evict}). The \gont{generated experience \gonx{is} stored} in the experience buffer (line~\ref{algo:store_exp}). \gonz{Once the experience buffer has $e_{EB}$ entries (line~\ref{algo:eb_full}), \gonx{\namePaper} trains the training network.} \gont{During training,} the training network samples a batch of experiences  from the \gor{ experience buffer} (line~\ref{algo:line:sample_experiences}) and updates its weights using stochastic gradient descent (SGD)~\cite{bottou2003stochastic} (line~\ref{algo:line:bellman_update}). \gonx{\namePaper} does \textit{not} perform a separate training step for the inference network.  \gon{Instead}, the training network weights are copied to the inference network (line~\ref{algo:weight_trans}), which removes the training
of the inference network from the critical path \gonx{of decision-making}.  
\setlength{\textfloatsep}{0.1cm}
\setlength{\floatsep}{0.00cm}
\begin{algorithm}[h]
\scriptsize
\setstretch{1}
 \caption{\namePaper's \gont{reinforcement learning-based} data placement \gon{algorithm}}
 \label{algo:rl_algo}
 \begin{algorithmic}[1]
 \State \textbf{Intialize:} \texttt{the experience buffer \textit{EB} to capacity \textit{$e_{EB}$}\label{algo:intial_eb}}
 \State \textbf{Intialize:}  \texttt{{the} {{training} network}  with random weights $\theta$\label{algo:intial_exp}}
 \State \textbf{Intialize:}  \texttt{{the} {inference} {network}  with random weights \textit{$\hat{\theta}$}\label{algo:intial_inf}}
 \State \textbf{Intialize:} \texttt{the observation vector $O_t$=$O(s_1)$ with storage request $s_1$=\{$req_{t}$\}, and host and  storage features\label{algo:intial_req}}


        \ForAll{\texttt{storage requests}     }
            \If{\texttt{(rand() $<\epsilon$)}} \Comment{\comm{with probability $\epsilon$, perform exploration}\label{algo:exploration1}}
                \State    \texttt{ random action $a_t$} \label{algo:exploration2}
            \Else \Comment{\comm{{with probability 1-$\epsilon$}{,} perform exploitation}} \label{algo:exploitation1}
                \State  \texttt{ $a_t=argmax_a Q_t(a)$}  \label{algo:exploitation2}
                \Comment{\comm{select action with the highest $Q_t$ value {from inference network}}}
            \EndIf
           \State \texttt{execute  $a_t$} \label{algo:act}
             \Comment{\comm{{place} the requested page {to}  fast or  slow storage}}
           \If{ no eviction}\label{algo:reward_if} 
                \State \texttt{$r_t \gets$  $\frac{1}{L_t}$\label{algo:reward_no_evict}} 
                \Comment{\comm{{reward, given} no eviction of a page from fast to slow storage}}
            \Else
              \State \texttt{$r_t \gets$ max(0,$\frac{1}{L_t}$-\textit{$R_p$})\label{algo:reward_evict}} 
              \Comment{\comm{{reward with} an eviction penalty in case of an eviction}}
              \EndIf
           \State \texttt{store {experience} $(O_t,a_t,r_t,O(t+1))$ in \textit{EB} \label{algo:store_exp}}
            \If{{\texttt{({num} requests \gonx{in \textit{EB}}==$e_{EB}$)}}\label{algo:eb_full}}   \Comment{\comm{train  {training} network when EB is full}}
               \State \texttt{sample random batch{es} of  experiences  from \textit{EB}, which are in format $(O_j,a_j,r_j,O(j+1))$} \label{algo:line:sample_experiences}\;
               \Comment{\comm{where $O_j$ represents an observation  at a time instant j from \textit{EB}}}
            \State \texttt{Perform stochastic gradient descent}
               \Comment{ \comm{update the {training} network weights \label{algo:line:bellman_update} }}
                  \State \texttt{$\hat{\theta} \gets \theta$ \label{algo:weight_trans}} \Comment{ \comm{{copy} the {training} network weights to the inference network }}
              \EndIf 

        \EndFor
\end{algorithmic}
   \end{algorithm}


\subsection{\gont{Detailed Design of \namePaper}}
\label{subsec:Sibyl/detail_design}

 \subsubsection{RL Decision Thread} In this thread, \namePaper~\gonx{makes} data placement decisions while storing \textit{experiences} in an experience buffer. 
 Sibyl extracts the observation vector \circled{1} from the attributes of the incoming request and the current system state (e.g., access \gonz{count}, remaining capacity in the fast storage) and uses \gonx{the} inference network \circled{2} to predict the Q-values for each \gonx{possible} action with 
 the given state vector.
 \gonz{While making data placement decisions, Sibyl balances the \gonx{random} \textit{exploration} of the environment (to \gonx{find a better} policy \gonzz{without getting stuck at a suboptimal one}) with the \textit{exploitation} of its current  policy (to \gonx{maximize its reward based on the current inference network weights}).} 

\head{Sibyl policy} For every storage request, 
Sibyl policy selects the action that leads to \gonx{the highest long-term} reward \circled{6}.
 We use \rbc{a} Categorical Deep Q-Network (also known as C51)~\cite{C51} to update  $Q(s,a)$.
C51's objective is to learn the distribution of Q-values, 
whereas other variants of Deep Q-Networks~\cite{sutton1999policy,baird1995residual,liang2016deep,mnih2013playing,silver2016mastering,silver2017mastering} aim to approximate \gonx{a single value for $Q(s,a)$}. This distribution helps \rbc{Sibyl} to capture more information from the environment 
to make \gonz{better} \gont{data placement} decisions~\cite{harrold2022data}.  

\gont{For \gonx{tracking} the state, we divide each feature into a small number of bins to reduce the state space \gonx{(see Section \ref{sec:Sibyl/rl_formulation})}, which directly affects the implementation overhead of \namePaper.} We \gonx{select the number of bins (Table~\ref{tab:Sibyl/state}) based on empirical} sensitivity analysis. 
Our state representation uses a more relaxed encoding of 40 bits \gont{(than using only 20 bits for the \gonx{observation vector})} 
to allow for future extensions (e.g., features with more bins). 
Similarly,  we use a relaxed 4\gonzz{-bit} encoding for the action to allow extensibility to a different number of storage devices. \gont{For the reward structure, we use a half-precision floating-point (16-bit) representation. }

\head{Experience buffer}  
 \gonx{\namePaper} stores \emph{experiences} \gonx{it} collect\gonx{s} while interacting with the HSS in an  \emph{experience buffer}~\cite{dqn}. The experience buffer is allocated in  the host \gonz{main memory (DRAM)}. 
To minimize \gonx{its} design overhead, we 
deduplicate data in the stored experiences. 
  To improve the training quality, we \gonx{perform batch training where each batch consists of randomly sampled experiences}. \gonzz{This technique of randomly sampling experiences from the experience buffer} is called  \textit{experience replay}~\cite{dqn}. 

\begin{figure}[h]
\centering
  \includegraphics[width=0.5\linewidth,trim={0cm 0.1cm 0cm 0cm},clip]{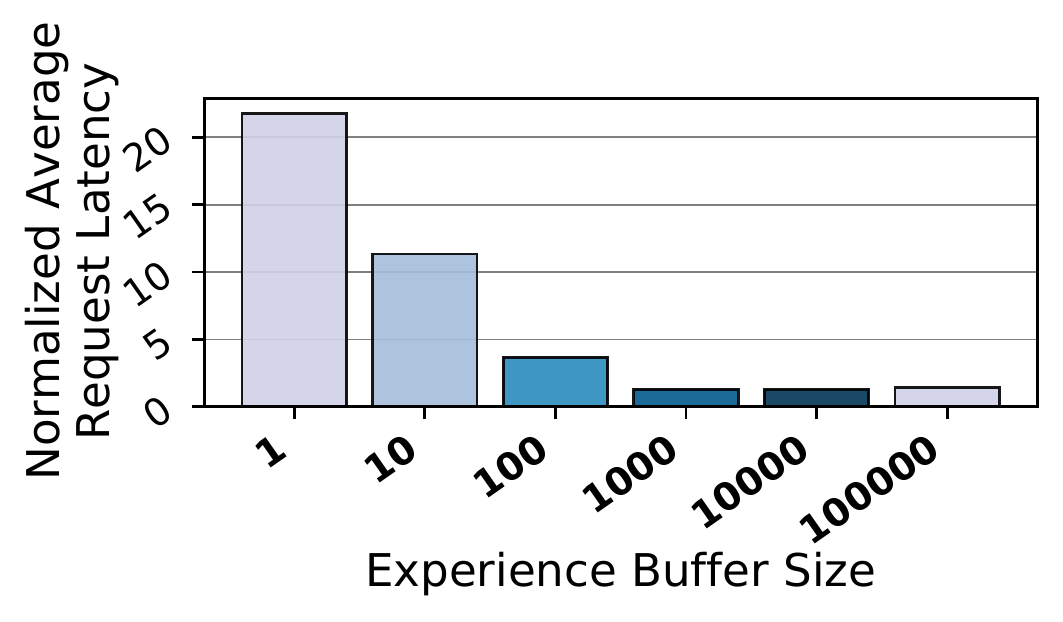}
  \vspace{-0.1cm}
\caption[Two numerical solutions]{Effect \gonx{of different experience buffer sizes on} the \gont{average request} latency (normalized to \fast) \label{fig:Sibyl/training_f}}
\end{figure}

Figure~\ref{fig:Sibyl/training_f} shows the effect \rbc{of} different \rbc{experience} buffer sizes \gonx{on \namePaper's performance} in the \hmssd configuration. 
We observe that \namePaper's performance saturates \gonx{at 1000 entries, which we select as \gonzz{the experience buffer} size}. 
{Since the size of our state representation is 40 bits, to store a single experience tuple, 
we  need 40-bit+\gonzz{4}-bit+16-bit+40-bit, i.e., 100 bits. 
\gont{In total, for 1000 experiences, the \gor{experience buffer} requires 100 KiB in the host DRAM.}}

\head{Exploration vs. exploitation}  An \gonx{RL} agent needs to \emph{explore} the environment to improve its policy 
\rbc{to maximize \gonx{its} long-term reward beyond  local maxima~\cite{sutton_2018}}.
At the same time, the agent \gonx{needs} to \emph{exploit} what it has already \emph{experienced} \gonx{so that it can take advantage of its learning so far}.   
\rbc{To balance exploration and exploitation,}
we use the $\epsilon$-greedy policy~\cite{tokic2011value}: the best-known action based on the agent's experience is selected with (1-$\epsilon$) probability, and \rbc{otherwise}, i.e., with $\epsilon$ probability, \rbc{another} action is chosen randomly. \gonx{Exploration allows \namePaper to experience states it may not otherwise get into~\cite{sutton_2018} and thus avoid missing higher long-term rewards.} 
To perform exploration, \namePaper randomly chooses to \gonx{place data} \gonx{to} the fast or the slow storage \juancr{device}, \gonx{so that it can} get more information about the HSS and the workload. Based on the received reward, \namePaper updates its \gonx{training network}. 
  \gonx{Such exploration} helps \namePaper to avoid making suboptimal data placement decisions \gonx{in the long run.}

\subsubsection{RL Training {Thread}}  
 \gont{This thread uses a} batch \rbc{of collected experiences} \circled{8} from the experience buffer \gonx{to train the training} network \circled{9}. 
The updated weights \gonx{of the training network} are transferred to the inference network after every 1000 requests \circleds{10}. 

\head{\gonz{Training} and inference \rbc{networks}} 
The training an\gonzz{d} inference network allows \gonzz{the} parallel execution of \gonx{decision} and training threads.
We use an identical neural network structure for the training and inference networks.
\gup{A \rbc{deep neural network can be} prohibitive due to the long time \gonx{it requires for} train\gonx{ing} and converge\gonx{nce}, preventing \namePaper to  adapt to new state-action pairs \gonx{in a timely manner}. Based on experiments, we find that a simple feed-forward network~\cite{bebis1994feed} with only two hidden layers~\cite{de1993backpropagation} \gonx{provides} good \gonzz{performance} 
for \gonx{\namePaper's} data placement task.} Figure~\ref{fig:Sibyl/model} shows the structure of our \rbc{\gonz{training}} network.\footnote{\rbc{The inference network is identical in shape to the \gonz{training} network.}} The network takes the observation vector $O_t$ as \gonx{its} input and produces a probability distribution of \rbc{Q-values} as \gonx{its} output. Before feeding the data to the network, we preprocess the data by normalizing and casting the data to low precision data types, which allows us to reduce memory in the experience buffer. 
Next, we apply two fully-connected \juangg{hidden} layers of 20 and 30 neurons, respectively. 
{We select these neurons based on our extensive design space exploration with different numbers of hidden layers and neurons per layer.} After the two hidden layers, we have an output layer of 2 neurons, one for each action. \gonx{Sibyl policy \circled{3} selects the action with the maximum Q-value.} All fully-connected layers use the swish {activation} function~\cite{ramachandran2017searching},  a non-monotonic function that
 outperforms ReLU~\cite{agarap2018relu}. 
 
 \gca{During the training of the \gonz{training} network, \gonzz{the} inference network's weights are fixed.}  \gont{ After every 1000 requests, \gonzz{the} weights of the \gonz{training} network are copied to the inference network, 
which removes the training of the inference network from the critical path. We set the number of requests to 1000 based on our empirical evaluation of the experience buffer size \gonx{(Figure~\ref{fig:Sibyl/training_f})}. Each training step is composed of \gont{8 batches of experiences from an experience buffer of 1000 experiences with a batch size of 128}. We perform the training on the host CPU rather than on \gonx{a dedicated hardware accelerator} because (1) the network size is small and the weights perfectly fit in on-chip caches of the CPU in our evaluated system, \gonzz{and}  (2) to avoid continuous weight transfer overhead between the host CPU and the \gonx{accelerator} over the external interface. }

\sloppy\head{Hyper-parameter tuning}
\label{subsec:Sibyl/lhs}
\gont{We improve \namePaper's accuracy by tuning its hyper-parameters.  Hyper-parameters are sets of RL algorithm variables that can be tuned to optimize the accuracy of the RL agent~\cite{paine2020hyperparameter,NAPEL}.} 
{For hyper-parameter tuning, we perform 
cross-validation~\cite{arlot2010survey} 
using different hyper-parameter values. During cross-validation, we randomly select one workload for hyper-parameter tuning and  \gonx{use the} other \gont{thirteen} workloads for validation. On the selected workload, we use different hyper-parameter configurations that we choose using \gonzz{the} \mbox{\textit{design of experiments}} (DoE)~\cite{montgomery2017design}. }DoE allows us  to minimize the number of experiments needed
{ to find the \gonx{best} hyper-parameter \grakesh{values}} without sacrificing the quality of the information gathered by the experiments. \gont{Unlike traditional supervised learning methods, we do \emph{not} train Sibyl \emph{offline} using a training dataset before deploying it for data placement. \gonx{All training happens online in \namePaper.} For every evaluated workload, Sibyl starts with no prior knowledge and gradually learns to make data placement decisions online by interacting with the hybrid storage system.  \namePaper  needs only one-time \gonx{offline} hyper-parameter tuning.}

Table~\ref{tab:Sibyl/hyperparameter} shows the hyper-parameters considered \gonx{in} \namePaper's design \gonx{as well as their chosen values after the tuning process.  
The discount  factor ($\gamma$) \gonx{determines the balance between} the immediate and future rewards. \gonx{At $\gamma$=0 ($\gamma$=1), \namePaper gives importance only to the immediate (long-term) reward.} The learning rate ($\alpha$) determines the rate at which neural network weights are updated. \gonzz{A lower $\alpha$ makes small updates to the neural network weights, which could take more training iterations to converge to an optimal policy. While a higher $\alpha$ results in large updates to the neural network weights, which could cause the model to converge too quickly to a suboptimal solution.} 
The exploration rate ($\epsilon$) balances exploration and exploitation for \gonx{\namePaper}. We also explore different batch sizes \gonx{(i.e., the number of samples processed in each training iteration)} and experience buffer sizes to train our \gonz{training} network.} 
\begin{table}[h]
\small
\begin{center}
 \caption{Hyper-parameters considered \gon{for} tuning }
    \label{tab:Sibyl/hyperparameter}
     \renewcommand{\arraystretch}{0.8}
\setlength{\tabcolsep}{10pt}
  \resizebox{1\linewidth}{!}{%
\begin{tabular}{l||l|l}
\hline
\textbf{Hyper-parameter}             & \textbf{Design Space} & \textbf{Chosen Value} \\  \hline

Discount factor   ($\gamma$)                           & 0-1  &0.9   \\
Learning rate     ($\alpha$)                  & $1e^{-5}-1e^{0}$ & $1e^{-4}$     \\
Exploration rate  ($\epsilon$)                           & 0-1  &0.001   \\
Batch size                          & 64-256   & 128     \\
\gonx{Experience buffer size ($e_{EB}$)}                         & 10-10000 & 1000      \\
\hline
\end{tabular}
}
\end{center}
\end{table}

\section{Evaluation Methodology} \label{sec:Sibyl/evaluation}
\head{Evaluation setup}
{
We evaluate  \namePaper~\gon{using}  real \gonf{systems with various HSS}   configurations.
The \gonf{HSS devices} \gon{appear} as a single flat {block device \rcamfix{that exposes} one contiguous logical block address space \gon{to the OS}, as depicted in \fig{\ref{fig:Sibyl/hybrid}}}. { We implement a lightweight custom block driver interface that manages the I/O requests to storage devices.  \tab{\ref{tab:Sibyl/devices}} {provides our system details, including the characteristics of the \gon{three}  storage devices \gon{we use}.}  
}  {To analyze the {sensitivity of our approach to} different device characteristics, we evaluate two different hybrid storage configurations (1) \em{performance-oriented HSS}: high-end device (\textsf{H})~\cite{inteloptane} and middle-end device (\textsf{M})~\cite{intels4510}, and (2) \em{cost-oriented HSS}: high-end device  (\textsf{H})~\cite{inteloptane} and low-end device (\textsf{L}) ~\cite{seagate}. 
\gca{We also evaluate two tri-hybrid \gon{HSS} configurations \gon{consisting of (1)}  \thold~and  \gon{(2)} \thnew~ devices.  }
{We \gon{run the} Linux Mint 20.1  operating system~\cite{linuxmint} with the  Ext3
file system~\cite{tweedie1998journaling}}. We use \gon{the} TF-Agents API~\cite{TFagents} to develop \namePaper. \gon{We evaluate \namePaper using two different metrics: \mbox{(1) \emph{average}} \emph{request latency}\rcamfix{, i.e.,} average \rcamfix{of the latencies of all storage \mbox{read/write} requests in a workload}, \gonzz{and} (2) \emph{request throughput (IOPS)}\rcamfix{, i.e., throughput of all storage requests in a workload in terms of completed I/O operations per second}. 
} 
}}
\begin{table}[h]
\begin{center}
\small
\setstretch{0.9}
\vspace{5pt}
\caption{{Host system and \gca{storage devices used in} hybrid storage configurations }}
 \label{tab:Sibyl/devices}
  \renewcommand{\arraystretch}{1} 
\setlength{\tabcolsep}{2pt} 
 \resizebox{1.0\columnwidth}{!}{%
\begin{tabular}{lll}
\hline
\multicolumn{1}{l|}{\textbf{Host System}}                  & 
\multicolumn{2}{l}{\begin{tabular}[c]{@{}l@{}}AMD Ryzen 7 2700G~\gon{\cite{amdryzen}}, 8-core\gon{s}@3.5 GHz, \\ 8$\times$64/32 KiB L1-I/D, 4 MiB L2, 8 MiB L3,  \\16 GiB RDIMM DDR4 2666 MHz
\end{tabular}}  
\\ \hline
\multicolumn{1}{l|}{\textbf{Storage \rcamfix{Devices}}}                  & \multicolumn{2}{l}{\textbf{Characteristics}}                                                                                                                                        \\ \hline
\multicolumn{1}{l|}{\textsf{H}: Intel Optane SSD P4800X~\cite{inteloptane}}     & \multicolumn{2}{l}{\begin{tabular}[c]{@{}l@{}}375 GB, PCIe 3.0 NVMe, SLC, R/W: 2.4/2 GB/s,\\ random R/W: 550000/500000 IOPS\end{tabular}} \\ \hline

\multicolumn{1}{l|}{\textsf{M}: Intel SSD D3-S4510~\cite{intels4510}} & \multicolumn{2}{l}{\begin{tabular}[c]{@{}l@{}}1.92 TB, SATA TLC (3D), R/W: 550/510 MB/s, \\ random R/W: 895000/21000 IOPS\end{tabular}}           \\ \hline
\multicolumn{1}{l|}{\textsf{L}: Seagate \ghpca{HDD} ST1000DM010~\cite{seagate} } & \multicolumn{2}{l}{\begin{tabular}[c]{@{}l@{}}1 TB, SATA 6Gb/s 7200 RPM \\ \rakesh{Max. Sustained Transfer Rate: 210 MB/s}\end{tabular}}   \\      \hline
\multicolumn{1}{l|}{\textsf{L$_{SSD}$}: ADATA SU630 SSD ~\cite{adatasu630} } & \multicolumn{2}{l}{\begin{tabular}[c]{@{}l@{}}960 GB, SATA 6 Gb/s, TLC,   \\ Max R/W: 520/450 MB/s\end{tabular}}      
\\ \hline                          \multicolumn{1}{l|}{\textbf{HSS \gon{Configurations}}}              & \multicolumn{1}{l|}{\textbf{Fast \ghpca{Device}}  }                                                            & \textbf{Slow \ghpca{Device}}                                                             \\ \hline
\multicolumn{1}{l|}{\textsf{H\&M} (Performance-oriented) }                       & \multicolumn{1}{l|}{high-end (\textsf{H})}                                                                   & middle-end (\textsf{M})                                                                \\
\multicolumn{1}{l|}{\textsf{H\&L} (Cost-oriented) }                          & \multicolumn{1}{l|}{high-end (\textsf{H})}                                                                   & low-end (\textsf{L})                                                                   \\ \hline                                           
\end{tabular}
}
\end{center}
\end{table}

{{\head{Baselines}} \label{subsec:Sibyl/eval:baselines} {We compare \namePaper against \rcamfix{two } state-of-the-art heuristic-based \gon{HSS} data placement techniques, (1) cold data eviction (\cde) \cite{matsui2017design} and (2) history-based \gon{page selection} (\hps)~\cite{meswani2015heterogeneous}, \gon{(3)} a state-of-the-art supervised learning-based technique (\arcc)~\cite{ren2019archivist}, \gon{and (4)} a recurrent neural network (RNN)-based data placement technique (\kleio), adapted from \rakeshisca{Kleio~\cite{doudali2019kleio}, a data placement technique for hybrid memory systems}. \gup{\kleio 
provides a state-of-the-art ML-based data placement \gon{baseline}.} 
We compare the above policies with three extreme \gon{baselines}:
(1) \slow{}, {where all data resides in \gon{the} slow \gon{storage (i.e., there is no fast storage)}}, (2) \fast, {where all data resides in \gon{the}  fast \gon{storage}}, {and (3) \textsf{Oracle}}
\cite{meswani2015heterogeneous}, which exploits \gon{complete} {knowledge of} future I/O-access patterns  
{to perform data placement and to }\js{select victim data blocks for eviction from} the fast device.

{\head{Workloads}} 
We use fourteen different \js{block-I/O traces} from {the} MSRC benchmark suite~\cite{MSR} that are collected from real enterprise server workloads. 
{We carefully select the fourteen traces to have {distinct} I/O-access patterns, as shown in \tab{\ref{tab:Sibyl/workload}}, in order to study a diverse set of workloads with different \rcamfix{randomness} and  \rcamfix{hotness} \gon{properties} (see Figure~\ref{fig:Sibyl/apps}). 
We quantify a workload’s randomness using the average request size of the workload; the higher (lower) the average request size, the more sequential (random) the workload. \gon{The average access \rcamfix{\gonz{count}} provides the average \rcamfix{of the access counts of all pages in a workload;}
the higher (lower) the average access \rcamfix{\gonz{count}}, the hotter (colder) the workload.  
\rcamfix{\tab{\ref{tab:Sibyl/workload}} also  shows}  \rcamfix{the number of} unique requests in a workload.} To demonstrate {\namePaper}'s ability to generalize and provide performance gains across \gon{\emph{unseen traces}}, \gon{i.e., traces} that are \emph{not} used to tune the \gon{hyper-parameters of} \namePaper, we evaluate {\namePaper} using \gon{four} additional workloads from FileBench~\cite{tarasov2016filebench}.} 

\begin{table}[h]
\caption{Characteristics of 14 evaluated workloads}
 \label{tab:Sibyl/workload}
 \centering
 \tiny
 \setstretch{0.85}
   \renewcommand{\arraystretch}{0.85} 
\setlength{\tabcolsep}{2pt} 
\resizebox{0.9\linewidth}{!}{%
\begin{tabular}{@{}lrrrrr@{}}
\hline
 \textbf{Workload}        & \multicolumn{1}{l}{\textbf{Write}} & \multicolumn{1}{l}{\textbf{Read}} & \multicolumn{1}{r}{\textbf{Avg. request}} & \multicolumn{1}{r}{\textbf{Avg. access}} & \textbf{No. of unique}  \\ 
                &         \%                  &    \%                      & \multicolumn{1}{r}{\textbf{size}}          & \multicolumn{1}{r}{\textbf{\gonz{count}}} & \textbf{requests} \\ 
         
         \hline
$hm\_1$    & 4.7\%                     & 95.3\%                   & 15.2                                  & 44.5                                  & 6265                     \\

$mds\_0$    & 88.1\%                     & 11.9\%                   & 9.6                                  & 3.5                                  &        31933              \\
$prn\_1$   & 24.7\%                    & 75.3\%                   & 20.0                                  & 2.6                                  & 6891                     \\
$proj\_0$  & 87.5\%                    & 12.5\%                   & 38.0                                  & 48.3                                 & 1381                     \\
$proj\_2$  & 12.4\%                    & 87.6\%                   & 42.4                                 & 2.9                                  & 27967                    \\
$proj\_3$  & 5.2\%                     & 94.8\%                   & 9.6                                  & 3.6                                  & 19397                    \\
$prxy\_0$  & 96.9\%                    & 3.1\%                    & 7.2                                  & 95.7                                 & 525                      \\
$prxy\_1$  & 34.5\%                    & 65.5\%                   & 12.8                                  & 150.1                                 & 6845                     \\
$rsrch\_0$ & 90.7\%                    & 9.3\%                    & 9.2                                 & 34.7                               & 5504                     \\
$src1\_0$  & 43.6\%                    & 56.4\%                   & 43.2                                 & 12.7                                  & 13640                    \\
$stg\_1$   & 36.3\%                    & 63.7\%                   & 40.8                                 & 1.1                                  & 3787                     \\
$usr\_0$   & 59.6\%                    & 40.4\%                   & 22.8                                  & 19.7                                  & 2138                     \\
$wdev\_2$  & 99.9\%                    & 0.1\%                    & 8.0                                  & 17.7                                 & 4270                     \\
$web\_1$   & 45.9\%                    & 54.1\%                   & 29.6                                  & 1.2                                  & 6095                     \\ \hline
\end{tabular}
}
\end{table}


 \begin{figure*}[t]
\centering
\begin{subfigure}[h]{.85\textwidth}
  \centering
  \includegraphics[width=\linewidth,trim={0.2cm 0.2cm 0cm 0.2cm},clip]{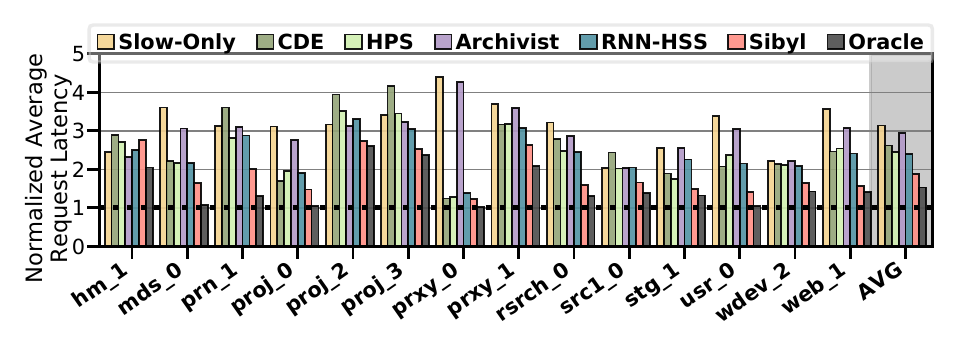}
  \vspace{-0.6cm}
   \caption{\textsf{H\&M} \gon{HSS} configuration\label{fig:Sibyl/perf_1}}
\end{subfigure}%

\begin{subfigure}[h]{.85\textwidth}
  \centering
  \includegraphics[width=\linewidth,trim={0.2cm 0.2cm 0.2cm 0.2cm},clip]{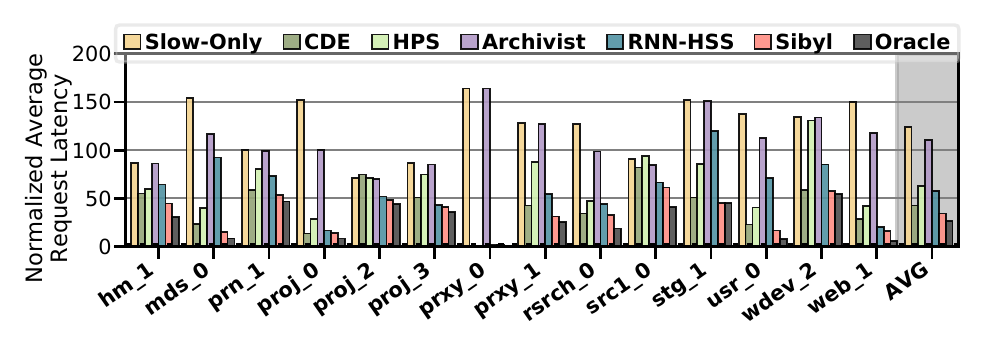}
    \vspace{-0.6cm}
   \caption{\textsf{H\&L} \gon{HSS} configuration \label{fig:Sibyl/perf_2}}
\end{subfigure}
\caption[Two numerical solutions]{\gon{Average request latency} under two different hybrid storage configurations (normalized to \fast
)
\label{fig:Sibyl/perf}}
\end{figure*}
 
\begin{figure*}[h]
\centering
\begin{subfigure}[h]{.85\textwidth}
  \centering
  \includegraphics[width=\linewidth,trim={0.2cm 0.2cm 0cm 0.2cm},clip]{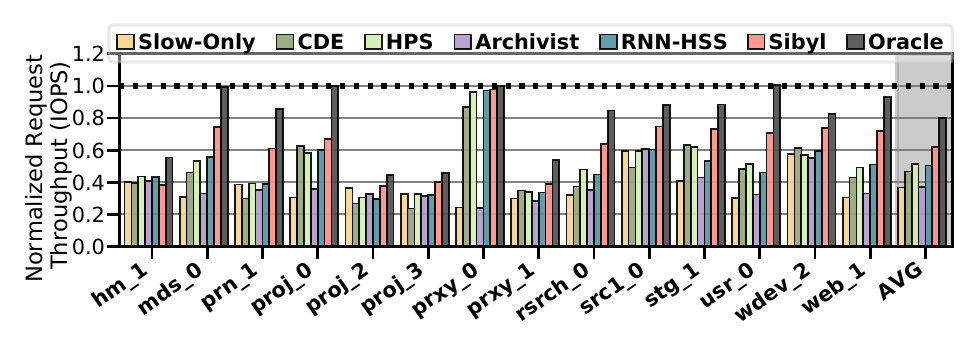}
  \vspace{-0.6cm}
\caption{\textsf{H\&M} \gon{HSS} configuration 
  \label{fig:Sibyl/iops_1}}
\end{subfigure}%

\begin{subfigure}[h]{.85\textwidth}
  \centering
  \vspace{-0.04cm}
  \includegraphics[width=\linewidth,trim={0.2cm 0.2cm 0.2cm 0.2cm}, clip]{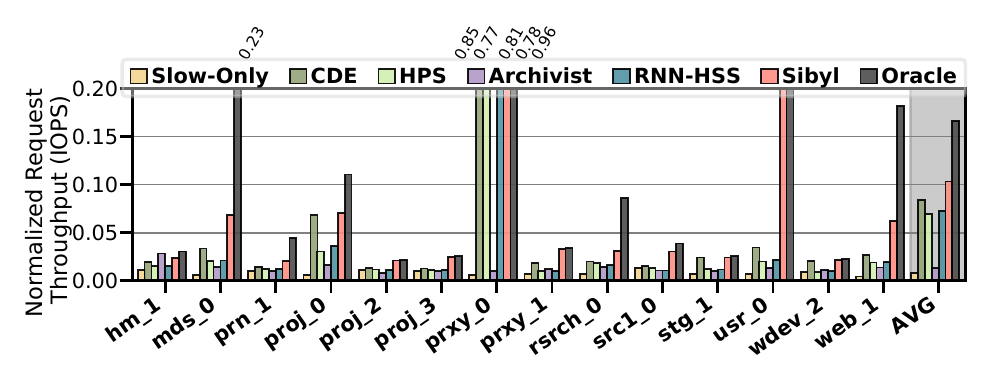}
  \vspace{-0.6cm}
  \caption{\textsf{H\&L} \gon{HSS} configuration 
  \label{fig:Sibyl/iops_2}}
\end{subfigure}
\caption{\gon{Request throughput (IOPS)} under two different hybrid storage configurations (normalized to \fast
)
\label{fig:Sibyl/iops}}
\end{figure*}

\section{Results}
\label{sec:Sibyl/results}
 \subsection{Performance Analysis}
 \label{subsection:Sibyl/Evaluation_perf}
Figure~\ref{fig:Sibyl/perf} compares the average \gon{request} latency of \namePaper against the baseline policies \gon{for \hmssd (Figure~\ref{fig:Sibyl/perf}(a)) and \hlssd (Figure~\ref{fig:Sibyl/perf}(b)) HSS configurations}. \js{All values are} normalized to \js{\fast}. 
We make \rcamfix{five major} observations. 
First, \namePaper consistently outperforms all the baselines for all the workloads in \hlssd and all but two workloads in \hmssd. \gon{In the} \js{\textsf{H\&M}} \gon{HSS} configuration (Figure~\ref{fig:Sibyl/perf}(a)), {where the latency difference between  two  devices  is  relatively  smaller than \textsf{H\&L},} \namePaper improves \gon{average} performance by {28.1\%, 23.2\%, 36.1\%, and 21.6\%} 
\gonzz{over} \cde, \hps, \arcc, and \kleio, respectively. In \gon{the} \js{\textsf{H\&L}} \gon{HSS} configuration (Figure~\ref{fig:Sibyl/perf}(b)), where there is a large difference between the latencies of the two {storage devices}, \namePaper~improves performance by 19.9\%, 45.9\%, 68.8\%, and 34.1\% 
\gonzz{over} \cde, \hps, \arcc and \kleio, respectively.  } 
\gor{
\rcamfix{We observe that the} larger the latency gap between HSS devices, the higher the expected benefits {of avoiding} the eviction penalty by placing only performance-critical pages in the fast storage.
Second, in \rcamfix{the} \textsf{H\&M} HSS configuration, \gon{\cde and \hps are ineffective for certain workloads (\textsf{hm\_1}, \textsf{prn\_1}, \textsf{proj\_2}, \textsf{proj\_3}, and \textsf{src1\_0}) even when compared to \slow.} 
\rcamfix{In contrast,} ~\namePaper~\gon{consistently \rcamfix{and significantly} outperforms \slow for all workloads because it} can learn the small latency difference between \js{the two storage devices in \textsf{H\&M}} and dynamically \gonzz{adapts its} data placement decisions, which is \rcamfix{difficult} \js{for \cde and \hps due to their inability to holistically take into account the underlying device characteristics.} \gsa{Third, \namePaper provides slightly lower performance than \gon{other} baselines in only two workloads: \slow, \hps, \arcc, and \kleio for \hmone and \cde and \hps for \proxy in the \hmssd HSS configuration. We observe that such workloads are write-intensive and have many random requests (in terms of \gon{both access pattern}  and request size). Therefore,  such  workload\gon{s} would benefit from  more frequent retraining of Sibyl’s \gonzz{training} network. We experimentally  show in Section \ref{subsection:Sibyl/mixed_workload} that using a lower learning rate \gonzz{during} the training of the \gonzz{training} network helps to improve \rcamfix{\namePaper's} performance for such workloads. \gon{ Fourth, \namePaper achieves, on average, 80\% of the performance of the \oracle, \rcamfix{which} has complete knowledge of future access patterns, \gont{across \hmssd and \hlssd}. Fifth, \kleio provides higher performance than heuristic-based policies (2.1\% and 8.9\% than \cde and \hps, respectively, in \hmssd and 9.8\% than \hps in \hlssd), \rcamfix{but \namePaper outperforms it by \gonz{27.9\%}}.
Unlike \namePaper, \rcamfix{the two machine learning-based policies,} \arcc and \kleio, do \emph{not} consider any system-level feedback, which leads to \rcamfix{their} suboptimal performance. }}

Figure~\ref{fig:Sibyl/iops} \gon{compares the request throughput (IOPS)} of {\namePaper} against other {baseline} policies. We make two observations. 
\rcamfix{First, in the \textsf{H\&M} ({\textsf{H\&L}}) HSS configuration (Figure~\ref{fig:Sibyl/iops}), {\namePaper} improves throughput by $32.6\%$ ($22.8\%$), $21.9\%$ ($49.1\%$), $54.2\%$ ($86.9\%$), and $22.7\%$ ($41.9\%$) \gonzz{over} {\cde}, {\hps}, {\arcc}, and {\kleio}, respectively.}
Second, \namePaper provides slightly lower performance than \slow, \cde, \hps, \arcc, and \kleio for \rcamfix{only} \hmone in \hmssd HSS configuration. We draw similar observations \rcamfix{for throughput results} as \rcamfix{we did for latency results} (Figure~\ref{fig:Sibyl/perf}) because as \rcamfix{\namePaper} consider\rcamfix{s} the request size in state features and \gonz{request} latency in the reward, \rcamfix{it} {also} indirectly captures throughput (size/latency).

\gon{ We conclude that \namePaper 
consistently provides higher performance than \rcamfix{all five} baselines and \rcamfix{significantly} improves both average request latency and request throughput.}}

\subsection{{Performance on Unseen Workloads}}
\label{subsection:Sibyl/perf_unseen}
To demonstrate {\namePaper}'s ability to generalize and provide performance gains across \textit{unseen} \rcamfix{workloads} that are \textit{not} used to tune the \rcamfix{hyper-parameters of the} data placement policy of \namePaper, we evaluate {\namePaper} using \gonzz{four} additional  workloads from FileBench~\cite{tarasov2016filebench}. \gon{No data placement policy \rcamfix{we evaluate}, including \namePaper, \rcamfix{is} tuned on these workloads. } Figure~\ref{fig:Sibyl/unseen} shows the performance of \rcamfix{these} unseen \rcamfix{workloads}. 
We observe the following observations. First, in \textsf{H\&M} (\textsf{H\&L}) \gon{HSS} configuration, {\namePaper}~outperforms {\kleio} and {\arcc}  by $46.1\%$ ($54.6\%$) and $8.5\%$ ($44.1\%$), respectively. Second, \mbox{\namePaper} may misplace some pages during the \rcamfix{online} adaptation period, but it provides significant performance benefit\rcamfix{s} over existing ML-based data placement techniques.} 
We conclude that {\namePaper} \gon{
provides high performance benefits on unseen \rcamfix{workloads} for which it has not been tuned.}

\begin{figure}[h]
    \centering
 \includegraphics[width=0.9\linewidth]{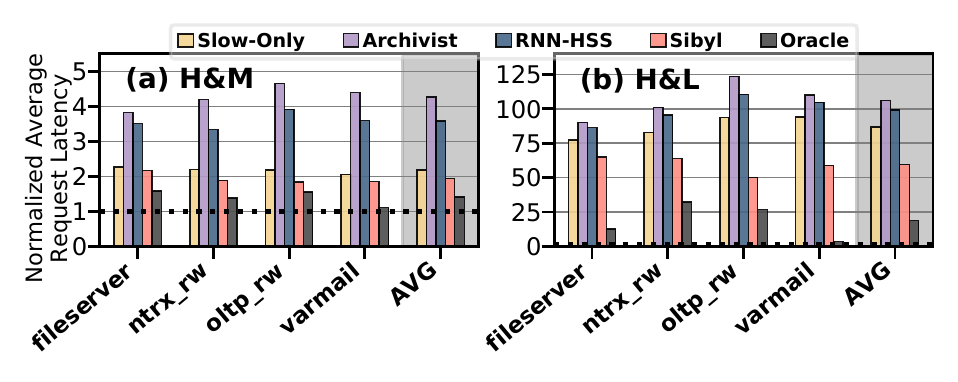}
    \caption{{\gon{Average request} latency on unseen workloads (normalized to \fast) under two HSS configurations}}
    \label{fig:Sibyl/unseen}
\end{figure}

\subsection{Performance on Mixed Workloads}
\label{subsection:Sibyl/mixed_workload}
We evaluate mixing  two or more workloads at the same time while \rcamfix{randomly varying} their relative start times.  Table~\ref{tab:Sibyl/mixed_workloads} describes the characteristics of these mixed workloads. These workloads are truly independent of each other, potentially creating more evictions from the fast storage device than a single workload.  Such a scenario \mbox{(1) leads} to unpredictable execution where requests arrive at different, unpredictable timesteps, (2) mimics distributed workloads, and (3) further tests the \gon{ability} of Sibyl to dynamically adapt its decision-making policy.

\begin{figure}[h]
    \centering
 \includegraphics[width=1\linewidth]{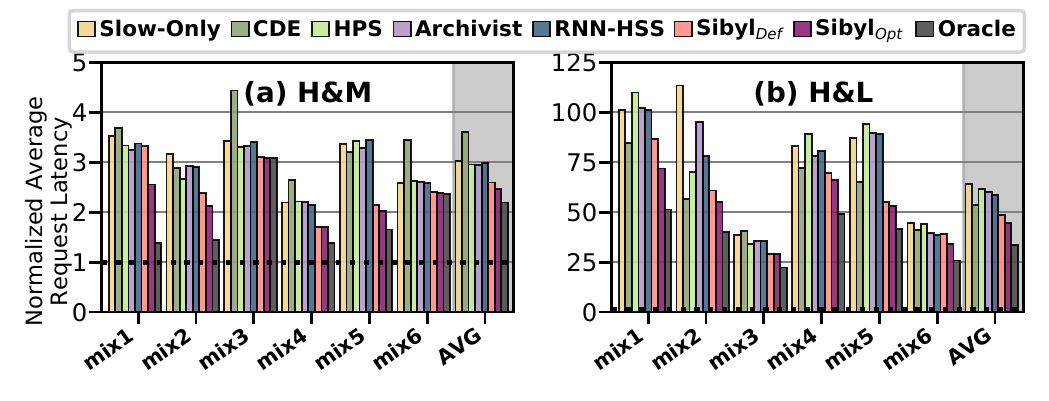}    
 \caption{{\gon{Average request} latency on mixed workloads (normalized to \fast) \rcamfix{and two HSS configurations}}}
  
    \label{fig:Sibyl/mixedWorkloads}
\end{figure}

Figure~\ref{fig:Sibyl/mixedWorkloads} shows \gonzz{average request latency} for mixed workloads. We use two different settings for \namePaper: (a) \sibyldef, where we use our default hyper-parameters (Section \ref{subsec:Sibyl/lhs}), and (b) \sibylopt, where we optimize the hyper-parameters \rcamfix{for} these mixed workloads and use a lower learning rate ($\alpha$) of $1e^{-5}$. A lower learning rate performs small\rcamfix{er} updates to the \gonz{training}  network’s weights in each training \rncam{iteration}, thus requiring more training to converge to an optim\gon{al} solution.
\gsa{{
\begin{table}[h]
 \caption{Characteristics of mixed workloads}
    \label{tab:Sibyl/mixed_workloads}
\centering
\normalsize
 \setstretch{0.8}
   \renewcommand{\arraystretch}{0.8} 
\setlength{\tabcolsep}{2pt}
  \resizebox{0.8\linewidth}{!}{%
\begin{tabular}{c||c|l}
\hline
\textbf{\rcamfix{Mix}}             & \textbf{\rcamfix{Workloads}} & \textbf{Description} \\ 
\hline
\hline
\textbf{mix1}                     &  $prxy\_0$~\cite{MSR} and $ntrx\_rw$~\cite{tarasov2016filebench} & \begin{tabular}{l}    
 Both $prxy\_0$ and \\ $ntrx\_rw$ are write-intensive
\end{tabular}
    \\
\hline
\textbf{mix2}                      &  $rsrch\_0$~\cite{MSR} and $oltp\_rw$~\cite{tarasov2016filebench}  &
\begin{tabular}{l} 
$rsrch\_0$ is  write-intensive and \\ $oltp\_rw$ is read-intensive
\end{tabular}
\\
\hline
\textbf{mix3}                      &  $proj\_3$~\cite{MSR} and $YCSB\_C$~\cite{cooper2010benchmarking}& 
\begin{tabular}{l}
Both $proj\_3$ and \\ $YCSB\_C$ are read-intensive
\end{tabular}
\\
\hline
\textbf{mix4}                      & $src1\_0$~\cite{MSR} and fileserver~\cite{tarasov2016filebench} &
\begin{tabular}{l}
Both $src1\_0$ and \\ fileserver have nearly equal \\ numbers of reads and writes
\end{tabular} \\
\hline
\textbf{mix5}                     & 
\begin{tabular}{l}
$prxy\_0$~\cite{MSR}, $oltp\_rw$~\cite{tarasov2016filebench} and \\ fileserver~\cite{tarasov2016filebench}
\end{tabular}
&
\begin{tabular}{l}
$prxy\_0$ is write-intensive, \\ $oltp\_rw$ is read-intensive, and \\fileserver has nearly equal \\numbers of reads and writes
\end{tabular}
\\
\hline
\textbf{mix6}                     &   
\begin{tabular}{l}
$src1\_0$~\cite{MSR}, $YCSB\_C$~\cite{cooper2010benchmarking} and \\ fileserver~\cite{tarasov2016filebench}  
\end{tabular}
&
\begin{tabular}{l}
$src1\_0$ and fileserver have\\nearly equal numbers \\of reads and writes while \\$YCSB\_C$ is read-intensive
\end{tabular}
\\
\hline
\end{tabular}
}
\end{table}
}
}

We make two observations.   First, \sibyldef consistently outperforms \cde, \hps, \arcc, and \kleio~ \gon{by} 27.9\%, 12.2\%, 12.1\%, and 12.9\%,  
 respectively, in \rcamfix{the} \textsf{H\&M} HSS \gon{configuration} and 9.4\%, 21.3\%, 19.4\%, and 17.1\%, respectively, in \textsf{H\&L} HSS \gon{configuration}. Second,  with a lower learning rate \rcamfix{and optimized hyper-parameters}, \sibylopt provides  5.2\% (9.3\%) higher average performance for \textsf{H\&M} (\textsf{H\&L}) \gon{HSS} configuration than \sibyldef.
 \gon{Third}, for \mix,  \hps provides comparable performance to \sibyldef in \textsf{H\&M},
 and \cde provides slightly  better performance  in \textsf{H\&L}.
 As discussed in Section \ref{subsection:Sibyl/Evaluation_perf}, \proxy  is write-intensive and has  random requests (with an average request size of 7.2) \gon{within} every 1000 requests, which is the experience buffer size to train the \gonz{training}  network. Such a workload requires more frequent retraining of Sibyl’s \gonz{training}  network to achieve higher performance. \gon{We conclude that \namePaper ~\gonz{can} \rcamfix{effectively} adapt its \rcamfix{data placement} policy \rcamfix{online} to highly dynamic workloads.}


\gon{
\subsection{Performance with Different Features}
{Figure~\ref{fig:Sibyl/ablation} compares the use of some of the most \rcamfix{useful} features for the \emph{state} of \namePaper in our \textsf{H\&L} \gon{HSS} configuration.  \textsf{All} represents using all the six features \gca{in Table~\ref{tab:Sibyl/state}}. \namePaper~\gca{autonomously} decides which features are important to maximize the performance of \gon{the running} workload. 

\begin{figure}[h]
  \centering
    \includegraphics[width=1\linewidth,trim={0cm 0.35cm 0cm 0cm},clip]{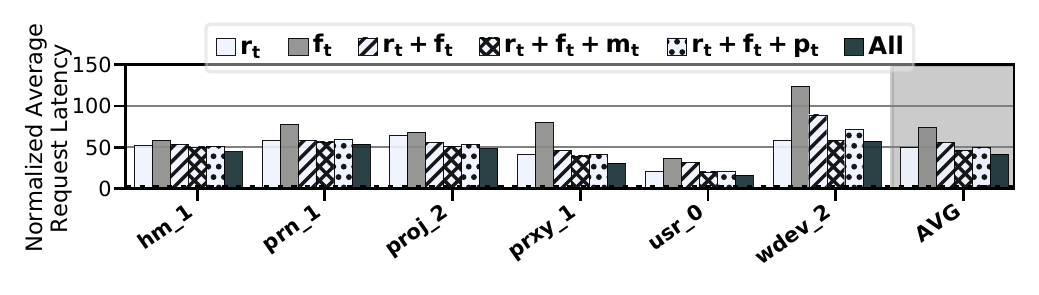}
  \caption{\rcamfix{Average request latency when using} different features (see Table~\ref{tab:Sibyl/state}) for the state space of \namePaper in the \textsf{H\&L} \gon{HSS configuration}  (normalized to \fast) }
\label{fig:Sibyl/ablation}
 \end{figure}

We make two {key} observations from Figure~\ref{fig:Sibyl/ablation}. First, \namePaper consistently achieves the lowest latency \gup{(up to 43.6\% lower)} by using all the features {mentioned in Table~\ref{tab:Sibyl/state}} (\textsf{All} in {Figure~\ref{fig:Sibyl/ablation}}). Second, by using the same features as in \rcamfix{baseline} heuristic-based polic\gonz{ies}, \namePaper is able to perform better data placement decisions. For example, \rcamfix{\textsf{$r_t$} and \textsf{$f_t$}  configurations of} \namePaper \rcamfix{~(in Figure~\ref{fig:Sibyl/ablation})} use \rcamfix{only one} feature, \rcamfix{just like} {\cde and \hps} do. 
\rcamfix{These two \namePaper configurations outperform {\cde and \hps}  policies by 4.9\% and 5.5\%, respectively (\textit{ref.} {Figure~\ref{fig:Sibyl/perf}(b)}). Using the same features as a heuristic-based policy, Sibyl autonomously finds a \rncam{higher-performance} dynamic policy that can maximize the reward function, which heuristic-based policies cannot possibly do.}
We conclude that {\namePaper}~uses a richer set of features 
that can capture multiple aspects of a storage request to \gon{\gonz{m}ake \gonz{better}}  {data} placement decisions than a heuristic-based policy. \gup{RL reduces 
\juangg{the design burden on system architects,} as \namePaper autonomously learns to use \gon{the provide\rcamfix{d}} features to achieve the highest cumulative reward.} In contrast, traditional heuristic-based policies use \rcamfix{features to make rigid data placement decisions without any system-level feedback,} \gonz{and thus they underperform \rncam{compared to \namePaper}.}
} 
}

\subsection{Performance with Different Hyper-Parameters}

\begin{figure}[h]
\begin{subfigure}[t]{.45\linewidth}
  \includegraphics[width=\linewidth,trim={0cm 0cm 0cm 0cm},clip]{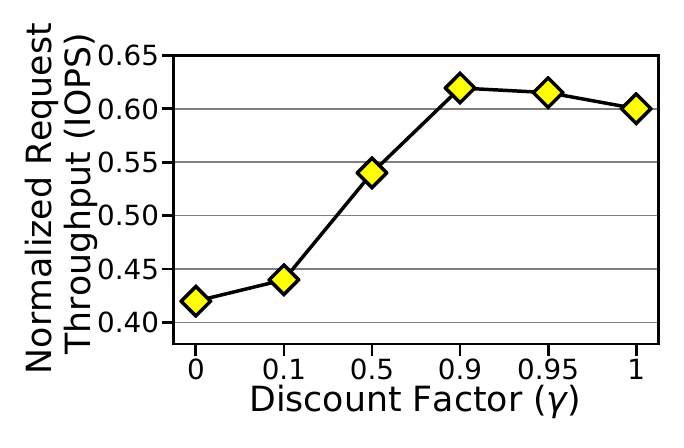}
   \vspace{-3.8cm}
  \caption{ \hspace{-2cm}
  \label{fig:Sibyl/discount}}
\end{subfigure}%
\begin{subfigure}[t]{.45\linewidth}
  \includegraphics[width=\linewidth,trim={0cm 0cm 0cm 0cm},clip]{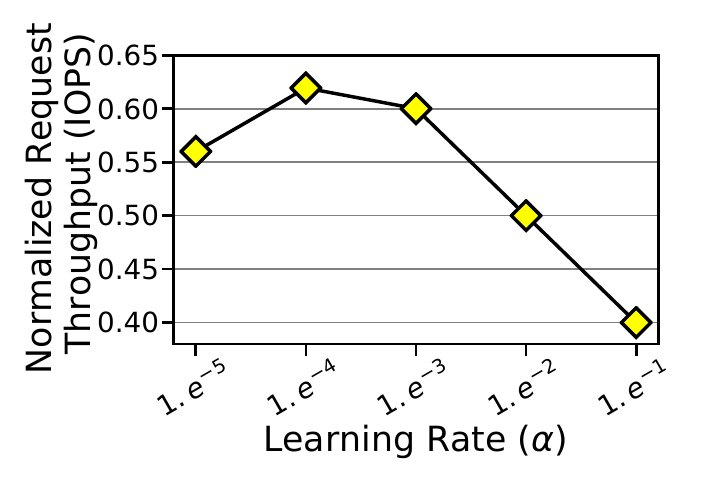}
  \vspace{-3.9cm}
   \caption{\hspace{-2cm}
  \label{fig:Sibyl/learning_rate}}
\end{subfigure}
\begin{subfigure}[t]{.45\linewidth}
\hspace{4cm}
  \includegraphics[width=\linewidth,trim={0cm 0cm 0cm 0cm},clip]{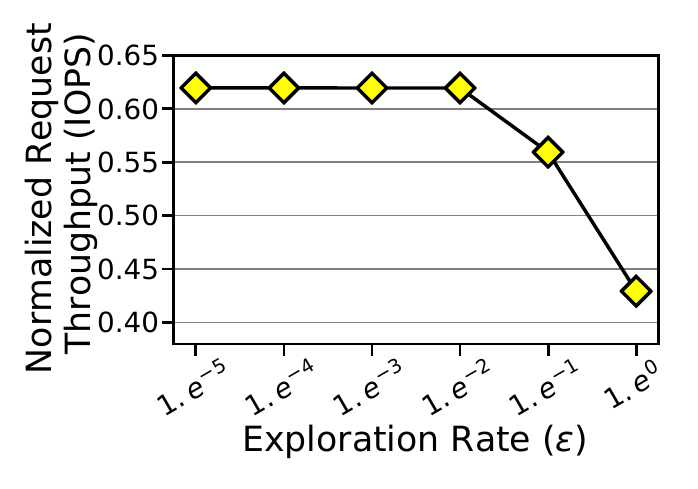} 
  \vspace{-3.9cm} 
   \caption{\hspace{-8cm}
  \label{fig:Sibyl/epsilon}}
\end{subfigure}
\caption{Sensitivity of \namePaper~\rcamfix{throughput} to: (a) the discount factor ($\gamma$), (b) the learning rate ($\alpha$), (c) the exploration rate ($\epsilon$), \rcamfix{averaged across 14 workloads (normalized to Fast-Only)} \label{fig:Sibyl/hyper_senstivity}}
\end{figure}

Figures~\ref{fig:Sibyl/hyper_senstivity}(a), \ref{fig:Sibyl/hyper_senstivity}(b), and \ref{fig:Sibyl/hyper_senstivity}(c) show the effect \gon{of}  three critical hyper-parameters (discount factor, learning rate, and exploration rate) \rcamfix{on \namePaper's throughput in \hmssd HSS configuration}. 
Figure~\ref{fig:Sibyl/hyper_senstivity}(a) \rcamfix{shows} that \namePaper's \rcamfix{throughput} drops sharply at $\gamma=0$.
At $\gamma=0$, \namePaper gives importance \gonz{only} to the immediate reward \gonz{and not} \rncam{at all to} the long-term reward, \gca{leading to} lower performance. We use $\gamma=0.9$, where \namePaper ~\rcamfix{is more forward-looking, giving enough weight to} long-term rewards.
\rcamfix{Figure~\ref{fig:Sibyl/hyper_senstivity}(b) shows} that at \rcamfix{a learning rate of} $\alpha=1e^{-4}$, \namePaper~\gon{provides} the best performance. \rcamfix{The learning rate determines the rate at which \gonz{training} network w\gonz{e}ights are updated.} \gonz{Both too slow and too fast} updates are detrimental \gonz{for}  adaptive learning and \gonz{stable \rncam{exploitation} of a learned policy}, \rncam{respectively}.
Third, Figure~\ref{fig:Sibyl/hyper_senstivity}(c)  shows  that the performance of \mbox{\namePaper} drops sharply if it performs exploration \rncam{too frequently} \rcamfix{(i.e., $\epsilon=1e^{-1}$)} and \rcamfix{thus} does not \gonz{sufficiently} exploit its learned policy. \namePaper achieves \rcamfix{the highest} performance improvement\rcamfix{s} for \rcamfix{$ 1e^{-5}\leq ~\epsilon \leq 1e^{-2}$}. 



{\subsection{{Sensitivity to Fast Storage Capacity}}
{Figure~\ref{fig:Sibyl/sensitivity} shows the \gon{average request latency} of \namePaper and baseline policies \gon{as we vary}  the available capacity in the fast  \gon{storage}. \gonzz{The} \rcamfix{\rncam{x}-axis denotes a range of fast storage device sizes available for data placement and represented in terms of percentages of the entire fast storage device capacity, \gonz{where 100\% represents \rncam{the size where} all pages of a workload can fit in the fast storage}}. 
\begin{figure}[h]
    \centering
 \includegraphics[width=1\linewidth]{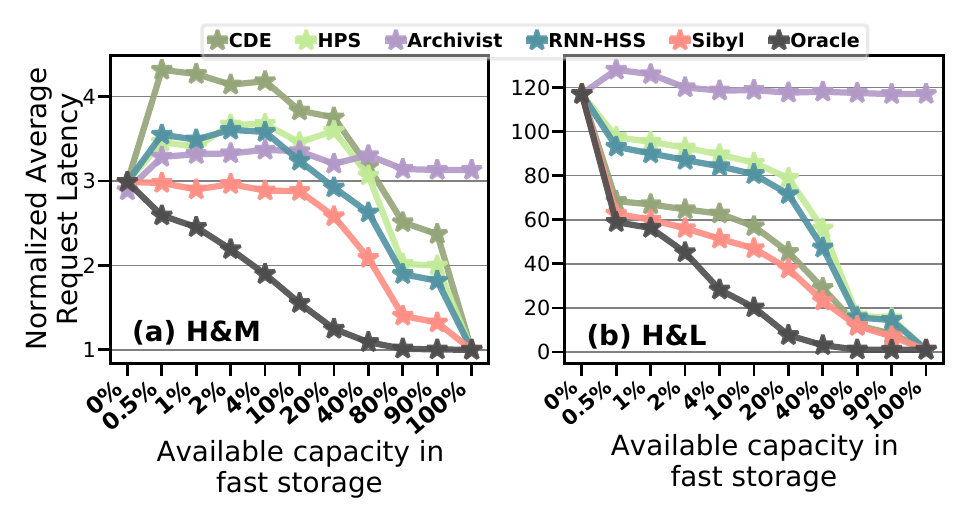}
    \caption{\gon{Average request latency for various fast storage \gonz{device} sizes (normalized to \fast)}}
    \label{fig:Sibyl/sensitivity}
\end{figure} 
}

\gonzz{We make two observations.} First, \rcamfix{for all fast storage sizes}, \namePaper performs better than  the baseline heuristic- 
and supervised learning-based policies for both \textsf{H\&M} and \textsf{H\&L} HSS configurations. 
\rcamfix{Even when the fast storage size is as small as 1\%,}~\namePaper outperforms \cde, \hps, \arcc, \kleio by 47.2\% (11.5\%), 17.3\% (58.9\%), 12.3\% (110.1\%), 21.7\% (50.2\%), respectively, in \hmssd (\hlssd)
}. Second, at a larger 
({smaller}) \rcamfix{fast storage device size}, the 
\rcamfix{performance approaches that of} \gca{the} \fast (\slow) policy, \gf{ except \rcamfix{for} \arcc. \arcc classifies pages \gca{as} hot or cold \gca{at} the beginning of an epoch and does not change its placement decision throughout the execution of that epoch. It does not perform any promotion or eviction of data. We observe that \arcc often mispredicts the target device for a request and classifies the same number of requests for the fast and slow storage device under different fast storage sizes.

As we vary the \gonz{size} of \gonz{the} \rcamfix{fast storage device}, a dynamically adaptable data placement policy is required, which considers features from both the running workload and the underlying storage system. \gon{ We conclude that \namePaper can provide scalability by dynamically \rcamfix{and effectively} adapting \rcamfix{its policy} to the available storage size to achieve high performance. 
}} 

\subsection{{Tri-Hybrid Storage Systems}}
\label{subsec:Sibyl/trihybrid}
\begin{figure}[h]
\centering
\begin{subfigure}[h]{.85\textwidth}
  \centering
  \includegraphics[width=\linewidth,trim={0.2cm 0.2cm 0cm 0.2cm},clip]{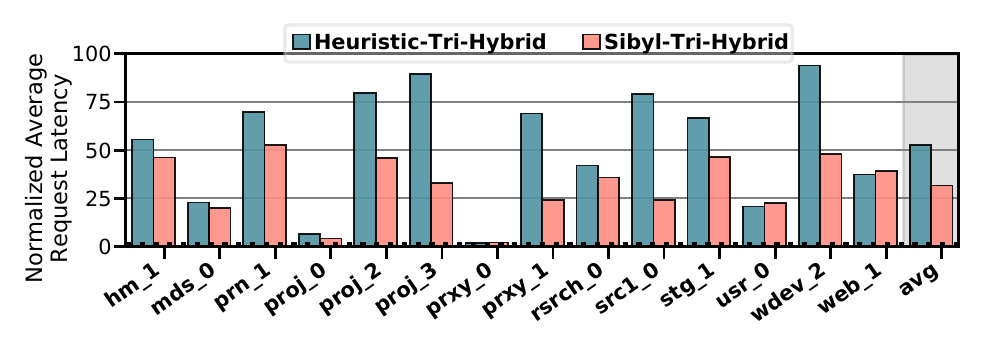}
  \vspace{-2em}
   \caption{\thold~configuration\label{fig:Sibyl/triybrid1}}
\end{subfigure}%

\begin{subfigure}[h]{.85\textwidth}
  \centering
  \includegraphics[width=\linewidth,trim={0.2cm 0.2cm 0.2cm 0.2cm},clip]{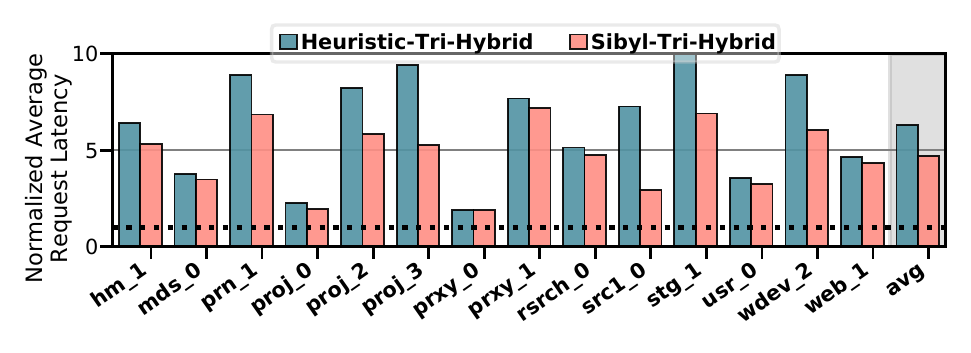}
    \vspace{-2em}
   \caption{\thnew ~configuration \label{fig:Sibyl/triybrid2}}
\end{subfigure}
  \caption{\gor{\gon{Average request} latency for \rcamfix{the tri-hybrid HSS} (normalized to \fast) }}
\label{fig:Sibyl/triybrid}
\end{figure}

We evaluate two \gca{different} tri-HSS configurations, \thold~and \thnew ~(Table \ref{tab:Sibyl/devices}), implemented as a single flat block device. The \thnew \ configuration has a low-end SSD (L$_{SSD}$), 
\gon{whose} performance \gon{is} lower than \rcamfix{the} \textsf{H} and \textsf{M} devices but higher than \rcamfix{the} \textsf{L} device. We restrict the available capacity of \textsf{H} and \textsf{M} 
\rcamfix{to 5\% and 10\%, respectively,}
of the working set size of \gon{a given workload}. This ensures data eviction from \textsf{H} and \textsf{M} devices once \gon{they are} full. We compare the performance of \namePaper on a tri-hybrid system with a \gon{state-of-the-art} heuristic-based policy~\cite{matsui2017design, matsui2017tri} that divides data into \emph{hot}, \emph{cold}, and \emph{frozen} and places \gon{them} \rcamfix{respectively} into \textsf{H}, \textsf{M}, and \textsf{L} devices.\footnote{\gon{\cde, \hps, \arcc, and \kleio do \textit{not} consider more than two devices and \gonz{are not easily adaptable to a tri-hybrid HSS}.}} Figure~\ref{fig:Sibyl/triybrid} shows the performance \gon{of the} heuristic-based and \namePaper data placement policies. 

We observe that  \namePaper outperforms the heuristic-based policy by, on average, \gor{43.5\% (48.2\%) and 23.9\% (25.2\%) for \mbox{\thold} (\mbox{\thnew})}. \rcamfix{This is because Sibyl is much more dynamic and adaptive to  \gonz{the storage} system configuration \gonz{due to its RL-based \rncam{decision-making}} than the baseline heuristic-based policy, which is rigid in its decision-making.}
To extend \namePaper for three storage devices, we had to only (1) add a new action in \namePaper's action space, and (2) add the remaining capacity in the \textsf{M} device as a state feature. \gon{We conclude that \namePaper provides \gca{ease of extensibility} \rcamfix{to new storage \gonz{system} configurations, which} reduces the system architect's burden in designing sophisticated data placement mechanisms.}  

 
\section{Explainability Analysis}
 \label{sec:Sibyl/explanability}
\begin{figure}[h]
  \centering
    \includegraphics[width=1\linewidth,trim={0.2cm 0.2cm 0.2cm 0.2cm},clip]{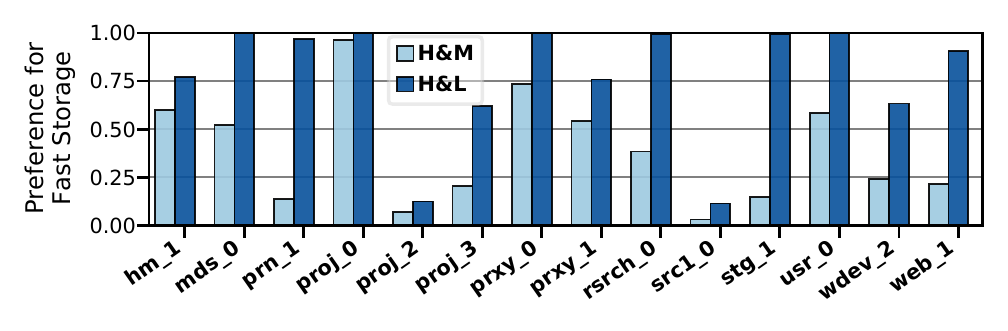}
  \caption{\namePaper's preference for the fast {storage device} under different \gon{HSS} configurations}
\label{fig:Sibyl/explain}
\vspace{-0.1cm}
 \end{figure}

We perform an explainability analysis to understand our results further and explain Sibyl’s decisions. 
We extract \namePaper's actions for different workloads under \hmssd and \hlssd~ \gon{HSS configurations and analyze the page placements for each workload}. 
Figure~\ref{fig:Sibyl/explain} shows \gon{\namePaper's preference for \rcamfix{the} fast storage \rcamfix{device} over \rcamfix{the} slow storage \rcamfix{device}, measured as the} ratio of the  number of fast storage placements to the \gont{sum of the number of placements in \rcamfix{both} fast and slow storage devices} \gon{(i.e., $\text{Preference=}\frac{\# \text{fast placements}}{\# \text{fast}+\# \text{slow placements}}$)}.

  \begin{figure*}[h]
\centering
\begin{subfigure}[h]{.85\textwidth}
  \centering
  \includegraphics[width=\linewidth]{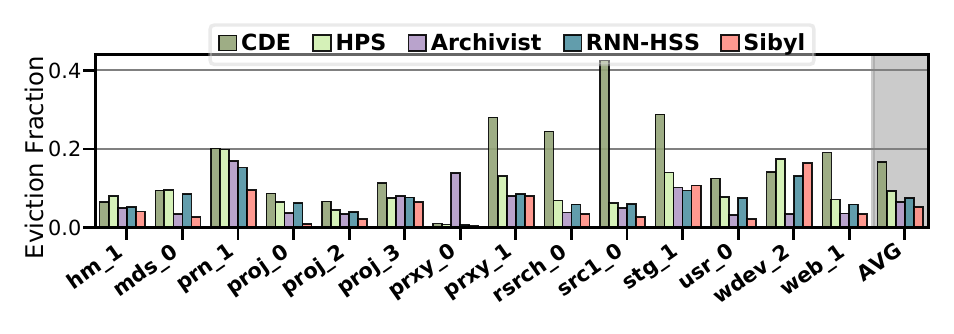}
   \vspace{-0.8cm}
  \caption{\textsf{H\&M} \gon{HSS configuration}
  \label{fig:Sibyl/evict_1}}
\end{subfigure}%

\begin{subfigure}[h]{.85\textwidth}
  \centering
  \includegraphics[width=\linewidth,trim={0cm 0cm 0cm 0cm},clip]{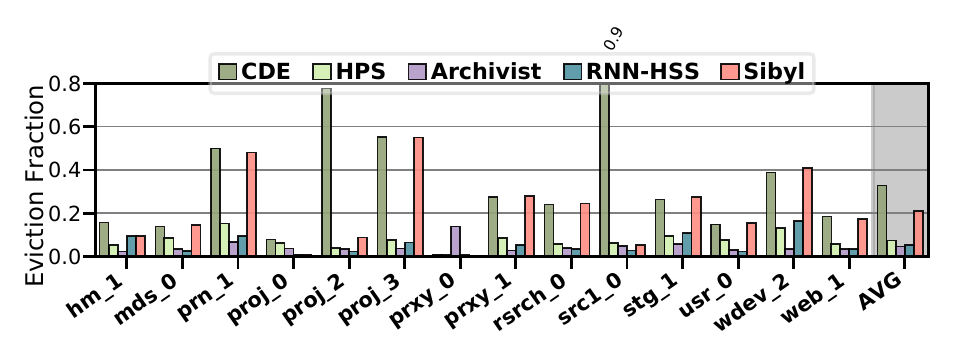}
   \vspace{-0.8cm}
   \caption{\textsf{H\&L}  \gon{HSS configuration}
  \label{fig:Sibyl/evict_2}}
\end{subfigure}
\caption[Two numerical solutions]{Comparison of evictions from the fast storage \gon{to the slow storage} (normalized to the total number of \gon{storage requests}) \label{fig:Sibyl/evict}}
\end{figure*}

We make the following four observations. 
First, in the \js{\textsf{H\&L}} configuration, where the latency difference is large {between the two storage devices}, \namePaper~ \gon{prefers} to place \gon{more} data in the fast \gon{storage} device. \namePaper learns that despite the eviction penalty, the benefit of serving more requests from the fast \gon{storage} device is significant. On the other hand, in the \js{\textsf{H\&M}} device configuration, where the latency difference between two devices is smaller compared to \rakesh{\textsf{H\&L}}, \namePaper places only performance-critical pages in the faster \rcamfix{storage device} to avoid the eviction penalty.

Second, \rcamfix{in the} {\textsf{H\&M}} configuration, \namePaper~shows less preference to place pages from  \texttt{mds\_0}, \texttt{prn\_1}, \texttt{proj\_2}, \texttt{proj\_3}, \texttt{src1\_0}, \texttt{stg\_1}, and \texttt{web\_1}  in \juan{the} fast storage \gon{device}. These workloads \gonz{are} cold and sequential (Table~\ref{tab:Sibyl/workload}) \gonz{and thus} \gon{are less suitable for} the fast storage \gon{device}. Therefore, \gon{for such workloads}, \namePaper shows more preference for  \gon{the} slow storage \gon{device}. \rcamfix{In contrast,} for hot and random workloads (\texttt{prxy\_0} and \texttt{prxy\_1}), \namePaper shows \rcamfix{more} preference to place \rcamfix{pages} in the fast \gon{storage} device.


Third, for \texttt{rsrch\_0}, \texttt{wdev\_2}, and \texttt{web\_1}, \namePaper~ \gonzz{places} \gonff{$\le$40\%} \gon{of pages} in the fast storage \gon{device}. Such requests have random access patterns, while pages with cold and sequential accesses are \gonzz{placed in the} slow storage. 

\gont{Fourth},  in the \js{\textsf{H\&L}} setting, \namePaper shows more preference to \gonzz{place} requests in the fast \gon{storage} device, except for \texttt{proj\_2} and \texttt{src1\_0}  workloads. {We observe that these two workloads \gonff{are} \gon{highly random with} a low average access \gonz{count} (Table~\ref{tab:Sibyl/workload}). Therefore,  aggressive placement in \gonzz{the}  fast storage is not beneficial for  long-term performance.  }


\js{We also measure} the number of evictions \gonff{(as a \gonz{fraction} of all storage requests)} that occur while using \namePaper~and other baseline policies, as shown in  Figure~\ref{fig:Sibyl/evict}. 
We make two observations. First, \gon{in \gonff{the} \hmssd HSS configuration, \namePaper~\gonff{leads to} 68.4\%, 43.2\%, 19.7\%, and 29.3\% \gonff{fewer} evictions from the fast storage than \cde, \hps, \arc, and \kleio, respectively. }
Second, \cde places more data in the fast \gon{storage}, which leads to a \js{large} number of evictions in both \gon{HSS} configurations. However, if the latency difference between the two devices is large (e.g., \js{\textsf{H\&L}} configuration), \gon{\cde} provides higher performance than other baseline policies (see Figure~\ref{fig:Sibyl/perf}(b)). 
Therefore, in the \js{\textsf{H\&L} HSS configuration}, we observe \gonzz{that}~\namePaper follows a similar policy, leading to more evictions compared to other baselines. 



\section{Overhead Analysis}
\label{sec:Sibyl/overhead_analysis}

\subsection{Inference and Training Latencies} 
The input layer \gon{of the} \gonz{training}  and inference network\gont{s} consists of \js{six} neurons, \gont{equal} to the number of features \js{listed} in Table~\ref{tab:Sibyl/state}. \gont{Each feature is} normalized to transform the \gont{value} range of different features to a common scale. { The size of one state entry is 40 bits (32 bits for state features and 8 bits for the counter used for tracking \gonzz{the remaining capacity in the} fast storage device).  We make use of two hidden layers with 20 and 30 neurons each. The final output layer has neurons equivalent to our action space, \gca{ i.e., two for \gont{dual-HSS} configurations and three for \gont{the} tri-HSS configurations}. 

\head{Inference \gonff{latency}} Our inference network has 52 \gont{inference} neurons (20+30+2)  with 780 \gon{weights} (6$\times$20+20$\times$30+30$\times$2). 
{As a result, {\namePaper} requires 780 MAC operations per inference (1$\times$6$\times$20+1$\times$20$\times$30 +1$\times$30$\times$2). On our evaluated \gont{CPU}, we can perform these operations in $\sim$10ns, which is several orders of magnitude smaller than the I/O read latency of even a high-end SSD ($\sim$10us)~\cite{inteloptane, samsung2017znand}.} 
\gca{\gont{\namePaper's inference} computation can also be performed in the SSD controller.}

\gor{{\noindent{\head{Training \gonff{latency}}}  For each training step, \mbox{\namePaper} needs to compute 1,597,440 MAC operations, where each batch requires 128$\times$6$\times$20+ 128$\times$20$\times$30+128$\times$30$\times$2 MAC operations. 
This computation takes $\sim$2us on our evaluated \gont{CPU}. 
\gonff{This} training latency does not affect the \gonff{benefits} of \mbox{\namePaper}
\rcamfix{because (1)}
training occurs asynchronously \rcamfix{with} inference, \gonzz{and}
\rcamfix{(2)} training latency is \gonff{$\sim$5$\times$} smaller than the I/O read latency of even a high-end SSD.  }}

\noindent\gonff{We conclude that Sibyl's performance benefits come at small latency overheads that are easily realizable in existing CPUs.}


\vspace{-0.35cm}
\subsection{Area Overhead} 

\noindent{\head{Storage cost}} We use a half-precision floating-point format for the weights \gon{of the \gonz{training} and the inference network\rcamfix{s}}. With 780 16-bit weights, \rcamfix{each} neural network requires 12.2 KiB of memory. 
Since we use 
\rcamfix{the same} network architecture for the \gonz{two} networks, we need 24.4 KiB of memory. In total, with an experience buffer of 100 KiB (Section \ref{subsec:Sibyl/detail_design}), \namePaper require\rcamfix{s} \gonzz{124.4} KiB of \gont{DRAM} overhead, \js{which is negligible compared to {the} memory size of modern computing systems}.  } 

\noindent{\head{Metadata cost}} HSSs need to maintain the address mapping information for the underlying storage devices~\cite{tsukada2021metadata}.  {\namePaper} \gonff{requires} 40 bits to store  \gont{state information} (\gonff{i.e., the per-page state features;} see Table~\ref{tab:Sibyl/state}). This overhead is $\sim$0.1\% of the total storage capacity when using a 4-KiB data placement granularity (5-byte per 4-KiB data). 

\noindent\gon{We conclude that \namePaper~\gon{has \gonzz{a} very} modest cost in \gon{terms of} \gon{storage capacity} overhead \gon{in \gonz{main memory (DRAM)}}.}

\section{Discussion}
\label{sec:Sibyl/discussion}

\gsa{\head{Cost of generality} 
\gon{We identify two main limitations of using RL for data placement. First, currently, RL is largely a \emph{black-box policy}.  Our explainability analysis (Section \ref{sec:Sibyl/explanability}) tries to provide intuition into Sibyl’s internal mechanism. However, providing rigorous explainability to reinforcement learning-based mechanisms is an active field of research~\cite{verma2018programmatically,liu2018toward,juozapaitis2019explainable,madumal2020explainable,sequeira2020interestingness,puiutta2020explainable}, a problem that is beyond the scope of this paper. Perfectly finding worst-case workloads against an RL policy is, therefore, very difficult, in fact, impossible, given the state-of-the-art in reinforcement learning. There are many dynamic decisions that the agent performs, which cannot be easily explained or modeled in human-understandable terms.}   
Second, \gonff{\namePaper} requires \emph{engineering effort} to (1) \gonff{thoroughly} tune the RL hyper-parameters, and (2) implement and integrate \namePaper components into the host OS's storage management layer. 
This second limitation is not specific to Sibyl and applies to any ML-based storage management technique. \gonff{As quantified in  Section \ref{sec:Sibyl/overhead_analysis}, \namePaper's storage and latency overheads are small. }   }

\head{\namePaper's implications} \gor{\gont{\namePaper} (1) provides performance improvements on a wide variety of workloads and system configurations (our evaluations \gonff{in} Section \ref{sec:Sibyl/results} show that \namePaper outperforms all evaluated state-of-the-art data placement policies \gonff{under all system configurations}),  (2) provides extensibility by {reducing} the designer burden when extending data placement policies to multiple devices \gonff{and different storage configurations}, and (3) \gonff{enables reducing } the fast storage {device} \gonff{size} by taking \gonff{better} advantage of \rcamfix{the} fast-yet-small storage device and large-yet-slow storage device to deliver high storage capacity at low latency.}

\head{Adding more features and optimization objectives}
\gsa{An RL-based approach simplifies adding new features (such as bandwidth utilization) \gont{in} \gonzz{the} RL state and optimization objectives (such as endurance) using \gonzz{the} RL reward function. This flexibility allows an RL-based mechanism to self-optimize and adapt its decision-making policy to achieve an objective without \gont{the designer} explicitly defining \emph{how} to achieve it. We demonstrate and evaluate example implementations of Sibyl using a reward scheme that is a function of  request latency and eviction latency. We \gont{find} that \gonx{request} latency in the reward structure best encapsulates system conditions since latency could vary for each \gon{storage} request \gont{based on complex system conditions}. To optimize for a different device-level objective, one needs to define a new reward function with appropriate state features, e.g., to optimize for endurance, one might use the number of writes to an endurance-critical device in the reward function. Another interesting research direction would be to perform multi-objective optimization, \gont{e.g., optimizing for \gonx{both}} performance and energy. We leave \gont{the study of different objectives and features to} future \gont{work}.
}

\head{Necessity of the reward} \gsa{RL training is highly dependent upon the quality of the reward function and state features. Using an incorrect reward or improper state features could lead to severe performance degradation. }Creating the right reward is a human-driven effort that could benefit from design insights. We tried two other reward structures to achieve our objective to \grakesh{\rcamfix{improve}  \gonff{system} performance}:
\begin{itemize}[leftmargin=*, noitemsep, topsep=0pt]
\item \textbf{Hit rate of the fast storage device:} Maximizing the hit rate of \ghpca{the} fast \gon{storage} device is another \gonff{potentially plausible objective}. However,  
\rcamfix{if we use} \gonff{the} hit rate as \gonzz{a} reward, \namePaper (1) tries to aggressively place data in \gonzz{the} fast storage \rcamfix{device}, \gon{which leads to unnecessary evictions}, \gonzz{and} (2)  cannot capture the asymmetry in the latencies  present in \gonx{modern} storage devices \gonff{(e.g., due to  \mbox{read/write} latencies, latency of garbage collection, queuing delays, error handling latencies, \gonzz{and} write buffer state)}.

\item \textbf{High negative reward for eviction:} We also tried a negative reward for eviction and \gonz{a} zero \rcamfix{reward} in other cases. We observe that such a reward structure \gon{provides suboptimal performance because \gonff{\namePaper places more pages in the slow device to avoid evictions. Thus, \gonx{with such a reward structure,} \namePaper is not able to effectively utilize the fast storage.}} 

\end{itemize}
  \noindent We conclude that our chosen  reward structure works well for a wide variety of workloads \gonff{Section \ref{sec:Sibyl/results}}, as \gonff{\gonx{reinforced} by} our generality studies using unseen workloads in \mbox{Section \ref{subsection:Sibyl/perf_unseen}}.   

\head{Managing hybrid main memory using RL} \gca{The key idea of \namePaper can be adapted for managing hybrid main memory architectures. However, managing data placement at different levels of the memory hierarchy
has its own set of challenges~\cite{li2017utility_HMM,agarwal2015page_HMM,agarwal2017thermostat_HMM,goglin2016exposing,ham2013disintegrated_HMM,lin2016memif,malladi2016dramscale,pavlovic2013data,pena2014toward,qureshi2009scalable,yoon2012row,meza2013case, meza2012enabling, ren2015thynvm} that \namePaper would need to adapt \rcamfix{to}, such as the low latency decision-making \gonff{and control} requirements in  main memory. 
Even with the \gonx{use} of hybrid main memories, many systems \rcamfix{continue to} benefit from using hybrid storage devices \gonff{due to much lower cost\gonx{-}per\gonx{-}bit of storage, which accommodates increasingly larger datasets.} 
 Therefore,  we focus on hybrid storage systems and leave it \rcamfix{to} future \rcamfix{work} to \rcamfix{\gonff{study} RL to manage} hybrid main memories. 
}

\section{Related Work}

\gagan{To our knowledge, this is the first work to propose \rcamfix{a reinforcement learning}-based data placement technique for hybrid storage systems. \rcamfix{\namePaper can} continuously learn from and adapt to the \rcamfix{running} application and \rcamfix{the} storage \rcamfix{configuration and} device characteristics. }\gca{We \rcamfix{briefly} discuss closely-related prior works that propose data management techniques for hybrid memory/storage systems and RL-based system optimizations.}

\head{Heuristic-based data placement}
Many prior works~\cite{matsui2017design,sun2013high,heuristics_hyrbid_hystor_sc_2011,vasilakis2020hybrid2,lv2013probabilistic,li2014greendm,guerra2011cost,elnably2012efficient,heuristics_usenix_2014,li2017utility_HMM,agarwal2015page_HMM,agarwal2017thermostat_HMM,ham2013disintegrated_HMM, bu2012optimization, krish2016efficient, lin2011hot, tai2015sla, zhang2010automated, wu2012data, iliadis2015exaplan, lv2013hotness, matsui2017tri, feng2014hdstore,salkhordeh2015operating, pavlovic2013data, meza2013case, yang2017autotiering, chiachenchou2015batman, kim2011hybridstore, wang2019panthera, ramos2011page, liu2019hierarchical,luo2020optimal,doudali2021cori} propose heuristic-based techniques to perform data placement. 
These techniques rely on statically-chosen design features that \rcamfix{usually} favor certain workload\gonzz{s} and/or device characteristics, \rcamfix{leading to \gonx{relatively} rigid policies}.
In Section \ref{sec:Sibyl/motivation_limitations} and Section \ref{sec:Sibyl/results}, we  show \gca{that} \namePaper outperforms two \rcamfix{state-of-the-art} works, CDE~\cite{matsui2017design} and HPS~\cite{meswani2015heterogeneous}.

\head{ML-based data placement}
\rcamfix{Several works~\cite{doudali2019kleio,ren2019archivist,cheng2019optimizing,shetti2019machine,sen2019machine} propose ML-based techniques for data placement in hybrid memory/storage systems. These  works 1) are based on supervised learning techniques that require frequent \rcamfix{and very costly} retraining to adapt to changing workload and device characteristics, and 2) have not been evaluated on a real system.} \gonn{\rcamfix{We evaluate} \kleio, which is  inspired by the state-of-the-art data placement technique in hybrid main memory~\cite{doudali2019kleio}. It uses sophisticated recurrent neural networks (RNNs) for data placement \rcamfix{and} shows promising results \rcamfix{compared to heuristic-based techniques}. However, it has \gonx{two major} limitations \gonx{that make it impractical or difficult to implement}: \juangg{it} (1) trains  an RNN for each page, which \rcamfix{leads to large} computation\gonx{, storage, and training time \gonzz{overheads}}, and (2) requires offline application profiling.}
\gagan{Our evaluation (ref. Section \ref{subsection:Sibyl/Evaluation_perf}) shows that \namePaper outperforms \rcamfix{two state-of-the-art ML-based data placement techniques}, \kleio~\cite{doudali2019kleio} and \arcc~\cite{ren2019archivist}, \rcamfix{across a wide variety of workloads.}}
\head{{\rcamfix{RL-based techniques in storage systems}}}
Recent works \gon{(e.g., \cite{liu2019learning,yoo2020reinforcement,wang2020reinforcement,rl_GC_TECS, kang2018dynamic})} propose the use of RL-based approaches for managing different aspects of storage systems. These works cater to use cases and objectives that are \gonx{very} different \rcamfix{from} \namePaper's.
Specifically, {Liu}~\etal~\cite{liu2019learning} (1) \rcamfix{propose} data placement in cloud systems and \emph{not} \gf{hybrid storage \rcamfix{systems}, (2) consider devices with unlimited capacity, sidestepping the capacity limitations, (3) \emph{emulate} a data center network rather than \gon{use} a real system \gon{for design and evaluation}, and (4) focus only on data-analytics workloads. }
{Yoo}~\etal~\cite{yoo2020reinforcement} do \emph{not} focus on data placement; they instead deal with dynamic storage resizing 
based on workload characteristics using a trace-based simulator.  Wang~\etal~\cite{wang2020reinforcement} (1) focus on cloud \gon{systems} to predict the data storage consumption, and 
(2) do \emph{not} consider hybrid storage systems. 
\namePaper is the first RL-based mechanism \gon{for data placement in hybrid storage systems.} 
 

\head{RL-based system optimizations}
Past works~\ghpca{\cite{lin2020deepNOC,ipek2008self,liu2020imitation,rl_NOC_AIDArc_2018,rl_voltage_scaling_TC_2018,mirhoseini2021chip,mutlu2021intelligent_DATE, pythia, peled2015semantic, martinez2009dynamic, multi_scheduler_HPCA_2012, jain2016machine,zheng2020agile, pd2015q}} propose RL-based methods for various system optimizations, such as memory scheduling~\cite{ipek2008self, multi_scheduler_HPCA_2012}, data prefetching~\cite{pythia,peled2015semantic}, cache replacement~\cite{liu2020imitation}, and network-on-chip arbitration~\cite{rl_NOC_AIDArc_2018,lin2020deepNOC}. \gca{Along with \namePaper\rcamfix{, designed} for efficient data placement in hybrid storage systems, \rcamfix{this body of work} demonstrate\rcamfix{s} that RL is a promising approach to designing high-performance\rcamfix{, and highly-adaptive} self-optimizing computing systems. }


\section{Conclusion}

We introduce \namePaper, the first reinforcement learning-based mechanism for  data placement in hybrid storage systems.
Our extensive \gagan{real-system} {evaluation} demonstrate\rcamfix{s} that \namePaper~\gon{provides adaptivity and extensibility by} continuously learning from and autonomously adapting to the workload characteristics\rcamfix{, storage configuration and device characteristics,} and system-level feedback to maximize the overall long-term performance of a hybrid storage system. {We interpret \gca{\namePaper}'s policy through our explainability analysis} and {conclude that \namePaper provides an \rcamfix{effective} and robust approach \gonx{to} data placement in current and future hybrid storage systems.} 
\rcamfix{We hope that \namePaper and our open-sourced implementation of it~\cite{sibylLink} inspire future work and ideas in self-optimizing storage and memory systems.}

\chapter{Conclusions and Future Directions}
\label{chapter:conclusion}

In this dissertation, the goal was twofold. First, overcome the data movement bottleneck by following a data-centric approach of bringing processing close to the memory  and ensure that system components are not overwhelmed by data; thus, enabling high-performance in an energy-efficient way. Second, leverage the enormous amount of data to devise computer architecture mechanisms that can assist us in architectural decisions or optimizations. To this end, \dd{the dissertation} presented five contributions to effectively handle and leverage the vast amount of data for future computing systems. Table~\ref{tab:intro/methods_end} shows nine architectural methods that we use across our five contributions. 

\begin{table}[h]
\caption[List of contributions and methods used to achieve the thesis statement]{We highlight across five contributions nine different methods (concepts) used in this dissertation to achieve the thesis statement of handling data well. We make use of two guiding principles: (1) \textit{data-centric} (DC) is bringing processing closer to where data resides and ensuring that data does not overwhelm system components; (2) \textit{data-driven} (DD) is leveraging data to perform architectural decisions or predictions.
\label{tab:intro/methods_end}}
\centering
\HUGE
\renewcommand{\arraystretch}{0.8}
\setlength{\arrayrulewidth}{2pt}
\setlength{\tabcolsep}{10pt}
\resizebox{1\linewidth}{!}{
\begin{tabular}{l|l|l|Hl|l|l}
\toprule 
\rule{0pt}{32pt}& \multicolumn{5}{c}{\textbf{CONTRIBUTIONS}} \\ \cmidrule{2-7} 
\textbf{METHODS} & \textbf{NERO~\cite{singh2020nero}}                                                                                                   & \textbf{Low Precision~\cite{singh2019low}}                                                                                        & \textbf{PreciseFPGA~\cite{not_published_arpan2021preciseNN}}                                                                                      & \textbf{NAPEL~\cite{NAPEL}}                                                                                       & \textbf{LEAPER~\cite{singh2021leaper_fpga}}                                                                                                                 & \textbf{Sibyl~\cite{not_published_singh2021qrator}}                                                                                   \\ \hline 
\rowcolor[HTML]{DCDCDC} 
\mystrut  
\textbf{Specialization} (DC)                                                                          &  \begin{tabular}[c]{@{}l@{}}FPGA-based \\ accelerator\end{tabular}    &                                             \begin{tabular}[c]{@{}l@{}}Implementation on\\an FPGA for fixed-\\point and floating-\\point representation\end{tabular}                                                                                                                 & \begin{tabular}[c]{@{}l@{}}FPGA-based \\ accelerator\end{tabular}                                         &                                                                                                      & \begin{tabular}[c]{@{}l@{}}FPGA-based \\ accelerator\end{tabular}                                                               &                                                                                                   \\\hline 
\rowcolor[HTML]{DCDCDC} 
\mystrut  
\begin{tabular}[c]{@{}l@{}}\textbf{Revisit memory}\\\textbf{hierarchy} (DC)       \end{tabular}                                                                & \begin{tabular}[c]{@{}l@{}}Scratchpad-based\\ hybrid memory\end{tabular}                                 &                                                                                                               &                                                                                                           &                                                                                                      &                                                                                                                                 & \begin{tabular}[c]{@{}l@{}}Tiered hybrid\\storage system\end{tabular}                             \\\hline 
\rowcolor[HTML]{DCDCDC} 
\mystrut  
\begin{tabular}[c]{@{}l@{}}\textbf{Reducing copies}\\\textbf{between host and}\\\textbf{accelerator} (DC)\end{tabular} & \begin{tabular}[c]{@{}l@{}}Shared memory \\space    \end{tabular}                                                                                             & Quantize data                                                                                                 & Quantize data                                                                                             &                                                                                                      &                                                                                                                                 &                                                                                                   \\\hline 
\rowcolor[HTML]{DCDCDC} 
\mystrut   
\begin{tabular}[c]{@{}l@{}}\textbf{Dataflow} \\ \textbf{architecture} (DC)     \end{tabular}                                                                                   & Task pipelining                                                                                                 &                                                                                                               &                                                                                                           &                                                                                                      &                                                                                                                                 &                                                                                                   \\\hline 
\rowcolor[HTML]{DCDCDC} 
\mystrut  
\begin{tabular}[c]{@{}l@{}}\textbf{Near-memory}\\\textbf{computing} (DC)    \end{tabular}                                                               & \begin{tabular}[c]{@{}l@{}}{Processing near-}\\{high-bandwidth} \\{memory}   \end{tabular}                                                                                            &                                                                                                               &                                                                                                           & \begin{tabular}[c]{@{}l@{}}Processing near-\\3D stacked memory \end{tabular}                           &                                                                                                                                 &                                                                                                   \\\hline 
\rowcolor[HTML]{DCDCDC} 
\mystrut   
\begin{tabular}[c]{@{}l@{}}\textbf{Reducing memory}\\ \textbf{footprint} (DC)     \end{tabular}                                                          &                                                                                                          \begin{tabular}[c]{@{}l@{}}Single and half\\floating-point\\precision \end{tabular}        & \begin{tabular}[c]{@{}l@{}}Different number \\ representations--\\ posit, fixed-point,\\and floating-point\end{tabular}                           & \begin{tabular}[c]{@{}l@{}}Automatic\\ quantization of \\ fixed-point\\ representation\end{tabular} &                                                                                                      &                                                                                                                                 &                                                                                                   \\\hline 
\rowcolor{gray!45} 
\mystrut  
\textbf{Static ML} (DD)                                                                 &                                                                                                                 &                                                                                                               &                                             \begin{tabular}[c]{@{}l@{}}Learns resource\\ and power\\utilization\end{tabular}                                                                &\begin{tabular}[c]{@{}l@{}}Learns application\\ performance and\\ energy consumption\end{tabular}                                                                                   & \begin{tabular}[c]{@{}l@{}}Learns resource\\ utilization and\\ performance\\ models\end{tabular}                                    &         
\\\hline 
\rowcolor{gray!45} 
\mystrut  
\begin{tabular}[c]{@{}l@{}}\textbf{Speed up design} \\ \textbf{space exploration}\\  (DD)\end{tabular}         &    \begin{tabular}[c]{@{}l@{}}Auto-tuning\\for data transfer\\window size\end{tabular}                                                                                                                      &                                                                                                               & Supervised ML                                                                                                  & \begin{tabular}[c]{@{}l@{}}Supervised ML,\\Design of \\ experiment\end{tabular}                                      & \begin{tabular}[c]{@{}l@{}}Few-shot\\learning\end{tabular}                                                                                                           &                           \begin{tabular}[c]{@{}l@{}}Design of\\experiment\end{tabular}                                                                            \\\hline 
\rowcolor{gray!45}
\mystrut  
\textbf{Dynamic ML} (DD)                                                          &                                                                                                                 &                                                                                                               &                                                                                                           &                                                                                                      &                                                                                                                                 & \begin{tabular}[c]{@{}l@{}}Reinforcement\\ Learning\end{tabular}                                                                             \\ \hline 
\mystrut  
\textbf{Goal}                                                                                & {\begin{tabular}[c]{@{}l@{}}Overcome memory\\bottleneck of\\weather prediction\\ application\end{tabular}} & {\begin{tabular}[c]{@{}l@{}}Investigate\\ computationally\\ cheaper number\\ representations\end{tabular}} &   {\begin{tabular}[c]{@{}l@{}}Automate DSE\\ for fixed-point\\ representation\end{tabular}}                                                                                  & {\begin{tabular}[c]{@{}l@{}}Quick performance\\ and energy\\ estimates of new \\ applications\end{tabular}} & {\begin{tabular}[c]{@{}l@{}}Quick area and\\ performance\\ estimates on\\ new high-end \\FPGA-based \\platforms\end{tabular}} & {\begin{tabular}[c]{@{}l@{}}Efficient and \\ high performance\\data placement \\mechanism\end{tabular}} \\ \bottomrule
\end{tabular}
}
\end{table}

\noindent \textbf{A data-centric architecture:} In \textbf{Chapter~\ref{chapter:nero}}, we designed NERO, a data-centric accelerator for a real-world weather prediction application. NERO overcomes the memory bottleneck {of weather {prediction} stencil kernels} by {exploiting near-memory {computation} capability on} specialized field-programmable gate array (FPGA) accelerators with high-bandwidth memory (HBM) {that are attached} {to the host CPU}. {Our} experimental results showed that {\nameNERO} outperforms a 16 core POWER9 {system} by $4.2\times$ and $8.3\times$ when running two {different} compound stencil kernels. {\nameNERO} {reduces the energy consumption by} $22\times$ and $29\times$ {for the same two kernels {over the POWER9 system}} {with an energy efficiency of 1.5 GFLOPS/Watt and 17.3 GFLOPS/Watt}. {We concluded that} employing near-memory acceleration solutions for weather prediction {modeling}~{is~promising} {as a means to achieve {both} high performance and {high} energy efficiency}.

\noindent \textbf{Precision tolerance and alternate number system:} In \textbf{Chapter~\ref{chapter:low_precision_stencil}}, we  explored the applicability of different number formats and searched for the appropriate bit-width for complex stencil kernels, which are one of the most widely used scientific kernels. Further, we leveraged the arbitrary fixed-point precision capabilities of an FPGA to demonstrate these kernels' achievable performance on state-of-the-art hardware that includes IBM POWER9 CPU with an FPGA board connected via CAPI interface. Thus, this chapter filled the gap between current hardware capabilities and future systems for stencil-based scientific applications.


\noindent \textbf{Data-driven alternative to system simulation:} In \textbf{Chapter~\ref{chapter:napel}},  we proposed NAPEL, a machine learning-based application performance and energy prediction framework for data-centric architectures. NAPEL uses ensemble learning to build a model that, once trained for a fraction of programs on a number of architecture configurations, can predict the performance and energy consumption of {different} applications. Our inexpensive performance model can provide, on average, an additional 10$\times$ reduction in performance evaluation time with an error rate lower than 15\% compared to a computationally-intensive state-of-the-art NMC simulator.

\noindent   \textbf{Data-driven modeling of FPGA-based systems:} In \textbf{Chapter~\ref{chapter:leaper}}, we presented LEAPER, the first use of \textit{few-shot learning} to transfer FPGA-based computing models across different hardware platforms and applications. Experimental results showed that our approach delivers, on average, 85\% accuracy  {when we use our transferred model for prediction in a cloud environment} with \textit{5-shot learning} and reduces {design-space exploration} time by 10$\times$, from days to {only} a few hours. These machine learning-based mechanisms follow \textit{data-driven} techniques by using the vast amount of data to provide fast and accurate performance prediction results.

\noindent \textbf{Data-driven mechanism for data-placement in hybrid storage systems}: In \textbf{Chapter~\ref{chapter:sibyl}}, we proposed Sibyl, a  reinforcement learning (RL)-based data-placement technique for a hybrid storage system (HSS). Sibyl observes different features \gon{of} the \gon{running workload}  \gon{as well as the}  storage devices to make system-aware data placement decisions. For every decision \gon{it makes}, Sibyl receives a reward from the system that it uses to evaluate the long-term \gon{performance} impact of its decision and continuously optimizes its data placement policy online. Our real system evaluation results show that  
 Sibyl \gon{ provides \gonzz{21.6\%/19.9\%} performance improvement in a performance-oriented\gonn{/cost-oriented} HSS configuration 
 compared to the best previous data placement technique.} 
Our evaluation using an HSS configuration with three different storage devices shows that Sibyl outperforms \gonn{the state-of-the-art} \gonzz{data placement} 
policy by 23.9\%-48.2\%\gonn{,} while significantly reducing the system architect's burden in designing a data placement mechanism that can simultaneously incorporate three storage devices. We show that Sibyl  achieves 80\% of the performance of an {oracle} policy that has \gon{complete} knowledge~of~future access patterns while incurring \gon{a \gonn{very modest} storage}  overhead of only \gonzz{124.4} KiB. 

We conclude that the mechanisms proposed by this dissertation provide promising solutions to handle data well by following a \textit{data-centric} approach and further demonstrates the importance of leveraging data to devise \textit{data-driven} policies.


\section{Outlook and Future Directions}

Throughout this dissertation, we considered data as a paramount resource and provided various mechanisms to effectively handle and leverage the vast amount of data. We identify topics for future research both in data-centric computing and data-driven optimization techniques.

\subsection{Data-Centric Computing}
In data-centric computing, we ensure that our computing systems are not overwhelmed by data. There are two different approaches to enable data-centric computing. First, near-memory computing (\acrshort{nmc})~\cite{nair2015active,teserract,7446059,ahn2015pim,hsieh2016accelerating,tom,ke2020recnmp,gomezluna2021upmem,cali2020genasm,fernandez2020natsa,singh2020nero} adds processing capabilities close to the existing memory architectures. Second, computation-in memory (\acrshort{cim})~\cite{seshadri2013rowclone,chang2016low,rezaei2020nom,gao2019computedram,seshadri2019dram,seshadri2017ambit,li2016pinatubo,computecache,li2017drisa} exploits the memory architecture and intrinsic properties of emerging technologies to perform operations using memory itself. We identify the following key topics for future research that
we regard as essential to unlock the full potential of data-centric computing.

\noindent\textbf{Architecture:} It is unclear which emerging memory technology best supports data-centric computing; for example, much research is going into new 3D stacked DRAM and non-volatile memories such as PCM and ReRAM. The future of these new technologies relies heavily on advancements in endurance, reliability, cost, and density. Moreover, these memories have different memory attributes, such as latency, bandwidth, cost, and energy consumption. We believe a hybrid approach of complementing two different technologies, as we show in Chapter~\ref{chapter:sibyl}, can revolutionize our current systems. Huang~\etal~\cite{qian2015study} evaluate one such architecture that tightly integrates CPU, DRAM, and a flash-based NVM to meet the memory needs of big data applications, i.e., larger capacity, smaller delay, and wider bandwidth. Processing near-heterogeneous memories is a new research topic with high potential (could provide the best of both worlds), and in the future, we expect much interest in this direction. \gthesis{Further, we see a trend towards processing near caches~\cite{computecache,neuralcache2018}, and we expect more works in this direction. The right level of cache where to place the processing unit would be an interesting future direction. }

The interplay of NMC units with the emerging interconnects standards like CXL~\cite{cxlwhitepaper}, CCIX~\cite{ccixwhitepaper}, and CAPI~\cite{openCAPI} could be vital in improving the performance and energy efficiency of big data workloads running on NMC enabled servers. More quantitative exploration is required for interconnecting networks between the near-memory compute units and between the host and near-memory compute systems.  Besides the design of the near-memory computing device itself, the integration of such architectures into the overall computing system and how multiple near-memory computing devices can work together to scale to larger data volumes are critical challenges to solve.

\noindent\textbf{Software:} Most of the evaluated architectures focus on the compute aspect. Few architectures focus on providing coherency and virtual memory support. As highlighted in Section~\ref{sec:nmc_review/challenges}, lack of coherency and virtual memory support makes programming difficult and obstructs the adoption of the NMC paradigm. 
 At the application level, algorithms need to provide code and data co-location for efficient processing. For example, in NMC, algorithms should prevent excessive movement of data between vaults (as in 3D stacked memory, see Figure~\ref{fig:nmc_review/HMCLayout}) and across different memory modules. Whenever it is not possible to avoid an inter-vault data transfer, we should provide light-weight data migration mechanisms. 
 


\noindent\textbf{Tool support and benchmarks:} The field requires a generic set of open-source tools and techniques for these novel systems. Often researchers have to spend a significant amount of time and effort in building the needed simulation environment. Application characterization tools should add support for static and dynamic decision support for offloading processing and data to near-memory systems. NMC-specific metrics are required to assist in the offloading decision to assess whether an application is suitable for these architectures. These tools could provide region-of-interest (or hotspots) in an application that should be offloaded to an NMC system. Besides, a standard benchmark set is missing to gauge different architectural proposals in this domain.

\subsection{Data-Driven System Optimization}

We are seeing an enormous amount of data being generated in different applications domains. Rather than discarding the available data, our machines should leverage this data to understand inherent characteristics or patterns to make better architectural decisions or predictions. To this end, machine learning-based approaches provide an attractive tool that can assist our computer architecture in various aspects. In this dissertation, we use this tool for performance and energy prediction to reduce the time for simulation overhead (Chapter~\ref{chapter:napel}), resource and application performance prediction to overcome the disadvantages of the slow FPGA downstream mapping process (Chapter~\ref{chapter:leaper}), effective data-placement (Chapter~\ref{chapter:sibyl}), and perform design space exploration for the fixed-point precision (Appendix~\ref{chapter:precise_fpga}). These mechanisms are likely just the beginning of a paradigm shift in computer architecture design and use.  Below we highlight some aspects of computer architecture where data-driven mechanisms can greatly assist us.

\noindent \textbf{Architecture:} Past works have demonstrated the applicability of using supervised learning-based techniques in hardware such as prefetching~\cite{liao2009machine}, branch prediction~\cite{jimenez2001dynamic}, and cache replacement~\cite{shi2019applying}. However, hardware feasibility is one of the fundamental challenges in the adoption of these techniques. In the future, we need to make these approaches hardware friendly with low cost and area overhead by using techniques such as pruning, quantization, model compression, etc. 

\noindent \textbf{Software:} The operating system is a ripe place to use ML-based approaches as  software-based mechanisms can tolerate higher latency and implementation overhead than latency-critical aspects of computer architecture such as prefetching and cache-replacement. We can apply modern ML techniques, such as deep reinforcement learning and natural language processing, on problems like software caching, task scheduling, power management, virtual memory management, etc. 

\noindent \textbf{Explainability:} Explainability is providing an explanation of the internal mechanisms of a machine learning model that led the model to a certain prediction or decision. The machine learning community has started to look into the explainability to understand these models' inner workings. However, we still use these tools as mere black-box models in the computer architecture community without making them as white-box models. Explainability can lead to certain insights that have been unknown to our current community, and learn from it to develop better human-driven policies. Since implementing ML algorithms in the hardware can be costly, leveraging insights from ML algorithms can help us better optimize our computing systems.\\

We hope that the ideas, analyses, methods, and techniques presented in this dissertation will enable the development of energy-efficient data-intensive computing systems and drive the exploration of new mechanisms to improve the performance and energy efficiency of future computing systems.





\renewcommand\bibname{Bibliography}

\appendix

\chapter[\texorpdfstring{Review of Near-Memory Data-Centric Architectures}{Review of Near-Memory Data-Centric Architectures}
]{\chaptermark{header} Review of Near-Memory Data-Centric Architectures}
\chaptermark{}
\label{chapter:appendixA}
In the past, many near-memory computing (NMC) architectures have been proposed and/or designed. This appendix summarizes a large selection of NMC architectures that are either targeting main memory (Appendix~\ref{sec:nmc_review/procmainmem}) or storage class memory (Appendix~\ref{sec:nmc_review/procstorage}). We couple different memory technologies with three broad classes of processing units: programmable unit, fixed-functional unit, and reconfigurable unit. The classification and evaluation metrics are detailed in Section~\ref{sec:nmc_review/classificationmodel}.
\section{Processing Near-Main Memory}
\label{sec:nmc_review/procmainmem}
Processing near-main memory can allow us to reduce the data movement bottleneck by circumventing memory-package pin-count limitations.  We describe some of the notable architectures that process close to main memory. All solutions discussed in this section are summarized in Table~\ref{tab:nmc_review/arcClassification}.

\subsection{Programmable Unit}
\label{subsec:nmc_review/programmableunitmm}

\textbf{NDC (2014)} Pugsley~\etal~\cite{6844483} focus on Map-Reduce workloads, characterized by localized memory accesses and embarrassing parallelism. The architecture consists of a central multi-core processor connected in a daisy-chain configuration with multiple 3D-stacked memories. Each memory houses many ARM cores that can perform efficient memory operations without hitting the memory wall. However, they were not able to fully exploit the high internal bandwidth provided by HMC. The NMC processing units need careful redesigning to saturate the available bandwidth.

\textbf{TOP-PIM (2014)} Zhang~\etal\cite{zhang2014top} propose an architecture that consists of an accelerated processing unit (APU) coupled with 3D-stacked memory. Each APU consists of a GPU and a CPU on the same silicon die. The authors focus on providing code portability with ease-of-programmability.  The kernels analyzed span from graph processing to fluid and structural dynamics. However, the authors use traditional coherence mechanisms based on restricted memory regions that restrict data placement restriction.

\textbf{AMC (2015)} Nair~\etal\cite{nair2015active} develop an architecture called active memory cube (AMC), which is built upon the HMC-based memory. They add several processing elements to the vault of HMC, which they refer to it as \emph{lanes}. Each lane has a computational unit, a dedicated register file, and a load/store unit that performs read and write operations to a dedicated part of AMC memory. The host processor coordinates the communication between AMCs. However, such an approach requires low communication overhead to avoid performance degradation and undoing benefits of processing near-memory.

\textbf{PIM-enabled (2015)} Ahn~\etal~\cite{ahn2015pim} leverage an existing programming model so that the conventional architectures can exploit the NMC concept without changing the programming interface. They add compute-capable commands and specialized instructions to trigger the NMC computation. NMC processing units are composed of computation logic (e.g., adders) and an SRAM operand buffer. The authors place these processing units in the logic layer of an HMC-based memory. Offloading at the instruction level, however, could lead to significant overhead. In addition, the proposed solution requires substantial changes on the application side, hence reducing application readiness and can be a hurdle for wide adoption.

\textbf{TESSERACT (2015)} Ahn~\etal~\cite{teserract} focus on graph processing applications.
Their architecture consists of a host processor connected to an HMC-based memory. Each HMC vault has an out-of-order processor mapped to it. These cores can see only their local data partition, but they can communicate with each other using a message-passing protocol. The host processor has access to the entire address space of the HMC. The authors also use prefetching of data to exploit the high available memory bandwidth in their systems. However, the performance benefits largely depend upon the efficient distribution of graphs to the vaults in HMC-based memory, which the authors do not consider in this work.

\textbf{TOM (2016)} Hsieh~\etal~\cite{tom} propose an NMC architecture consisting of a host GPU interconnected to multiple 3D-stacked memories with small, light-weight GPU cores. They develop a compiler framework that automatically identifies possible offloading candidates. Code blocks are marked as beneficial to be offloaded by the compiler if the saving in memory bandwidth during the offloading execution is higher than the cost to initiate and complete the offload. A runtime system takes the final
decision as to where to execute a block. Furthermore, the framework uses a mapping scheme that ensures data and code co-location. The cost function proposed for code offloading makes use of static analysis to estimate the bandwidth saving. However, static analysis may fail on code with indirect memory accesses.

\textbf{Pattnaik\footnote{Architecture has no name, first author's name is shown.\label{noAuthor}} (2016) }Pattnaik~\etal~\cite{7756764}, similar to~\cite{tom}, develop an NMC-assisted GPU architecture. An affinity prediction model decides where to executed a kernel while a scheduling mechanism tries to minimize the application execution time. The scheduling mechanism can overrule the decision made by the affinity prediction model. The author proposes to invalidate the L2 cache of the host GPU after each kernel execution to keep memory consistency between the host GPU and the near-memory cores.  However, their affinity prediction model is trained for a specific set of applications and architecture that would have low prediction performance for a new, unknown application or architecture. 

\gcheck{
\textbf{MONDRIAN (2017)} Drumond~\etal~\cite{de2017mondrian} demonstrate that a hardware/software co-design approach is required to achieve efficiency and performance for NMC systems. In particular, they show that the current optimization of data-analytic algorithms heavily relies on random memory accesses while the NMC system prefers sequential memory accesses to saturate the huge bandwidth available. Based on this observation, the authors propose an architecture that consists of a mesh of HMC with tightly connected ARM cores in the logic layer.}

\textbf{MCN (2018)} Alian~\etal~\cite{alian2018application} use a light-weight near-memory processing unit in the buffered DRAM DIMM. The memory channel network (MCN) processor runs an OS with network software layers essential for running a distributed computing framework. The most striking feature of MCN is that the authors demonstrate unified near-data processing across various nodes using ConTutto FPGA~\cite{ConTutto_2017_MICRO} with IBM POWER8. However, supporting the entire TCP/IP stack on the near-memory accelerator requires a complex accelerator design. Current trends in the industry, however, are pushing for a simplified accelerator design shifting the complexity on the host core's side~\cite{openCAPI}).

\textbf{DNN-PIM (2018)} Liu~\etal~\cite{liu2018processing} propose heterogeneous NMC architecture
for the training of deep neural network (NN) models. The logic layer of 3D-stacked memory comprises programmable ARM cores and large fixed-function units (adders and multipliers). They extend the OpenCL programming model to accommodate the NMC heterogeneity. Both fixed-function NMC and programmable NMC appear as distinct compute devices. A runtime system dynamically maps and schedules NN
kernels on a heterogeneous NMC system based on online profiling of NN kernels.

\textbf{Boroumand\textsuperscript{\ref{noAuthor}} (2018)} Boroumand \etal\cite{googleWorkloads} evaluate NMC architectures for
Google workloads. They observe that: (1) many Google workloads spend a considerable amount of energy on data movement; (2) simple functions are responsible for a significant fraction of data movement. Based on their observations, they propose two NMC architectures, one with a general-purpose processing core and the other with a fixed-function accelerator coupled with HBM-based memory. While accelerating the Google workloads, the authors take into account the low area and power budget in consumer devices. They evaluate the benefits of the proposed NMC architectures.

\textbf{FIMDRAM (2021)} Kwon~\etal~\cite{fimdram2021ISSCC} make the following two observations. First, ML has become a driving factor in the advancement of various technologies. Second, the improvements in DRAM bandwidth could only be delivered with a significant increase in power consumption. Based on the above observations, the authors' goal is to develop an energy-efficient DRAM architecture for ML-based workloads that can deliver high bandwidth with low power consumption. To this end, the authors design an NMC architecture with 16-way floating-point16 (FP16) programmable compute units with an access granularity as the host processor, which simplifies various system-level aspects such as address mapping and interleaving. The experimental results demonstrate that FIMDRAM can provide $2\times$ performance and 70\% less energy consumption than a GPU+HBM implementation for DeepSpeech2 workload. FIMDRAM shows promising results for ML-based workloads, however, the efficacy for other data-intensive workloads such as weather prediction, genomics, and databases is still an open question.

\subsection{Fixed-Function Unit}
\label{subsec:nmc_review/ffunit}

\textbf{JAFAR (2015) }Xi~\etal\cite{xi2015beyond} embed an accelerator in a DRAM module to
implement database select operations. The key idea is to use the near-memory accelerator to scan and filter data directly in the memory while only the relevant data will be moved to the host CPU. Thus, having a significant reduction in data movement. The authors suggest using memory-mapped registers to read and write via application program interface (API) function calls to control the accelerator. Even though JAFAR shows promising potential in database applications, its evaluation is quite limited as it can handle only filtering operations. More complex operations fundamental for the database domain such as sorting, indexing and compression are not considered.

\textbf{IMPICA (2016)} Hsieh~\etal~\cite{hsieh2016accelerating} accelerate pointer chasing operations, ubiquitous in data structures. They propose adding specialized units that decouple address generation from memory accesses in the logic layer of 3D-stacked memory. These units traverse through the linked data structure in memory and return only the final node found to the host CPU. They also propose to completely decouple the page table of IMPICA from the host CPU to avoid virtual memory-related issues. The
memory coherence is assured by demarking different memory zones for the accelerators and the host CPU. This design provides a state-of-the-art technique for address translation in NMC for pointer chasing.

\textbf{Vermij\textsuperscript{\ref{noAuthor}} (2017)} Vermij~\etal~\cite{vermij2017sorting} propose a system for sorting algorithms where phases having high temporal locality are executed on the host CPU, while algorithm phases with poor temporal locality are executed on an NMC device. The architecture proposed consists of a memory technology-agnostic controller located at the host CPU side and a memory-specific controller tightly coupled with the NMC system. The NMC accelerators are placed in the memory-specific controllers and are assisted by an NMC manager. The NMC manager also supports cache coherency, virtual memory management, and communications with the host processor.

\textbf{GraphPIM (2017)} Nai~\etal~\cite{nai2017graphpim} map graph workloads in the HMC by exploiting its inherent atomic\footnote{An atomic instruction such as compare-and-set is an indivisible instruction where the CPU is not interrupted when performing such operations.} functionality. As they focus on atomics, they can offload at instruction granularity. Notably, they do not introduce new instructions for NMC and use the host instruction set to map to NMC atomics through an uncacheable memory region.  Similar to~\cite{ahn2015pim}, offloading at instruction granularity can have significant overhead. Besides, the mapping to NMC atomics instruction requires the graph framework to allocate data on particular memory regions via custom \textit{malloc}. This custom allocation requires changes on the application side, reducing the application readiness.

\textbf{GRIM-Filter (2018)} Kim~\etal~\cite{kim2018grim} develop a seed location filter for the read mapping stage of the genomics pipeline that exploits the high memory bandwidth and near-memory processing capabilities of 3D-stacked DRAM to improve the performance of DNA read mappers.  The authors demonstrate a $5.6\times-6.4\times$ lower false-negative rate with end-to-end performance improvement of $1.8\times-3.7\times$ over a state-of-the-art DNA read mapper. GRIM-Filter targets only the
short reads (i.e., segments on the order of several hundred base pairs long), while accelerating long reads is an important problem~\cite{whylongread_2018}.

\textbf{RecNMP (2020)} Ke~\etal~\cite{ke2020recnmp} overcome the memory bottleneck of personalized recommendation systems. A personalized recommendation is a fundamental part of services like a search engine, social network, etc. After characterizing production-recommendation models, the authors observe that the recommendation model's sparse embedding step leads to a memory bottleneck. To this end, the authors propose RecNMC, which
uses light-weight processing cores to accelerate embedding operations in the DIMM of existing standard DRAM. RecNMP accelerates the shows $9.8\times$ memory latency speedup and $45.9\%$ memory energy savings.

\textbf{GenASM (2020)} Cali~\etal~\cite{cali2020genasm} accelerate the \emph{approximate string matching }(ASM) step of genome analysis. ASM is a computationally-expensive step as it usually uses dynamic programming (DP)-based algorithms. The authors modify the underlying ASM algorithm to increase its parallelism and reduce its memory footprint significantly. Their accelerator design follows a systolic-array-based architecture that they place in the logic layer of HMC-based memory. The authors demonstrate the flexibility of their accelerator design in accelerating multiple steps of genome analysis.

\textbf{NATSA (2020)} Fernandez~\etal~\cite{fernandez2020natsa} design a near-memory accelerator for time series analysis. Specifically, the authors implement a matrix profile algorithm that has a low arithmetic intensity and operates on a large amount of data. As a result, this algorithm is memory-bound and performs poorly on multi-core systems. NATSA's processing element consists of floating-point arithmetic units that they place next to HBM-based 3D stacked memory.
NATSA improves performance by $9.9\times$ on average and reduces energy by $19.4\times$ on average over a multi-core implementation. However, it considers HBM as a stand-alone unit without a host-system. Such a scenario leads to practical concerns such as processing orchestration and data-mapping.

\subsection{Reconfigurable Unit}
\label{subse:nmc_review/reconfigunit}
\textbf{Gokhale\textsuperscript{\ref{noAuthor}} (2015)} Gokhale~\etal~\cite{gokhale2015near} propose to place a data rearrangement engine (DRE) in the logic layer of the HMC to accelerate data accesses while still performing the computation on the host CPU. The authors target cache unfriendly applications with high memory latency due to irregular access patterns, e.g., sparse matrix multiplication.  Each of the DRE engines consists of a scratchpad, a simple controller processor, and a data mover. To make use of the DRE units, the
authors develop an API that supports several operations. Each operation is issued by the main application running on the host and served by a control program loaded by the OS on each DRE engine. Similar to~\cite{7756764}, the authors propose to invalidate the CPU caches after each fill and drain operation to keep memory consistency between the near-memory processors and the main CPU. This approach can introduce a significant overhead. Furthermore, the synchronization mechanism between the CPU and the near-memory processors is based on polling. Therefore, the CPU wastes clock cycles waiting for the near-memory accelerator to complete its operations. On the other hand, a light-weight synchronization mechanism based on interrupts could be a more efficient alternative.

\textbf{HRL (2015)} Gao~\etal~\cite{7446059} propose a reconfigurable logic architecture called heterogeneous reconfigurable logic (HRL) that consists of three main blocks: fine-grained configurable logic blocks (CLBs) for control unit, coarse-grained functional units (FUs) for basic arithmetic and logic operations, and output multiplexer blocks (OMBs) for branch support. Each memory module follows HMC like technology and houses multiple HRL devices in the logic layer. The central host processor is responsible for data partition and synchronization between NMC units. As in the case of~\cite{zhang2014top}, to avoid consistency issues and virtual-to-physical translation, the authors propose a memory-mapped non-cacheable memory region that puts restrictions on data placement.

\textbf{NDA (2015)} Farmahini~\etal~\cite{7056040} propose three different NMC architectures using coarse-grained reconfigurable arrays (CGRA) on commodity DRAM modules. This proposal requires minimal change to the DRAM architecture. However, programmers should identify which code would run close to memory. This dependence leads to increased programmer effort for demarking compute-intensive code for execution. Also, it does not support direct communication between NMC stacks.


\section{Processing Near-Storage Class Memory}
\label{sec:nmc_review/procstorage}
NAND flash-based non-volatile memories (NVM) are trying to fill the latency gap between DRAMs and disks are termed as storage-class memories (SCM)~\cite{7151782}. SCM, like NVRAM, is even touted as a future replacement for DRAM~\cite{ranganathan2011microprocessors}. Moving computation in SCM has some of the similar benefits to DRAM concerning savings in bandwidth, power, latency, and energy but also because of the higher density it allows to work on much larger data-sets as compared to DRAM~\cite{quero2015self}.

\subsection{Programmable Unit}
\label{subsec:nmc_review/programmableunitscm}

\textbf{XSD (2013)} Cho~\etal~\cite{cho2013xsd} propose a solid-state drive (SSD)-based architecture that integrates graphics processing unit (GPU) close to the memory. They provide an API based on the MapReduce framework that allows users to express parallelism in their application and exploit the parallelism provided by the embedded GPU. They develop a performance model to tune the SSD design. The experimental results show that the proposed XSD is approximately $25\times$ faster than an SSD model incorporating a high-performance embedded CPU. However, the host CPU instruction set architecture (ISA) needs to be modified to launch the computation on the GPU embedded inside the SSD.

\textbf{WILLOW (2014)} Seshadri~\etal~\cite{seshadri2014willow} propose a system that has programmable processing units referred to as storage processor units (SPUs). Each SPU runs a small operating system (OS) that maintains and enforces security. On the host-side, the Willow driver creates and manages a set of objects that allow the OS and applications to communicate with SPUs. The programmable functionality is provided in the form of \textit{SSD Apps}. Willow enables programmers to augment and extend SSD semantics with application-specific features without compromising file system protection. The programming model based on remote procedure call (RPC) supports the concurrent execution of multiple SSD Apps and trusted code execution. However, it neither supports dynamic memory allocation nor allows users to load their tasks to run on the SSD dynamically.

\textbf{SUMMARIZER (2017)} Koo~\etal~\cite{koo2017summarizer} design APIs that can be used by the host application to offload filtering tasks to the inherent ARM-based cores inside an SSD processor. This approach reduces the amount of data transferred to the host and allows the host processor to work on the filtered result. They evaluate static and dynamic strategies for dividing the work between the host and SSD processor. However, sharing the SSD controller processor for user applications and SSD firmware can lead to performance degradation due to interference between I/O tasks and in-storage compute tasks.

\textbf{CompStor (2018) } Torabzadehkashi~\etal~\cite{torabzadehkashi2018compstor} propose an architecture that consists of NVMe over PCIe SSD and FPGA-based SSD controller coupled with in-storage processing subsystem (ISPS) based on the quad-core ARM A53 processor.  They modify the SSD controller hardware and software to provide high bandwidth and low latency data path between ISPS and the flash media interface. Fully isolated control and data paths ensure concurrent data processing and storage functionality without degradation in either one's performance. The architecture supports porting a Linux operating system. However, homogeneous processing cores in the SSD are not sufficient to meet the requirement of complex modern applications~\cite{torabzadehkashi2019catalina}.

\subsection{Fixed-Function Unit}
\label{subsec:ffunit2}

\textbf{Smart SSD (2013)} Kang~\etal~\cite{kang2013enabling} implement the Smart SSD features in the firmware of a Samsung SSD and modify the Hadoop core and MapReduce framework to use task-lets as a map or a reduce function. To evaluate the prototype, they used a micro-benchmark and log analysis application on both the device and the host. Their SmartSSD is able to outperform host-side processing drastically by utilizing internal application parallelism. Likewise, Do~\etal~\cite{park2014query} extend Microsoft SQL Server to offload database operations onto a Samsung Smart SSD. The selection and aggregation operators are compiled into the firmware of the SSD. Quero~\etal~\cite{quero2015self} modify the Flash Translation Layer (FTL) abstraction and implement an indexing algorithm based on the B++tree data structure to support sorting directly in the SSD. The approach of modifying the SSD firmware to support the processing of certain functions is fairly limited and can not support the wide variety of workloads~\cite{torabzadehkashi2019catalina}.

\textbf{ProPRAM (2015)} Wang~\etal~\cite{wang2015propram} observe that NVM is often supporting built-in logic like for data comparison, write or flip-n-write module. Therefore, the authors propose to exploit the existing resources inside NVM-based memory chips to accelerate the simple non-compute intensive functions in emerging big data applications. They expose the peripheral logic to the application stack through ISA extension. Like~\cite{kang2013enabling,park2014query,quero2015self}, this approach cannot support the diverse workload requirements.

\textbf{BISCUIT (2016)} Gu~\etal~\cite{gu2016biscuit} present a near-memory computing framework that allows programmers to write a data-intensive application to run in a distributed manner on the host and the NMC-capable storage system. The storage hardware incorporates a pattern matcher IP designed for NMC.  The authors evaluate their approach by accelerating MySQL-based application. However, the NMC system acts as a slave to the host CPU, and all data is controlled by the host CPU. Therefore, this architecture needs to manage communications messages from the host effectively.

\subsection{Reconfigurable Unit}
\label{subse:reconfigunit2}

\textbf{BlueDBM (2015) }Jun~\etal~\cite{jun2015bluedbm} present a flash-based platform, called BlueDBM, built of flash storage devices augmented with an application-specific FPGA-based in-storage processor. The data-sets are stored in the flash array and are read by the FPGA accelerators. Each accelerator implements a variety of application-specific distance comparators used in the high-dimensional nearest-neighbor search algorithms. They also use the same architecture exploration platform for graph analytics. However, the platform does not support dynamic task loading, similar to ~\cite{seshadri2014willow}, and has limited OS-level flexibility. 

\textbf{CARIBOU (2017) }Zsolt~\etal~\cite{istvan2017caribou} enable key-value store interface over TCP/IP socket to the storage node comprising of an FPGA connected to DRAM/NVRAM. They implement selection operators in the FPGA, which are parameterizable at runtime, both for structured and unstructured data, to reduce data movement and avoid the negative impact of near-data processing on the data retrieval rate. Like~\cite{jun2015bluedbm}, Caribou does not have OS-level flexibility, e.g., the file-system is not supported transparently.

\renewcommand{\namePaper}{PreciseFPGA\xspace} 
\chapter[\texorpdfstring{\namePaper: Low Precision Accelerator Search for FPGA}{\namePaper: Low Precision Accelerator Search for FPGA}
]{\chaptermark{header} \namePaper: Low Precision Accelerator Search for FPGA}
\chaptermark{\namePaper}
\label{chapter:precise_fpga}
\vspace{-0.2 cm}

In Chapter~\ref{chapter:low_precision_stencil}, we demonstrate the benefits of using lower precision datatypes on an FPGA. However, determining the optimal fixed-point precision in terms of accuracy and power on an FPGA is challenging because of the following two reasons: (1) the design space of choices makes it infeasible to perform an exhaustive manual-tuning; and (2) FPGAs have a slow downstream mapping process.  To this end, this appendix presents {\namePaper}, our preliminary work on building an automated solution for fixed-point precision evaluation for FPGA-based devices, which overcomes the aforementioned issues. 

\section{Introduction}

Fixed-point data types have been used for energy-efficient computing in various applications~\cite{guptaetal,cnn_approx,8bit_approx}. For a fixed-point configuration, the impact on accuracy, throughput, power, and resource utilization depends largely on its chosen configuration. A random selection of a fixed-point configuration can have a detrimental effect on the accuracy without significant gains in terms of area and power. Therefore, application-aware fixed-point configuration selection is critical to obtain an optimal solution in terms of resource/power consumption without compromising the accuracy.

An FPGA allows us to implement any arbitrary precision fixed-point configuration. However, the number of precision options leads to a very large design space that is infeasible to explore exhaustively on an FPGA because of its time-consuming design cycle. Recently, FPGA designers have started to adopt high-level synthesis (HLS)~\cite{xilinx_hls_Design_hub} to increase productivity with reduced time-to-market. HLS converts a C/C++ program into register-transfer logic (RTL) that is further synthesized and mapped to a target FPGA. This complete downstream process can take several hours for a single design point~\cite{dai2018fast}. Hence, it is impossible to enumerate all possible fixed-point precision choices and exhaustively search for the optimal design that offers low resource consumption while considering the hardware mapping constraints.

\textbf{Our key idea} is to develop an automated framework to obtain an application-aware optimal fixed-point configuration without exhaustively searching the entire design space while taking into account the mapping constraints of an FPGA. To this end, we propose \namePaper~- a resource and power estimation framework for arbitrary fixed-point data precision. 
\namePaper~requires HLS C-Synthesis results from {only two} fixed-point configurations (see Section~\ref{subsec:preciseFPGA/func_detec}) to provide a Pareto-optimal post-implementation solution with respect to power and error. \namePaper~uses these two C-Synthesis utilization reports to obtain the relation between the operation-width of FPGA components and the fixed-point configuration. 
The extracted relation is extrapolated to obtain operation-width for all possible configurations. Compared to previous works~\cite{o2018hlspredict,Pyramid,linAnalyzer,hls_scope,kao2020confuciux}, \namePaper~provides the following key capability. \namePaper is aware of the effect of FPGA-resource saturation, e.g., in case if we saturate digital signal processing (\acrshort{dsp}) units then our framework would map operands to a lookup table (\acrshort{lut}), which makes our approach more practical. We demonstrate our approach by quantizing weights of a neural network (\acrshort{nn})-based accelerator.
\vspace{-0.15 cm}
\section{Motivation}\label{preciseFPGA/motivation}

In fixed-point representation (see Section~\ref{sec:low_precision/methodology}), 
the feasible fixed-point configuration  $<x,y>$ for bit-width $\leq N$ can be represented by Equation~\ref{preciseFPGA/eq1}, where $x$ is the total \bitwidth including $y$ integer bits.  In general, the total number of fixed-point configurations can be given by Equation~\ref{preciseFPGA/eq2}. 

\begin{equation}\label{preciseFPGA/eq1}
x \leq N \quad \forall x \in [2, N] \quad \forall y \in [1, x-1] 
\end{equation}
\vspace{-0.5 cm}
\begin{equation}\label{preciseFPGA/eq2}
 S = \frac{N(N-1)}{2}
\end{equation}

 As we are interested in reduced precision fixed-point configurations, we restrict the maximum \bitwidth to the default \bitwidth of our baseline single-precision floating-point representation, i.e., 32-bit. An exhaustive search to obtain an optimal fixed-point configuration requires post-implementation resource and power estimates for all possible configurations. Therefore, with a \bitwidth of 32, we would need 496 FPGA downstream mapping runs (using Equation~\ref{preciseFPGA/eq2}) to complete an exhaustive search. We have discussed different number systems, which include fixed-point, floating-point, and posit, in Chapter~\ref{chapter:low_precision_stencil}.

To motivate the need for \namePaper, we perform three experiments to study the behavior of an NN model with different fixed-point precision configurations.

\subsection{Effect on Power Consumption}
\begin{figure}[h!]
  \centering
\includegraphics[width=0.8\linewidth,trim={0cm 0cm 0cm 0.7cm},clip]{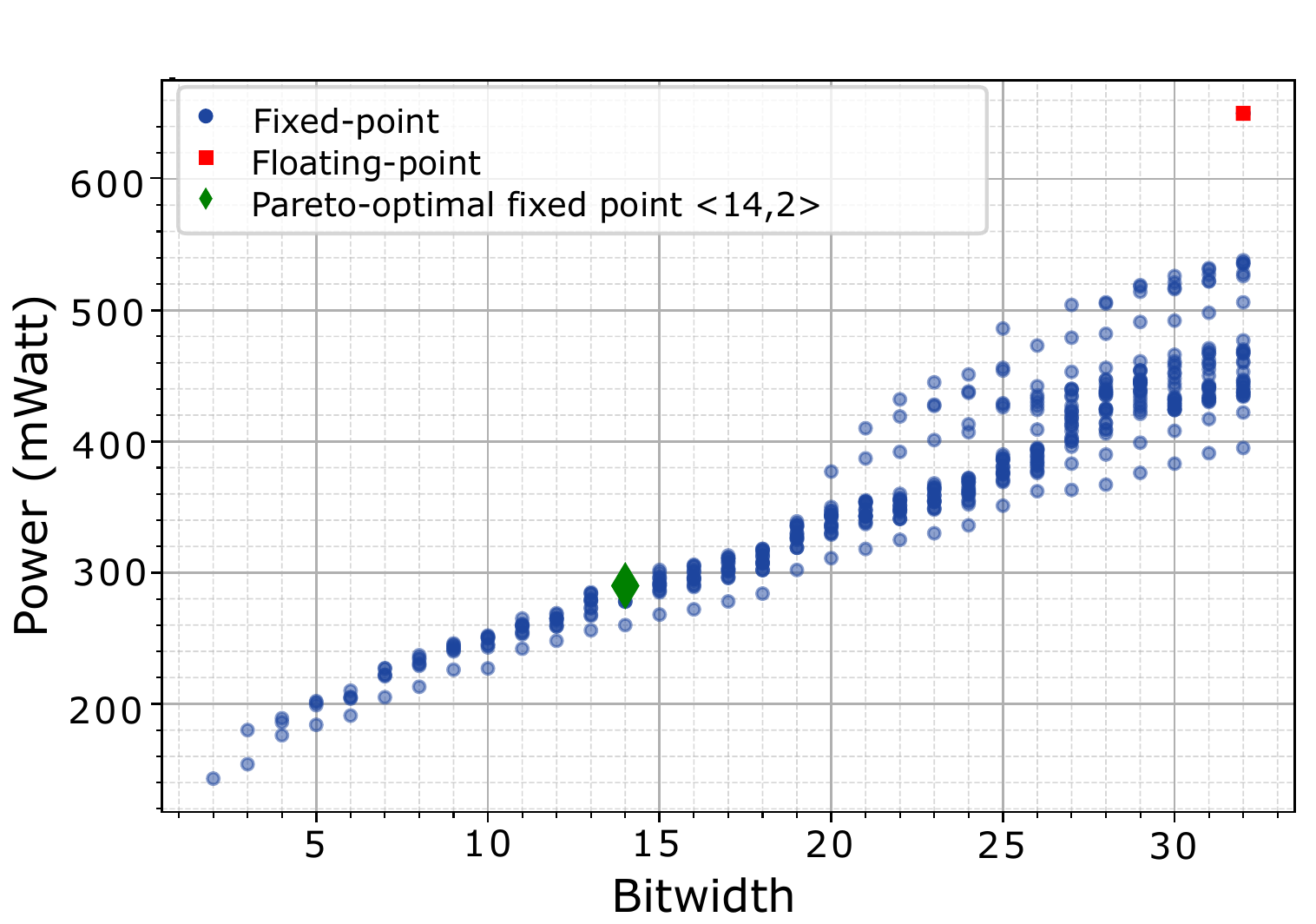}
\caption{Power consumption with change in \bitwidth for LeNet-5~\cite{lecun2015lenet, lenet_code}.}\label{preciseFPGA/fig_power}
 \end{figure}

{Figure~\ref{preciseFPGA/fig_power}} shows the power consumption of the LeNet-5~\cite{lecun2015lenet, lenet_code} convolutional  neural network (CNN) model using different \bitwidth. We analyze the impact on the post-implementation power consumption using all possible fixed-point configurations (\tikzcircle[fill=navyblue]{3pt}). We use Equation~\ref{preciseFPGA/eq1} to enumerate all possible fixed-point configuration options. We draw two conclusions from Figure~\ref{preciseFPGA/fig_power}. First, the power consumption increases with increasing \bitwidth. Second, the maximum power consumption is given by floating-point (\crule[red]{0.2cm}{0.2cm})  with a \bitwidth of 32.

\subsection{Effect on Inference Accuracy}\label{preciseFPGA/motivation_inference_accuracy_loss}
\begin{figure}[h!]
  \centering
\includegraphics[width=0.8\linewidth,trim={0cm 0cm 0cm 0.6cm},clip]{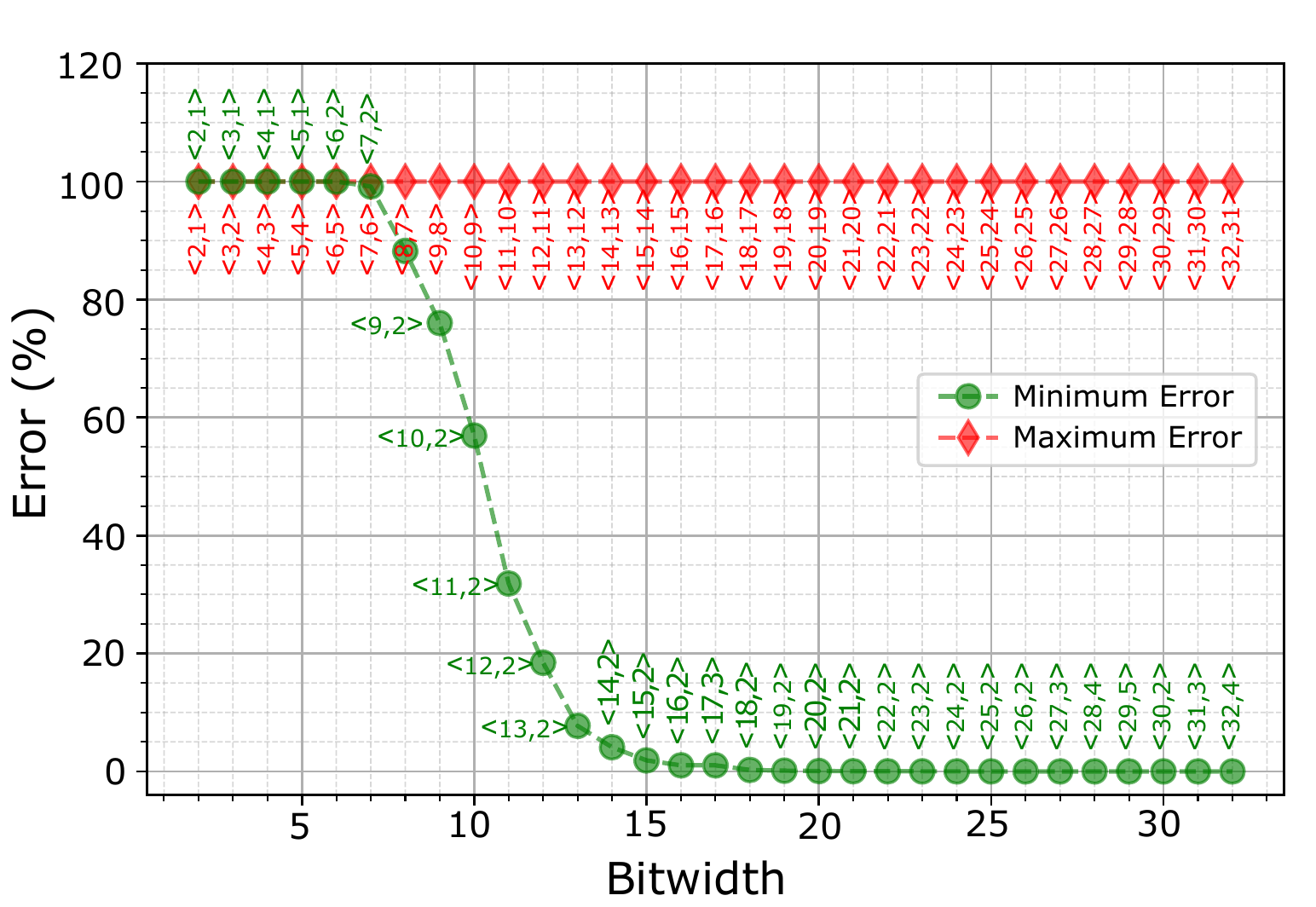}
\caption{Error in the inference accuracy of LeNet-5~\cite{lecun2015lenet, lenet_code}. To find the optimum \bitwidth that minimizes the error without a drastic increase in power consumption, we need to determine both the \integer and the \fractional part of a fixed-point configuration.}\label{preciseFPGA/fig_error}
 \end{figure}

We also studied the impact of different fixed-point configurations on inference accuracy compared to our baseline 32-bit floating-point by following the approach in Section~\ref{sec:low_precision/methodology}. We use the percentage accuracy drop of a fixed-point configuration with respect to floating-point implementation as an error metric. We calculated the error for all possible 496 fixed-point configurations using the C-simulation results.  We plot the maximum ($\mathbin{\color{red}\blacklozenge}$) and the minimum  ($\tikzcircle[fill=forestgreen]{3pt}$) error (\%) of the fixed-point configurations with respect to \bitwidth in Figure~\ref{preciseFPGA/fig_error}.
We draw two conclusions from Figure~\ref{preciseFPGA/fig_error}. First, increasing the \bitwidth only does not reduce the error. To find the right \bitwidth, we need to find both the appropriate \integer and the \fractional part. Second, increasing the \bitwidth beyond a certain point does not significantly improve accuracy but drastically increases the power consumption. For LeNet-5, we found the optimum \bitwidth, in terms of power consumption and accuracy, at $<14,2>$. 

\subsection{Design Space Exploration Time}
While designing an efficient FPGA-based implementation, our goal is to find a Pareto-optimal configuration with respect to power and accuracy. The Pareto-Optimal configuration can vary according to the power budgets and accuracy requirements of a model.  In Figure~\ref{preciseFPGA/fig_power}, we observe that the optimal configuration is obtained at $<14,2>$ ($\mathbin{\color{forestgreen}\blacklozenge}$), and using a \bitwidth beyond 14 does not add much to the accuracy (see Figure~\ref{preciseFPGA/fig_error}) but increases the power consumption linearly. This Pareto-optimal point enables $\sim$55\% savings in terms of power with only a 4\% loss in accuracy compared to a full-precision floating-point implementation. 

However, to achieve this Pareto-optimal design point, we have to perform an exhaustive search for all design options. In the case of LeNet-5 architecture, it took us nearly five days to get the post-implementation results for all possible 496 configurations. This runtime is impractical, and it prohibits us from an effective design-space exploration. Therefore, we require a fast and accurate framework to explore the optimum fixed-point configuration that does not require post-implementation results for all possible configurations. This motivated us to create \namePaper, which can predict the post-implementation resource and power consumption for fixed-point configurations using only C-Synthesis results of two fixed-point configurations.  

\vspace{-0.1 cm}
\section{\namePaper}\label{preciseFPGA/overview}
\namePaper~is a resource and power estimation framework for fixed-point precision arithmetic targeting Xilinx HLS~\cite{xilinx_hls_Design_hub} designs. \namePaper~uses two realistic assumptions. First, the change in precision does not affect the control path, which means the number of operations in the application is not varied with the change in precision. Second, the operand-width associated with the precision arithmetic changes linearly with the change in the fixed-point configuration. In this section, we describe the main components of \namePaper. First, we give an overview (Section~\ref{subsection:preciseFPGA/overview}) of our framework. Second, we mention the HLS-based features that vary with change in fixed-point configuration (Section~\ref{preciseFPGA/hls_report}). Third, we describe the two most important blocks of \namePaper: (1) function detector and feature predictor (Section~\ref{subsec:preciseFPGA/func_detec}) and (2) resource and power predictor (Section~\ref{subsec:preciseFPGA/resource_predic}). 
\vspace{-0.2 cm}
\begin{figure}[!htp]
    \centering
    \includegraphics[width=0.9\textwidth]{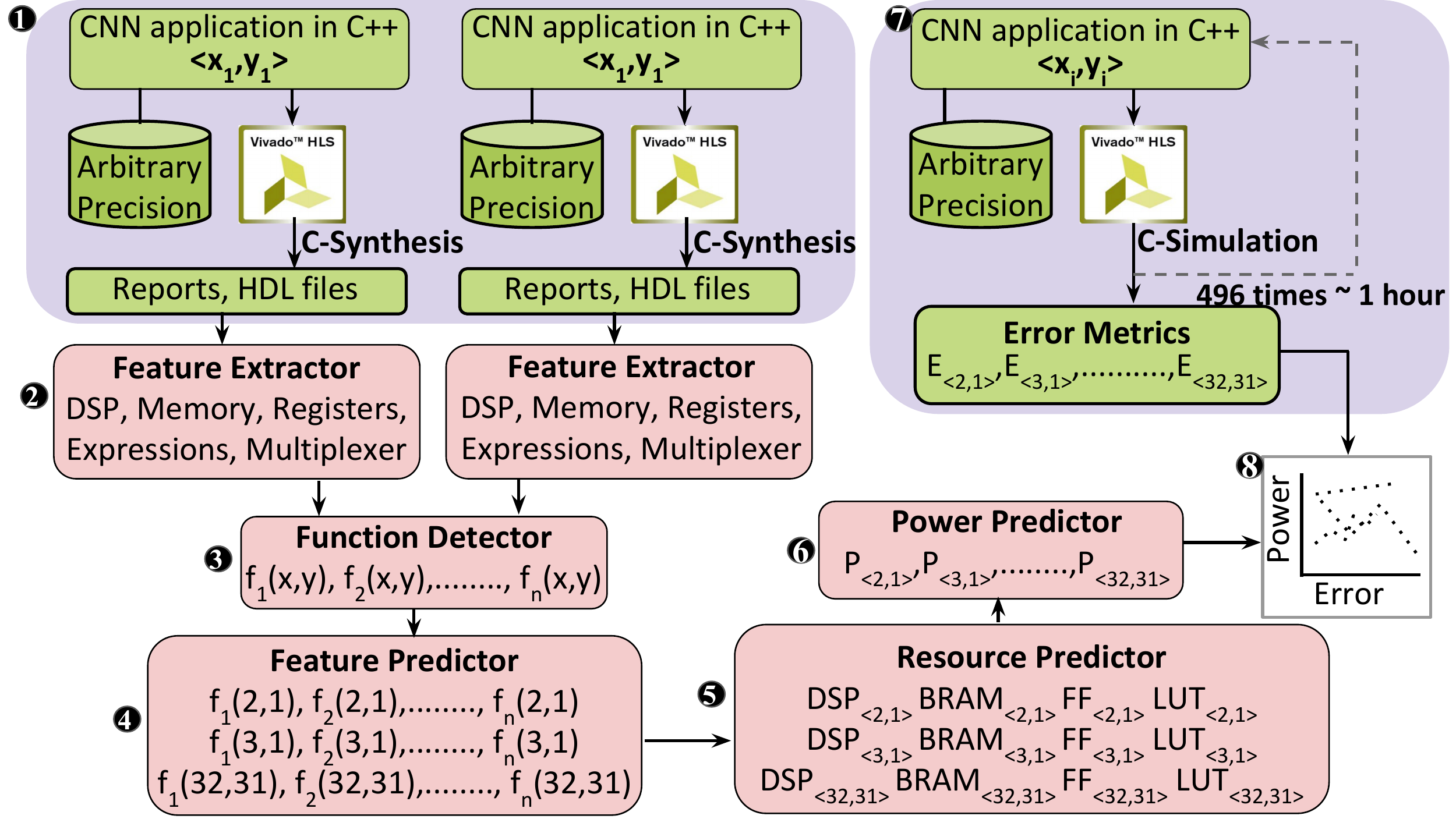}
    \caption{Overiew of \namePaper.}
    \label{preciseFPGA/block_diagram}
\end{figure}
\vspace{-0.7 cm}
\subsection{Overview}
\label{subsection:preciseFPGA/overview}
The C-Synthesis results of a C++ application obtained for two fixed-point precision configurations (\circled{1} in Figure \ref{preciseFPGA/block_diagram}) are fed through a \emph{feature extractor} block (\circled{2} in Figure \ref{preciseFPGA/block_diagram}). We extract all resource and timing-based features from the synthesis and the implementation reports. 
The \emph{function detector} (\circled{3} in Figure \ref{preciseFPGA/block_diagram}) uses the features extracted from the {feature extractor} to calculate offsets $\mathbf{\hat{x}}$, $\mathbf{\hat{y}}$, and $\mathbf{\hat{x}-\hat{y}}$ of each instance~\footnote{Instance refers to the functions or operators in a design.} for each of the two fixed-point configurations. We compare the calculated offsets of the two configurations to find a common offset function. This common function represents the relation of \bitwidth with an instance.

The \emph{feature predictor} block (\circled{4} in Figure~\ref{preciseFPGA/block_diagram}) uses the identified common functions to generate the feature values for all instances of all possible fixed-point configurations $<x_i,y_i>$. The {resource predictor} block (\circled{5} in Figure \ref{preciseFPGA/block_diagram}) uses the features generated from the {feature predictor} to predict the post-implementation resource utilization for all possible fixed-point configurations. The \emph{power predictor} block (\circled{6} in Figure \ref{preciseFPGA/block_diagram}) uses the resource estimates obtained from the resource predictor to obtain the post-implementation power prediction using a regression model called the Support Vector Regression (SVR)\cite{xyz}. The {power predictor} output along with the C-Simulation (\circled{7} in Figure \ref{preciseFPGA/block_diagram}) is used to obtain a Pareto-optimal fixed-point configuration (\circled{8} in Figure \ref{preciseFPGA/block_diagram}) with respect to power and error in an inference task. We further explain the above steps in the following sections.

\vspace{-0.1 cm}
 
\subsection{HLS-based Features} \label{preciseFPGA/hls_report}
The Xilinx HLS tool transforms a C/C++ specification into a register transfer level (RTL) implementation using a C-Synthesis step. Our framework uses the reports and the RTL source code (generated in the C-Synthesis step of the FPGA design cycle) to predict the post-implementation resource and power consumption. 

We mention the important components of the HLS report in Table-\ref{preciseFPGA/table_hls_report}. A C-Synthesis report captures the variables mapped to FPGA resources under five main categories (``Comp''). First, \textbf{Memory} represents instances mapped to block RAM (BRAM) or slice logic (i.e., FF and LUT) and is reported along with its three features, i.e., depth or words, width or bits, and banks. \gthesis{The bank depends on the memory ports.}  Second, 
\textbf{Expression (Expr)} represents the variables responsible for logical and arithmetic operations and lists their operand widths and operation type.  Third, \textbf{Multiplexer (MUX)} includes the selector instances with the number of inputs and input \bitwidth. Fourth, \textbf{DSP} describes multiply-accumulate (MAC) and multiplication (MUL) operation instances above a certain operand-width that are mapped to DSP slices. The operand-width of a DSP instance is extracted from the respective RTL source code generated from the C-Synthesis step. Fifth, \textbf{Register (Reg)} is implemented using flip-flop and stores the results of a look-up table.
\begin{table}[!htp] 
 \caption{Components and their features extracted from the C-Synthesis report.}
           \label{preciseFPGA/table_hls_report}
\centering
 \resizebox{1\linewidth}{!}{%
\begin{tabular}{ c | c | c | c | c | c |c |c|c}
\toprule
            \multicolumn{1}{c|}{\textbf{Comp}} & {\textbf{Feature1}} & {\textbf{Feature2}} &{\textbf{Feature3}} &{\textbf{Status}} &{\textbf{FF}}&{\textbf{LUT}}&{\textbf{BRAM}}&{\textbf{DSP}}\\ \hline
            Memory & depth & width & bank  &\xmark, \cmark, \xmark &\cmark &\cmark &\cmark &\xmark \\ \hline
            Expr & op1 width & op2 width & op type  &\cmark, \cmark, \xmark &\xmark &\cmark &\xmark &\cmark\\ \hline
            MUX & ip size & bits & total bits  &\xmark, \cmark, \cmark &\xmark &\cmark &\xmark &\xmark \\ \hline
            DSP & op1 width & op2 width & op3 width  &\cmark, \cmark, \cmark &\cmark &\cmark &\xmark &\cmark \\ \hline
            Reg & variable bits & const bits & N/A  &\cmark, \xmark, \xmark &\cmark &\xmark &\xmark &\xmark \\ \bottomrule
           \end{tabular}
           }
        \end{table}
        
\textbf{Precision dependent features:} 
The features expected to change with change in precision type are marked with \cmark~and others with \xmark~in the \emph{Status} column of Table-\ref{preciseFPGA/table_hls_report}. 
For example, as mentioned above, for the {Memory} component, HLS reports define memory in terms of depth, width, and banks. 
A change in the fixed-point precision would affect the memory width only because the depth depends on the array size and banks depend on the access type (i.e., a single-port or a double-port). Thus, the \emph{Status} has \xmark~for depth and bank, and \cmark~for width. In FPGA, we can use \acrshort{bram} or LUT and FF to implement the memory component. Thus, these components have a \cmark~ while DSP is marked with \xmark~as we cannot use DSP to implement a memory.

\vspace{-0.15 cm}
\subsection{Function Detector and Feature Predictor}
\label{subsec:preciseFPGA/func_detec}
%
We require C-Synthesis reports of only two  fixed-point configurations (for example $<x_1,y_1>$ and $<x_2,y_2>$) to determine if a feature is fixed or varies with a change in precision. As mentioned before, an instance that is a part of the control-flow, such as loop count, is expected to remain constant, whereas an instance involving fixed-point operands would change with a change in data width precision. 

The features extracted for an application with precision configurations $<x_1,y_1>$ and $<x_2,y_2>$ are given as an input to the {function detector} block (\circled{3} in Figure \ref{preciseFPGA/block_diagram}). The function detector uses two steps. In the first step, the offset between the feature \emph{value}\footnote{Value refers to HLS reported output of an instance's feature, which belongs to a specific component (see Table~\ref{preciseFPGA/table_hls_report}) i.e, memory, expression, multiplexer, DSP, or register. } and \texttt{bitwidth}, \texttt{integer}, and \texttt{fraction} part of a fixed-point representation for both  $<x_1,y_1>$ and  $<x_2,y_2>$ configurations is calculated separately.  This offset captures the relation between feature value and different components of the number representation, i.e., \texttt{bitwidth}, \texttt{integer}, and \texttt{fraction}. The calculated offsets are mentioned under columns $\mathbf{\hat{x}, \hat{y}}$, and $\mathbf{\hat{x}-\hat{y}}$ for each of the two configurations. We obtain these offsets by subtracting \texttt{bitwidth}, \texttt{integer}, and \texttt{fraction} part of a fixed-point representation  from the obtained \emph{Value} of an instance's feature, respectively.   

To understand the relationship between an instance of an application and the \texttt{bitwidth, integer}, and \texttt{fraction} part of a fixed-point representation, we need to ensure that the two chosen fixed-point configurations follow the conditions given by Equation~\ref{preciseFPGA/unique}
\vspace{-0.2 cm}
\begin{equation}\label{preciseFPGA/unique}
    x_{1}\not=x_{2} \quad \text{and} \quad y_{1}\not=y_{2} \quad \text{and} \quad (x_{1}-y_{1})\not=(x_{2}-y_{2})
\end{equation}
\vspace{-0.7 cm}

In the second step, the offsets obtained in the first step for two different fixed-point precision configurations are compared to find a common matching offset function, which we refer to it as \emph{function detected}. In the case of a match, the \emph{function detected} is replaced by the common offset function plus the offset's calculated value. In case there is no match, the \emph{function detected} is replaced with 1, which implies that an instance's feature is independent of the chosen data-width and, therefore, remains constant for all the fixed-point configurations.

\textbf{Example:} In Table~\ref{preciseFPGA/table_func_generate_new} we explain the above step using the instances captured from the HLS C-Synthesis report for $<17,8>$ and $<16,10>$ fixed-point configurations for LeNet-5. We report the values achieved by HLS for different features under the \emph{Value} column for each precision. In the case of the \emph{Expression} component with instance name $out\_V\_d0$, we have two features, op1-width and op2-width. First, we calculate the offsets for each feature of the \emph{Expression} component. For each fixed-point precision configuration, we mention the calculated offsets under $\mathbf{\hat{x},  \hat{y}}$, and $\mathbf{\hat{x}-\hat{y}}$. Second, we compare the offsets obtained for the two fixed-point precision configurations to find a common function. In the case of \emph{op1-width}, the offsets obtained for $<x_1,y_1>$ are different than the offsets obtained for $<x_2,y_2>$. However, for op2-width, $\mathbf{\hat{x}}$  with a value of -1 is common between the offsets of the two precision configurations. Thus, the function detected for op1-width and op2-width features are 1 and $\mathbf{\hat{x}-1}$, respectively.

\begin{table}[h]
\HUGE
 \renewcommand{\arraystretch}{0.95}
  \caption{Example of an HLS C-Synthesis report with outputs from function-detector and feature-predictor.}
   \centering
 \label{preciseFPGA/table_func_generate_new}
\setlength{\tabcolsep}{6pt}
 \resizebox{1\linewidth}{!}{%
\begin{tabular}{|c|c|c|c|c|c|c|c|c|c|c|c|c|c|c|c|c|c|}
\toprule
\multirow{2}{*}{\textbf{\begin{tabular}[c]{@{}l@{}}Comp\end{tabular}}} & \multirow{2}{*}{\textbf{\begin{tabular}[c]{@{}l@{}}Instance\\ Name\end{tabular}}} & \multirow{2}{*}{\textbf{\begin{tabular}[c]{@{}l@{}}Feature\\ Name\end{tabular}}} & \multicolumn{4}{c|}{\textbf{\textless{}x=17, y=8\textgreater{}}} & \multicolumn{4}{c|}{\textbf{\textless{}x=16, y=10\textgreater{}}} & \multirow{2}{*}{\textbf{\begin{tabular}[c]{@{}l@{}}Function \\ Detected\end{tabular}}} & \multicolumn{6}{c|}{\textbf{Feature Predicted}}                                                                                                                  \\ \cline{4-11} \cline{13-18} 
&                                                                                   &                                                                                  & \textbf{Value}    & \textbf{$\hat{x}$}   & \textbf{$\hat{y}$}   & \textbf{$\hat{x}-\hat{y}$}   & \textbf{Value}    & \textbf{$\hat{x}$}    & \textbf{$\hat{y}$}   & \textbf{$\hat{x} - \hat{y}$}   &                                                                                        & \textbf{\textless{}2,1\textgreater{}} & \textbf{\textless{}3,1\textgreater{}} & \textbf{\textless{}3,2\textgreater{}} & \textbf{\textless{}4,1\textgreater{}} & \textbf{\textless{}4,2\textgreater{}} & \textbf{\textless{}4,3\textgreater{}} \\ \hline
Memory & C2\_weights\_V\_0\_U & width   & 9       & -8   & +1   & 0   & 6   & -10      & -4     & 0       & $\hat{x} - \hat{y}$         & 1         & 2     & 1     & 3        & 2          & 1       \\ \hline
\multirow{2}{*}{Exr}      & \multirow{2}{*}{out\_V\_d0}                                 & op1 width                                                           & 1                 & -16          & -7           & -8             & 1                 & -15           & -9           & -5             & 1                                      & 1                                     & 1                         & 1                & 1      & 1         & 1                          \\ \cline{3-18} 
             &                           & op2 width                           & 16                & -1           & +8           & +7             & 15                & -1            & +5           & +9             & $\hat{x}$-1                                                                                    & 1                                     & 2                                     & 2                                     & 3                                     & 3                                     & 3                                     \\ \hline
MUX                                                                        & p\_Val2\_40\_reg                                                                  & bits                                                                             & 17                & 0            & +9           & +8             & 16                & 0             & +6           & +10            & $\hat{x}$                                                                                      & 2                                     & 3                                     & 3                                     & 4                                     & 4                                     & 4                                     \\ \hline
\multirow{2}{*}{DSP}                                                               & \multirow{2}{*}{nnet\_mac\_muladd}                                                & op1 width                                                                        & 16                & -1           & +8           & +7             & 15                & -1            & +5           & +9             & $\hat{x}$-1                                                                                    & 1                                     & 2                                     & 2                                     & 3                                     & 3                                     & 3                                     \\ \cline{3-18} 
                                                                                   &                                                                                   & op2 width                                                                        & 10                & -7           & +2           & +1             & 7                 & -9            & -3           & +1             & $\hat{x}-\hat{y}$ +1                                                                                  & 2                                     & 3                                     & 2                                     & 4                                     & 3                                     & 2                                     \\ \hline
Reg                                                                           & C2\_weights\_V\_0\_reg                                                            & bits                                                                             & 9                 & -8           & +1           & 0              & 6                 & -10           & -4           & 0              & $\hat{x}-\hat{y}$                                                                                    & 1                                     & 2                                     & 1                                     & 3                                     & 2                                     & 1                                     \\ \bottomrule
\end{tabular}
}
\end{table}

In the \emph{feature prediction} phase, we use the \emph{function detected} to find the value of an instance's feature for all possible fixed-point precision configurations. In Table~\ref{preciseFPGA/table_func_generate_new}, we demonstrate the feature prediction for six different fixed-point configurations (<2,1>, <3,1>, <3,2>, <4,1>, <4,2>, and <4,3>). For each of these configurations, to obtain the feature value of an instance, we replace <x,y> values of a configuration into the \emph{function detected}.

\subsection{Resource and Power Predictor}
\label{subsec:preciseFPGA/resource_predic}
The {resource predictor} step uses the feature values generated from the {feature predictor} (\circled{4} in Figure \ref{preciseFPGA/block_diagram}) to predict the post-implementation DSP, BRAM, FF, and LUT resource utilization for all possible fixed-point configurations. For each FPGA resource type, we employ a different estimation technique \gthesis{based on the feature complexity}. 

\noindent\textbf{BRAM:} 
Our target Xilinx FPGA board~\cite{xilinx_zynq} has block RAM resources that store up to 36-Kbits of data and can be configured as either two independent 18-Kbits RAM or one 36-Kbit RAM. Each 18-Kb block RAM can be configured as a 16K x 1, 8K x2 , 4K x 4, 2K x 9, 1K x 18 or 512 x 36 in dual-port mode. For a given  memory instance with width, depth, and banks, our BRAM resource predictor finds the possible BRAM utilization for all possible combinations of the configuration. The BRAM configuration requiring the lowest cascade depth and the lowest BRAM utilization is selected. This process is repeated for all the BRAM instances for the particular fixed-point configuration. Total BRAM count is obtained as the sum over all the BRAM instances. 

\noindent\textbf{FF and LUT}:  FF and LUT estimates vary non-linearly with the feature values. Therefore, for the prediction of the FF and LUT, we use SVR (Support Vector Machine)~\cite{xyz}, which is a simple non-linear machine learning algorithm, to map HLS estimates to post-implementation FF and LUT count. We use the calculated FF, and LUT estimates from the feature predictor step as features to train our SVM model.

\noindent\textbf{DSP:}\label{preciseFPGA/resource_prediction_dsp}
As discussed in Section~\ref{preciseFPGA/hls_report}, MAC and MUL operations above a certain threshold are mapped to the dedicated DSP slices on an FPGA. This threshold value is extracted from the RTL generated in the C-Synthesis step of the HLS design cycle.  There is a linear mapping between the number DSP slices a particular MAC, or MUL operation requires and the operand-width. Therefore, we use operand-width to estimate post-implementation DSP utilization. 

\textbf{Power Prediction:} The power prediction step (\circled{6} in Figure \ref{preciseFPGA/block_diagram}) uses the resource estimates obtained from the resource predictor to obtain the post-implementation power prediction for a certain fixed-point configuration. We again use a simple SVM~\cite{xyz} model for power prediction because of the non-linearity between the resource utilization and power consumption. We train a separate SVM-based model for each FPGA resource type, i.e., BRAM, FF, LUT, and DSP.



\vspace{-0.1 cm}
\section{Evaluation}

We evaluate our proposed method with LeNet-5~\cite{lecun2015lenet, lenet_code}-based CNN architecture. A CNN model consists of a combination of convolution layers, activation layers, fully-connected layers, and pooling layers~\cite{CNN}.   We vary the number of layers and dimensions of each layer of a CNN model to obtain 8 different NN architectures that have an impact on all the FPGA resource types, i.e., FF, LUT, BRAM, and DSP. Table-\ref{table_results} shows the number of parameters of each CNN model used in our evaluation. We use the Xilinx Vivado HLS 2019.1~\cite{xilinx_hls_Design_hub} tool to implement CNN designs targeted for the Zynq-7z020~\cite{xilinx_zynq} edge FPGA-based device. To predict the post-implementation resources and power utilization of a particular model, we use that specific model as the test set and exclude that model from the training and validation set. From the remaining set of 7 models, we randomly choose 6 models for training and 1 model for validation.


\subsection{Resource and Power Prediction}\label{results_pred}
For each model, we obtain C-Synthesis results for only two fixed-point configurations. We feed these results to our \namePaper~framework and obtain power prediction results for all the possible configurations of a model.

In Table~\ref{table_results}, we mention the total runtime and error results. The time taken to predict power and resource utilization for all possible fixed-point configurations using \namePaper~and Xilinx HLS~\cite{xilinx_hls_Design_hub} are mentioned under the columns \textit{Total Runtime (\namePaper)} and \textit{Total Runtime (HLS)}, respectively. Total time taken by \namePaper~consists of the prediction time of making predictions for all possible fixed-point configurations (496 configurations) and runtime for C-Synthesis~\cite{xilinx_hls} for two distinct precision configurations.  We calculate the \textit{speedup} in time obtained using \namePaper~ instead of the exhaustive search (using the HLS tool) in the column \textit{Total Runtime (Speedup)}. We also mention the average prediction error for resources (FF, BRAM, LUT, and DSP) and power estimates. We calculate the prediction error by averaging the relative error with respect to HLS post-implementation estimates.  

We achieve an average prediction error of $4.40\%, 4.30\%, 4.04\%,$ and $4.21\%$ for FF, LUT, BRAM, and DSP, respectively. We observe \namePaper is, on average, $674\times$ faster than an exhaustive search using HLS. For power prediction, we achieve an error of less than 4\% for all the considered models.

\begin{table}[h]
\Huge
 \centering
 \caption{Model runtime and resource prediction results.}
 \label{table_results}
 \resizebox{1\linewidth}{!}{%
 \begin{tabular}{|c|c|c|c|c|c|c|c|c|c|c|}
\toprule
 \multicolumn{1}{|c|}{}& & \multicolumn{3}{c|}{\textbf{Total Runtime (hrs)}} &  \multicolumn{5}{c|}{\textbf{Average Relative Prediction Error (\%)}}\\
 \cline{3-10}
 \multirow{-2}{*}{\textbf{Model}}  &\multirow{-2}{*}{\textbf{\# Parameters}}&{HLS}&{\namePaper} &{Speedup}&{FF}&{LUT}&{BRAM}&{DSP} &{Power}\\
\hline
1 & 16278 & 223 & 0.205 & 1087 & 3.93 & 4.93 & 3.18 & 4.73 &2.71
\\
\hline
2 & 42704 & 107 & 0.213 & 502 & 4.26 & 3.96 & 2.83 & 4.99 &3.07	
\\
\hline
3 & 14758 & 103 & 0.204 & 505 & 4.65 & 4.24 & 5.44 & 2.89 &2.63\\
\hline
4 & 14822 & 115 & 0.219 & 525 & 4.42 & 3.80 & 5.91 & 3.25 &	2.42	 \\
\hline
5 & 9222 & 111 & 0.221 & 502 & 3.54 & 4.19 & 6.53 & 3.43 &2.64 \\
\hline
6 & 36004 & 132 & 0.227 & 581 & 4.04 & 4.19 & 2.84 & 4.95 &	3.41	\\
\hline
7 & 24114 & 119 & 0.232 & 512 & 6.07 & 4.14 & 2.75 & 4.52 &3.68 \\
\hline
8 & 16498 & 239 & 0.203 & 1177 & 4.24 & 4.92 & 2.89 & 4.92 &	3.72 \\
\hline
\multicolumn{4}{|c|}{\textbf{Average}} &\textbf{673.8} &\textbf{4.40} &\textbf{4.30}		&\textbf{4.04}		&\textbf{4.21} &\textbf{3.03
}
\\
\bottomrule
 \end{tabular}%
 }
\end{table}

\subsection{Pareto Curve Generation}
\begin{figure}[h]
  \centering
  \includegraphics[width=\textwidth]{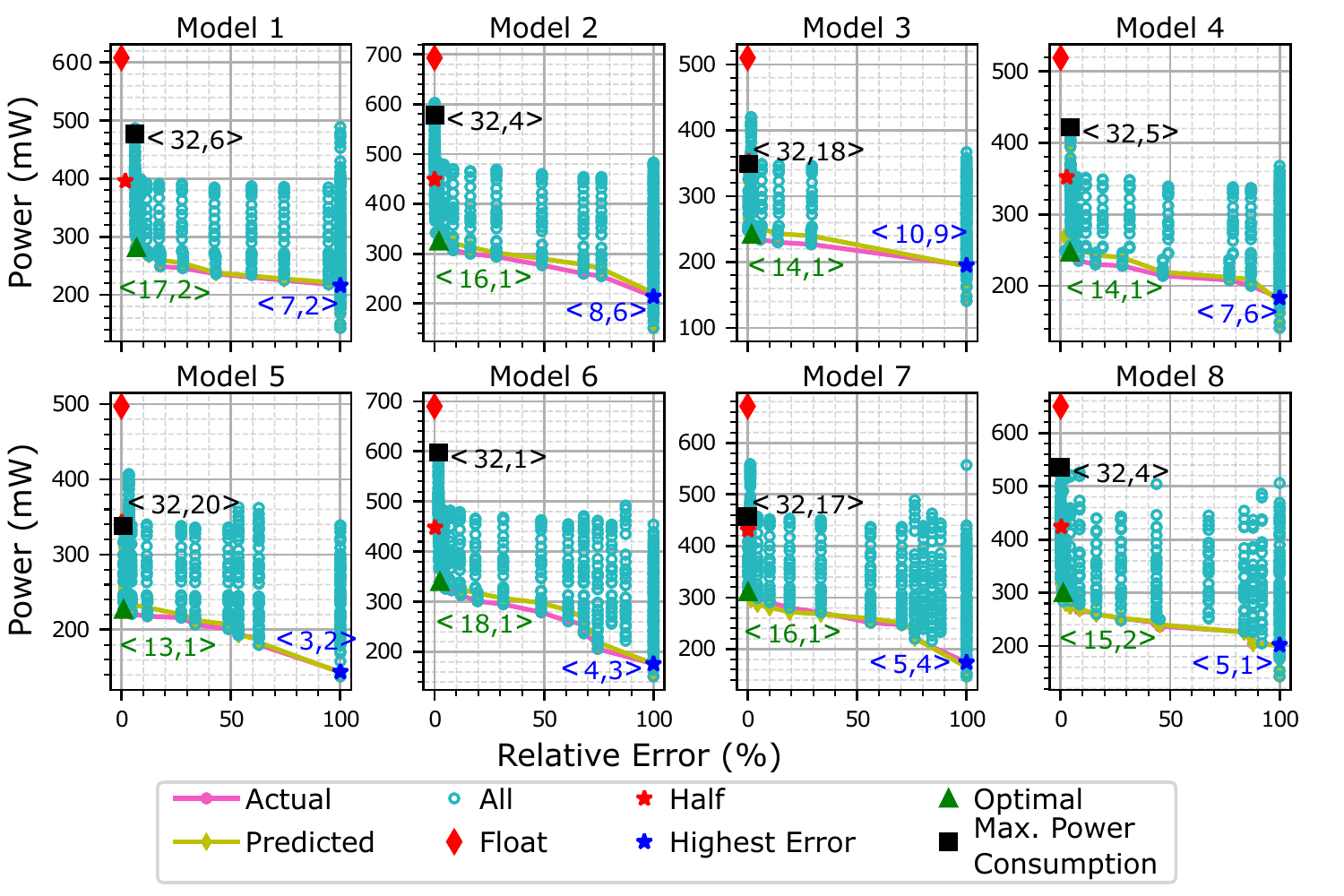}
  \vspace{-0.4 cm}
  \caption{Pareto-Optimal plot of power vs relative error for 8 different architectures of LeNeT-5-based~\cite{lecun2015lenet, lenet_code} CNN model.}
  \label{plot_pareto}
\end{figure}

We use the MNIST~\cite{mnist} dataset to obtain the error in a CNN model's inference task under test while using 32-bit floating-point baseline and 496 possible fixed-point configurations. Figure~\ref{plot_pareto} shows the plot for the predicted power using \namePaper and the relative error for each model.  For the prediction error, we calculate the change in relative accuracy of fixed-point implementation compared to the baseline floating-point implementation. The power consumption estimated by \namePaper~ and HLS tool is shown as \textit{Predicted} and \textit{Actual}, respectively. We also show the performance of baseline 32-bit floating-point (\textit{Float}) as well as 16-bit floating point (\textit{Half}) in the plot. The figure also highlights the entire search space (496 configurations) using the points labeled as \textit{All}. 

From the Pareto curve, we make three observations. First, the {\textit{Predicted} Power} faithfully follows the trend of the {\textit{Actual} Power}. Second, after attaining optimal configuration, in terms of power consumption and error, an increase in \bitwidth does not improve the error but drastically increases the power consumption. Third, the optimal configuration enables greater saving in power consumption than the floating-point implementation and higher \bitwidth fixed-point configurations. 

\vspace{-0.15 cm}
\section{Related Work}
In this section, we discuss related work that makes use of prediction and analytic-based techniques to overcome the productivity issue with FPGAs. 

\noindent\textbf{ML-based.} HLSPredict~\cite{o2018hlspredict} estimates pre-implementation resource utilization using program-counters (PC) available in off-the-shelf CPU-based processors. As the PC of a processor is used, arbitrary precision cannot be evaluated using this method. Additionally, pre-implementation resource estimation deviates dramatically compared to post-implementation results~\cite{dai2018fast}. Works such as
~\cite{dai2018fast}, Pyramid\cite{Pyramid}, and Minerva~\cite{Minerva} predict post-implementation resource utilization using an ML framework. However, these frameworks require C-Synthesis~\cite{xilinx_hls} run results for all possible fixed-point configurations, which is a time-consuming process. 

\noindent\textbf{Analytic-based.} Frameworks such as Lin-Analyzer~\cite{linAnalyzer} and HLSCope+~\cite{hls_scope} predict performance metrics for applications using an analytic-based approach. These works do not take into the effect of different fixed-point precision configurations. FINN~\cite{FINN} targets quantized NNs to generate dataflow style architectures for the network. However, these frameworks do not predict post-implementation resource utilization or power prediction for all possible fixed-point configurations. 

ApproxFPGAs~\cite{approxfpgas} is an ML framework that analyzes ASIC-based approximate circuits (ACs) to determine a set of Pareto-optimal FPGA-based designs with respect to power and performance. This framework is specialized only for ASIC-based ACs. Hence, it cannot be generalized for other applications. Additionally, this framework does not predict post-implementation DSP, BRAM, LUT, and FF utilization.

Compared to \namePaper, none of the above frameworks incorporates the resource bound of a particular FPGA device. For example, if BRAM or DSP is exhausted,  how can it  be mapped to the slice logic is a question not answered by those frameworks. Hence, to our knowledge, \namePaper~is the only framework that: (1) predicts post-implementation power and resource utilization for all fixed-point configurations with only two C-Synthesis~\cite{xilinx_hls} run results, and (2) \namePaper~takes into account the heterogeneity of FPGA by mapping the logic to other possible components when any of the components is exhausted.
\section{Conclusion}

In this appendix, we propose \namePaper, a resource and power prediction framework for the exploration of  fixed-point representation on an FPGA.  \linebreak \namePaper~uses high-level synthesis results for only two fixed-point configurations to predict the post-implementation power and resource utilization for all possible fixed-point configurations. While predicting the post-implementation results, \linebreak\namePaper~takes into account the effect of resource saturation and provides a Pareto-optimal configuration in terms of power consumption and error.  We show that  \namePaper can provide accurate estimates of resource utilization and power consumption with up to three orders of magnitude reduction in design space exploration time. In the future, we aim to extend \namePaper to quantize the activations of a neural network and further examine our approach for other application domains.



\chapter[\texorpdfstring{Other Works of the Author}{Other Works of the Author}
]{\chaptermark{header} Other Works of the Author}
\chaptermark{}
\label{chapter:appendixB}
In addition to the works presented in this thesis, I have also contributed to several other research works done in collaboration with students and researchers at ETH, IBM, CMU, and TUe. In this section, I briefly overview these works.

In a recent work~\cite{not_published_singh2021micro}, we demonstrate the capability of a data-centric near-HBM FPGA-based accelerator for pre-alignment filtering step in genome analysis. Compared to our previous work on weather prediction, we see that genome analysis is even more memory-bounded and can significantly benefit from near-memory acceleration.  We map the pre-alignment filtering algorithm to an HBM-based FPGA architecture and create a heterogeneous memory hierarchy using on-chip URAM, BRAM, and on-package HBM.  We perform in-depth scalability analysis for both HBM and DDR4-based FPGA boards and show the memory bottleneck of the pre-alignment phase of genome analysis on a state-of-the-art IBM POWER9 system. Our analysis shows that our HBM-based pre-alignment filtering design provides a $20.1\times$ higher speedup and $115\times$ higher energy efficiency than a 16 cores IBM POWER9 system.

Casper~\cite{not_published_alain2021stencil} designs a \emph{near-cache} accelerator that improves the performance and energy efficiency of stencil computations by eliminating the need to transfer data to the processor for computation while minimizing the unnecessary data movement within the cache hierarchy as well. To this end, we propose Casper, a novel hardware/software codesign approach specifically targeted at stencil computations. We minimize data movement by placing a set of stencil processing units (SPUs) near the last-level cache (LLC) of a traditional CPU architecture and provide novel mechanisms to introduce data mapping changes and support unaligned loads needed for high-performance stencil computations. Computation is mapped to {SPU} so that each SPU operates on the data that is located in the closest LLC slice. Thus, we can reduce the overall data access latency and energy consumption while matching the compute performance to the peak bandwidth of the LLC.

In our work~\cite{not_published_keerthana2021weather}, we propose a highly efficient and flexible interconnect network capable of running the entire COSMO model on a CGRA-based accelerator. The central part of the COSMO model (called dynamical
core or \textit{dycore}) performs $\sim$80 compound stencil computations.  Our  algorithm is able to route the connections in the weather stencil kernels while reducing the delay of the longest wire segment. Further, we propose a post routing optimization to improve the placement of the weather stencil kernels' operations in the COSMO CGRA-based accelerator to improve the routing results with minimum time complexity. Our evaluation shows that the proposed techniques achieve 50\% lower delay and 12\% lower power than a baseline CGRA switch box-interconnect network.

TDO-CIM~\cite{published_vadivel2020tdo} proposes a compiler for computation-in memory (CIM). Computation in-memory is a promising non-von Neumann approach aiming at completely diminishing the data transfer to and from the memory subsystem. In recent years, several CIM-based architectures have been proposed. However, the compiler support for such architectures is still lagging. In this work, we {close} this gap by proposing an end-to-end compilation flow for CIM based on the LLVM compiler infrastructure.  Our compiler {automatically} and {transparently} invokes a CIM accelerator without any user intervention. Therefore, enabling legacy code to exploit CIM-based acceleration. We develop a {light-weight run-time library} for data allocation, transfer, and execution of computational tasks on a CIM device.  We evaluate our approach using an open-source simulation environment based on the Gem5-simulator, where we model a host CPU connected to a CIM accelerator.

Agile auto-tuning~\cite{published_diamantopoulos2020agile} explores the possibility of automatically guiding the auto-tuning of an overlay architecture for different transprecision settings by leveraging knowledge from hardware experience. By adopting the concept of agile development, we built a pipeline of engineering tasks that support the auto-tuning process. Instead of eliminating the overlay hardware design space with pruning techniques, we propose a technique that builds a prediction model to quantify the impact of a hardware design choice towards an optimization goal. We show that the features with the highest impact differ for different precisions.



\begingroup
\setstretch{0.9}
\setlength\bibitemsep{0.1pt}
\printbibliography[heading=bibintoc,title={Bibliography}]
\endgroup

\begingroup
\setstretch{1}
\setlength\bibitemsep{0.2pt}
\begin{refsection}
\nocite{singh2020nero_own}
\nocite{singh2019narmada_own}
\nocite{singh2019napel_own}
\nocite{singh2019near_own}
\nocite{singh2019low_own}
\nocite{singh2018review_own}

\defbibnote{myPrenote}{
 {\bf First Author:}
}

\begin{refcontext}[sorting=ydndt]
\printbibliography[heading=bibintoc,  title={List of Publications},
    prenote=myPrenote
  ]
\end{refcontext}
\end{refsection}

\begin{refsection}
\nocite{diamantopoulos2020agile}
\nocite{vadivel2020tdo}
\nocite{corda2019platform}
\nocite{corda2019memory}
\nocite{chris2019DATE}

\defbibnote{myPrenote}{
 {\bf Co-Author:}
}
\begin{refcontext}[sorting=ydndt]
\printbibliography[
   heading=none, 
    prenote=myPrenote
  ]
\end{refcontext}
\end{refsection}

        

\begin{refsection}
\nocite{poster-singh2021leaper}
\nocite{poster-singh2021ISC}
\defbibnote{myPrenote}{
 {\bf Poster:}
}

\printbibliography[sorting=none,
   heading=none, 
    prenote=myPrenote
  ]

\end{refsection}

    \begin{refsection}
    
        \nocite{patent}
        
        \defbibnote{myPrenote}{
         {\bf Patent:}
        }
        \printbibliography[sorting=none,
           heading=none, 
            prenote=myPrenote
          ]
    
    \end{refsection}

\endgroup
\thispagestyle{plain}

\chapter*{Curriculum Vitae}
\addcontentsline{toc}{chapter}{Curriculum Vitae}

Gagandeep was born in Chandigarh, India, in 1992. He received a joint M.Sc. degree with distinction in Integrated Circuit Design from Technische Universität München (TUM), Germany, and Nanyang Technological University (NTU), Singapore in 2017. He joined Eindhoven University of Technology, Netherlands, in 2017 to pursue a Ph.D. degree as a part of the Marie Sklodowska-Curie EID (Ph.D.) Program under the supervision of Prof. Henk Corporaal. 
From June 2018 to January 2020, he was a Predoctoral Researcher at IBM Research Zurich, Switzerland, in the group of Dr. Christoph Hagleitner. Since January 2020, he has been an Academic Guest in Prof. Onur Mutlu's group at ETH Zurich, Switzerland.  He is passionate about FPGA design, computer architecture, and applied machine learning with skills in both hardware and software design.

\end{document}